\documentclass[%
 aps,
 pra,
 reprint,
 superscriptaddress,
 longbibliography,
 amsfonts,		
 amssymb,
 amsmath,		
 preprintnumbers,
 floatfix,
 showpacs,
 letterpaper,
 byrevtex,
 citeautoscript%
]{revtex4-1}

\usepackage[autoload,graphics]{svn-multi}
\svnidlong
 {$HeadURL: file:///Repository/SVN/Articles/AngularMomentumRadio/branches/PRA/main.tex $}
 {$LastChangedDate: 2015-05-28 19:52:28 +0200 (Thu, 28 May 2015) $}
 {$LastChangedRevision: 98 $}
 {$LastChangedBy: bt $}
\svnid{$Id: main.tex 98 2015-05-28 17:52:28Z bt $}

\usepackage{nameref}
\usepackage{enumitem}
\usepackage[font=itshape]{quoting}
\usepackage{comment}
\usepackage{microtype}
\usepackage{booktabs}
\usepackage{xcolor}
\usepackage{ifpdf}
 \ifpdf%
  \usepackage[pdftex]{graphicx}
 \else%
  \usepackage{graphicx}
\fi
\usepackage{color}
\usepackage[american]{babel}
\usepackage[T1]{fontenc}
\usepackage[scaled=0.92]{helvet}
\renewcommand*{\rmdefault}{ptm}
\usepackage{mathtools}
\usepackage{mathptmx}
\usepackage{siunitx}
\usepackage{braket}
\RequirePackage{bm}

\usepackage{mymacros}

\begin{document}
\renewcommand*{\th}{\vartheta}
\renewcommand*{\d}[1]{\mathrm{d}#1}

\preprint{SVN revision \svnauthor-\svnrev}

\title{The physics of angular momentum radio}

\author{B.~Thid\'e}
\email[Electronic address: ]{bt@irfu.se}
\homepage[Home page: ]{www.physics.irfu.se/~bt}
\altaffiliation[Also at: ]{\UNIVPDScuolaG.}
\affiliation{\IRF}

\author{F.~Tamburini} 
\email[Electronic address: ]{fabrizio.tamburini@unipd.it}
\altaffiliation[Also at: ]{\UNIVPDviaMar.}
\affiliation{\TWISTOFF}

\author{H.~Then} 
\affiliation{\UNIVBRmath}

\author{C.\,G.~Someda} 
\affiliation{\TWISTOFF}

\author{R.\,A.~Ravanelli} 
\affiliation{\SIAE}

\date{\svndate}

\svnidlong
 {$HeadURL: file:///Repository/SVN/Articles/AngularMomentumRadio/branches/PRA/abstract.tex $}
 {$LastChangedDate: 2015-05-28 19:23:04 +0200 (Thu, 28 May 2015) $}
 {$LastChangedRevision: 97 $}
 {$LastChangedBy: bt $}
\svnid{$Id: abstract.tex 97 2015-05-28 17:23:04Z bt $}

\begin{abstract}

To this day, wireless communications, as well as radio astronomy and
other radio science and technology applications, have made predominantly
use of techniques built on top of the electromagnetic linear momentum
physical layer.  As a supplement and/or alternative to this conventional
approach, techniques rooted in the electromagnetic angular momentum
(moment of momentum) physical layer have been advocated, and promising
results from proof-of-concept laboratory and real-world angular momentum
wireless and fiber communication experiments were recently reported.
A physical observable in its own right, albeit much more sparingly
used than other electromagnetic observables such as energy and linear
momentum, the angular momentum exploits the rotational symmetry of
the electromagnetic field and the rotational (spinning and orbiting)
dynamics of the pertinent charge and current densities.  Here we present
a review of the fundamental physical properties of the electromagnetic
angular momentum observable, derived from the fundamental postulates
of classical electrodynamics (the Maxwell-Lorentz equations), and put
it into its context among the other Poincar\'e symmetry invariants of
the electromagnetic field. The multi-mode quantized character and other
physical properties that sets the classical electromagnetic angular
momentum apart from the electromagnetic linear momentum are pointed
out. We give many of the results both in terms of the electric and
magnetic field vectors and in terms of the complex Riemann-Silberstein
vector formalism (Majorana-Oppenheimer representation of classical
electrodynamics) and discuss formal parallels with first quantization
formalism.  We introduce generalized and symmetrized extensions of
the Jefimenko integral formulas for calculating exact expressions
for the electric and magnetic fields directly from arbitrary given
electric and/or magnetic source distributions and use them for deriving
expressions for the volumetric density of the electromagnetic angular
momentum radiated from any given arbitrary localized distribution of
charges and currents. Our results generalize earlier results, obtained by
other authors for certain specific configurations only, and facilitate
the modeling and design of angular momentum transducers. They show
that the volume integrated angular momentum density, \ie, the total
angular momentum emitted by the sources, always tends asymptotically
to a constant when the distance from the source volume tends to
infinity, proving that a radiation arrow of time exists also for
angular momentum, as it does for linear momentum (the volume integrated
Poynting vector). Consequently, angular momentum physics (torque action)
offers an alternative or a supplement to linear momentum physics (force
action) as a means of transferring information wirelessly over very long
distances. This finding opens possibilities for, among other things,
a more flexible utilization of the radio frequency spectrum and paves
the way for the development of new information transfer techniques.
We discuss implementation aspects and illustrate them by examples based
on analytic and numerical analyses. References are made to experiments
demonstrating the feasibility of using angular momentum in real-world
applications, and how the vortical beams can be shaped and their
divergence controlled. A scenario with angular momentum transducers
of new types, including optomechancial ones and those based on the
interplay between charge, spin and orbital angular degrees of freedom,
heralded by recent advances in spintronics/orbitronics, condensed-matter
skyrmion and quantum ring physics and technology, is briefly delineated.
Our results can be readily adapted and generalized to other symmetry
related radiation phenomena and information transfer scenarios.

\end{abstract}

\maketitle
\tableofcontents
\svnidlong
 {$HeadURL: file:///Repository/SVN/Articles/AngularMomentumRadio/branches/PRA/introduction.tex $}
 {$LastChangedDate: 2015-05-27 11:03:09 +0200 (Wed, 27 May 2015) $}
 {$LastChangedRevision: 96 $}
 {$LastChangedBy: bt $}
\svnid{$Id: introduction.tex 96 2015-05-27 09:03:09Z bt $}

\section{Introduction}

Transferring information wirelessly by means of electromagnetic fields
amounts to generating such fields within a spatial volume of finite
extent, encoding information onto physical observables carried by
these fields, radiating the observables as volumetric densities into the
surrounding space, converting them remotely into observables
by using volume integrating sensors, and feeding these observables
into information-decoding receivers. The volumetric density of each
observable is second order (quadratic, bilinear) in the fields, falls off
asymptotically in space as the square of the distance from the source,
and fulfills a conservation law.  Of all physical observables carried by
the electromagnetic field, only the linear momentum (volume integrated
Poynting vector, ``power'') is fully exploited in present-day radio
science and communication applications.

A fundamental physical limitation of the linear momentum vectorial
observable, which exploits only the \emph{translational} symmetry and the
pertinent degrees of freedom of the electromagnetic field, is that it is
single-mode in the sense that a linear momentum radio communication link
comprising one single transmitting and one single receiving transducer
of a conventional type (\eg, a linear half-wave dipole antenna), known
as a single-input-single-output (SISO) link, provides only one physical
channel per carrier frequency.

In contrast, electromagnetic angular momentum, the pseudo-vectorial
physical observable that manifests the \emph{rotational} symmetry
of the electromagnetic field, is multi-mode, being a superposition of
discrete, mutually independent classical electromagnetic angular momentum
eigenmodes, each uniquely identified by the number of $2\pi$ that the
phase of the field varies azimuthally during one oscillation period,
\ie, by its topological charge.  This allows a frequency reuse in the
sense that several independent angular momentum channels can co-exist
on the same carrier frequency.

Hence, a wireless communication link designed so that the transmitting and
receiving ends each are equipped with a single, monolithic transducer,
allowing the dynamics of the charges to be two- or three-dimensional,
thus enabling the generation and detection of angular momentum modes
directly, would constitute an angular momentum SISO link system.
The multi-mode property of the angular momentum allows such a system to
accommodate multiple physical information transfer channels on a single
carrier frequency and within the same frequency bandwidth as that of
a single-channel linear momentum SISO system.  The actual number of
physical angular momentum channels that can be realized in a given
application and/or setup is limited classically only by engineering
and implementation issues.

The multi-mode capability of systems based on electromagnetic angular
momentum is a consequence of the fact that different angular momentum
eigenmodes are topologically distinct and linearly independent by
virtue of their mutual functional orthogonality in a Hilbert space of
denumerably infinite dimension. Even in situations when the different
eigenmodes overlap in space, time and frequency, the individual angular
momentum eigenmodes are physically independent and separable and can be
extracted by using efficient angular momentum mode filtering techniques.

In order to put angular momentum into its proper physical context and
on a firm theoretical basis, we first review the conservation/symmetry
properties of the electromagnetic field that make wireless information
transfer possible.  Then we continue with a theoretical presentation
of the physics itself, touching upon first quantization quantum-like
aspects of classical electrodynamics and its observables.

We derive generic expressions for the angular momentum density and prove
that for any system described by the Maxwell equations, the magnitude of
the electromagnetic angular momentum always tends to a constant far away
from an arbitrary source. This has earlier been shown only for specific
source configurations or in certain approximations.

Furthermore, we discuss concepts and issues related to the exploitation
of the angular momentum physical layer in practical implementations,
both formally and by way of examples. To supplement and support this
theoretical approach, we also report experimental results.  Both those
obtained by using conventional radio and antenna technologies, as well as
results from innovative experiments showing how one can shape of angular
momentum carrying beams and control their spatial divergence.  We also
delineate the possibilities for completely new concepts in transducer
design and technology that are offered by recent advances in nano-
and optomechanics, spintronics and orbitronics.

A theoretical description of electromagnetic angular momentum in the
context of the ten Poincar\'e symmetry invariants and the associated
conservation laws is presented in an Appendix where, among other things,
exact, gauge-invariant expressions, valid for sources for which the
magnetic vector potential falls off spatially sufficiently rapidly, are
derived for both the linear and angular momenta. In another Appendix we
derive exact expressions, valid in all space, for the fields from one or
more electric Hertzian dipoles, and for second-order field quantities
that enters the expressions for the linear and angular momentum densities,
as well as other locally conserved quantities, carried by these fields.

\svnidlong
 {$HeadURL: file:///Repository/SVN/Articles/AngularMomentumRadio/branches/PRA/background.tex $}
 {$LastChangedDate: 2015-05-28 19:23:04 +0200 (Thu, 28 May 2015) $}
 {$LastChangedRevision: 97 $}
 {$LastChangedBy: bt $}
\svnid{$Id: background.tex 97 2015-05-28 17:23:04Z bt $}

\section{Background}
\label{sect:background}

Transferring information electromagnetically is made possible by the
fundamental property of the electromagnetic field $(\E,\B)$ to be able
to carry physical observables, \ie, directly measurable intrinsic
and extrinsic electromagnetic quantities, over long distances through
free space, solids, fluids, gases, and plasmas. These observables are
generated within a volume of finite spatial extent (the source volume
$V'$; \eg, a transmitting antenna), propagated in the form of volumetric
densities to a remotely located volume of likewise finite spatial extent
(the observation volume $V$; \eg, a receiving antenna) over which they
are volume integrated into observables, allowing the information carried
by them to be subsequently extracted and decoded.

The volumetric density of every electromagnetic observable carried
by the fields $\E,\B$ is a linear combination of quantities that are
second order (quadratic, bilinear) in the fields, such as $\E\bdot\E$,
$\B\bdot\B$, $\E\cross\B$, $\x\cross(\E\cross\B)$, and/or of derivatives
of the fields. These quantities fulfill local equations of continuity,
and are therefore conserved. As shown by \textcite{Noether:NGWG:1918},
this conservation is a manifestation of a symmetry of the Maxwell theory~%
\cite{%
Poincare:RCMP:1906,%
Fushchich&Nikitin:JPA:1992,%
Ibragimov:AAM:2008%
}.  In classical physics terms the observables, and their volumetric
densities, are viewed as the result of dynamical changes in the charged
particle distribution in the source volume $V$ and, causally, in the
observation volume $V'$, as discussed in Appendix~\ref{app:Poincare}.

In classical electrodynamics the interplay between the microscopic
electric charge and current densities $\rhoe(t,\x)$ and $\je(t,\x)$
and the electric and magnetic fields, $\E(t,\x)$ and $\B(t,\x)$, as well
as between the fields themselves, are described by the Maxwell-Lorentz
equations.  Allowing for magnetic charge and current densities,
represented by the pseudoscalar $\rhom(t,\x)$ and the pseudovector
$\jm(t,\x)$, respectively, reflecting the dual symmetry of the theory~%
\cite{%
Dirac:PRSLA:1931,%
Schwinger:S:1969,%
Schwinger&al:Book:1998%
},
these equations can be written in the symmetrical form~%
\cite{%
Jackson:Book:1998,%
Thide&al:Incollection:2011,%
Thide:Book:2011%
}
\begin{subequations}
\label{eq:Maxwell_micro}
\begin{gather}
\label{eq:Maxwell_divE}
 \div\E = \frac{\rhoe}{\epz}
\\
\label{eq:Maxwell_divB}
 \div\B = \frac{\rhom}{\epz c^2}
\\
\label{eq:Maxwell_curlE}
 \curl\E + \pdd{t}\B = -\frac{1}{\epz c^2}\jm
\\
\label{eq:waves:Maxwell_curlB}
 \curl\B - \frac{1}{c^2}\pdd{t}{\E} = \frac{1}{\epz c^2}\je
\end{gather}
\end{subequations}
where $c$ is the speed of light in free space and $\epz$ is the
dielectric permittivity of free space.  This set of linear, inhomogeneous,
coupled, partial differential equations are the postulates of classical
electrodynamics from which all properties of the classical electromagnetic
field, its observables, and interactions can be derived. These postulates
are valid for radio and optical wavelengths alike inasmuch as they are
scale invariant and are therefore valid for classical wireless information
transfer at all frequencies.

\begin{table}
\caption{\label{tab:RS}%
 Various second-order products of the Riemann-Silberstein vector
 ${\RS=\E+\im c\B}$ with itself and their meaning in terms of quadratic
 and bilinear products of the fields $\E$ and $\B$.  An asterisk ($\cc{}$)
 denotes complex conjugate, and $\ox$ denotes the tensor (Kronecker,
 dyadic) product operator.%
}
\begin{ruledtabular}
\begin{tabular}{ll}
 \textbf{Complex notation}
 & \textbf{Real notation}
\\
\midrule
  $\RS\bdot\RS$
 &
  $\E\bdot\E - c^2\B\bdot\B + 2\im c\E\bdot\B$
\\
  $\RS\bdot\cc{\RS}
  = \SQRT{(\RS\bdot\RS)\cc{(\RS\bdot\RS)}}$
 &
  $\E\bdot\E + c^2\B\bdot\B \equiv E^2 + c^2B^2$
\\
  $\RS\cross\RS$
 &
  $\0$
\\
 $\cc{\RS}\cross\RS = -\RS\cross\cc{\RS}$
 &
  $2\im c(\E\cross\B)$
\\
  $\RS\ox\RS$
 &
  $\E\ox\E - c^2\B\ox\B + \im c(\E\ox\B + \B\ox\E)$
\\
  $\cc{\RS}\ox\cc{\RS}=\cc{(\RS\ox\RS)}$
 &
  $\E\ox\E - c^2\B\ox\B - \im c(\E\ox\B + \B\ox\E)$
\\
  $\RS\ox\cc{\RS}$
 &
  $\E\ox\E + c^2\B\ox\B - \im c(\E\ox\B - \B\ox\E)$
\\
  $\cc{\RS}\ox\RS=\cc{(\RS\ox\cc{\RS})}$
 &
  $\E\ox\E + c^2\B\ox\B + \im c(\E\ox\B - \B\ox\E)$
\end{tabular}
\end{ruledtabular}
\end{table}

It is often convenient to combine the real-valued fields $\E$ and $\B$ into
the complex-valued Riemann-Silberstein vector~%
\cite{%
Weber:Book:1901,%
Silberstein:AP:1907a,%
Silberstein:AP:1907b,%
Silberstein:Book:1914,%
Bateman:Book:1915%
}
\begin{align}
\label{eq:RS}
 \RS(t,\x) = \E(t,\x) + \im{}c\B(t,\x)\;,\;\im^2=-1
\end{align}
some properties of which are listed in Table~\ref{tab:RS}. Expressing
the charge and current densities in their complex forms
\begin{subequations}
\label{eq:rho&j}
\begin{gather}
\label{eq:rho}
 \rho(t,\x) = \rhoe(t,\x) + \frac{\im}{c}\rhom(t,\x)
\\
\label{eq:j}
 \j(t,\x) = \je(t,\x) + \frac{\im}{c}\jm(t,\x)
\end{gather}
\end{subequations}
we see that Dirac's symmetrized Maxwell-Lorentz
equations~\eqref{eq:Maxwell_micro}
become
\begin{subequations}
\label{eq:Maxwell_G}
\begin{gather}
 \div\RS = \frac{\rho}{\epz}
\\
 \curl\RS = \frac{\im}{c}\pdd{t}{\RS} + \frac{\im}{\epz c}\j
\end{gather}
\end{subequations}

Following
\textcite{%
Bialynicki-Birula&Bialynicka-Birula:OC:2006,%
Bialynicki-Birula&Bialynicka-Birula:PRA:2009%
}%
, who instead of $\RS$ use a differently normalized complex vector, \viz\
$\F=\RS\SQRT{\epz/2}$, we first introduce the linear momentum operator for
the electromagnetic field,
\begin{align}
\label{eq:op_pfield}
 \Op{p}\field
  = \sum_{i=1}^3 \xunit_i\op{p}_i\field
  = \sum_{i=1}^3 \xunit_i\bigg(-\im\hslash\pdd{x_i}{}\bigg)
  = -\im\hslash\grad
\end{align}
where $\hslash$ is the normalized Planck constant. Secondly, we
introduce the vector-like symbolic construct
\begin{subequations}
\label{eq:Svec}
\begin{align}
 \vectens{S}
  \equiv \sum_{i=1}^3\matr{S}_i\xunit_i
\end{align}
with components
\begin{align}
\label{eq:maths:S_i}
\matr{S}_1 =
 \begin{pmatrix*}[c]
  0 & 0 & 0 \\*
  0 & 0 & -\im \\*
  0 & \im & 0 \\*
 \end{pmatrix*}
\quad
\matr{S}_2 =
 \begin{pmatrix*}[c]
  0 & 0 & \im \\*
  0 & 0 & 0 \\*
  -\im & 0 & 0 \\*
 \end{pmatrix*}
\quad
\matr{S}_3 =
 \begin{pmatrix*}[c]
  0 & -\im & 0 \\*
  \im & 0 & 0 \\*
  0 & 0 & 0 \\*
 \end{pmatrix*}
\end{align}
\ie, spin-1 matrices fulfilling the angular momentum commutation rule
\begin{gather}
 [\matr{S}_i,\matr{S}_j] = -\im\eijk\matr{S}_k, 
\shortintertext{where}
\eijk =
\begin{cases} 
  1 & \text{if $i,j,k$ is an even permutation of 1,2,3} \\*
  0 & \text{if at least two of $i,j,k$ are equal} \\*
 -1 & \text{if $i,j,k$ is an odd permutation of 1,2,3}
\end{cases} 
\end{gather}
\end{subequations}
is the totally antisymmetric rank-3 pseudotensor (the Levi-Civita symbol).
Thirdly, we introduce the field helicity operator~%
\begin{align}
\label{eq:helicity_op}
 \op{\Lambda}
  = \frac{\vectens{S}\bdot\Op{p}\field}{\abs{\Op{p}\field}}
  \equiv \frac{\sum_{i=1}^3\matr{S}_i\op{p}_i\field}{\abs{\Op{p}\field}}
\end{align}
Using Eqns.~\eqref{eq:op_pfield}, \eqref{eq:Svec}, and
\eqref{eq:helicity_op}, we can write the Maxwell-Lorentz
equations~\eqref{eq:Maxwell_micro} in free space and other domains where
$\rhoe=\rhom=0$ and ${\je=\jm=\0}$ in the form
\begin{subequations}
\label{eq:Maxwell_G_freespace}
\begin{gather}
\label{ex:free_Dirac:transversality}
 \Op{p}\field\bdot\RS = 0
\\
\label{ex:free_Dirac:G_waveequation}
 \im\hslash\pdd{t}{\RS} = \op{H}\field\RS
\end{gather}
\end{subequations}
where 
\begin{align}
 \op{H}\field
  = c\abs{\Op{p}\field}\op{\Lambda}
  = c\vectens{S}\bdot{\Op{p}\field}
  \equiv c{\sum_{i=1}^3\matr{S}_i\op{p}_i\field}
\end{align}

Recalling the quantal relation $\pfield=\hslash\veck$,
where $\veck=k\kunit$ is the wave vector, $k=2\pi/\lambda$,
and $\lambda$ is the wavelength, we see that the first
equation, Eqn.~\eqref{ex:free_Dirac:transversality}, is the
transversality condition, $\veck\perp\RS$, for electromagnetic
waves propagating in free space, whereas the second equation,
Eqn.~\eqref{ex:free_Dirac:G_waveequation}, describes the dynamics of
the field and takes the form of a Schr\"{o}dinger/Pauli/Dirac-like
equation where $\op{H}\field$ behaves as a Hamilton operator for
the free electromagnetic field. The electromagnetic theory that
corresponds to classical mechanics is the geometrical optics (a.k.a.\
eikonal or ``ray optics'') approximation of the Maxwell-Lorentz
equations~\cite{Kobe:FP:1999}.

The solutions of Eqn.~\eqref{ex:free_Dirac:G_waveequation} can, in certain
aspects, be viewed as first quantization photon wave functions that obey
the superposition principle~%
\cite{%
Majorana:ResearchNotes:2009,%
Oppenheimer:PR:1931,%
Laporte&Uhlenbeck:PR:1931,%
Archibald:CJP:1955,%
Good:PR:1957,%
Hammer&Good:PR:1957,%
Moses:PR:1959,%
Kursunoglu:JMP:1961,%
Good&Nelson:Book:1971,%
Mignani&al:LNC:1974,%
Barut:Book:1980,%
Giannetto:LNC2:1985,%
Sipe:PRA:1995,%
Bialynicki-Birula:APP:1994,%
Bialynicki-Birula:PO:1996,%
Esposito:FP:1998,%
Schwinger&al:Book:1998,%
Gersten:FPL:1999,%
Kobe:FP:1999,%
Keller:PHR:2005,%
Khan:PS:2005,%
Dragoman:JOSAB:2007,%
Smith&Raymer:2007,%
Tamburini&Vicino:PRA:2008,%
Bogush&al:RM:2009,%
Wang&al:PRA:2009,%
Mohr:APNY:2010,%
Aste:JGSP:2012,%
Redkov&al:AACA:2012,%
Yamamoto&al:JPSJ:2012,%
Bialynicki-Birula&Bialynicka-Birula:JPA:2013,%
Barnett:NJP:2014,%
Dressel&al:ARXIV:2014%
}%
. It should be noted, however, that
\textcite[p.~204]{Cohen-Tannoudji&al:Book:1997} argue against ``\ldots
suggesting (in spite of warnings) the false idea that the Maxwell waves
are the wave functions of the photon.'' See also~%
\textcite{%
Li:OE:2013,%
Andrews:JN:2014%
}.

For any given, prescribed set of charge and current densities, every
electromagnetic observable can be calculated with, in principle,
arbitrary accuracy from the first order fields $\E$ and $\B$ that
are solutions of Eqns.~\eqref{eq:Maxwell_micro}.  As was pointed out
by \textcite{Silberstein:AP:1907a,Silberstein:AP:1907b}, it is very
convenient to calculate these second order quantities in a formalism
based on the use of the complex Riemann-Silberstein vector~\eqref{eq:RS},
known as the Majorana-Oppenheimer formalism.  This formalism is also
convenient for canonical (second) quantization of the electromagnetic
field \cite{Keller:PHR:2005,Bialynicki-Birula&Bialynicka-Birula:OC:2006}.

However, the fields themselves are abstractions in the sense that they
cannot be measured directly but have to be estimated from measurements
of physical observables that are quadratic or bilinear in the fields. To
quote \textcite{Dyson:Maxwell:1999}:

\begin{quote}

 ``The modern view of the world that emerged from Maxwell's theory is
 a world with two layers. The first layer, the layer of the fundamental
 constituents of the world, consists of fields satisfying simple linear
 equations. The second layer, the layer of the things that we can directly
 touch and measure, consists of mechanical stresses and energies and
 forces. The two layers are connected, because the quantities in the
 second layer are quadratic or bilinear combinations of the quantities
 in the first layer.
 \ldots
 The objects on the first layer, the objects that are truly fundamental,
 are abstractions not directly accessible to our senses. The objects that
 we can feel and touch are on the second layer, and their behavior is
 only determined indirectly by the equations that operate on the first
 layer. The two-layer structure of the world implies that the basic
 processes of nature are hidden from our view.''

 ``The unit of electric field-strength is a mathematical abstraction, chosen
 so that the square of a field-strength is equal to an energy-density that can
 be measured with real instruments.
 \ldots
 It means that an electric field-strength is an abstract quantity,
 incommensurable with any quantities that we can measure directly.''

 ``It may be helpful for the understanding of quantum mechanics to stress
 the similarities between quantum mechanics and the Maxwell theory.''

\end{quote}

From these considerations it should be clear that it is essential to
analyze not only the properties of the fields $\E$ and $\B$ as such but
also the properties of the electromagnetic observables, which are second
order in these fields or derivatives thereof, in order to satisfactorily
assess and optimally exploit the information-carrying capability of the
classical electromagnetic field, In Sect.~\ref{sect:physics} we will show
that for any conceivable distribution of sources of the electromagnetic
field, the physical observables (the volume integrated densities) tend to
a constant value when they propagate in free space and the distance from
the radiation source tends to infinity.  This means that there exists no
finite distance from the source at which any of these observables tends
to zero or beyond which it vanishes globally.  Consequently, they can all
carry information over arbitrary large distances in free space, albeit
subject to engineering limitations (spatial and temporal distribution,
transducer geometry, detector sensitivity, integration time, \ldots).

Of all physical observables that may potentially be used for wireless
information transfer, present-day radio science and engineering
implementations almost exclusively make full use only of the linear
momentum vectorial observable.  For instance, a conventional receiving
antenna is typically wire-like, \ie, the observation volume $V$ is 
a very thin cylinder where current-carrying electrons are constrained
to move in one dimension, resulting in a scalar, dissipative conduction
current flowing along the cylinder axis. The electrons can be viewed as
moving relative to a fixed neutralizing background of heavy ions with
negligible recoil.  Such an antenna, typically with a length $L$ of the
order half a wavelength $\lambda$, \ie, ${L\sim\lambda/2=\pi/k}$, senses
a projection of the field linear momentum density vector $\pfield$ and
integrates the component of the electric conduction current density vector
$\je(t,\x)$ over $L$ along the axis of the thin, cylindrical antenna
into a time-varying, spatially averaged scalar conduction current. This
scalar current is fed to the receiver equipment (usually an electronics
device) where it is processed and the wirelessly transferred information
is extracted. In the process, the vector property of the current density
is not exploited and information may therefore be wasted.

Linear momentum wireless information transfer techniques have the
advantage of being simple, robust, and easy to implement, but
have the drawback that they can be wasteful when it comes to the use
of the frequency spectrum. This is because electromagnetic linear
momentum is single-mode, permitting only one independent linear-momentum
physical channel of information transfer per carrier frequency (per
wave polarization, per antenna).  The ensuing over-crowding of the
radio frequency bands, often referred to as the ``spectrum crunch'',
has become a serious problem in radio communications.  It also imposes
limitations on radio astronomy and other radio science applications as
well as on fiber optics communications. It is doubtful whether further
evolutionary refinements of conventional techniques and methods built on
the linear momentum density (\ie, Poynting vector) physics layer alone
will, in the long term, suffice to remedy these problems.

This situation has led to the idea of exploiting so far underutilized
physical observables of the electromagnetic field that do not suffer
from the inherent information-carrying limitation of present-day radio
techniques. The development and exploitation of a new physical layer
that provides additional information-carrying capabilities already at the
fundamental physics level, before any multi-transducer or other spectral
density enhancing techniques are applied at the engineering application
layer, will set the stage for new radio paradigms that may enable a wiser
exploitation of the radio frequency spectrum resource and for new types
of transducers.

Specifically, it has been proposed that the electromagnetic angular
momentum (moment of momentum) pseudovectorial observable, discussed
already by Maxwell in his writings in the 1860's~\cite{Siegel:Book:2003},
studied by other authors since over a century~%
\cite{%
Sadovskii:1899,%
Poynting:PRSL:1909,%
Abraham:PZ:1914,%
Bateman:PR:1926}%
,
and nowadays treated in standard physics textbooks~%
\cite{%
Heitler:Book:1954,%
Bogolyubov&Shirkov:Book:1959,%
Messiah:Book:1970,%
Eyges:Book:1972,%
Landau&Lifshitz:Book:1975,%
Berestetskii&al:Book:1989,%
Mandel&Wolf:Book:1995,%
Cohen-Tannoudji&al:Book:1997,%
Schwinger&al:Book:1998,%
Jackson:Book:1998,%
Griffiths:Book:1999,%
Rohrlich:Book:2007%
},
be fully exploited in radio science and technology~%
\cite{%
Gibson&al:OE:2004,%
Thide&al:PRL:2007,%
Franke-Arnold&al:LPR:2008,%
Thide&al:Incollection:2011,%
Tamburini&al:NPHY:2011,%
Tamburini&al:APL:2011,%
Tamburini&al:NJP:2012,%
Tamburini&al:NJP:2012a,%
Tamburini&al:RS:2015%
}.
The rationale being that it has been demonstrated in experiments at
microwave and optical wavelengths that by exploiting the electromagnetic
angular momentum, one may enhance the information-carrying capability
of wireless communications substantially~%
\cite{%
Gibson&al:SPIE:2004,%
Wang&al:NPHO:2012,%
Litchinitser:S:2012,%
Tamburini&al:NJP:2012,%
Tamburini&al:NJP:2012a,%
Bozinovic&al:S:2013,%
Mahmouli&Walker:IEEEWCL:2013,%
Willner:MOC:2013,%
Ren&al:GLOBECOM:2014,%
Xie&al:GC:2014,%
Yan&al:ICC:2014,%
Yan&al:NCOM:2014,%
Willner&al:AOP:2015%
}.

This enhancement is mainly due to the fact that angular momentum is
discretized (quantized) already at the classical level. The electric
and magnetic fields in a beam that carries a non-integer amount of
electromagnetic angular momentum is namely a superposition of discrete
angular momentum eigenmodes, \ie\ vortices~%
\cite{%
Papanicolaou&Tomaras:NPB:1991,%
Zagrodzinski:PC:2002%
}%
. Each eigenmode is characterized by a unique integer that denotes its
azimuthal quantum number (mode index), also known as the topological charge.

As an illustration, let us consider a radio or light beam that propagates
in free space.  Expressed in cylindrical coordinates $(\rh,\ph,z)$
and in complex notation, an individual electric field vector $\E(t,\x)$
of the beam is assumed to be given by the expression
\begin{align}
\label{eq:E_varsep}
 \E(t,\rh,\ph,z)
 = R(\rh)\Phi(\ph)Z(z)T(t)\E_0
\end{align}
where
\begin{align}
 \E_0
  = E_0\eunit
  = \abs{E_0}\e^{\im\delta_0}\eunit
\end{align}
Here $E_0$ and $\delta_0$ are real-valued scalar constants, but we
allow the unit vector $\eunit$ to be complex, \eg, a linear combination
of the helical base vectors
\begin{align}
 \xunit_\pm = \frac{1}{\SQRT{2}}\big(\xunit\pm\im\yunit\big)
\end{align}
that represent circularly polarized wave fields.

We assume that the field $\E$ has a strongly peaked longitudinal wave
vector spectrum containing essentially only one single longitudinal wave
wave vector component $k_z$, corresponding to the linear momentum $z$
component $p_z=\hslash{}k_z$ carried by each photon of the beam, and
that the $z$ dependence of the field therefore can be written
\begin{subequations}
\begin{align}
 Z(z)
 = Z_0(\Omega z)\e^{\im{k_zz}}
\end{align}
where $Z_0(\Omega z)$ is a slowly varying real-valued envelope function of $z$,
as indicated by the slowness parameter $\Omega$. Likewise, assuming also
an essentially single temporal mode, we write
\begin{align}
 T(t) = T_0(\Omega t)\e^{-\im\w{t}}
\end{align}
where $T_0(\Omega t)$ is a slowly varying real-valued function of time $t$, and
$\w$ is the angular frequency, corresponding to the energy $\hslash\w$
carried by each photon of the beam.

The radial part $R(\rh)$ describes the variation of $\E$ in the
radial, $\rh$, direction, perpendicular to the $z$ direction. Assuming
well-behavedness in the usual manner, we can represent $R$, which is
real-valued, as a Fourier integral
\begin{gather}
\label{eq:R_Fourier}
 R(\rh) = \int\D{k_\rh}\,R_{k_\rh}\e^{\im k_\rh\rh}
\shortintertext{where the complex Fourier amplitudes are}
 R_{k_\rh} = \frac{1}{2\pi}\int\D{\rh}\,R(\rh)\e^{-\im k_\rh\rh}
\end{gather}
and $k_\rh$ is the radial ($\rh$) component of the wave vector.

The azimuthal part $\Phi(\ph)$ describes the variation of $\E$ in the
direction perpendicular to both $z$ and $\rh$. If $\Phi$ is well-behaved
enough that it can be represented by a (complex) Fourier series [\cf\
\textcite[Eqn.~(2)]{Yao&al:OE:2006},
\textcite[Eqns.~(1)--(2)]{Jha&al:PRA:2008}, and
\textcite[Eqns.~(1)--(2)]{Jack&al:NJP:2008}]
\begin{align}
\label{eq:Phi_Fourier}
 \Phi(\ph) = \sum_{m=-\infty}^\infty c_m\,\e^{\im m\ph}
\end{align}
\end{subequations}
where $c_m$ are (constant) Fourier amplitudes, then the electric
field $\E$ is a vector sum of discrete angular momentum modes
\begin{subequations}
\label{eq:E_sum_m}
\begin{gather}
 \E(t,\x) = \sum_{m=-\infty}^\infty \E_m(t,\rh,\ph,z)
\shortintertext{where}
 \E_m
  = E_0Z_0(\Omega z)T_0(t\Omega )R(\rh)
   \e^{\im(m\ph+k_zz-\w t+\delta_0)}\eunit
\\
  = E_0Z_0(\Omega z)T_0(\Omega t)\int\D{k_\rh}R_{k_\rh}
    \e^{\im(k_\rh\rh + m\ph + k_zz-\w t+\delta_0)}\eunit
\end{gather}
\end{subequations}
In the last step, we used Eqn.~\eqref{eq:R_Fourier}.

The $\exp{(\im{m}\phi)}$ phase factor of the $\E_m$ mode gives the
beam a helical phase front and causes a screw dislocation along the $z$
axis. This singularity results in a vanishing linear momentum density
(Poynting vector) on the $z$ axis ($\rh=0$), leading to an annular shape
of the linear momentum density.  For $m=0$ the phase front is flat and
the field has no phase singularity and the distribution of linear momentum
density is not annular.

The smallest radius of the annulus of a beam carrying an OAM
${L_z=\hslash{m},m=\pm1,\pm2,\ldots}$ eigenmode that can be effectively
observed is $(\abs{m}-1/2)\lambda/(2\pi)$ as given by the Bohr-Sommerfeld
quantization condition.

If only the $\ph$ dependence with $m\neq0$ is taken into account,
then the linear momentum density of the beam would diverge
spatially and the radius of the ring-shaped locus of maximum linear
momentum density would increase with increasing $m$ as discussed
by~\textcite{Padgett&al:NJP:2015}. However, as can be seen in
Eqns.~\eqref{eq:E_sum_m}, the structure of the amplitude and the phase
of $\E$ is determined not only by the azimuthal part $\Phi$, but also by
the radial part $R$ and the longitudinal part $Z$.  Hence, by modifying
$R$ and $Z$, which can be done if one uses two- or three-dimensional
transducers and/or reflectors, or phased arrays of linear antennas
that approximate such transducers, one can structure an OAM-carrying beam.
In fact, as has been demonstrated experimentally by
\textcite{%
Chen&al:OL:2013,%
Ostrovsky&al:OL:2013,%
Garcia-Garcia&al:OL:2014,%
Vaity&Rusch:OL:2015,%
Li&al:OSA:2015%
},
it is possible to shape OAM carrying beams so that the radii of the
linear momentum annulii do not scale with $m$.

The quantization of the orbital part of the classical electromagnetic
angular momentum is a consequence of the single-valuedness of the
electromagnetic fields upon their winding an integer number of $2\pi$
in azimuth angle $\ph$ around the beam axis $z$, resulting in a
Bohr-Sommerfeld type quantization condition~%
\cite{%
Bohr:PM:1913,%
Sommerfeld:Book:1921,%
Messiah:Book:1970,%
Bohm:Book:1989,
Nienhuis:Incollection:2008%
}.
Mathematically the quantization can be understood in terms of the
well-known functional space orthogonality condition
\begin{align}
 \frac{1}{2\pi}
  \int_0^{2\pi}\!\D{\ph}\,\cc{\big(\e^{\im{}m_1\ph}\big)}\e^{\im{}m_2\ph}
 = \delta_{m_1m_2}
\end{align}
where $\delta_{m_1m_2}$ is the Kronecker delta
\cite{%
Berry:JOA:2004c,%
Gibson&al:OE:2004,%
Celechovsky&Bouchal:NJP:2007,%
Barreiro&al:NPHY:2008,%
Pors&al:PRL:2008,%
Martelli&al:EL:2011,%
Kumar&al:OL:2011,%
Wang&al:NPHO:2012%
}%
. This mutual orthogonality ensures that different spatially and
temporally overlapping $L_z\field$ eigenmodes can co-exist on the same
carrier frequency $\w$ and in the same polarization state $\sigma$
without interfering each other, allowing a simultaneous transfer
of different information streams independently, without the need
for extra frequency bandwidth.  In other words, OAM spans a state
space of dimension $N=1,2,3,\ldots$, and can therefore be regarded
as an ``azimuthal (tangential) polarization'' with arbitrarily many
states~\cite{Litchinitser:S:2012}. This makes it possible, at least
in principle, to use OAM to physically encode an unlimited amount
of information onto any part of an EM beam, down to the individual
photon~\cite{Mair&al:N:2001,Watson:S:2002}.

Guided by Eqns.~\eqref{eq:pfield_ave}, \eqref{eq:Lfield},
and~\eqref{eq:Lfield_cart_z}, we define the expectation value of an
observable $O\field$ carried by the classical electromagnetic field as
\begin{align}
\label{eq:expval}
 \langle{O\field}\rangle
 = \bra{\E(t,\x)}\op{O}\field\ket{\E(t,\x)}
 = \bra{\RS(t,\x)}\op{O}\field\ket{\RS(t,\x)}
\end{align}
The operator for the $z$ component of the orbital angular momentum (OAM)
around the $z$ axis is given by Eqn.~\eqref{eq:Lfield_cart_z}.
Applying this to the electric field given by
expressions~\eqref{eq:E_sum_m}, we see that
\begin{align}
 \op{L}_z\field\E(t,\x)
 = -\im\hslash\pdd{\ph}{}\E(t,\x)
 = \sum_{m=-\infty}^\infty c_m m\hslash\E_m(t,\rh,\ph,z)
\end{align}
from which we conclude that in general $\E$ is a weighted vector sum of
fields carrying different integer $z$ components OAM.  In particular,
for a beam that is in an OAM $z$ component eigenstate $m$,
\begin{align}
 \op{L}_z\field \E_m = m\hslash\E_m
\end{align}
which means that the $z$ component of the orbital angular momentum
of each photon in this beam is $m\hslash$.

\begin{figure}
 \resizebox{1.\columnwidth}{!}{%
  \input{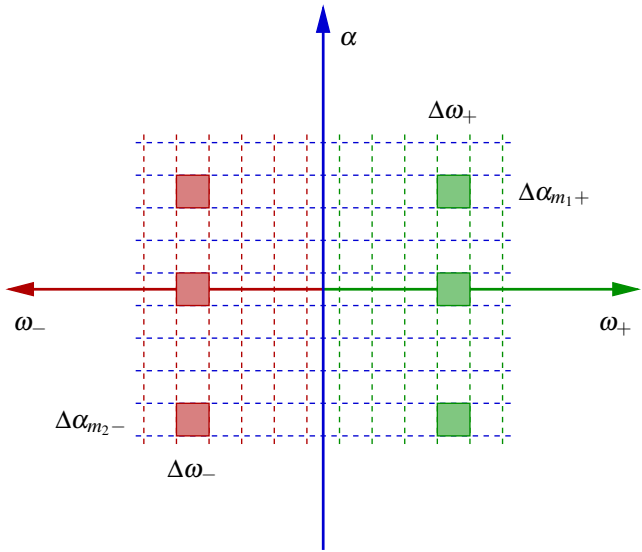}
 }
 \caption{\label{fig:freqplane}%
  The two-dimensional oscillation/rotation plane spanned by the temporal
  phase oscillation rate, \ie, the frequency $\w$ (horizontal axis),
  and the angular phase rotation rate $\alpha$, \ie, the $z$ component
  of the orbital angular momentum (OAM) $L_z$ (vertical axis) in units
  of Planck's normalized constant $\hslash$.  The right-hand, green part
  of the horizontal axis represents right-hand circular polarization
  (spin), and the left-hand, red part of the axis represents the
  opposite, left-hand circular polarization (spin).  The $\alpha$
  axis is rasterized into rotational frequency bins of azimuthal mode
  numbers $m$ because of the single-valuedness of the fields that causes
  a quantization of $L_z$, and the $\w$ axes are rasterized in terms of
  bandwidth segments.  The solid green squares in the right-hand plane
  represent different transmission channels on the same carrier frequency
  $\w_+$, with the same frequency bandwidth $\Delta\w_+$, and in the
  same spin (polarization) state $\sigma=+1$, but in different rotational
  frequency bins $\alpha_{m_1}$ and $\alpha_{m_2}$, corresponding to two
  different OAM azimuthal quantum numbers $m_1$ and $m_2$, respectively.
  The solid red squares in the left-hand plane represent transmission
  channels with the same carrier frequency ($\w_-=\w_+$) and frequency
  bandwidth ($\Delta\w_-=\Delta\w_+$) as for the right-hand (green)
  plane, but in the opposite spin state $\sigma=-1$, still allowing a more
  optimum use of the frequency spectrum.  All possible squares in the
  $(\w_\pm,\alpha)$ plane constitute unique, independent information
  transfer eigenmode channels $(\w,\sigma,\alpha_m)$.  For instance,
  the six solid squares in the figure use the same carrier frequency and
  occupy the same frequency bandwidth.  This is therefore an example of
  a six-fold frequency re-use.%
 }
\end{figure}

In the particular case that the azimuthal part in Eqn.~\eqref{eq:E_varsep}
is
\begin{align}
\label{eq:Phi(phi)}
\Phi(\ph)
 = \e^{\im k_\ph\rh\ph}
 \equiv \e^{\im\alpha\ph}
\end{align}
where $k_\ph$ is the $\ph$ (azimuthal) component of the wave vector,
and $\alpha$ has a non-integer value~%
\cite{%
Gotte&al:OE:2008,%
Huang&al:OC:2012%
},
the $m$th Fourier amplitude \eqref{eq:Phi_Fourier} is~%
\cite{%
Berry:JOA:2004c,%
Tamburini&al:APL:2011,%
Fadeyeva&al:JOSAB:2014%
}
\begin{align}
 c_m = \frac{\exp(\im\pi\alpha)\sin(\pi\alpha)}{\pi(\alpha-m)}
\end{align}
and the electric field can be written
\begin{gather}
 \E(t,\x)
 = E_0Z_0(\Omega z)T_0(\Omega t)R(\rh)\e^{\im\beta}\,\unit{e}
\shortintertext{where}
 \beta
 = \alpha\ph + k_zz - \w t + \delta_0
 = k_\ph\rh\ph + k_zz - \w t + \delta_0
\end{gather}

In real-valued notation, the electric field mode under consideration
becomes
\begin{align}
 \E(t,\x)
 = E_0Z_0(\Omega z)T_0(\Omega t)R(\rh)(\cos\beta\Re{\unit{e}} - \sin\beta\Im{\unit{e}})
\end{align}
which, of course, is the physically correct form to use when calculating
second order quantities such as the densities of the electromagnetic
observables that we are discussing here.  Alternatively, a description
in terms of the Zernike polynomials may be assumed~%
\cite{%
Zernike:P:1934,%
Dholakia&Cizmar:NPHO:2011,%
Rogel-Salazar&al:JOSAB:2014%
}.

Clearly, an electromagnetic beam carrying an arbitrary amount
${\hslash\alpha=\hslash{}k_\ph\rh}$ of OAM is a superposition of
discrete classical OAM eigenmodes $\E_m$ that are mutually orthogonal
in a function space sense and propagate independently of each other in
ordinary spacetime \cite{Molina-Terriza&al:PRL:2002,Torner&al:OE:2005}.
Such a beam therefore exhibits a discrete Fourier spectrum in orbital angular
momentum domain $\alpha$ where each angular momentum eigenmode component
occupies a bin
\begin{align}
 \alpha_m \in \{\alpha\in\Rone;\,m-1/2\leq\alpha<m+1/2\},
 m=0,\pm1,\pm2,\ldots
\end{align}
This was shown experimentally by \textcite{Leach&al:NJP:2004},
thus confirming the prediction by \textcite{Berry:JOA:2004c}; see
also Fig.~\ref{fig:freqplane}, 
\textcite{Molina-Terriza&al:JEOS:2007},
\textcite{Jack&al:NJP:2008},
\textcite{Lavery&al:JO:2011},
\textcite{Huang&al:OC:2012},
and
\textcite{Tamburini&al:APL:2011},
who used ${\alpha=-1.12}$.

Orbital angular momentum adds, as it were, an extra dimension of periodic
variation, effectively augmenting the one-dimensional ordinary oscillation
frequency axis to a two-dimensional frequency plane as outlined
graphically in Fig.~\ref{fig:freqplane}.  The horizontal axis represents
the usual temporal phase oscillation rate, \ie, the angular frequancy
$\w=\partial{\beta}/\partial{t}$, whereas the vertical axis represents
the angular phase rotation rate, \ie, the orbital angular momentum
$\alpha=\partial{\beta}/\partial\ph$ in a plane perpendicular to
the propagation axis of the beam.  Since this additional dimension
in the frequency is the result of making use of a hitherto largely
unexploited physical property of the electromagnetic field, wireless
communication based on angular momentum is fundamentally different
from present-day wireless engineering techniques. In particular,
it does not require arrays of multiple, spaced antennas or other
transducers as is the case for spatial division multiplexing
techniques, such as the multiple-input-multiple-output (MIMO)
technique~\cite{Vucetic&Yuan:Book:2003}.  Nor does it require
massive, power-consuming digital post-processing and has therefore
been described as a cost-effective and ``green'' alternative
to MIMO~\cite{Boffi&al:SPIE:2013} where the algorithms require
increasing overheads and the required processing resources scales
roughly with $M^2$, where $M$ is the mode number~\cite{Yu:OE:2015}.
The properties of electromagnetic/photon OAM have been studied for
frequencies ranging from low-frequency radio~\cite{Thide&al:PRL:2007} to
X-ray~\cite{Sasaki&al:PRL:2008}. It has been shown that in the optical
regime it is possible to generate OAM with an azimuthal mode number of
at least $m=10000$ in free space~\cite{Courtial&al:OC::1997}.

From a more fundamental physical point of view, the enhancements
made possible by invoking the angular momentum physics layer, compared
to systems that are based solely on the linear momentum physics layer,
is consistent with the fact that all classical fields carry angular
momentum~%
\cite{%
Belinfante:P:1940,%
Soper:Book:1976,%
Cohen-Tannoudji&al:Book:1997%
}
and that the dynamics of a physical system is not completely specified
unless both its total linear momentum and its total angular momentum are
specified at all instants~\cite{Truesdell:Book:1968}.  Specifying only
one of them is not sufficient. Therefore, exploiting only the linear
momentum and not the angular momentum of a system of charges, currents
and pertinent electromagnetic fields is not exploiting such a system
to its full capacity.

From the discussion above, it should be clear that in order to make
optimal use of an electromechanical system for radio science and
technology, including radio communications, one must, in addition to
the one-dimensional translational degrees of freedom represented by the
system's linear momentum, invoke also the mechanical and electromagnetic
angular momenta of the system~%
\cite{%
Heitler:Book:1954,%
Bogolyubov&Shirkov:Book:1959,%
Messiah:Book:1970,%
Eyges:Book:1972,%
Landau&Lifshitz:Book:1975,%
Berestetskii&al:Book:1989,%
Ribaric&Sustersic:Book:1990,%
Mandel&Wolf:Book:1995,%
Cohen-Tannoudji&al:Book:1997,%
Schwinger&al:Book:1998,%
Jackson:Book:1998,%
Allen&al:Book:2003,%
Rohrlich:Book:2007,%
Andrews:Book:2008,%
Torres&Torner:Book:2011,%
Yao&Padgett:AOP:2011,%
Thide:Book:2011%
}%
, \ie, the rotational degrees of freedom that are necessary to completely
and correctly specify the system's two- and three-dimensional dynamics~%
\cite{%
Beth:PR:1935,%
Beth:PR:1936,%
Holbourn:N:1936,%
Carrara:N:1949,%
Allen:AJP:1966,%
Carusotto&al:NC:1968,%
Chang&Lee:JOSAB:1985,%
Vulfson:USP:1987,%
Kristensen&al:OC:1994,%
He&al:PRL:1995,
Friese&al:PRA:1996,%
Then&Thide:ARXIV:2008,%
Helmerson&al:Topologica:2009,%
Padgett&Bowman:NPHY:2011,%
Ramanathan&al:PRL:2011,%
Elias:AA:2012,%
Emile&al:PRL:2014%
}%
.

As is the case for electromagnetic linear momentum, electromagnetic
angular momentum can propagate not only in free space, but also in---and
interact with---various propagation media~%
\cite{%
Torner&al:OE:2005,%
Molina-Terriza&al:JEOS:2007%
},
including plasmas~%
\cite{%
Thide:PPCF:2007,%
Mendonca&al:ARXIV:2008,%
Mendonca&al:PRL:2009,%
Tamburini&al:EPL:2010,%
Tamburini&Thide:EPL:2011,%
Thide&al:Incollection:2011,%
Mendonca:PPCF:2012%
}%
. This includes waveguides \cite{Ibanescu&al:S:2000} and optical fibers~%
\cite{%
Alexeyev&al:SPQEOE:1998,%
Bozinovic&al:S:2013,%
Brunet&al:OE:2014%
},
where the decay rate, due to absorption, is the same for the propagation
of angular momentum as for linear momentum since they are both bilinear
in the same electric and magnetic fields.

\svnidlong
 {$HeadURL: file:///Repository/SVN/Articles/AngularMomentumRadio/branches/PRA/physics.tex $}
 {$LastChangedDate: 2015-05-28 19:23:04 +0200 (Thu, 28 May 2015) $}
 {$LastChangedRevision: 97 $}
 {$LastChangedBy: bt $}
\svnid{$Id: physics.tex 97 2015-05-28 17:23:04Z bt $}

\section{Physics}
\label{sect:physics}

In this section we introduce symmetrized integral formulas for
calculating electromagnetic fields generated by an arbitrary, given
distribution of electric and magnetic charges and currents in a volume
of finite spatial extent. We then use the results to study the physical
properties of the electromagnetic observables carried by these fields.
Specifically, we analyze those properties of the electromagnetic angular
momentum observable that sets it apart from the electromagnetic linear
momentum observable when it comes to information-carrying capacity,
within a fixed given frequency bandwidth around a fixed given carrier
frequency, for a point-to-point communication link system equipped with
single monolithic transducers (``antennas'') at each end of the link.
The general formulas derived for the fields are presented in their
Majorana-Oppenheimer form and the fields are consistently approximated
in both the paraxial and the far-zone approximations, allowing us to
derive correctly truncated expressions for the angular momentum density
valid in both these approximations.

\subsection{Electromagnetic fields and observables}

In classical electrodynamics, different methods are routinely used for
calculating electromagnetic fields, and physical observables derived from
them, once the sources are given. Commonly the methods amount to solving
the Maxwell-Lorentz differential equations \eqref{eq:Maxwell_micro}
in one way or another, or to find the potentials, choose a gauge and
then perform the calculations.

Over the past decades, integral equations that allow the
immediate calculation of the fields directly from given charge
and current distributions have gained increasing attraction;
see~\textcite{Souza&al:AJP:2009} and references cited therein. In
Chapter VIII of his~\citeyear{Stratton:Book:1941} textbook,
\textcite{Stratton:Book:1941} calculates, without the intervention
of potentials, temporal Fourier transform expressions for the
retarded electric and magnetic fields, $\E(t,\x)$ and $\B(t,\x)$,
respectively, generated by arbitrary distributions of charge and
current densities at rest relative to the observer. In Chapter~14
of the second edition of their textbook on electrodynamics,
published in \citeyear{Panofsky&Phillips:Book:1962},
\textcite{Panofsky&Phillips:Book:1962} present a variant form
of these expressions, and give them also in ordinary space-time
coordinates. Four years later, \citeauthor{Jefimenko:Book:1966} published
his electrodynamics textbook~\cite{Jefimenko:Book:1966} where the Panofsky
and Phillips expressions for the retarded $\E$ and $\B$ fields were
given in Chapter~15. These expressions are sometimes referred to as the
Jefimenko equations~\cite{Heald&Marion:Book:1995,Jackson:Book:1998},
but we propose that they should rather be called the
Stratton-Panofsky-Phillips-Jefimenko or the SPPJ equations, a name we
will use throughout.

Let us consider a finite (source) volume $V'$ that is located in otherwise
empty space (free space) and let $V'$ contain an arbitrary distribution
of electric and magnetic charge and current densities $\rho(t',\x')$
and $\j(t',\x')$, given in complex form in Eqns.~\eqref{eq:rho&j};
in conventional radio, $V'$ is the volume in space that is occupied by
the transmitting antenna (typically a narrow, straight, electrically
conducting cylinder), and $\je(t',\x')=\Re{\j(t',\x')}$ is the
dissipative electric conduction current constrained to oscillate along
this essentially one-dimensional transmitting antenna.  The fields
produced by the sources in $V'$ propagate to a remotely located observer
who uses a sensing volume $V$ to integrate (average) the densities
of the observables carried by the field in the local volume element
$\dV'$ around the coordinate $\x$ in $V$. In conventional radio, $V$
comprises the receiving antenna, and $\je(t,\x)=\Re{\j(t,\x)}$ is the
dissipative electric conduction current constrained to oscillate along
the receiving antenna.

\begin{figure}
 \resizebox{1.\columnwidth}{!}{%
  \input{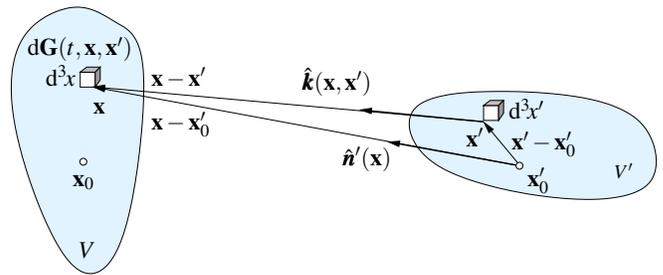}
 }
 \caption{\label{fig:vols}%
  The electric and/or magnetic charges and currents located in the
  small volume element $\cramped{\dV'}$ around the point~$\x'$ in
  the source volume $V'$, with its fixed reference point~$\x_0'$, give
  rise to elemental fields in the volume element $\cramped{\dV}$
  around the observation point (field point) $\x$ in the
  measurement volume $V$ with its fixed reference point $\x_0$.
  In the Figure the volume elements are drawn as small cubes and
  the fields in $\cramped{\dV}$ are given in Majorana-Oppenheimer
  representation by the complex Riemann-Silberstein elemental field
  vector ${\D\RS(t,\x,\x')=\D\E(t,\x,\x')+\im{c}\D\B(t,\x,\x')}$.
  The elemental fields propagate along the unit vector of $\xx'$
  that connects~$\x'$ with~$\x$, denoted $\kunit(\x,\x')$. If all
  individual $\kunit(\x,\x')$ from all individual source points $\x'$
  in~$V'$ that reach~$\x$ are almost parallel, for instance if $V'$ has
  a small lateral extent as seen from~$\x$ (as illustrated), all these
  unit vectors can be approximated by the~$\x'$ independent unit vector
  ${\nunit'(\x)\equiv\kunit(\x,\x_0')}$ that connects~$\x_0'$ with~$\x$.
  This is the paraxial approximation~\cite{Heald&Marion:Book:1995}.%
 }
\end{figure}

The primary elemental fields $\D\E$ and $\D\B$ that are set up at time
$t$ in $\dV$ originate from the local dynamics at the retarded time $t'$
of the charges and currents in $\dV'$; see Fig.~\ref{fig:vols}. These
elemental fields were emitted at the retarded time
\begin{align}
\label{eq:tret}
 t' = \tret'(t,\x,\x') = t - \frac{\abs{\x(t)-\x'(t')}}{c}
\end{align}
and propagated to $\x$ along the local unit vector
\begin{align}
 \kunit(\x,\x') = \frac{\xx'}{\abs{\xx'}}
\end{align}

Choosing a suitable gauge, one can readily calculate exact, closed-form
expressions for the electrodynamic scalar and magnetic vector potentials
from the electric and magnetic charges and currents, for a generic source
distribution.  From these potentials it is possible to derive exact
expressions for the corresponding electromagnetic elemental fields---as
well as the fields themselves---directly, without the intermediate step
of solving the Maxwell-Lorentz equations from first principles or even
calculating the explicit potentials themselves~%
\cite{%
Stratton:Book:1941,%
Panofsky&Phillips:Book:1962,%
Jefimenko:Book:1966,%
Heald&Marion:Book:1995,%
Jackson:Book:1998%
}%
.

If the sources in $V'$ are at rest (\ie, have negligible bulk motion)
relative to the observation/field point $\x$ (assumed to be at rest in
the lab system),
it is possible to express the elemental form of the Riemann-Silberstein
vector \eqref{eq:RS} in terms of a non-covariant complex field vector
density $\vecTheta$ as follows:
\begin{multline}
\label{eq:dG}
 \D\RS(t,\x,\x')
 = (\D\x\bdot\grad){\RS} 
 =\D\E(t,\x,\x') + \im c\D\B(t,\x,\x')
\\
 = \Big(
    \vecTheta(t,\x,\x')
    +\frac{\im}{c}\kunit(\x,\x')\cross\cc{\vecTheta}(t,\x,\x') 
   \Big)\dV'
\end{multline}
where
\begin{align}
\label{eq:vecTheta}
 \vecTheta(t,\x,\x')
 = \vecThetae(t,\x,\x') + \frac{\im}{c}\vecThetam(t,\x,\x')
\end{align}
and the electric charge $\vecThetae$ and its magnetic charge dual
$\vecThetam$ can be written as vector sums~%
\cite{%
Thide&al:ARXIV:2010,%
Thide:Book:2011,%
Thide&al:SPIE:2014%
}
\begin{subequations}
\label{eq:vecThetaem}
\begin{align}
 \vecThetaem(t,\x,\x') = \sum_{j=1}^4\vecbem_j(t,\x,\x')
\end{align}
where
\begin{gather}
\label{eq:bem1}
 \vecbem_1(t,\x,\x')
 = \frac{\rhoem\Big(t-\frac{\abs{\xx'}}{c},\x'\Big)}
    {4\pi\epz\abs{\xx'}^2}\,\kunit(\x,\x')
\\
\label{eq:bem2}
 \vecbem_2(t,\x,\x')
 = \frac{\jem\Big(t-\frac{\abs{\xx'}}{c},\x'\Big)\bdot\kunit(\x,\x')}
    {4\pi\epz c\abs{\xx'}^2}\,\kunit(\x,\x')
\\
\label{eq:bem3}
 \vecbem_3(t,\x,\x')
 = \frac{\jem\Big(t-\frac{\abs{\xx'}}{c},\x'\Big)
    \cross\kunit(\x,\x')}{4\pi\epz c\abs{\xx'}^2}
    \cross\kunit(\x,\x')
\\
\label{eq:bem4}
 \vecbem_4(t,\x,\x')
 = \frac{\pdd{t}{\jem\big(t-\frac{\abs{\xx'}}{c},\x'\big)}
    \cross\kunit(\x,\x')}{4\pi\epz c^2\abs{\xx'}}
    \cross\kunit(\x,\x')
\end{gather}
\end{subequations}

An appealing feature of this representation is that the expressions for
the four vectors $\vecbem_j$ do not involve any spatial derivatives.
As pointed out by~\textcite{Heras:AJP:1994}, this is an advantage
as this avoids some mathematical subtleties associated with the
retardation that mixes space and time [see Eqn.~\eqref{eq:tret}].
Furthermore, $\abs{\vecbem_j},\,j=1,2,3$, all have a $1/\abs{\xx'}^2$
spatial fall-off, whereas $\abs{\vecbem_4}$ falls off as $1/\abs{\xx'}$
and therefore dominates over the other three at very long distances
$\abs{\xx'}$ from the source volume $V'$.

Formulas~\eqref{eq:dG} are generalized (and symmetrized) elemental field
versions of the SPPJ equations~%
\cite{%
Stratton:Book:1941,%
Panofsky&Phillips:Book:1962,%
Jefimenko:Book:1966,%
Heald&Marion:Book:1995,%
Jackson:Book:1998%
}
and clearly show that interesting geometric relations exist between
the elemental fields. For instance, for purely electric sources and for
purely magnetic sources $\D\E$ is always and everywhere perpendicular
to $\D\B$. The new representation \eqref{eq:dG} has also proved to
be economical and in other ways advantageous to use in practical
computations.

\subsubsection{Purely electric sources}

From now on, we consider only electric charges and currents, \ie we set
$\vecThetam=\0$.  However, as is well known, magnetic currents
that manifest the duality of electrodynamics~\cite{Bliokh&al:NJP:2013} and
are useful in practical applications such as modeling electric slot
antennas.  Formulas for this case can be readily derived following the
recipes given below.

We shall frequently decompose a vector $\vv$ into its component
$\vv_\parallel$ parallel to a unit vector $\eunit$, and its component
$\vv_\perp$ transverse to $\eunit$, according to the following scheme:
\begin{subequations}
\label{eq:v_decomposition}
\begin{gather}
 \vv = \vv_\parallel + \vv_\perp
\shortintertext{where}
 \vv_\parallel = v_\parallel\eunit = (\vv\bdot\eunit)\eunit
\\
 \vv_\perp = \vv - \vv_\parallel = \vv - (\vv\bdot\eunit)\eunit
\end{gather}
\end{subequations}
We shall use this decomposition several times in the following and start
with $\vecThetae$ that we write as a sum of one longitudinal component,
$\vecThetae_\parallel$, parallel to the unit vector $\kunit(\x,\x')$,
and one transverse component, $\vecThetae_\perp$, perpendicular to
$\kunit(\x,\x')$, \ie,
\begin{subequations}
\label{eq:vecThetae_parallel+vecThetae_perp}
\begin{gather}
 \vecThetae = \vecThetae_\parallel + \vecThetae_\perp
\shortintertext{where}
 \vecThetae_\parallel 
 = \vecbe_1 + \vecbe_2
\\
  \vecThetae_\perp
 = \vecbe_3 + \vecbe_4
\end{gather}
\end{subequations}
This representation of $\vecThetae$ is very convenient when one
wishes to calculate and analyze electromagnetic field quantities that
are second order in the field vectors.

The local field vectors at a field point $\x$ in the observation
volume $V$, $\E(t,\x)$ and $\B(t,\x)$, are obtained by integrating
the contributions from the entire source volume $V'$ to the field
vector density $\vecThetae$, \ie, from Eqns.~\eqref{eq:dG} and
\eqref{eq:vecThetaem} for purely electric charges and currents, taking
vector perpendicularity into account.  The result is
\begin{subequations}
\label{eq:E&B}
\begin{align}
\begin{split}
 \E(t,\x) &= \int_{V'}{\D\E}
  = \intV{'}{\vecThetae}
  = \intV{'}{(\vecThetae_\parallel+\vecThetae_\perp)}
\\&
  = \sum_{j=1}^4\intV{'}{\vecbe_j(t,\x,\x')}
 \end{split}
\\
 \begin{split}
 \B(t,\x) &= \int_{V'}{\D\B}
  = \frac{1}{c}\intV{'}{(\kunit\cross\vecThetae)}
  = \frac{1}{c}\intV{'}{(\kunit\cross\vecThetae_\perp})
\\&
  = \frac{1}{c}\sum_{j=3}^4\intV{'}{\kunit(\x,\x')\cross\vecbe_j(t,\x,\x')}
 \end{split}
\end{align}
\end{subequations}
Interestingly, $\vecThetae$ behaves as a volumetric
super-density, that, when integrated, yields exact expressions for
the fields from which one can, in turn, calculate exact expressions
for volumetric densities of the observables that are generated by the
sources in $V'$.

\subsection{Useful approximations}

The vector density $\vecThetae$ as given by
Eqns.~\eqref{eq:vecThetae_parallel+vecThetae_perp} has one 
component, represented by the vectors $\vecbe_1$ and $\vecbe_2$ defined by
Eqns.~\eqref{eq:bem1} and~\eqref{eq:bem2}, respectively. This component
is parallel to the unit vector $\kunit(\x,\x')$ from a point $\x$
in the source volume $V'$ to a point $\x'$ in the observation volume
$V$ over which the observables are integrated (the longitudinal field
component). The other component, represented by the vectors $\vecbe_3$ and
$\vecbe_4$ in Eqns.~\eqref{eq:bem3} and~\eqref{eq:bem4}, is perpendicular
to $\kunit(\x,\x')$ (the transverse field component).

\subsubsection{The paraxial approximation}

In a configuration where all the source points $\x'$ in $V'$ are observed
at $\x$ in $V$ such that $\kunit(\x,\x')$ are all nearly parallel,
the paraxial approximation, illustrated in Fig.~\ref{fig:vols}, can be
applied. This amounts to setting
\begin{align}
 \kunit(\x,\x') \approx \kunit(\x,\x_0') \equiv \nunit'(\x)
\end{align}
\ie, approximating the unit vector from $\x'$ to $\x$ with the $\x$-ward
pointing normal unit vector of a sphere centered on a fixed reference
point $\x_0'$ in $V'$. When we introduce this approximation into the exact
expressions for the vectors $\vecbe_j$ in Eqns.~\eqref{eq:vecThetaem},
we obtain the paraxially approximate vectors
\begin{subequations}
\begin{align}
\label{eq:c1-c4_paraxial}
 {\vecbe_1}\parax(t,\x,\x')
&= \frac{\rhoe\Big(t-\frac{\abs{\xx'}}{c},\x'\Big)}
    {4\pi\epz\abs{\xx'}^2}\,\nunit'(\x)
\\ 
 {\vecbe_2}\parax(t,\x,\x')
&= \frac{\je\Big(t-\frac{\abs{\xx'}}{c},\x'\Big)
     \bdot\nunit'(\x)}{4\pi\epz c\abs{\xx'}^2}\,\nunit'(\x)
\\ 
 {\vecbe_3}\parax(t,\x,\x')
&= \frac{\je\Big(t-\frac{\abs{\xx'}}{c},\x'\Big)
    \cross\nunit'(\x)}{4\pi\epz c\abs{\xx'}^2}\cross\nunit'(\x)
\\ 
 {\vecbe_4}\parax(t,\x,\x')
&= \frac{\pdd{t}{\je\big(t-\frac{\abs{\xx'}}{c},\x'\big)}
    \cross\nunit'(\x)}{4\pi\epz c^2\abs{\xx'}}
    \cross\nunit'(\x)
\end{align}
\end{subequations}
These expressions can be further simplified with the help of standard
vector identities, with the result that the field vector density given by
Eqns.~\eqref{eq:vecThetae_parallel+vecThetae_perp} can be approximated by
\begin{align}
 {\vecThetae}\parax 
&= {\vecThetae_\parallel}\parax
 + {\vecThetae_\perp}\parax
\end{align}
where
\begin{subequations}
\label{eq:vecThetae_paraxial}
\begin{align}
\begin{split}
 {\vecThetae_\parallel}\parax
&= \frac{c\rhoe\nunit'}{4\pi\epz c\abs{\xx'}^2}
 + \frac{(\je\bdot\nunit')\nunit'}{4\pi\epz c\abs{\xx'}^2}
\\
&\equiv \frac{c\rhoe\nunit'}{4\pi\epz c\abs{\xx'}^2}
 + \frac{\je_\parallel}{4\pi\epz c\abs{\xx'}^2}
\end{split}
\\
\begin{split}
  {\vecThetae_\perp}\parax
&= \frac{(\je\bdot\nunit')\nunit' -\je}{4\pi\epz c\abs{\xx'}^2}
 +\frac{(\Dotje\bdot\nunit')\nunit' - \Dotje}{4\pi\epz c^2\abs{\xx'}}
\\
&\equiv -\frac{\je_\perp}{4\pi\epz c\abs{\xx'}^2}
   -\frac{\Dotje_\perp}{4\pi\epz c^2\abs{\xx'}}
\end{split}
\end{align}
\end{subequations}
For compactness, introduced the overdot ($\,\Dot{}\,$) notation
\begin{align}
 \Dotje
 \equiv \bigg(\pdd{t}{\je}\bigg)_{\!t=t'}
 = \pdd{t}{\je\big(t-\frac{\abs{\xx'}}{c},\x'\big)}
\end{align}
in the last equation and also suppressed obvious arguments.

Using the above expressions \eqref{eq:vecThetae_paraxial}, we conclude
that in the paraxial approximation the formulas~\eqref{eq:E&B} can be
approximated by
\begin{subequations}
\label{eq:E&B_paraxial}
\begin{multline}
 \E\parax(t,\x) 
 = \E_\parallel\parax(t,\x) + \E_\perp\parax(t,\x)
\\
 = \intV{'}{{\vecThetae_\parallel}\parax(t,\x,\x')}
     + \intV{'}{{\vecThetae_\perp}\parax(t,\x,\x')}
\end{multline}
\begin{multline}
\label{eq:B_paraxial}
 \B\parax(t,\x)
 = \B_\perp\parax(t,\x)
 = \frac{1}{c}\nunit'(\x)\cross\intV{'}{{\vecThetae_\perp}\parax(t,\x,\x')}
\\
 = \frac{1}{c}\nunit'(\x)\cross\E_\perp\parax(t,\x)
 = \frac{1}{c}\nunit'(\x)\cross\E\parax(t,\x)
\end{multline}
\end{subequations}
Whereas $\E\parax$ has two mutually perpendicular components, $\B\parax$
has only one component.  This component is perpendicular both to $\nunit$
and to the component of $\E\parax$ that, in turn, is perpendicular
to $\nunit$. These geometrical properties always hold in the paraxial
approximation, even if the far-zone approximation is not applicable.
On the other hand, sufficiently far away from a source volume of
finite extent, such as an antenna array, the paraxial approximation
is always applicable.

A consistently truncated paraxial approximation expression for the electromagnetic
linear momentum density is obtained by inserting the paraxially valid
approximate expressions~\eqref{eq:E&B_paraxial} for the fields into the
exact expression~\eqref{eq:gfield}. The result is
\begin{subequations}
\label{eq:gfield_parax}
\begin{gather}
 {\gfield}\parax(t,\x)
 = {\gfield_\perp}\parax(t,\x)
 + {\gfield_\parallel}\parax(t,\x)
\shortintertext{where}
 {\gfield_\perp}\parax(t,\x)
 = \epz\E_\parallel\parax(t,\x)\cross\B_\perp\parax(t,\x)
\\
 {\gfield_\parallel}\parax(t,\x)
 =  \epz E_\perp\parax(t,\x) B_\perp\parax(t,\x)\nunit'(\x)
\end{gather}
\end{subequations}

The angular momentum density carried by $(\E,\B)$ with
respect to the fixed moment point $\xm$ is given by the exact
expression~\eqref{eq:hfield}.  Inserting the paraxially valid approximate
expressions~\eqref{eq:E&B_paraxial} for the fields into \eqref{eq:hfield},
we obtain
\begin{multline}
\label{eq:hfield_parax_xm}
 {\hfield}\parax(t,\x;\xm)
 = (\x-\xm)\cross{\gfield}\parax(t,\x)
\\
 = {\hfield}\parax(t,\x;\0) - \xm\cross{\gfield}\parax(t,\x)
\end{multline}
where the first term in the second RHS is the angular momentum around
the origin $\x=\0$
\begin{align}
\label{eq:hfield_parax_0_raw}
\begin{split}
 {\hfield}\parax(t,\x;\0)
 ={}& \epz\x\cross\big(
    {\gfield_\perp}\parax(t,\x) + {\gfield_\parallel}\parax(t,\x)
   \big)
\\
 ={}& \epz\big(\B_\perp\parax(t,\x)\bdot\x\big)\E_\parallel\parax(t,\x)
\\
 &- \epz\big(\E_\parallel\parax(t,\x)\bdot\x\big)\B_\perp\parax(t,\x)
\\
 &+ \epz E_\perp\parax(t,\x) B_\perp\parax(t,\x)\x\cross\nunit'(\x)
\end{split}
\end{align}
as follows from Eqns.~\eqref{eq:gfield_parax} and the `bac-cab' vector
identity.  Recalling that, by definition, the radius vector $\x$ from the
source, located at the origin, to the field point is $\x=\abs{\x}\nunit'$,
we note that $\B_\perp\bdot\x=0$ and $\x\cross\nunit'=0$, which simplifies
Eqn.~\eqref{eq:hfield_parax_0_raw} to
\begin{align}
\label{eq:hfield_parax_0}
\begin{split}
 {\hfield}\parax(t,\x;\0)
&= -\epz \abs{\x}{E_\parallel\parax}{\B_\perp\parax}
\end{split}
\end{align}
which, in a spherical polar coordinate system $(r,\th,\ph)$ where $\x=r\runit$
and $\runit=\nunit'$, becomes
\begin{align}
\label{eq:hfield_parax_0_spherical}
\begin{split}
 {\hfield}\parax(t,r,\th,\ph;\0)
&= -\epz rE_r\parax\B_\perp\parax
\end{split}
\end{align}
This shows that in cases when the paraxial approximation can be
applied, the angular momentum density with respect to the origin is
always proportional to the product of the distance $r$ from the source,
the amplitude of the near-zone longitudinal (radial) component of the
electric field vector, which goes like $r^{-2}$, and the amplitude of
the transverse component of the magnetic field vector, which goes like
$r^{-1}$, so that the whole expression goes as $r^{-2}$ as it should. For
a single frequency component $\w$, the angular momentum density
(pseudo)vector has one DC component and one component that oscillates
along the magnetic field (pseudo)vector at twice the oscillation
frequency $\w$.  The cycle average of the latter component vanishes.

Using the paraxial relation \eqref{eq:B_paraxial} and simple vector
algebra, we can express the result in Eqn.~\eqref{eq:hfield_parax_0} entirely
in terms of the electric field in the following way:
\begin{align}
\label{eq:hfield_parax_0_alt}
\begin{split}
 {\hfield}\parax(t,\x;\0)
&= \frac{\epz}{c}\big(
    \x\bdot\E_\parallel\parax(t,\x)\ox\E_\perp\parax(t,\x)
   \big)\cross\nunit'(\x)
\end{split}
\end{align}
where $\ox$ is the dyadic product operator.  When the paraxial
approximation is not applicable, there exists a certain angular spread in
the wave vectors that may require corrections to the geometrical factors.

\subsubsection{The far-zone approximation}

In the far-zone approximation, the
integrands~\eqref{eq:vecThetae_parallel+vecThetae_perp} and
\eqref{eq:vecThetae_paraxial} can be consistently approximated in order to
further simplify the integrals~\eqref{eq:E&B} and \eqref{eq:E&B_paraxial},
respectively. With reference to Fig.~\ref{fig:vols}, and with the choice
of the reference point $\x_0'$ as the origin of the spherical polar
coordinate used, implying that
\begin{subequations}
\begin{gather}
 \xx_0' = \vecr = r\runit
\\
 \x'-\x_0' = \vecr' = r'\runit'
\end{gather}
\end{subequations}
the identity
\begin{align}
\begin{split}
 \frac{1}{\abs{\xx'}} 
 = \frac{1}{\abs{(\xx_0')-(\x'-\x_0')}}
 = \frac{1}{\abs{\rr'}}
\end{split}
\end{align}
holds.  For configurations such that the shortest distance between
$V'$ and $V$ is large, or, more precisely, when
\begin{align}
 \sup{r'}
 = \sup\abs{\x'-\x_0'}
 \ll \inf\abs{\xx_0'}
 = \inf{r}
\end{align}
the Taylor expansion
\begin{align}
\begin{split}
 \frac{1}{\abs{\xx'}} &= \frac{1}{\abs{\rr'}}
\\&
 = \frac{1}{r}
 - \big(\vecr'\bdot\grad\big)\frac{1}{r} 
 - \frac{1}{2}\big(\vecr'\bdot\grad\big)^2\frac{1}{r} 
 + \ldots 
\\&
 = \frac{1}{r}
 + \frac{\vecr'\bdot\runit}{r^2}
 + \frac{1}{2}
  \frac{\runit\bdot\big(3\vecr'\ox\vecr'-\tunity{r'}^2\big)\bdot\runit}{r^3} 
 + \Ordo{r^{-4}}
\end{split}
\end{align}
converges rapidly. This allows us to set ${1/\abs{\xx'}\approx1/r}$
in Eqns.~\eqref{eq:vecThetae_paraxial}.  As a result, the un-truncated
expressions for the longitudinal and transverse components of the field
density vector in the combined paraxial and far-zone approximations can
be written
\begin{subequations}
\label{eq:vecThetae_paraxial_far}
\begin{align}
 {{\vecThetae_\parallel}\parax}\far
 &= \frac{c\rhoe\nunit'}{4\pi\epz cr^2}
 + \frac{\je_\parallel}{4\pi\epz cr^2}
\\
  {{\vecThetae_\perp}\parax}\far
 &= -\frac{\je_\perp}{4\pi\epz cr^2}
   -\frac{\Dotje_\perp}{4\pi\epz c^2r}
\end{align}
\end{subequations}
Accordingly, the integrals in Eqns.~\eqref{eq:E&B_paraxial} simplify
considerably.

Following~%
\textcite{%
Thide&al:ARXIV:2010,%
Tamburini&al:NJP:2012a%
},
we introduce the following shorthand notations:
\begin{subequations}
\label{eq:integrals}
\begin{align}
\label{eq:qe}
 \qe(t') &\equiv \intV{'}{\rhoe(t',\x')}
\\
 \vIe(t') &\equiv \intV{'}{\je(t',\x')}
\\
\begin{split}
 \DotIe(t') &\equiv \intV{'}{\Dotje(t',\x')}
\\
  &\equiv \intV{'}{\pdd{t}{\je\big(t-\frac{\abs{\xx'}}{c},\x'\big)}}
\end{split}
\end{align}
\end{subequations}
Then, under the assumption of paraxial as well as far-zone approximation,
we can write the perpendicular and parallel components of the fields
as follows:
\begin{subequations}
\label{eq:E&B_paraxial_far}
\begin{gather}
 {\E_\perp\parax}\far(t,\x)
 = -\frac{\vIe_\perp(t')}{4\pi\epz cr^2}
  -\frac{\DotIe_\perp(t')}{4\pi\epz c^2r}
\\
 {\E_\parallel\parax}\far(t,\x)
 = \frac{\qe(t')\nunit'(\x)}{4\pi\epz r^2}
  +\frac{\vIe_\parallel(t')}{4\pi\epz cr^2}
\\
 {\B_\perp\parax}\far(t,\x)
 = \frac{\vIe_\perp(t')\cross\nunit'(\x)}{4\pi\epz cr^2}
  +\frac{\DotIe_\perp(t')\cross\nunit'(\x)}{4\pi\epz c^2r}
\end{gather}
\end{subequations}

Evaluating the linear momentum density formula~\eqref{eq:gfield} in the
paraxial approximation, keeping all terms and approximating each one
of them in a consistent way, we see that far away from the source the
linear momentum density is
\begin{subequations}
\label{eq:gfield_parax_far}
\begin{align}
 {{\gfield}\parax}\far
 = {{\gfield_\perp}\parax}\far + {{\gfield_\parallel}\parax}\far
\end{align}
where
\begin{gather}
\label{eq:gfield_perp_parax_far}
 {{\gfield_\perp}\parax}\far
 = \frac{1}{16\pi^2\epz c^2r^3}
 \big(c\qe+I\Sup{e}_\parallel\big)
 \bigg(\frac{\I_\perp\Sup{e}}{r} + \frac{\DotIe_\perp}{c}\bigg)
\\
\label{eq:gfield_parallel_parax_far}
 {{\gfield_\parallel}\parax}\far
 = \frac{1}{16\pi^2\epz c^2r^2}
  \bigg({\frac{\vIe_\perp}{r}+\frac{\DotIe_\perp}{c}}\bigg)\bdot
  \bigg({\frac{\vIe_\perp}{r}+\frac{\DotIe_\perp}{c}}\bigg)\nunit'
\end{gather}
\end{subequations}

If we express this result in terms of the explicit integrals in
Eqns.~\eqref{eq:integrals}, we find that as the distance $r$ from
the source tends to infinity, the linear momentum density vector is
given by the consistently truncated expression
\begin{multline}
\label{eq:gfield_far_int}
 {{\gfield}\parax}\far(t,\x) =
\\
  \frac{1}{16\pi^2\epz c^4r^3}
  \intV{'}{\Big(c\rhoe+j\Sup{e}_\parallel
 + \Ordo{r^{-1}}
   \Big)}
  \intV{'}{\Dotje_\perp}
\\
 +\frac{1}{16\pi^2\epz c^4r^2}
  \Big(\intV{'}{\Dotje_\perp}\bdot\intV{'}{\Dotje_\perp}
 + \Ordo{r^{-1}}
  \Big)\nunit'
\end{multline}
To leading order in $r$ the magnitude of this vector is the scalar
\begin{align}
\begin{split}
\label{eq:abs_gfield_far_int}
 \abs*{{{\gfield}\parax}\far(t,\x)}
 ={}& \frac{1}{16\pi^2\epz c^4r^2}
   \abs*{\intV{'}{\Dotje_\perp}}^2
  +\Ordo{r^{-3}}
\end{split}
\end{align}
that exhibits the well-known $r^{-2}$ asymptotic fall-off.

Starting from the exact expression for the angular momentum density,
Eqn.~\eqref{eq:hfield}, and using the un-truncated paraxially approximate
expressions \eqref{eq:gfield_parax_far}, noticing that $\x=r\nunit'$,
we see that the un-truncated paraxially approximate expression for the
angular momentum density is
\begin{align}
\label{eq:hfield_parax_far_0}
 {{\hfield}\parax}\far(t,\x;\0) 
  = \frac{c\qe(t')+I\Sup{e}_\parallel(t')}{16\pi^2\epz c^2r^2}
 \nunit'\cross\bigg(\frac{\I_\perp\Sup{e}}{r} + \frac{\DotIe_\perp}{c}\bigg)
\end{align}
Expressing this result explicitly in the integrals in
Eqns.~\eqref{eq:E&B_paraxial}, and truncating consistently, we obtain
\begin{multline}
\label{eq:hfield_parax_far_0_int}
 {{\hfield}\parax}\far(t,\x;\0) 
 =- \frac{1}{16\pi^2\epz c^3r^2}
  \intV{'}{\rhoe}\intV{'}{\Dotje_\perp}\cross\nunit'
\\
  -\frac{1}{16\pi^2\epz c^4r^2}
   \intV{'}{j\Sup{e}_\parallel}\intV{'}{\Dotje_{\!\!\perp}}\cross\nunit'
  +\Ordo{r^{-3}}
\end{multline}
In the first integral in the first term we may, with the help of the continuity
equation, formally express $\rhoe$ in $\je$ as follows
\begin{align}
 \intV{'}{\rhoe(t',\x')} = -\int^{t'}\!\D{t'}\intV{'}{\divprime\je(t',\x')}
\end{align}
which, by introducing the shorthand notation
\begin{align}
 \tvje(\x') = \int^{t'}\!\D{t'}\je(t',\x')
\end{align}
allows us to write
\begin{subequations}
\begin{align}
 \intV{'}{\rhoe(t',\x')} 
 &= -\intV{'}{\divprime\tvje(\x')}
\\
 &= -\ointSvecdot{'}{\tvje(\x')}
\end{align}
\end{subequations}
where, in the last step, the divergence theorem was applied. We note that
we can, in the same vein, formally write Eqn.~\eqref{eq:qe} as
\begin{subequations}
\begin{align}
 \qe(t') &\equiv -\intV{'}{\divprime\int^{t'}\!\D{t'}\je(t',\x')}
\\
 &= -\int^{t'}\!\D{t'}\intV{'}{\divprime\je(t',\x')}
\end{align}
\end{subequations}

Recalling formula~\eqref{eq:hfield_parax_xm} and using
expression~\eqref{eq:gfield_far_int}, we conclude that for an arbitrary
source distribution $(\rhoe,\je)$ located in a volume $V'$ of finite
extent, the angular momentum density with respect to an arbitrary fixed
moment point $\xm$ in the paraxial, far-zone approximation is
\begin{subequations}
\label{eq:hfield_parax_far_xm_int}
\begin{multline}
 {{\hfield}\parax}\far(t,\x;\xm)
  =
\\
  \frac{1}{16\pi^2\epz c^4r^2}
  \intV{'}{\big(c\divprime\tvje-j\Sup{e}_\parallel\big)}
  \intV{'}{\Dotje_\perp}\cross\nunit'
\\
 +\frac{1}{16\pi^2\epz c^4r^2}
  \abs*{\intV{'}{\Dotje_\perp}}^2
  \xm\cross\nunit'
  +\Ordo{r^{-3}}
\end{multline}
or, since $\nunit'\bdot\tvje=\tje_\parallel$,
\begin{multline}
 {{\hfield}\parax}\far(t,\x;\xm)
  =
  \frac{1}{16\pi^2\epz c^3r^2}
  \ointS{'}{\tje_\parallel}
  \intV{'}{\Dotje_\perp}\cross\nunit'
\\
  -\frac{1}{16\pi^2\epz c^4r^2}
   \intV{'}{j\Sup{e}_\parallel}\intV{'}{\Dotje_{\!\!\perp}}\cross\nunit'
\\
 +\frac{1}{16\pi^2\epz c^4r^2}
  \abs*{\intV{'}{\Dotje_\perp}}^2
  \xm\cross\nunit'
  +\Ordo{r^{-3}}
\end{multline}
\end{subequations}
Sufficiently far away from $V'$, the paraxial approximation is trivially
fulfilled.  Therefore, Eqns.~\eqref{eq:hfield_parax_far_xm_int}
show that the magnitude of the angular momentum density always
has an $r^{-2}$ asymptotic fall-off. See also Table~\ref{tab:g&h},~%
\textcite{%
Thide&al:ARXIV:2010,%
Tamburini&al:NJP:2012a%
}).
Eqns.~\eqref{eq:hfield_parax_far_xm_int} generalize the proof,
published by \textcite{Abraham:PZ:1914} for the special case that the source
is a single electric Hertzian dipole and the moment point is the origin,
to hold for any conceivable combination of charge and current densities,
located inside a source volume $V'$ of finite extent, and for an arbitrary
moment point $\xm$.

We note from Eqns.~\eqref{eq:E&B} for $\E$ and $\B$ that far away from
$V'$, the dominating contribution to the magnitude of the linear momentum
density (the Poynting vector) comes from the asymptotic approximation
of the far-zone $\vecbe_4$ term, \ie, the transverse component of
the time rate of change of the current density, whereas the dominating
contribution to the magnitude of the angular momentum in the same region
comes from both the \emph{far-zone} $\vecbe_4$ term for $\B$ and
the \emph{near-zone} $\vecbe_1$ and $\vecbe_2$ terms for $\E$,
\ie, from both the perpendicular component of the time rate change of 
the current density vector and the (time integrated) divergence and the
longitudinal component of the current density itself. Hence, the far-zone
$\E$ field has in general no direct, primary influence on the far-zone
angular momentum; see, however, formula~\eqref{eq:hfield_parax_0_alt}.
This may at first seem like a paradox but is a fact that was
established more than a century ago \cite{Abraham:PZ:1914}; see also
\textcite{Then&Thide:ARXIV:2008,Thide&al:ARXIV:2010,Tamburini&al:NJP:2012a}.

Equations~\eqref{eq:hfield_parax_far_xm_int} can be used for designing
transducers that can generate beams with a coherent superposition of
angular momentum eigenmodes that can be sensed, resolved and analyzed far
away from the source volume $V'$.  It is important that the volume $V$
used for integrating the angular momentum density is chosen appropriately
and that the angular momentum measurements are made in an optimum way,
so that the $r^{-2}$ fall-off is judiciously exploited.

\svnidlong
 {$HeadURL: file:///Repository/SVN/Articles/AngularMomentumRadio/branches/PRA/implementation.tex $}
 {$LastChangedDate: 2015-05-28 19:23:04 +0200 (Thu, 28 May 2015) $}
 {$LastChangedRevision: 97 $}
 {$LastChangedBy: bt $}
\svnid{$Id: implementation.tex 97 2015-05-28 17:23:04Z bt $}

\section{Implementation}
\label{sect:implementation}

For many reasons, not least to mitigate the problems of growing
frequency congestion and increasing energy consumption in radio science,
radar, and wireless and fiber optic communication applications,
it is desirable to strive for a more resource-conserving, efficient
and flexible use of the electromagnetic field than is offered by the
prevalent linear-momentum techniques. In this Section we discuss physical
issues that are essential to consider when developing and implementing
angular momentum techniques in the radio domain. These results are
of interest not only for radio but also for other wavelengths, since,
as was pointed out by~\textcite{Thide&al:PRL:2007}, it is often more
convenient to perform fundamental studies of electromagnetic radiation,
including those addressed in this article, in radio than in optics.
One reason being that it is relatively easy to vary the frequency and
to control the phase of radio waves and beams made from them.  They can
therefore be produced with a very high degree of coherence and spectral
purity. Furthermore, the fact that it is possible to use with digitally
controlled transmitters, receivers and antenna arrays that enables
experiments to be completely software defined, adds considerable
flexibility and versatility compared to the typical experiment at
optical wavelength, something that further the serendipity factor.

Let us consider a spatially limited source volume $V'$ from which
an electromagnetic beam is emitted into free space where there is no
creation or annihilation of neither linear nor angular field momentum. The
emission takes place during a finite time interval $\Delta{t_0}$,
thus forming a signal ``pulse'' that after a sufficiently long time
$t_0$ after the end of the emission will be propagating radially
outward with speed $c$ into the surrounding free space as discussed in
Sect.~\ref{sect:physics}.  This signal ``pulse'' will then be located
in a finite volume $V_0$ between two concentric spherical shells, one
with radius $r_0=ct_0$ relative to the reference point $\x_0'$ in $V'$
(see Fig.~\ref{fig:vols}), and another with radius $r_0+\Delta{}r_0$
where ${\Delta{r_0}=c\Delta{t_0}}$ \cite{Schwinger&al:Book:1998}.
According to the conservation laws~\eqref{eq:conservation_law_lin_mom}
and \eqref{eq:conservation_law_ang_mom}, the cycle averaged field momentum
$\pfield$ and angular momentum $\Jfield$ contained in this spatially
limited volume do not fall off asymptotically at large distances from
the source, but tend to constant values.

Consequently, both linear and angular electromagnetic field momenta
can propagate---and be used for information transfer---over in
principle arbitrarily long distances. Of course, the magnitude and
angular distribution of the respective momentum densities depend on
the specific spatio-temporal properties of the actual radiating and
receiving transducers (``antennas'') used. Some transducers, such as
one-dimensional linear dipole antennas used in radio today, are effective
radiators and sensors of linear momentum, whereas well-defined coherent
superpositions of angular momentum eigenmodes are more optimally radiated
and sensed by transducers that make full use of two- or three-dimensional
current distributions.

\subsection{Wireless information transfer with linear momentum}

If we identify the individual terms in the global law of conservation of
linear momentum in a closed volume, \eqref{eq:conservation_law_lin_mom},
we can rewrite this equation as a balance equation between the time
rate change of mechanical linear momentum, \ie, the force on the matter
(the charged mechanical particles) in $V$, the time rate change of field
linear momentum in $V$, and the flow of field linear momentum into $V$,
across the surface $S$ bounding $V$, as follows:
\begin{align}
\label{eq:lin_mom_int}
 \underbrace{\intV{}{\f}}_{
  \mathclap{\text{Force on the matter}}}
 + \ddt{}\underbrace{\intV{}{\epz(\E\cross\B)}}_{
    \mathclap{\text{Field momentum}}}
 + \underbrace{\ointSvecdot{}{\tens{T}}}_{\mathclap{\text{Linear momentum flow}}}
 = \0
\end{align}
Hence, electric charges that are subject to an oscillating electromotive
force, \eg, a current fed into an antenna from a transmitter via a
transmission line, experience a mechanical force that sets the charges
into translational oscillatory motion and therefore gives
rise to a time-varying linear conduction current density $\je(t',\x')$
in $V'$. The conservation law~\eqref{eq:conservation_law_lin_mom} also
shows that $\je(t',\x')$ is accompanied by the simultaneous emission
of a time-varying electromagnetic field linear momentum $\pfield(t)$
that propagates in free space away from $V'$ in the form of the linear
momentum density $\gfield(t,\x')$, as described in earlier Sections and
in Appendix~\ref{app:Poincare}.

For illustrative purposes, let us consider an electromechanical system
comprising an electromagnetic field that interacts with point charges
$q_i$, $i=1,2,\ldots,j,\ldots,n$, located at $\x_i$ at time $t_i$. All
charges have the same charge $q$ and same mass $m$.  The charge density
$\rhoe_j$ representing the \emph{j}th point charge $q_j$ at $\x_j$
can be described in the form of the Dirac delta distribution
\begin{align}
\label{eq:qdelta}
 \rhoe_j=q\delta(\xx_j)
\end{align}
According to the balance equation~\eqref{eq:lin_mom_int},
the Lorentz force acting on $q_j$ due to the imposed linear momentum
flow (the Maxwell stresses) is
\begin{align}
\label{eq:Lorentz_force}
\begin{split}
 \F_j \equiv \F(t_j,\x_j) &= \intV{}\f = \intV{}{\rhoe_j(\E+\vmech\cross\B)}
\\
 &= q\E(t_j,\x_j) + q\vmech(t_j,\x_j)\cross\B(t_j,\x_j)
\\
 &\equiv q\E_j + q\vmech_j\cross\B_j
 = q\E_j + \je_j\cross\B_j
\end{split}
\end{align}
Expressed in the mechanical linear
momentum of particle $j$, $\pmech_j$, this can be written
\begin{align}
 \dd{t}{\pmech_j} - \frac{q}{m}\pmech_j\cross\B_j = q\E_j
\end{align}
This illustrates that the process of estimating the fields from
the measured current in a linear antenna of finite one-dimensional
extent where the charges experience a holonomic constraint due to their
cylindrically symmetric axial confinement, is difficult and can only be
projective.  The situation gets worse if we also include self-interaction
effects, re-radiation, pre-acceleration and related complications. For
a detailed discussion see~\textcite[Supplement]{Rohrlich:Book:2007}.

In a remote sensing volume $V$, the conservation
law~\eqref{eq:conservation_law_lin_mom} shows that the reverse process
takes place in that (a part of) the flow of field linear momentum density
$\gfield(t,\x)$ emitted from $V'$ is integrated  in $V$ into a field
linear momentum $\pfield(t)$ that is accompanied by a mechanical linear
momentum $\pmech(t)$. This gives rise to a translational motion of
the charges and, hence, a translational (non-rotational) conduction
current.  If $V$ is a very thin cylindrical conductor, as in typical
radio communications scenarios, this is a one-dimensional, scalar,
(Ohmic) dissipative conduction current, known as the antenna current,
that is fed to the receiving equipment. Clearly, a single translational
symmetric receiving antenna cannot sense the electromagnetic angular
momentum, manifesting rotational symmetry, carried by the same beam.

Let us calculate the amount of linear momentum $\pfield$ carried by
the electromagnetic field generated by an arbitrary source that is
localized in a volume $V'$, \eg, a typical transmitting antenna. This is
most readily done by evaluating the integral in the right-hand member
of Eqn.~\eqref{eq:pfield} that expresses the field linear momentum
in a volume away from $V'$, in a spherical polar coordinate system
$(r,\th,\ph)$ with its origin at the center of the source region, \eg,
the antenna phase center. One finds that the total linear momentum
carried by such an electromagnetic ``pulse'' of finite duration
${\Delta{t}=\Delta{r_0}/c}$ is
\begin{align}
\label{eq:pfield_far}
 \pfield(t)
 = \int_{r_0}^{r_0+\Delta{r_0}}\D{r}\,r^2
   \int_0^{2\pi}\D{\ph}
   \int_0^\pi\D{\th}\sin\th\,{\gfield(t,r,\th,\ph)}
\end{align}
when it propagates in free space \cite{Schwinger&al:Book:1998}.

The integration over the angular domain (the two last
integrals) yields a function of $r$ (and $t$) that, for very
large $r_0=ct_0$, becomes proportional to $r^{-2}$ as shown by
expressions~\eqref{eq:gfield_parax_far} and~\eqref{eq:gfield_far_int}.
Taking into account that this function shall in the remaining (\ie,
first) integral in the right-hand member of Eqn.~\eqref{eq:pfield_far}
be first multiplied by $r^2$ and thereafter integrated over the finite
radial interval ${[r_0,r_0+\Delta{r_0}]\equiv[ct_0,c(t_0+\Delta{t_0})]}$,
we see that $\abs{\pfield}$ tends to a constant when $r_0$ tends to
infinity. This asymptotic independence of distance from a localized
source, allowing the field linear momentum generated by this source
to be transported all the way to infinity without radial fall-off
and therefore be irreversibly lost there, is the celebrated arrow of
radiation asymmetry (see \textcite[pp.~328--329]{Eddington:Book:2010},
\textcite[Chap.~6]{Jackson:Book:1998}, \textcite{Zeh:Book:2001},
\textcite[Chap.~9]{Rohrlich:Book:2007}, and
\textcite{Wheeler&Zurek:Book:2014}).

The angular distribution in the far zone of the magnitude of the
linear momentum density, $\abs{\cramped{\cramped{\gfield}}\far}$
and, consequently, of the Poynting vector $\abs{\S}$ [see
relation~\eqref{eq:Planck_rel}], is commonly referred to as ``the
radiation pattern'' or, in the case of antenna engineering, ``the antenna
pattern'' or ``antenna diagram''.

\subsection{Wireless information transfer with angular momentum}

As the global law of conservation of angular momentum in a closed volume,
Eqn.~\eqref{eq:conservation_law_ang_mom}, shows, the electric charges
in a source volume $V'$ that are subject to a time-varying surface
integrated angular momentum flux $\tens{M}$ suffer a change in their
total mechanical (matter) angular momentum $\hmech$, \ie, they experience a
mechanical torque $\vtau\mech$. This torque sets the charges into two-
or three-dimensional rotational motion that causes the emission of
a superposition of electromagnetic angular momentum eigenmodes that
propagates in free space away from $V'$ in the form of the angular
momentum density $\hfield$ as described in earlier Sections and in
Appendix~\ref{app:Poincare}.

The angular momentum conservation law~\eqref{eq:conservation_law_ang_mom}
can be written in the form of a balance equation between the time
rate change of mechanical torque on the matter (the charged mechanical
particles) in $V$, the time rate change of field angular momentum and
the flow of angular momentum across the surface $S$, which encloses $V$,
as follows:
\begin{multline}
\label{eq:ang_mom_int}
 \underbrace{\intV{}{\vtau(\xm)}}_{
  \mathclap{\text{Torque}}}
 + \ddt{}\underbrace{\intV{}{(\xxm)\cross\epz(\E\cross\B)}}_{
    \mathclap{\text{Field angular momentum}}}
\\
 +\underbrace{
  \ointSvecdot{}{\tens{M}(\xm)}
  }_{\mathclap{\text{Angular momentum flow}}}
 = \0
\end{multline}

For illustrative purposes let us consider the \emph{j}th charge $q_j$ in
a system of $n$ point charges $q_i, i=1,2,\ldots,j,\ldots,n$, each
carrying the charge $q$. The charge density can be represented by the
Dirac delta distribution as given by Eqn.~\eqref{eq:qdelta}.  Then we find
that the mechanical torque on $q_j$ is
\begin{align}
\label{eq:Lorentz_torque}
\begin{split}
 \vtau_j(\xm) \equiv \vtau(t_j,\x_j;\xm)
  &= \intV{}{(\xxm)\cross\f}
\\
  &= \intV{}{(\xxm)\cross(\rhoe_j\E+\je\cross\B)}
\\
  &= (\x_j-\xm)\cross\F_j
\end{split}
\end{align}
where $\F_j$ is the Lorentz force on $q_j$, given by
Eqn.~\eqref{eq:Lorentz_force}. 

In a remote observation volume $V$ where the charges are not constrained
to move along a line but can move freely in two or three dimensions,
the reverse process takes place in that (a sufficient amount of) the flow
of field angular momentum density emitted from $V'$ is intercepted and
integrated into a mechanical angular momentum $\Jmech$. This gives
rise to a rotational motion of the charges and, hence, to a two- or
three-dimensional rotational (non-translational) current density in
$V$. Applications that exploit the rotational symmetry of the fields,
\ie, make use of the angular momentum degrees of freedom, must therefore
be implemented in techniques that are based on the angular momentum
physical layer.

The field angular momentum about the origin, $\Jfield(\0)$, contained
in the ``pulse'' emitted from the same localized source and volume as
described by equation \eqref{eq:pfield_far}, is
\begin{align}
\label{eq:Jfield_int}
 \Jfield(t,\0)
 = \int_{r_0}^{r_0+\Delta{r_0}}\D{r}\,r^2
   \int_0^{2\pi}\D{\ph}
   \int_0^\pi\D{\th}\sin\th\,{\hfield(t,r,\th,\ph)}
\end{align}
The expressions~\eqref{eq:hfield_parax_far_0} and
\eqref{eq:hfield_parax_far_0_int} explicitly show that the angular
momentum density $\hfield$ emitted from any electromagnetic radiation
source has precisely the same asymptotic $r^{-2}$ radial fall-off in all
directions as the linear momentum density $\gfield$. Hence, it is clear
that also $\abs{\Jfield}$ tends asymptotically to a constant value and
is irreversibly lost when the ``pulse'' approaches infinity. This is
the angular momentum analog of the well-known linear momentum arrow
of radiation.

\begin{figure}
 \resizebox{1.\columnwidth}{!}{%
  \input{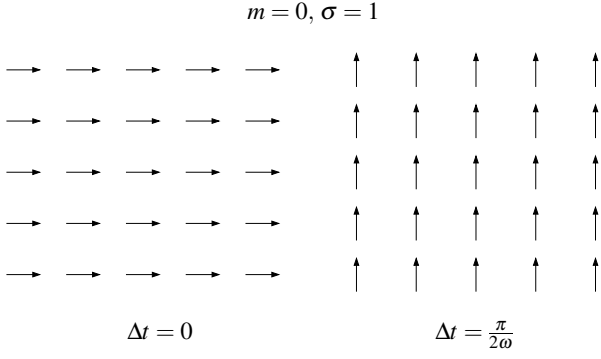}
 }
 \caption{\label{fig:fields_circ_pol}%
  Schematic plot of the instantaneous directions of the field vectors
  in a right-hand circular polarized cylindrical beam (SAM $\sigma=1$),
  projected onto the phase plane perpendicular to the $z$ axis (pointing
  out of the figure).  The right-hand panel shows the situation $1/4$
  period later than the left-hand panel.  The beam does not carry OAM
  so the directions of the field vectors vary only with time and rotate
  in unison.%
 }
\end{figure}

We point out that for one and the same radiating device (``antenna''),
the angular distribution of the radiated \emph{angular} momentum
density $\abs{\hfield}$ is different from the angular distribution
of the radiated \emph{linear} momentum density, $\abs{\gfield}$,
\ie, the ``antenna pattern''; see Eqns.~\eqref{eq:gfield_parax_far}
and~\eqref{eq:hfield_parax_far_0}, \textcite{Thide&al:Incollection:2011},
and \textcite{Then&Thide:ARXIV:2008}.

\subsubsection{Spin angular momentum (SAM) \vs\ orbital angular momentum
(OAM)}

\begin{figure}
\centering
 \resizebox{1.\columnwidth}{!}{%
  \input{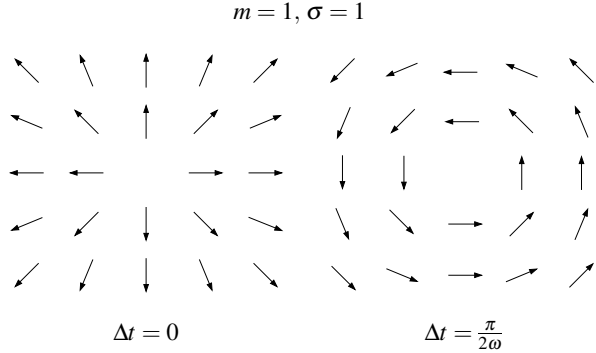}
 }
 \caption{\label{fig:fields_circ_pol_OAM}%
  Same as Fig.~\ref{fig:fields_circ_pol}, but for the case that the beam
  also carries a $z$ component of OAM, $L_z$, with azimuthal quantum
  number (mode index) $m=1$, corresponding to a phase variation $\e^{\im
  \ph}$ of the field where $\ph$ is the azimuth angle around the $z$
  axis.  As a result, the instantaneous direction of the field vector
  depends on $\ph$ in addition to varying with time due to SAM.%
 }
\end{figure}

It is shown in Subsection~\ref{subsect:angular_momentum} in
Appendix~\ref{app:Poincare}, that if the vector potential $\A$ fulfills
the criteria for Helmholtz decomposition, it is possible to separate
the analytic expression for the total field angular momentum, $\Jfield$,
into two exact, gauge independent expressions.  One that is not dependent
of the moment point, $\Sfield$, \viz, expression~\eqref{eq:Sfield_total},
and one that is, $\Lfield$, \viz, expression~\eqref{eq:Lfield_total}. In
radio communication scenarios these expressions can be identified with
the spin and orbital angular momentum, respectively, of the radio beam.

Figure~\ref{fig:fields_circ_pol} illustrates how the intrinsic spin
part of the angular momentum (SAM), $\Sfield$, manifests itself as
left- or right-hand circular polarization, \ie, as a unison rotation
in time of the field vectors such that they all have one and the same
direction across the phase front of the beam at given time $t$, and
all have rotated into another direction (modulo $2\pi$ in oscillation
frequency phase $\w{t}$) in each point at time $t+\Delta{t}$.  Hence,
the measurement of SAM (polarization) requires temporal coherence.

A common technique to observe and exploit SAM in radio is to use two
co-located linear dipole antennas arranged orthogonally to each other
and to determine the polarization state by comparing the phases of the
antenna currents induced in the two antennas by using interferometry. In
antenna engineering such antennas are known as turnstile antennas
\cite{Brown:E:1936,Vulfson:USP:1987,Emile&al:PRL:2014}. Another
technique is to use longitudinal mode helical antennas; see
\textcite[example~12.7.1, p.~461]{Someda:Book:2006}.

In contrast to SAM, the orbital part of the electrodynamic angular
momentum, the orbital angular momentum (OAM) $\Lfield$, is an extrinsic
(coordinate dependent) property.  Figure~\ref{fig:fields_circ_pol_OAM}
illustrates how OAM manifests itself as a rotation in space of the field
vectors such that they at any given instant have different directions
(modulo $2\pi$ in oscillation frequency phase $\w{t}$) for different
azimuth angles positions $\ph$ and $\ph+\Delta{\ph}$ around the $z$
(beam) axis in the phase plane. Hence, the measurement of OAM requires
spatial coherence.

We emphasize that whereas the spin angular momentum (SAM) is an intrinsic
property of the photon, both its linear and orbital angular momenta are
extrinsic properties.

\subsubsection{Angular momentum transducers}

As the conservation law~\eqref{eq:conservation_law_ang_mom} shows, angular
momentum modes cannot be radiated and measured by single, monolithic
transducers that work solely on the principle of force, \eg, are based on
(Ohmically) dissipative conduction currents flowing along a line, as is
the case for linear dipole-type antennas commonly used today.  Instead,
these modes should be radiated and measured by transducers (actuators,
sensors, ``antennas'') that work on the principle of rotational degrees
of freedom and torque, thus being capable of exploiting the angular
momentum directly
\cite{%
Beth:PR:1935,%
Beth:PR:1936,%
Holbourn:N:1936,%
Carrara:N:1949,%
Allen:AJP:1966,%
Carusotto&al:NC:1968,%
Chang&Lee:JOSAB:1985,%
Vulfson:USP:1987,%
Kristensen&al:OC:1994,%
He&al:PRL:1995,%
Friese&al:PRA:1996,%
Then&Thide:ARXIV:2008,%
Helmerson&al:Topologica:2009,%
Padgett&Bowman:NPHY:2011,%
Ramanathan&al:PRL:2011,%
Elias:AA:2012,%
Coles&al:LPR:2013,%
Emile&al:PRL:2014,%
Williams&al:PRL:2014%
}.

In general, angular momentum transducers must support---and be able to
control and sense---the three-dimensional dynamics of the charges.  If,
however, the rotation is confined to a single plane, it suffices that
the transducers are effectively two-dimensional. Such transducers can be
realized by equipping single linear-momentum antennas with two-dimensional
azimuthally dependent reflectors, lenses or phase plates~%
\cite{%
Gagnon&al:IEEETAP:2010,%
Tamburini&al:APL:2011,%
Salem&Caloz:METAMATERIALS:2013,%
Cheng&al:SR:2014,%
Gao&al:APL:2014,%
Schemmel&al:OE:2014,%
Yan&al:NCOM:2014,%
Yu&Capasso:NMAT:2014,%
Hui&al:IEEEAWPL:2015%
}
or by employing helicoidal parabolic antennas
\cite{Tamburini&al:NJP:2012,Trevino&al:OE:2013}.  Other examples are
the flat plate transducer developed by \textcite{Bennis&al:EuCAP:2013}
and the screen detector used by \textcite{Krenn&al:NJP:2014}.

Interesting possibilities for innovative monolithic transducers
are offered by developments in nano- and optomechanics~%
\cite{
Bochmann&al:NPHY:2013,%
Shi&Bhattacharya:JPB:2013,%
Shi&Bhattacharya:JMO:2013,%
Andrews&al:NPHY:2014,%
Aspelmeyer&al:JOSAB:2010,%
Aspelmeyer&al:RMP:2014,%
Bagci&al:N:2014,%
Emile&al:PRL:2014,%
Zhang&al:PRL:2015%
},
and by atomic Bose-Einstein condensates~%
\cite{%
Mondal&al:PRA:2014%
}.
Recent research into spintronics and orbitronics herald the arrival of
transducers that are based on the interplay between charge, spin and
orbital degrees of freedom and that may be realized with dissipation-less
non-charge currents~%
\cite{%
Wolf&al:S:2001,%
Murakami&al:S:2003,%
Sinova&al:PRL:2004,%
Zutic&al:RMP:2004,%
Deac&al:NPHY:2008,%
Bernevig&al:PRL:2005,%
Dussaux&al:NCOM:2010,%
Kirilyuk&al:RMP:2010,%
Tokatly:PRB:2010,%
Verma:EFY:2011,%
Maekawa&al:2013,%
Vannucchi&al:APL:2013,%
Yan&al:PRB:2013,%
Brataas&Hals:NNAN:2014,%
Kuschel&Reiss:NNAN:2015%
},
including antennas realized with condensed-matter skyrmions (magnetic vortex
structures)~%
\cite{%
Papanicolaou&Tomaras:NPB:1991,%
Vogel&al:APL:2011,%
Mochizuki:PRL:2012,%
Pulecio&al:NCOM:2014%
}
and quantum rings~%
\cite{%
Quinteiro&Beradkar:OE:2009,%
Koksal&Berakdar:PRA:2012,%
Nieminen&al:NJP:2015%
}.

However, monolithic multi-dimensional ``antennas'' that make direct use
of the angular momentum degree of freedom for the radio frequency range
are not yet readily available.  Until they are, one may use existing
phased arrays of linear-momentum antennas of the type that are used
in today's radio astronomy and wireless communications. This is because
these multi-transducer arrays, including MIMO arrays, sample a radio beam
with transducers, essentially in the form of a finite set of integrating
cylindrical volumes ${V_n,n=1,2,\ldots,M}$ deployed at of discrete
points distributed over a certain region in space, thereby approximating,
within the limitations and constraints set by the sampling theorem,
\cite{%
Nyquist:AIEET:1928,%
Shannon:PIRE:1949,%
Jerri:PIEEE:1977%
},
the properties of a single, monolithic two- and three-dimensional transducer
that samples the field in a continuous manner~%
\cite{%
Thide&al:PRL:2007,%
Then&al:ARXIV:2008,%
Then&Thide:ARXIV:2008,%
Sjoholm&Palmer:MSc:2009,%
Mohammadi&al:IEEETAP:2010,%
Thide&al:Incollection:2011,%
Edfors&Johansson:IEEETAP:2012,%
Huang&al:NL:2012,%
Tennant&Allen:EL:2012,%
Bai&al:LAPC:2013,%
Barbuto&al:APSURSI:2013,%
Deng&al:IJAP:2013,%
Gao&al:JO:2013,%
Li&al:EuMC:2013,%
Weber:OPTIK:2013,%
Bai&al:EL:2014,%
Barbuto&al:PIER:2014,%
Cano&al:LAPC:2014,%
Opare&Kuan:2014,%
Bai&al:IJAP:2015,%
Opare&al:ACCT:2015,%
Liu&al:IEEEAWPL:2015%
}.
The fact that it is \emph{possible} to use arrays of conventional
linear-momentum antennas (\eg, dipole antennas) to exploit (a subset of
approximate) angular momentum modes, does of course not imply that it is
\emph{necessary} to use such arrays. Maxwell's equations allow any number
and combination of sources $\rho$ and $\j$ with any spatio-temporal and
topological properties in the source volume~$V'$.  And the currents do not
have to be conduction currents in the usual sense~\cite{Bagci&al:N:2014}.

Within the limitations just mentioned, it is possible to use arrays of
linear antennas and currently available radio equipment and techniques
for experimental investigations of certain properties of a finite subset
of (approximate) OAM eigenmodes and coherent superpositions of them.
This will provide useful insights into the properties of OAM information
transfer and therefore pave the way for the development of future
innovative angular momentum radio concepts based on single transducers.

These experimental investigations typically amount to measuring the
phase of the currents sampled at the feedpoints of spatially distributed
individual 1D antenna elements, \ie, a spatial sampling of one component
of the linear momentum of the charges, and then performing a phase
gradient analysis~%
\cite{%
Nienhuis:Incollection:2008,%
Roichman&al:PRL:2008,%
Berkhout&al:PRL:2010,%
Mohammadi&al:RS:2010,%
Thide&al:Incollection:2011,%
Lavery&al:JO:2011,%
Yao&Padgett:AOP:2011,%
Lavery&al:NJP:2013,%
Bennis&al:EuCAP:2013%
}
and post-processing the data so that the approximate (possibly aliased)
spectrum of the $z$~component of the angular momentum, $L_z$ can be
estimated. This way, approximations of individual OAM eigenmodes
may be extracted and an approximation of the correct spectrum of
individual OAM eigenmodes and the information they carry be decoded.
It should be emphasized that this way of estimating $L_z$ is based on the
application of the operator defined by formula~\eqref{eq:Lfield_cart_z}
on the electric (or magnetic) field.  However, as was discussed in
Sect.~\ref{sect:background}, fields are, strictly speaking, not directly
observable.

Proof-of-concept experiments where existing conventional radio technology
was employed have shown that it is indeed possible to use the
total angular momentum, \ie, the  SAM+OAM physical observable, as a new
physical layer for radio science and technology exploitation. These
studies include numerical experiments~\cite{Thide&al:PRL:2007}
showing that it is feasible to utilize OAM in radio; controlled
laboratory experiments~%
\cite{Tamburini&al:APL:2011,%
Wang&al:ECOC:2011,%
Wang&al:NPHO:2012,%
Yan&al:NCOM:2014%
}
verifying that it is possible to generate and transmit radio beams
carrying non-integer OAM and to measure their OAM spectra in the form of
weighted superpositions of different integer OAM eigenmodes each with its
own quantum number (mode index), also known as topological charge; and
outdoor experiments~%
\cite{%
Gotte&al:OE:2008,%
Tamburini&al:NJP:2012,%
Tamburini&al:ARXIV:2013b%
}%
, verifying that in real-world settings different signals, physically
encoded in different OAM modes and overlapping each other in space
and time and subject to reflection, can be transmitted wirelessly to a
receiver located in the (linear momentum) far zone and be individually
extracted and decoded there.

We stress that for EM beams of the kind used in radio astronomy,
radar investigations, radio communications and similar applications,
OAM is distinctively different from---and independent of---SAM (wave
polarization). If such a beam is already $N$-fold OAM encoded, utilizing
also SAM will double the information transfer capacity by virtue of the
fact that the dimension of the state space then doubles from $N$ to $2N$.

\subsubsection{Electric dipoles}

An antenna type that is commonly used in today's radio is the
half-wavelength dipole antenna.  Many qualitative as well as quantitative
characteristics of this antenna are fairly accurately approximated by an
electric Hertzian dipole, \ie, an antenna whose length $L$ is much shorter
than the wavelength $\lambda$ of its emitted (nearly monochromatic)
field, $L\ll\lambda$, so that the phase variations  over $L$ of the
antenna current can be neglected, a condition that simplifies the
theoretical analysis considerably.  We have found it convenient for our
study to calculate analytically the linear and angular momentum densities
emitted by a single Hertzian dipole and combinations of such dipoles.
We choose a Cartesian coordinate system $(x_1,x_2,x_3)$ where the $x_3$
axis coincides with the polar axis of a spherical polar coordinate system
$(r,\th,\ph)$ and let the two systems have the same origin~$\x'=\0$.

\paragraph{A single Hertzian dipole along the polar axis}

The properties of Hertzian dipoles are described in many textbooks~%
\cite{%
Cohen-Tannoudji&al:Book:1997,%
Jackson:Book:1998,%
Thide:Book:2011%
}
and explicit formulas are derived in Appendix~\ref{app:Hertizan_dipoles}.

According to Eqn.~\eqref{eq:E3crossB3} for ${\E_3\cross\B_3}$ the
electromagnetic linear momentum density, ${\gfield_3=\epz\E_3\cross\B_3}$,
emitted from a single Hertzian dipole $\vd_3(\0)$ located at the origin
and directed along the $x_3$ axis, has nine different vector components.
Five of them are longitudinal ($\parallel\runit$) and four of them are
transverse ($\perp\runit$). It follows trivially from Planck's relation,
Eqn.~\eqref{eq:Planck_rel}, that the same is also true for the Poynting
vector~$\S$. However, the different components have different fall-offs,
implying that far away from the dipole the following approximation holds:
\begin{multline}
\label{eq:gfield_dipole_far}
 {\gfield_3}\far(t,r,\th,\ph)
\\
 =
 \frac{k^4d_3^2}{32\pi^2\epz c}
   \bigg(\frac{1 + \cos[2(kr-\w t+\delta_3)]}{r^2} + \Ordo{r^{-3}}\bigg)
    \sin^2\th\,\runit
\\
  + \frac{k^3d_3^2}{32\pi^2\epz c}\bigg(
   \frac{\sin[2(kr-\w t+\delta_3)]}{r^3} + \Ordo{r^{-4}}\bigg)
   \cos2\th\,\thunit
\end{multline}
To the same degree of approximation, the cycle (temporal) average of the
Poynting vector in the far zone is, to leading order,
\begin{align}
 \avet{\S_3\far}
 = c^2{\gfield_3}\far
  \approx%
 \frac{\w^4d_3^2}{32\pi^2\epz c^3}\bigg(\frac{1}{r^2} + \Ordo{r^{-3}}\bigg)
  \sin^2\!\th\,\runit
\end{align}
where we used the fact that $k=\w/c$ and that
\begin{align}
 \avet{\sin[2(kr-\w t+\delta_3)]} = \avet{\cos[2(kr-\w t+\delta_3)]} = 0
\end{align}
This is a well-known result that can be found in most textbooks.

To calculate the electromagnetic angular momentum density around $\xm=\0$
we use Eqn.~\eqref{eq:hfield}, Eqn.~\eqref{eq:E3crossB3}, and the fact
that $\x=r\runit$ and $\runit\cross\thunit=\phunit$ to obtain the
exact expression
\begin{multline}
 \hfield_3(t,r,\th,\ph;\0) = r\runit\cross\gfield_3
 = \frac{d_3^2}{8\pi^2\epz c}r\big[
    \fn{1}(\delta_3)\fn{2}(\delta_3)
\\
    - \fn{2}^2(\delta_3)
    + \fn{1}(\delta_3)\fn{3}(\delta_3) 
    - \fn{2}(\delta_3)\fn{3}(\delta_3)
   \big]\sin\th\cos\th\,\phunit
\end{multline}
where the functions $f_n$ are given by Eqns.~\eqref{eq:fn}.

At very long distances from a Hertzian dipole located at the origin
and oscillating along the $x_3$ axis we can use the approximation
in Eqn.~\eqref{eq:gfield_dipole_far} for $\gfield_3$.  Because of the
geometry, only the transverse part ($\perp\runit$) of the linear momentum
density, where the dominating part is $\Ordo{r^{-3}}$, contributes
to the linear momentum density in the far zone.  The radial component
($\parallel\runit$) of the linear momentum density (Poynting vector),
which is $\Ordo{r^{-2}}$ does not contribute to the angular momentum
density. If it would, then expression~\eqref{eq:Jfield_int} for
the total angular momentum at very large distances of a ``pulse''
would grow without bounds as the ``pulse'' propagates, thus
violating fundamental laws of physics. See the detailed, clarifying
discussion about this by~\textcite{Abraham:PZ:1914}; an English
translation of the essential part of this discussion is included
in~\textcite{Tamburini&al:NJP:2012a}. This almost paradoxical fact is
also discussed by~\textcite{Low:Book:1997}.  Therefore, in the far zone,
the angular momentum density is

\begin{multline}
 {\hfield_3}\far(t,r,\th,\ph;\0) \approx
  \frac{d_3^2}{16\pi^2\epz c}
  r\fn{1}(\delta_3)\fn{2}(\delta_3)\sin\th\cos\th\,\phunit
\\=
 -\frac{k^3d_3^2}{64\pi^2\epz r^2c}\sin[2(kr-\w t+\delta_3)]
  \sin2\th\,\phunit
\end{multline}
This result shows that ${\hfield_3}\far$ has the expected $r^{-2}$
fall off, has no $z$ component, and oscillates at $2\w$ along
$-\phunit$. This causes an electric charge in the far zone but not located
at ${\th=n\pi/2,n\in\mathbb{Z}}$, to rotate around its own center of
mass in the osculating plane normal to $\phunit$.  However, the temporal
(cycle) average of ${\hfield_3}\far$ is zero, which makes it nontrivial
to measure this rotational motion.  It is tempting to relate this and
similar double-frequency terms to photon \emph{Zitterbewegung}~%
\cite{%
Unal:FP:1997,%
Kobe:PLA:1999,%
Wang&al:PRA:2009,%
Leary&Smith:PRA:2014%
}.

\paragraph{Two co-located crossed Hertzian dipoles in the azimuthal plane}

We now want to calculate the field linear and angular momenta emitted from
a system of two co-located monochromatic Hertzian dipoles, one spatially
rotated from the other by $\pi/2$, fed at the same frequency but with
different temporal phase shifts.  For this purpose we chose the two
Hertzian dipoles $\vd_j,j=1,2$ with their generated fields described by
Eqns.~\eqref{eq:Edipole_fields}. Since the total field is a superposition
of the fields from the two individual dipoles, the total linear momentum
density $\gfield_{1\&2}$ emitted by the two dipoles is
\begin{align}
\label{eq:gfield_dipoles_1&2}
\begin{split}
 \gfield_{1\&2}
={}& \epz(\E_1+\E_2)\cross(\B_1+\B_2)
\\
={}&\epz\E_1\cross\B_1 + \epz\E_2\cross\B_2 
 + \epz\E_1\cross\B_2 + \epz\E_2\cross\B_1 
\\
&= \gfield_1 + \gfield_2 + \gfield_{12} + \gfield_{21}
\end{split}
\end{align}
We see that in addition to the vector sum of the two linear momentum
densities from the individual dipoles $\vd_1$ and $\vd_2$, the total
linear momentum contains two interference terms $\gfield_{12}$
and $\gfield_{21}$ which appear due to the fact that the total
electromagnetic field is the superposition of the fields from each
dipole. Using Eqns.~\eqref{eq:E1crossB1}--\eqref{eq:E2crossB2}
and \eqref{eq:E1crossB2}--\eqref{eq:E2crossB1}, we can calculate
the exact expression for $\gfield_{1\&2}$, and from that, using
Eqn.~\eqref{eq:hfield}, the exact expression for the total angular
momentum density $\hfield_{1\&2}$. If the two dipole moments have equal
amplitudes, $d_1=d_2=d$, and are in phase quadrature relative each other,
$\delta_1=0$ and $\delta_2=\pi/2$, we find that the exact expression
for the $z$ component of the angular momentum density $\hfield_{1\&2}$,
valid in all of space outside the dipoles, is
\begin{multline}
\label{eq:hfield_dipoles_1&2}
 h_{{1\&2}_z}\field(t,r,\th,\ph;\0) = 
 \frac{d^2(\0)k}{8\pi^2\epz c}\bigg(
   \frac{k^2}{r^2}\big\{
   \sin^2\ph\cos^2(kr-\w t)
\\
   +\cos^2\ph\sin^2(kr-\w t)
   +\sin\ph\cos\ph\sin[2(kr-\w t)]
  \big\}
\\
  +\frac{k}{r^3}\big\{
   2\sin\ph\cos\ph[\cos^2(kr-\w t)-\sin^2(kr-\w t)]
\\
   -(\sin^2\ph-\cos^2\ph)\sin[2(kr-\w t)]
\\
  +\frac{1}{r^4}\big\{
   \cos^2\ph\cos^2(kr-\w t)
  +\sin^2\ph\sin^2(kr-\w t)
\\
  -\sin\ph\cos\ph\sin[2(kr-\w t)]
  \big\}
  \bigg)\sin^2\th
\end{multline}

To turn the $z$ component of the angular momentum density,
$h_{{1\&2}_z}\field$, radiated from two co-located Hertzian dipoles,
arranged perpendicular to each other and oscillating in quadrature phase,
into the observable $z$ component of the total field angular momentum,
$J_z\field(\0)$, we must integrate \eqref{eq:hfield_dipoles_1&2} over an
observation volume $V$. Choosing $V$ as in Eqn.~\eqref{eq:Jfield_int},
\ie, as the region between two spherical shells with radii $r_0$
and $r_0+\Delta{r_0}$, representing the radial extent of the ``pulse''
emitted, we obtain
\begin{align}
\label{eq:Jfield_z_dipoles_1&2}
 J\field_{{1\&2}_z}(t,\0)
 = 
   \int_{r_0}^{r_0+\Delta{r_0}}\,\D{r}\,r^2
   \int_{\th_1}^{\th_2}\D{\th}\sin\th
   \int_{\ph_1}^{\ph_2}\D{\ph}\,h_{{1\&2}_z}\field(\0)
\end{align}
where $h_{{1\&2}_z}\field(\0)$ is given by
Eqn.~\eqref{eq:hfield_dipoles_1&2}.

If we integrate over all azimuth angles $\ph$ from $\ph_1=0$ to
$\ph_1=2\pi$, and also use the well-known identities
\begin{subequations}
\begin{gather}
 \int_0^{2\pi}\D{\ph}\,\sin^2\ph 
 = \int_0^{2\pi}\D{\ph}\,\cos^2\ph 
 = \pi
\\
 \int_0^{2\pi}\D{\ph}\,\sin\ph\cos\ph = 0
\\
 \sin^2(kr-\w t) + \cos^2(kr-\w t) = 1
\end{gather}
\end{subequations}
we obtain
\begin{align}
 J\field_{{1\&2}_z}(t,\0)
 = \frac{d^2(\0)k}{8\pi\epz c}
   \int_{r_0}^{r_0+\Delta{r_0}}\,\D{r}
   \bigg(k^2 +\frac{1}{r^2}\bigg)
   \int_{\th_1}^{\th_2}\D{\th}\sin^3\th
\end{align}
If we now integrate over all polar angles $\th$ from $\th_1=0$ to $\th_2=\pi$,
we obtain the explicit, closed-form expression
\begin{align}
\label{eq:Jfield_z_dipoles_1&2_shell}
 J\field_{{1\&2}_z}(t,\0)
 = \frac{d^2(\0)k}{6\pi\epz c}
   \bigg(k^2\Delta{r_0} +\frac{\Delta{r_0}}{k^2r_0(r_0+\Delta{r_0})}\bigg)
\end{align}
where the only time dependence comes from the duration of the emitted
``pulse'' $\Delta{t}=\Delta{r_0}/c$.  When the distance $r_0$ that the
``pulse'' has propagated is very large so that the $1/r_0^2$ contribution
can be neglected, the $z$ component of the field angular momentum can
be approximated by
\begin{align}
\label{eq:Jfield_dipoles_1&2_far}
 {J_{{1\&2}_z}\Sup{field}}\Sup{\!far}(t,\0)
 = \frac{d^2(\0)k}{6\pi\epz c}k^2\Delta{r_0}
 = \frac{d^2(\0)k^3}{6\pi\epz }\Delta{t}
\end{align} 
which propagates without radial fall-off as expected.

If the observation volume $V$, \ie, the angular momentum sensing
transducer, is a very thin circular ring with radius $r_0$, located in
the azimuthal ($xy$) plane ($\th=\pi/2)$, centered on the dipoles, and
subtending the small polar angular interval $\th_2-\th_2$ around the
azimuthal plane, the integrals in Eqn.~\eqref{eq:Jfield_z_dipoles_1&2}
are trivial and the result is
\begin{align}
\label{eq:Jfield_z_dipoles_1&2_ring}
 J\field_{{1\&2}_z}(t,\0)
 = \frac{d^2(\0)k}{8\pi\epz c}
   \bigg(k^2 +\frac{1}{r_0^2}\bigg)(\th_2-\th_1)
\end{align}
A torsion pendulum experiment that is a kind of radio analog of the
celebrated Einstein-de~Haas experiment~\cite{Einstein&deHaas:PKAWA:1915},
first suggested by \textcite{Vulfson:USP:1987}, was performed
in the microwave radio regime by \textcite{Emile&al:PRL:2014}.
The results was found to agree with the prediction in
Eqn.~\eqref{eq:Jfield_z_dipoles_1&2_ring}. Howerver, comparing this
equation with Eqn.~(4) in~\cite{Emile&al:PRL:2014}, shows, that the
latter is in error: there is a factor of $2\pi$ missing which explains
the factor of $\sim6$ discrepancy mentioned.

\begin{figure}
 \resizebox{1.\columnwidth}{!}{%
  \input{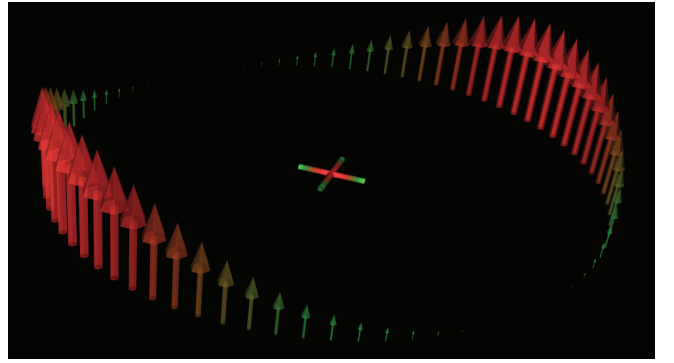}
 }
 \caption{\label{fig:h_crossed_dipoles}%
  Snapshot of the distribution of the angular momentum density
  pseudovector $\Jfield(t,\0)$ on a circle in the plane around the
  centers of two crossed half-wave dipole antennas, co-located at the
  origin $\x'=\0$, arranged perpendicular to each other, and fed with
  equal-amplitude currents at one an the same frequency but $\pi/2$
  out of phase with each other.%
 }
\end{figure}
\paragraph{Two co-located crossed half-wave dipole antennas}

The analytic results for two monochromatic crossed Hertzian dipoles in
quadrature phase derived above are in fair qualitative agreement with the
results obtained for the same configuration using two half-wave dipole
antennas.  Since the analytic treatment for the latter configuration
is a bit complicated, we resorted to a direct numerical evaluation of
the exact formulas \eqref{eq:E&B} and \eqref{eq:hfield}. We found that
our results were in excellent agreement with those obtained with two
enterprise-grade Maxwell solvers but with five times shorter compute time.
Graphical snapshots of the time evolution of the angular momentum density
$\hfield(\0)$ from the two crossed half-wave dipole antennas, fed $\pi/2$
out of phase relative to each other, measured along a circle in the
$x_1x_2$ plane, is displayed in Fig.~\ref{fig:h_crossed_dipoles}. We
see that $\hfield(\0)$ is positively directed along the entire circle.
The total field angular momentum, $\Jfield(\0)$, obtained by integrating
$\hfield(\0)$ along this circle, is therefore always directed in the
$x_3$ axis, representing rotation around this axis.  As mentioned above,
this mechanical rotation caused by the angular momentum carried by the
microwave beam was verified experimentally~\cite{Emile&al:PRL:2014}.

\subsection{Angular momentum and multi-transducer/space-division techniques}

The multi-antenna technique for enhancing the capacity of a link using
linear momentum that goes by the name multiple-input-multiple-output
(MIMO) is a space-division method that amounts to utilizing the
fact that the EM linear momentum $\pfield$ emitted from an ordinary
antenna---typically a linear dipole---is a vector quantity. Dividing
this EM linear momentum vector up into $M$ equal components such that
\begin{align}
 \pfield = \sum_{n=1}^M \pfield_n
\end{align}
letting each $\pfield_n$ be emitted from one of $M$ linear momentum
antennas and propagate along different paths to the receiving end where
the individual EM linear momenta $\pfield_n$ are combined vectorially
by digital signal post-processing, the spectral efficiency and/or the
signal-to-noise-ratio (SNR) can, under certain circumstances be increased.

\begin{figure}
\centering
 \resizebox{.85\columnwidth}{!}{%
  \input{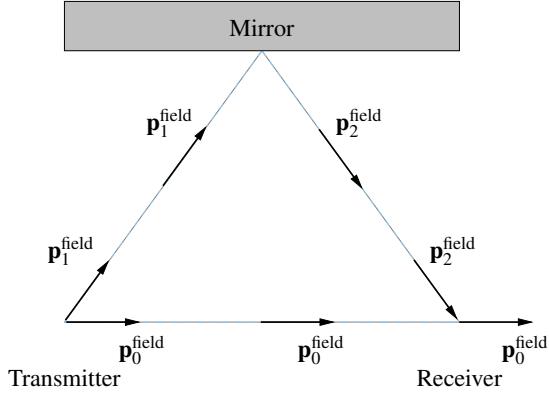}
 }
\caption{%
\label{fig:reflected_p}
 Communicating with electromagnetic linear momentum $\pfield$ in the
 presence of a reflective surface. In the two-dimensional plane of the
 figure, the vector $\pfield_0$ propagates directly along the straight
 path to the receiver, while the vector $\pfield_1$ is reflected off
 the mirror without recoil and produces the vector $\pfield_2$ at the
 receiver. Since $\pfield$ is a constant of the motion and preserves
 its parity upon reflection, the full 3D structure of the EM linear
 momentum can be obtained by post-processing. This is used in present
 MIMO schemes for enhancing the spectral density of linear-momentum
 radio communications in reflective environments.
}
\end{figure}
\begin{figure}
\centering
 \resizebox{.85\columnwidth}{!}{%
  \input{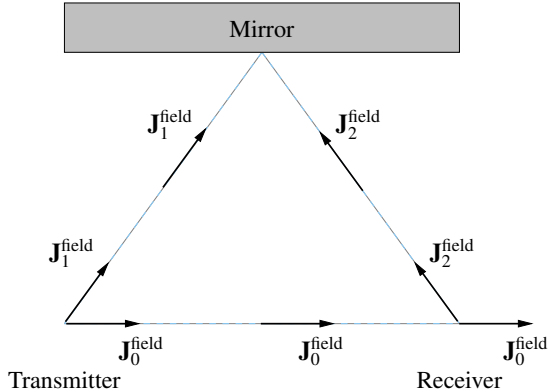}
 }
\caption{%
\label{fig:reflected_J}
 Communicating with electromagnetic angular momentum $\Jfield$ in the
 presence of a reflecting surface. Angular momentum is a pseudovector
 and therefore $\Jfield$ suffers a parity change when it is reflected
 off a dense medium. Taking this parity change into account properly,
 the reflected pseudovector can be used in MIMO-type schemes for enhancing
 the spectral density of angular-momentum radio communications, over and
 above the enhancement offered by the multi-state capability of angular
 momentum relative to the single-state linear momentum.
}
\end{figure}

A MIMO-like enhancement of the radio spectral capacity can be
achieved by the alternative technique of letting a beam of linear
momentum $\pfield$ emitted from an antenna be reflected off a wall,
say. This works because the linear momentum density (Poynting)
vector is an ordinary (polar) vector and therefore does not
suffer any parity change upon reflection. This is illustrated
schematically in Fig.~\ref{fig:reflected_p}; see also Fig.~1 in
\textcite{Andrews&al:N:2001} where the authors plot a vector which,
in physics terms, is the (approximate) far-field electric field vector
but the authors call ``polarization,''

For this to work, the Heisenberg uncertainty relation for
momentum must be fulfilled. If $\lambda$ denotes the wavelength,
then $p_n\Sup{field}=\hslash/\lambda$ is the $x$ component of the linear
momentum of each photon of the monochromatic radio beam emitted from the
$n$th antenna and this relation becomes
\begin{align*}
\label{eq:Heisenberg_p}
 \Delta{p_n\Sup{field}}\Delta{x}\ge\hslash/2
\end{align*}
Hence the spatial separation $\Delta{x}$ between the linear
momenta must be at least $\lambda/2$ in order to be statistically
independent. In practical radio implementations this means that the
distance between one (translational current carrying) antenna and
another one must fulfill the inequality
\begin{align*}
 \Delta{x}\ge\lambda/2
\end{align*}
Consequently, the space-time distance between each individual
linear-momentum MIMO antenna must be at least half a wavelength; in
practical implementations the separation must be much larger. This may
limit the usefulness of the MIMO technique in practice.  Furthermore,
MIMO often requires substantial post-processing and this may limit its
usefulness even further
\cite{%
Essiambre&Tkach:PIEEE:2012,%
Arik&al:JLWT:2013,%
Boffi&al:SPIE:2013%
}.

\begin{table*}
\caption{\label{tab:g&h}%
 Similarities and differences between the EM linear momentum
 density, $\gfield(t,\x)$, the Poynting vector $\S(t,\x)$, and the
 EM angular momentum density around a moment point $\x\Sub{m}$,
 $\hfield(t,\x,\x\Sub{m})$, carried by a classical electromagnetic
 field $[\E(t,\x),\B(t,\x)]$ in the presence of matter (particles)
 with mechanical linear momentum density $\gmech(t,\x)$ and mechanical
 angular momentum density $\hmech(t,\x,\x\Sub{m})$.%
}
\begin{ruledtabular}
\begin{tabular}{lcc}
  \textbf{Property}
 &
  \textbf{Linear momentum density}
 &
  \textbf{Angular momentum density}
\\
\midrule
 Definition 
 &
  $\gfield=\epz\E\cross\B=\S/c^2$
 &
  $\hfield=(\x-\x\Sub{m})\cross(\epz\E\cross\B)$
\\
 SI unit
 &
  \si{N.s.m^{-3}} (\si{kg.m^{-2}.s^{-1}})
 &
  \si{N.s.m^{-2}.rad^{-1}} (\si{kg.m^{-1}.s^{-1}.rad^{-1}})
\\
 Spatial fall off at large distances $r$
 &
  $\sim r^{-2} + \mathcal{O}(r^{-3})$ 
 &
  $\sim r^{-2} + \mathcal{O}(r^{-3})$ 
\\
 Typical phase factor
 &
  $\exp{\{\im(kz - \w t)\}}$
 &
  $\exp{\{\im(\alpha\ph + kz -\w t)\}}$ 
\\[.5ex]
 C (charge conjugation) symmetry
 &
 Even 
 &
 Even
\\
 P (spatial inversion) symmetry
 &
 Even (polar vector, ordinary vector)
 &
 Odd (axial vector, pseudovector)
\\
 T (time reversal) symmetry
 &
 Odd
 &
 Odd
\end{tabular}
\end{ruledtabular}
\end{table*}

As described earlier, when wireless communication is based on the angular
momentum physics layer, it is possible, in an ideal case, to $N$-tuple the
capacity within a fixed frequency bandwidth of a SISO system using single
monolithic transducers at both the transmitting and the receiving ends.
Furthermore, replacing the single transducer by $M$ transmitting and
receiving transducers in order to split the orbital angular momentum
(OAM) pseudovector into $M$ equal units $\Lfield_n$
\begin{align}
 \Lfield = \sum_{n=1}^M \Lfield_n
\end{align}
it should be possible to achieve an $N\times{}M$-fold increase in
capacity. Using, in addition to this OAM MIMO technique, also wave
polarization (spin angular momentum, SAM), the total capacity will
be $2N\times{}M$ higher than for a linear-momentum SISO channel. As
illustrated in Figure~\ref{fig:reflected_J}, the exploitation of the
angular momentum technique in a reflective environment requires that the
parity of $\Jfield$ can be taken properly into account.  Experiments have
shown that this is feasible~\cite{Tamburini&al:ARXIV:2013}.

\svnidlong
 {$HeadURL: file:///Repository/SVN/Articles/AngularMomentumRadio/branches/PRA/summary.tex $}
 {$LastChangedDate: 2015-05-25 23:33:01 +0200 (Mon, 25 May 2015) $}
 {$LastChangedRevision: 91 $}
 {$LastChangedBy: bt $}
\svnid{$Id: summary.tex 91 2015-05-25 21:33:01Z bt $}

\section{Summary}

We have presented a review of those fundamental physical principles
that comprise the foundation for radio science and technology.  We have
clarified that in electromagnetic experiments and applications such as
wireless communications, one has to rely on the existence and judicious
use of electromagnetic observables and have emphasized the fact that by
exploiting the angular momentum observable, in addition to the linear
momentum observable used today, many advantages can be achieved.

The similarities and differences between linear momentum and angular momentum
based techniques in radio science and communication applications are
summarized in Table~\ref{tab:g&h} and in the following list:
\begin{enumerate}
 
 \item
 The conventional techniques based on the linear momentum degrees of
 freedom of an electromechanical system exploits the translational
 degrees of freedom of the EM field and the associated particles.

  \begin{enumerate}
   
   \item
   The linear momentum $\pfield(t)$ can be measured remotely---and
   thereby be utilized in applications---by employing a conduction
   current based device that integrates the local linear momentum density
   $\gfield(t,\x)$ over a one-dimensional volume $V$, typically a very
   thin cylinder as is the case for the ubiquitous linear electric dipole
   antenna.

   From such a nonlocal measurement one can estimate the (weighted)
   average of the primary EM field over $V$ but not determine the exact
   field (within instrumental errors).

  \end{enumerate}

 \item
 Novel techniques based on the angular momentum degrees of freedom of an
 electromechanical system exploits the rotational degrees of freedom
 of the EM field and the associated particles.

  \begin{enumerate}

   \item
   The angular momentum $\Jfield(t)$ can be measured remotely---and thereby
   utilized in applications---by using a conduction current based device,
   or a spin- and/or orbit current based device, that integrates the local
   angular momentum density $\hfield(t,\x)$ over a two- or three-dimensional
   volume.
   
   \item
   In a beam situation typical for radio communication links,
   radio astronomy observations, and radar applications, the
   spin and angular parts of the angular momentum, $\Sfield$ and
   $\Jfield$ respectively, can be separated or, alternatively, both
   the helicity and the total angular momentum of the beam can be
   measured~\cite{Fernandez-Corbaton&al:PRA:2012}.  The main differences
   between the two parts of the angular momenta are:

   \begin{enumerate}

    \item
    The spin part of the angular momentum (SAM, polarization), $\Sfield$,
    manifests itself as a rotation in time of the field vectors such that
    they have different directions (oscillation phase) modulo $2\pi$
    at one position $\x$ in the phase front at different times $t$ and
    $t+\Delta{t}$. Hence, the measurement of SAM (polarization) requires
    temporal coherence.

    \item
    The orbital part of the angular momentum (OAM, twist), $\Lfield$,
    manifests itself as a rotation in space of the field vectors such
    that they have different directions (oscillation phase) modulo $2\pi$
    at different azimuthal angles $\ph$ and $\ph+\Delta{\ph}$ in the
    phase plane at a given instant $t$.  Hence, the measurement of OAM
    (twist) requires spatial coherence.

   \end{enumerate}


 \end{enumerate}


\end{enumerate}

\svnidlong
 {$HeadURL: file:///Repository/SVN/Articles/AngularMomentumRadio/branches/PRA/acknowledgments.tex $}
 {$LastChangedDate: 2015-05-28 19:23:04 +0200 (Thu, 28 May 2015) $}
 {$LastChangedRevision: 97 $}
 {$LastChangedBy: bt $}
\svnid{$Id: acknowledgments.tex 97 2015-05-28 17:23:04Z bt $}

\begin{acknowledgments}

The authors gratefully acknowledge useful discussions with professor
Iwo Bia{\l}ynicki-Birula, professor Anton Zeilinger, and professor 
Markus Aspelmeyer.
We also thank professor G\"oran F\"aldt, Uppsala, for reminding us
about the excellent treatment by \textcite{Truesdell:Book:1968} on the
fundamental properties of angular momentum, and professor Olle Eriksson,
Uppsala, for comments on skyrmions.

B.\,T. thanks the Institute for Quantum Optics and Quantum Information
(IQOQI), Vienna, where part of this work was done, for very generous
hospitality during repeated visits.

B.\,T. was financially supported by the Swedish National Space Board
(SNSB) and the Swedish Research Council (VR) under the contract number
2012-3297. H.\,T. acknowledges support from EPSRC grant EP/H005188/1 and,
for the early stage of the work, from the Centre for Dynamical Processes
and Structure Formation, Uppsala University, Sweden.

\end{acknowledgments}

\appendix
\svnidlong
 {$HeadURL: file:///Repository/SVN/Articles/AngularMomentumRadio/branches/PRA/poincare.tex $}
 {$LastChangedDate: 2015-05-26 20:56:10 +0200 (Tue, 26 May 2015) $}
 {$LastChangedRevision: 95 $}
 {$LastChangedBy: bt $}
\svnid{$Id: poincare.tex 95 2015-05-26 18:56:10Z bt $}

\section{The ten Poincar\'e invariants and the corresponding equations
of continuity}
\label{app:Poincare}

If all conserved quantities (constants of motion) of an
integrable physical system are known, the system dynamics can
be obtained trivially by formulating it in terms of action-angle
variables.  All in all the electromagnetic field has 84 constants of
motion~\cite{Fushchich&Nikitin:JPA:1992}. In this Appendix we discuss
the ten conserved electromagnetic observables that correspond to the
ten-dimensional Poincar\'e symmetry group $P(10)$, \ie, the ten Poincar\'e
invariants and their associated equations of continuity. They are the
scalar quantity energy, the three components of the linear momentum
vector, the three components of the angular momentum pseudovector,
and the three components of the energy center position vector.  Only a
small subset of these observables are currently used in radio science,
optics, and wireless communications.

Electromagnetic conservation laws can be viewed as constraints
on the field dynamics and, in this sense, every conserved quantity
is physically significant. The information contained in the conserved
quantities corresponds one-to-one to the information contained in the
Maxwell-Lorentz postulates~\cite{Philbin:PRA:2013}.

\subsection{Energy}

The energy density $\ufield(t,\x)$ carried by the fields generated by the
sources in $V'$ is
\begin{subequations}
\label{eq:ufield}
\begin{align}
 \ufield
&= \frac{\epz}{2}(\E\bdot\E + c^2\B\bdot\B)
\shortintertext{which, according to Eqns.~\eqref{eq:RS} and
Table~\ref{tab:RS}, can be written}
\label{eq:ufield_RS}
 \ufield
&= \frac{\epz}{2}\RS\bdot\cc{\RS}
 = \frac{\epz}{2}\abs{\RS}^2
 = \frac{\epz}{2}\SQRT{(\RS\bdot\RS)\cc{(\RS\bdot\RS)}}
\end{align}
\end{subequations}
Outside the source volume $V'$, the field energy in a certain observation
volume $V$ can be measured with an appropriate energy sensor, \eg,
a calorimeter or bolometer, that integrates $\ufield$ over $V$.

With the energy flux vector (Poynting vector) $\S(t,\x)$ given by the
formula
\begin{subequations}
\label{eq:S}
\begin{align}
 \S
&=  \epz c^2\E\cross\B
\shortintertext{or, using Eqns.~\eqref{eq:RS} and Table~\ref{tab:RS},}
\label{eq:S_RS}
 \S
&= \im\frac{\epz c}{2}\RS\cross\cc{\RS}
\end{align}
\end{subequations}
the energy density balance equation, which follows straightforwardly from
the Maxwell-Lorentz equations \eqref{eq:Maxwell_micro}, takes the form
\begin{align}
\label{eq:energy_diff}
 \pdd{t}{\ufield} + \div\S = -\je\bdot\E
\end{align}
\ie, a local continuity equation for the field energy density with a sink
(loss) term $\je\bdot\E$.  By using Eqns.~\eqref{eq:RS}, \eqref{eq:ufield_RS},
and \eqref{eq:S_RS}, we can write this as
\begin{align}
 \pdd{t}{(\RS\bdot\cc{\RS})} + \im c\div(\RS\cross\cc{\RS})
  = -\frac{1}{\epz}\je\bdot(\RS+\cc{\RS})
\end{align}

If $\varrho\mech(t§',\x')$ denotes the local mechanical mass density
field and $\vmech(t',\x')$ the local mechanical velocity field of the
charged particles at a point in $V'$, then, in the approximation where
statistical mechanics as well as relativistic and quantum physics effects
are negligible, the electric current density $\je(t',\x')$ carried by the
charged particles can be written as $\je=\rhoe\vecv\mech$ and the loss term
in Eqn.~\eqref{eq:energy_diff} can therefore be identified
as the increase in mechanical energy density of the particles~%
\cite{%
Schwinger&al:Book:1998,%
Jackson:Book:1998%
}%
:
\begin{align}
 \pdd{t}{\umech} = \je\bdot\E = \vecv\mech\bdot\rhoe\E
\end{align}
representing dissipation due to Ohmic losses, \ie, transfer of field energy
density to mechanical energy density of the current-carrying particles.

This identification makes it possible to introduce the total energy
density of the electromechanical system under study (charged particles
and their associated fields)
\begin{align}
 \usys(t,\x) = \ufield(t,\x) + \umech(t,\x) 
\end{align}
and to put Eqn.~\eqref{eq:energy_diff} in the form of an equation
of continuity where there are no sources or sinks (losses)~%
\cite{%
Schwinger&al:Book:1998,%
Jackson:Book:1998%
}%
:
\begin{align}
\label{eq:energy_diff_final}
 \pdd{t}{\usys} + \div\S = 0
\end{align}

We let $U(t)$ denote energy (volume integrated energy density).
Then, in obvious notation, the system energy is the sum of the energy of
the particles and and the energy of the field:
\begin{align}
 \Usys(t) = \intV{}{\usys(t,\x)}
  = \Ufield(t) + \Umech(t) 
\end{align}
Volume integrating the local continuity
equation~\eqref{eq:energy_diff_final} and using the divergence
theorem, the corresponding global continuity equation can be written
\begin{align}
\label{eq:energy_int}
 \dd{t}{\Usys} + \ointSvecdot{}{\S} = 0
\end{align}
This is the Poynting theorem, also known as the energy theorem in
Maxwell's theory.  It is utilized, for instance, in radiometry.
The field energy itself is utilized in simple, broadband ``on-off''
signaling as was used in Marconi's first radio experiments and also
in optical fiber communications. The conservation of energy as described
by~\eqref{eq:energy_int} is a manifestation of the temporal
translation invariance symmetry of electrodynamics and describes the
kinematics of the system.

\subsection{Linear momentum}

The electromagnetic linear momentum density $\gfield(t,\x)$
radiated from $V'$ is~%
\cite{%
Mandel&Wolf:Book:1995,%
Schwinger&al:Book:1998,%
Jackson:Book:1998,%
Thide&al:Incollection:2011%
}%
\begin{gather} 
\label{eq:gfield}
 \gfield
 = \epz\E\cross\B
\shortintertext{or, using Eqns.~\eqref{eq:RS},}
\label{eq:gfield_RS}
 \gfield
 = \im\frac{\epz}{2c}\RS\cross\cc{\RS}
\end{gather} 
and is, in free space, simply related to the Poynting vector~\eqref{eq:S}
by the Planck relation
\begin{align}
\label{eq:Planck_rel}
 \S = c^2\gfield
\end{align}

Let us introduce the electromagnetic linear momentum flux tensor
$\tens{T}(t,\x)$ as the negative of Maxwell's stress tensor~%
\cite{%
Schwinger&al:Book:1998,%
Thide&al:Incollection:2011%
}%
\begin{subequations}
\label{eq:T}
\begin{gather}
 \tens{T}
 = \ufield\tunity - \epz(\E\ox\E + c^2\B\ox\B) 
\shortintertext{or, using Eqns.~\eqref{eq:RS},}
\label{eq:T_RS}
 \tens{T}
 = \frac{\epz}{2}\big[\RS\bdot\cc{\RS}\tunity
  - (\RS\ox\cc{\RS} + \cc{\RS}\ox\RS)\big]
\end{gather}
\end{subequations}
where $\tunity$ is the rank-3 unit tensor and $\ox$ the tensor (Kronecker,
dyadic) product operator. Then the balance equation for linear
momentum density that follows from the Maxwell-Lorentz equations
\eqref{eq:Maxwell_micro} can be written~%
\cite{%
Schwinger&al:Book:1998,%
Jackson:Book:1998,%
Thide&al:Incollection:2011%
}%
\begin{subequations}
\label{eq:pddt_gfield}
\begin{gather}
 \pdd{t}{\gfield} + \div\tens{T} = -\f 
\shortintertext{where}
\label{eq:Lorentz_force_density}
 \f(t,\x) = \rhoe(t,\x)\E(t,\x) + \je\cross\B(t,\x)
\end{gather}
\end{subequations}
This polar vector $\f$ is the Lorentz force density\footnote{In certain
disciplines a non-standard convention is sometimes used in which the
term Lorentz force density denotes only the second term in the RHS of
Eqn.~\eqref{eq:Lorentz_force_density}. We follow the standard convention.}
that behaves like a loss of field linear momentum density.

By using Eqns.~\eqref{eq:RS}, \eqref{eq:gfield_RS}, and \eqref{eq:T_RS} we
can write this local equation of continuity for linear momentum density as
\begin{multline}
 \pdd{t}{(\RS\cross\cc{\RS})}
 - \im c\div\Big(\RS\bdot\cc{\RS}\tunity
   -\frac{1}{\epz}(\RS\ox\cc{\RS} + \cc{\RS}\ox\RS)\Big)
\\
 = -\im\frac{c}{\epz}\big[\rhoe(\RS + \cc{\RS})
   - {\epz}\je\cross(\RS - \cc{\RS})\big]
\end{multline}

In an electromechanical system of charged, massive
particles and pertinent fields, \eg, a transmitting antenna,
the mechanical linear momentum density the particles is
\begin{align}
 \gmech(t,\x) = \varrho\mech(t,\x)\vecv\mech(t,\x)
\end{align}
The local equation of continuity of linear momentum of an
electromechanical system, directly derivable from Maxwell's equations,
can therefore be written
\cite{%
Rohrlich:Book:2007,%
Schwinger&al:Book:1998,%
Jackson:Book:1998,%
Thide&al:Incollection:2011%
}%
\begin{align}
\label{eq:local_conservation_law_lin_mom}
 \pdd{t}{\gsys} + \div\tens{T} = \0
\end{align}
where
\begin{align}
 \gsys(t,\x) = \gfield(t,\x) + \gmech(t,\x) 
\end{align}
is the system linear momentum density.

The classical electromagnetic field linear momentum $\pfield$ that is
localized inside a volume $V$ embedded in free space is obtained by
volume integrating the linear momentum density $\gfield$ over $V$~%
\cite{%
Rohrlich:Book:2007,%
Jackson:Book:1998,%
Thide&al:Incollection:2011%
}
\begin{align}
\label{eq:pfield}
 \pfield(t)
  = \intV{}{\gfield(t,\x)}
  = \epz\intV{}{[\E(t,\x)\cross\B(t,\x)]} 
\end{align}
This integration can be done by using an appropriate volume integrating
linear momentum sensor, \eg, a linear dipole antenna.

The mechanical linear momentum in $V$ is
\begin{align}
\label{eq:pmech}
 \pmech(t)
  = \intV{}{\gmech(t,\x)}
  = \intV{}{\varrho\mech(t,\x)\vecv\mech(t,\x)}
\end{align}
which means that the total linear momentum of the system of particles
and fields in $V$ is 
\begin{align}
 \psys(t) = \pfield(t) + \pmech(t) 
\end{align}
It fulfills the global conservation law~%
\cite{%
Schwinger&al:Book:1998,%
Jackson:Book:1998,%
Thide:Book:2011%
}
\begin{align}
\label{eq:conservation_law_lin_mom}
 \dd{t}{\psys} + \ointSvecdot{}{\tens{T}} = \0
\end{align}
This is the linear momentum theorem in Maxwell's theory.  It demonstrates
that not only the mechanical particles (charges) but also the
electromagnetic field itself carries linear momentum (translational
momentum) that couples to the mechanical linear momentum of the matter
(charged particles) and thereby to the conduction current carried by
the charged particles.

\subsubsection{Gauge invariance}
\label{subsubsec:lin_mom_gauge_invariance}

It is common practice to investigate the properties of the
electromagnetic observables that we consider here, by expressing the
fields in terms of their potentials \cite{vanEnk&Nienhuis:JMO:1994}. If
the vector potential $\A(t,\x)$ is twice continuously differentiable in
space, it is well known that it fulfills the vector analytic identity
(see, \eg, \textcite{Thide:Book:2011})
\begin{align}
\label{eq:A_Helmholtz}
\begin{split}
 \A(t,\x) ={}& 
  -\grad\intV{'}{\frac{\del'\bdot\A(t,\x')}{4\pi\abs{\xx'}}}
\\ 
  &+\curl\intV{'}{\frac{\del'\cross\A(t,\x')}{4\pi\abs{\xx'}}}
\\ 
  &-\ointSvecdot{'}{\bigg(\frac{\A(t,\x')\ox(\xx')}{\abs{\xx'}^3}\bigg)}
\end{split}
\end{align}
The last integral, to be evaluated over a surface $S'$ enclosing the
entire volume $V'$ within which $\A\ne\0$ is localized, vanishes if it
is evaluated at very large distances and the expression within the big
parentheses falls off faster than $1/\abs{\xx'}^2=1/r^2$, or if it is
perpendicular to the normal unit vector $\nunit'$, or if it has certain
(a)symmetry properties \footnote{In fact, the surface integral may also
be a vector that is (at most) a function of time $t$.}. Then, Helmholtz's
decomposition is applicable, allowing us to write
\begin{align}
\label{eq:A_decomposed}
 \A(t,\x) &= \A\irrot(t,\x) + \A\rotat(t,\x)
\end{align}
where $\A\irrot$ is the irrotational part of $\A$, fulfilling
$\curl\A\irrot=\0$, and $\A\rotat$ is the rotational part, fulfilling
${\div\A\rotat=0}$. Since $\A\rotat$ is gauge invariant by definition,
it follows that in this case
\begin{align}
\label{eq:B_decomposed}
 \B(t,\x)
 = \curl\A(t,\x)
 = \curl\A\rotat(t,\x)
\end{align}

If we take \eqref{eq:pfield} as a starting point, and assume that
not only the vector potential $\A$ can be Helmholtz decomposed, but
also $\E$ is so well-behaved that it can be decomposed as
\begin{align}
\label{eq:E_decomposed}
 \E(t,\x) &= \E\irrot(t,\x) + \E\rotat(t,\x)
\end{align}
then the electromagnetic linear (translational) momentum can be
represented by the gauge invariant expression
\begin{align}
\begin{split}
\label{eq:pfield_int_irrot_rotat}
 \pfield(t)
 ={}& \epz\intV{}{\E\irrot\cross(\curl\A\rotat)}
\\
&+\epz\intV{}{\E\rotat\cross(\curl\A\rotat)}
\end{split}
\end{align}
Using straightforward vector analysis, we obtain the following
result~%
\cite{Thide:Book:2011}
\begin{align}
\begin{split}
\label{}
\label{eq:pfield_int_rotat}
 \pfield(t)
 ={}& \intV{}{\rhoe\A\rotat}
 - \epz\intV{}{\A\rotat\cross(\curl\E\rotat)}
\\
&- \epz\ointSvecdot{}{\E\rotat\ox\A\rotat}
\end{split}
\end{align}
Whenever Eqns.~\eqref{eq:A_decomposed} and~\eqref{eq:E_decomposed} hold,
this gauge invariant expression for $\pfield$ is exact. If the electric field
is such that it cannot be Helmholtz decomposed, every occurrence of $\E\irrot$
and $\E\rotat$ in Eqn.~\ref{eq:pfield_int_irrot_rotat}
and~Eqn.~\ref{eq:pfield_int_rotat} has to be replaced by $\E$ but would
still be exact and gauge independent.

If the expression within the big parentheses in the integrand of
the last integral in Eqn.~\eqref{eq:A_Helmholtz} is such that this
surface integral grows without bounds at very large distances,
it is not possible to decompose the vector potential as described
by Eqn.~\eqref{eq:A_decomposed} and $\A\rotat$ and $\E\rotat$ in
Eqn.~\ref{eq:pfield_int_irrot_rotat} and~Eqn.~\ref{eq:pfield_int_rotat}
would have to be replaced by $\A$ and $\E$.

\subsubsection{First quantization formalism}

If we restrict ourselves to consider a single temporal Fourier component
of the rotational (`transverse') component of $\E$ in a region where
$\rhoe=0$,  we find that in complex representation
\begin{align}
 \E\rotat = - \pdd{t}\A\rotat = \im\w\A\rotat
\end{align}
which allows us to replace $\A\rotat$ by $-\im\E\rotat/\w$, yielding,
\begin{align}
\begin{split}
 \Avet{\pfield}
 ={}& \Re{-\im\frac{\epz}{2\w}\intV{}{(\grad\ox\E\rotat)\bdot\cc{(\E\rotat)}}}
\\
  &{}+\Re{\im\frac{\epz}{2\w}\ointSvecdot{}{\E\rotat\ox\cc{(\E\rotat)}}}
\end{split}
\end{align}
If the tensor (dyadic) $\E\ox\cc{(\E\rotat)}$ is regular and falls off
sufficiently rapidly at large distances or if $\nunit\bdot\E=0$, we can
discard the surface integral term and find that the cycle averaged
linear momentum carried by the rotational components of the fields,
$\E\rotat$ and $\B\rotat\equiv\B=\curl\A\rotat$, is
\begin{align}
 \Avet{\pfield}
 = \Re{-\im\frac{\epz}{2\w}\intV{}{(\grad\ox\E\rotat)\bdot\cc{(\E\rotat)}}}
\end{align}
In complex tensor notation this can be written
\begin{align}
\label{eq:pfield_ave}
\begin{split}
 \Avet{\pfield}
 &= -\im\frac{\epz}{2\w}
  \sum_{i,j=1}^3\intV{}{\cc{(E\rotat_i)}\big(\xunit_j\pdj{E\rotat_i}\big)}
\\
 &= -\im\frac{\epz}{2\w}\sum_{i=1}^3\intV{}{\cc{(E\rotat_i)}\grad E\rotat_i}
\end{split}
\end{align}
Making use of the field linear momentum operator,
Eqn.~\eqref{eq:op_pfield}, we can write the expression for the linear
momentum of the electromagnetic field in terms of this operator as
\begin{align}
 \Avet{\pfield}
 = \frac{\epz}{2\hslash\w}
   \sum_{i=1}^3\intV{}{\cc{(E\rotat_i)}\,\Op{p}\field\,E\rotat_i}
\end{align}
If we introduce the vector
\begin{align}
\label{eq:Psi}
\boldsymbol{\Psi}
 = \sum_{i=1}^3\Psi_i\xunit_i
 = \SQRT{\frac{\epz}{2\hslash\w}}\E\rotat
 = \SQRT{\frac{\epz}{2\hslash\w}}\sum_{i=1}^3E\rotat_i\xunit_i
\end{align}
we can write 
\begin{align}
 \Avet{\pfield}
 = \sum_{i=1}^3\intV{}{\cc{\Psi}_i\,\Op{p}\field\,\Psi_i} 
\end{align}
or, if we assume Einstein's summation convention,
\begin{align}
 \Avet{\pfield}
 = \bra{\Psi_i}{\Op{p}\field}\ket{\Psi_i}
\end{align}
Thus we can represent the cycle (temporal) averaged linear momentum
carried by a monochromatic electromagnetic field as a sum of diagonal
quantal matrix elements (expectation values) where the rotational
(`transverse') component of the (scaled) electric field vector behaves
as a kind of vector wavefunction.

Methods based on electromagnetic linear momentum are used for radio science
and radar applications, and for wireless communications.

\subsection{Angular momentum}
\label{subsect:angular_momentum}

The volumetric density of the electromagnetic field angular momentum,
radiated from $V'$ is~%
\cite[Sect.~10.6]{Mandel&Wolf:Book:1995},
\cite{%
Schwinger&al:Book:1998,%
Thide:Book:2011%
}
\begin{align}
\label{eq:hfield}
\begin{split}
 \hfield(t,\x;\xm)
&= (\xxm)\cross\gfield
 = \epz(\xxm)\cross(\E\cross\B)
\end{split}
\end{align}
where $\xm=\sum_{i=1}^3{x_m}_i\xunit_i$ is a regular but otherwise arbitrary
point (the moment point). Using Eqns.~\eqref{eq:RS}, we can write this
in complex notation as
\begin{gather}
\label{eq:hfield_RS}
 \hfield(t,\x;\xm)
 = \im\frac{\epz}{2c}(\xxm)\cross(\RS\cross\cc{\RS})
\end{gather}

The mechanical angular momentum density of the charged particles
around $\xm$ is
\begin{align}
\label{eq:hmech}
\begin{split}
 \hmech(t,\x;\xm) 
  &= (\xxm)\cross\gmech(t,\x)
\\
  &= (\xxm)\cross\varrho\mech(t,\x)\vecv\mech(t,\x)
\end{split}
\end{align}

As emphasized by \textcite{Truesdell:Book:1968}, angular momentum
(also known as moment of momentum) is a physical observable in its
own right, in general independent of and not derivable from linear
momentum\footnote{See in particular Section V.5: ``Daniel Bernoulli and
Euler on the Dependence or Independence of the Law of Moment of Momentum
in 1744.''}.  Therefore, as is well known, the knowledge of one of these
two observables does not automatically imply a knowledge of the other.
This also follows from Noether's theorem~%
\cite{%
Noether:NGWG:1918,%
Neuenschwander:Book:2011%
}.

Let us introduce the electromagnetic angular momentum flux pseudotensor
\begin{align}
 \tens{M}(\xm)=(\xxm)\cross\tens{T}
\end{align}
where $\tens{T}$ is defined by formula~\eqref{eq:T}, and the physically
observable the Lorentz torque density pseudovector (axial vector) around
the moment point $\xm$ is
\begin{align}
\label{eq:Lorentz_torque_density}
 \vtau(t,\xxm) = (\xxm)\cross\f(t,\x)
\end{align}
where $\f$ is the Lorentz force density defined by
formula~\eqref{eq:Lorentz_force_density}.  Then the following local
conservation law can be derived from Maxwell's equations~%
\cite{%
Schwinger&al:Book:1998,%
Jackson:Book:1998,%
Thide:Book:2011,%
Bliokh&al:NJP:2014%
}
\begin{align}
\label{eq:pddt_hfield}
 \pdd{t}{\hfield(\xm)} + \div\tens{M}(\xm) = -\vtau
\end{align}
Hence, there is a loss of field angular momentum density to the mechanical
angular momentum density, in the form of Lorentz torque density $\vtau$,
showing that the angular momentum of the electromagnetic field and the
rotational dynamics of the charges and currents are coupled to each other.

By using Eqns.~\eqref{eq:RS}, \eqref{eq:hfield_RS}, and \eqref{eq:T_RS} we
can write this local equation of continuity for angular momentum density as
\begin{multline}
 \pdd{t}{[(\xxm)\cross(\RS\cross\cc{\RS})]}
\\
 - \im c\div\Big[(\xxm)\cross\Big(\RS\bdot\cc{\RS}\tunity
   -\frac{1}{\epz}(\RS\ox\cc{\RS} + \cc{\RS}\ox\RS)\Big)\Big]
\\
 = -\im\frac{c}{\epz}(\xxm)\cross\big[\rhoe(\RS + \cc{\RS})
   - {\epz}\je\cross(\RS - \cc{\RS})\big]
\end{multline}

Recalling the definition \eqref{eq:hmech} of the mechanical angular
momentum density, and that the angular momentum density of the system is
\begin{align}
 \hsys(t,\x;\xm) = \hfield(t,\x;\xm) + \hmech(t,\x;\xm) 
\end{align}
the continuity equation \eqref{eq:pddt_hfield} can be written
\begin{align}
\label{eq:local_conservation_law_ang_mom_dens}
 \pdd{t}{\hsys(t,\x;\xm)} + \div\tens{M}(\xm) = \0
\end{align}

This shows that a volume containing an arbitrary distribution of
charge density $\rhoe(t',\x')$ and current density $\je(t',\x')$ does
not only radiate linear momentum density (Poynting vector) but also
angular momentum density into the surrounding free space; see also
\cite{Thide&al:ARXIV:2010}.  However, since angular momentum describes
rotations, it cannot be sensed by a single linear dipole or
similar one-dimensional linear-momentum sensing antenna.

The total classical electromagnetic angular momentum around an arbitrary
moment point $\xm$ carried by the electromagnetic field in a volume $V$
is [see also~\textcite{Mandel&Wolf:Book:1995}]
\begin{align}
\begin{split}
 \Jfield(t;\xm) 
 ={}& \intV{}{\hfield(t,\x;\xm)}
\\
 ={}& \epz\intV{}{\x\cross[\E(t,\x)\cross\B(t,\x)]}
\\
  &{}- \xm\cross\epz\intV{}{[\E(t,\x)\cross\B(t,\x)]}
\\
 ={}& \Jfield(t;\0) -\xm\cross\pfield(t)
\end{split}
\end{align}
where, in the last step, Eqn.~\eqref{eq:pfield} was used.

The mechanical angular momentum in $V$ is
\begin{align}
\begin{split}
 \Jmech(t;\xm)
 &= \intV{}{\hmech(t,\x;\xm)}
\\
 &= \intV{}{(\xxm)\cross\varrho\mech(t,\x)\vecv\mech(t,\x)}
\\
 &= \Jmech(t;\0) -\xm\cross\pmech(t)
\end{split}
\end{align}
where, in the last step, Eqn.~\eqref{eq:pmech} was used.

Hence, the total angular momentum of the system of particles and
fields in $V$ is
\begin{align}
\begin{split}
 \Jsys(t;\xm)
&= \Jfield(t;\xm) + \Jmech(t;\xm) 
\\
&= \Jsys(t;\0) -\xm\cross\psys(t)
\end{split}
\end{align}
Clearly, if $\psys\ne0$, then it is always possible to find a moment
point $\xm$ that makes the total system angular momentum vanish.  If,
on the other hand, $\psys=\0$, then the total system angular momentum
is independent of moment point $\xm$.

The total angular momentum of the system of charges, currents and fields
fulfills the global conservation law
\cite{%
Schwinger&al:Book:1998,%
Rohrlich:Book:2007,%
Thide&al:Incollection:2011%
}
\begin{align}
\label{eq:conservation_law_ang_mom}
 \dd{t}{\Jsys(\xm)} + \ointSvecdot{}{\tens{M}(\xm)}
  = \0
\end{align}
\ie, a continuity equation without sources and sinks showing that for
a constant angular momentum flux (including zero) the system's angular
momentum is a constant of motion.  This is the angular momentum theorem
in Maxwell's theory. It shows that not only the mechanical particles
(charges) but also the electromagnetic field itself carries angular
momentum (rotational momentum) that couples to the mechanical angular
momentum of the matter (charged particles) and thereby to the conduction
current carried by the charged particles.

We see that the conservation law \eqref{eq:conservation_law_ang_mom}
describes the angular momentum analogs of the linear momentum processes
described by the conservation law \eqref{eq:conservation_law_lin_mom} and
shows that the use of single monolithic antenna devices based on torque
action will allow an alternative way of transferring information but
with the added benefit of higher information density offered by the
inherent quantized multi-state property of the classical electromagnetic
angular momentum.

\subsubsection{Gauge invariance}
\label{subsubsec:ang_mom_gauge_invariance}

If the vector potential can be decomposed as in
Subsubsect.~\ref{subsubsec:lin_mom_gauge_invariance} and if one uses
several vector analytic identities and employing partial integration,
it is possible to derive an exact, manifestly gauge invariant expression
for the total field angular momentum $\Jfield$ that is the sum of two
pseudovectors, one independent and one dependent of the moment point
$\xm$. The result is~\cite{Thide:Book:2011}
\begin{align}
 \Jfield(t,\xm) = \Sfield(t) + \Lfield(t,\xm) 
\end{align}
where 
\begin{subequations}
\begin{align}
\label{eq:Sfield_total}
 \Sfield(t) = \epz\intV{}\E(t,\x)\cross\A\rotat(t,\x)
\end{align}
and
\begin{multline}
\label{eq:Lfield_total}
 \Lfield(t,\xm)
  = \intV{}{(\xxm)\cross\rhoe(t,\x)\A\rotat(t,\x)}
\\
  + \epz\intV{}{(\xxm)\cross\big([\grad\ox\A\rotat(t,\x)]\bdot\E(t,\x)\big)}
\\
  - \epz\ointSvecdot{}{[\E(t,\x)\ox(\xxm)\cross\A\rotat(t,\x)]}
\end{multline}
\end{subequations}
The last integral vanishes if the integrand is regular and falls
off sufficiently rapidly, becomes perpendicular to $\nunit$, or
otherwise causes the surface integral to go to zero at very large
distances. However, if the vector potential $\A$ is not regular
and/or does not fall off faster than $1/r$ as $r$ tends to infinity,
the separation of $\Jfield$ into $\Sfield$ and $\Lfield$ is not gauge
invariant and, hence, not unique; \cf\ \textcite{Leader:PPN:2013} and
\textcite{Bialynicki-Birula&Bialynicka-Birula:JO:2011}.  See also
\textcite[Sect.~7]{Keller:PHR:2005}.

\subsubsection{First quantization formalism}

Let us now assume that $\A$ is so well-behaved that the conditions
for Helmholtz's decomposition~\eqref{eq:A_decomposed} are
fulfilled and, furthermore, that $\rhoe$ vanishes in the region
of interest.  Particularizing to a single temporal Fourier
component of the field $\propto\exp{(-\im\w{t})}$, we can then write
$\E=-\partial{\A}/\partial{t}=\im\w\A$ and obtain the following
expressions for the cycle averages of $\Sfield$ and $\Lfield$ in complex
notation:
\begin{subequations}
\begin{gather}
\label{eq:Sfield}
\Avet{\Sfield}
 = -\im\frac{\epz}{2\w} \intV{}(\cc{\E}\cross\E) 
\\
\label{eq:Lfield}
\begin{split}
\Avet{\Lfield(\xm)}
 ={}& -\im\frac{\epz}{2\w}
  \sum_{i=1}^3\intV{}{\cc{E}_i[(\xxm)\cross\grad]E_i}
\\
 ={}& \frac{\epz}{2\hslash\w}
  \sum_{i=1}^3\intV{}{\cc{E}_i(-\im\hslash\x\cross\grad)E_i}
\\
  &{}- \frac{\epz}{2\hslash\w}\xm\cross
  \sum_{i=1}^3\intV{}{\cc{E}_i(-\im\hslash\grad)E_i}
\\
 ={}& \Avet{\Lfield(\0)} - \xm\cross\Avet{\pfield}
\end{split}
\end{gather}
\end{subequations}
We note that if the cycle averaged electromagnetic field linear momentum
$\Avet{\pfield}\neq\0$, it is always possible to choose the moment point
$\xm$ such that $\Avet{\Lfield(\xm)}=\0$. But if $\Avet{\pfield}=\0$,
\ie, if the cycle average of the linear momentum of the monochromatic
field in question vanishes in a volume $V$, then the cycle averaged OAM,
$\Avet{\Lfield}$, is independent of $\xm$ in $V$.  In a beam geometry the
quantity $\Sfield$ can often to a good approximation be identified with
the spin angular momentum (SAM) carried by the field, and $\Lfield$
with its orbital angular momentum (OAM).

The fact that we can express the temporal averages $\Avet{\pfield}$ and
$\Avet{\Lfield}$ in terms of expectation values with the quantal operators
for linear and angular momentum, $\Op{\p}\field=-\im\hslash\grad$ and
$\Op{\L}\field=-\im\hslash\x\cross\grad$, respectively, hints to the
first-quantization character of the Maxwell-Lorentz equations.

As is well known from quantum mechanics, but true also for classical fields,
the Cartesian components of $\Op{\L}$ expressed in cylindrical coordinates
$(\rho,\ph,z)$ are
\begin{subequations}
\label{eq:Lfield_cart}
\begin{gather}
\label{eq:Lfield_cart_x}
\op{L}_x\field = -\im\hslash\bigg[
 \sin\ph\Big(z\pdd{\rho}{}-\rho\pdd{z}{}\Big)
 +\frac{z}{\rho}\cos\ph\pdd{\ph}{}
 \bigg]
\\
\label{eq:Lfield_cart_y}
\op{L}_y\field = -\im\hslash\bigg[
 \cos\ph\Big(z\pdd{\rho}{}-\rho\pdd{z}{}\Big)
 -\frac{z}{\rho}\sin\ph\pdd{\ph}{}
 \bigg]
\\
\label{eq:Lfield_cart_z}
\op{L}_z\field = -\im\hslash\pdd{\ph}{}
\end{gather}
\end{subequations}
duly recalling that $\hslash$ should be divided out; \cf\
formula~\eqref{eq:Lfield}.  The magnitude squared of the orbital angular
momentum operator is
\begin{align}
 \abs{\Op{\L}\field}^2 
  = \big(\op{L}_x\field\big)^2 
  + \big(\op{L}_y\field\big)^2 
  + \big(\op{L}_z\field\big)^2 
\end{align}
with eigenvalues $\hslash^2{}l(l+1)$.

Orbital angular momentum is utilized in photonics and in optical trapping~%
\cite{%
Allen&al:PO:1999,%
Franke-Arnold&al:NJP:2004,%
Molina-Terriza&al:NPHY:2007,%
Torres&Torner:Book:2011%
}
but has not yet been exploited fully in wireless communications. The
conservation of angular momentum is a consequence of the rotational
symmetry of electrodynamics.

\subsection{Boost momentum}

Due to the invariance of Maxwell's equations under Lorentz
transformations, the first spatial momentum of the energy density
\eqref{eq:ufield}, and the linear momentum density
\eqref{eq:gfield} multiplied by the elapsed time
\begin{align}
 \vecxi(t,\x)
 = 2(\xxm)\ufield - (\tt_0)\frac{1}{\epz}\gfield
\end{align}
fulfills a local (differential) conservation law.  The integrated quantity
\begin{align}
 \xce(t) = \frac{\intV{}{\vecxi(t,\x)}}{\Ufield}
\end{align}
is the center of energy~\cite{Boyer:AJP:2005}. Apparently this physical
observable has not yet been exploited to any significant extent in radio
or optics.

\svnidlong
 {$HeadURL: file:///Repository/SVN/Articles/AngularMomentumRadio/branches/PRA/hertzian.tex $}
 {$LastChangedDate: 2014-10-27 17:50:01 +0100 (Mon, 27 Oct 2014) $}
 {$LastChangedRevision: 53 $}
 {$LastChangedBy: bt $}
\svnid{$Id: hertzian.tex 53 2014-10-27 16:50:01Z bt $}

\section{Hertzian dipoles}
\label{app:Hertizan_dipoles}

When the oscillations of an electric charge distribution $\rhoe(t',\x')$
are so small that their maximum amplitude (and the largest extent
of the smallest volume $V'$ that fully encloses the charge distribution)
is much smaller than the wavelength of the emitted fields, we can use
the multipole expansion.  The lowest order contribution to the fields
comes from the electric monopole (charge) scalar
\begin{align}
\label{eq:app:qe}
 \qe(t') = \intV{'}\rhoe(t',\x')
\end{align}
The second lowest contribution comes from the electric (Hertzian) dipole
moment vector with respect to an arbitrary moment point $\x'_0$,
\begin{align}
\label{eq:rad:d}
 \vd(t',\x'_0)
 &= \intV{'}{(\x'-\x'_0)\,\rhoe(t',\x')}
\\
 &= \intV{'}{\x'\,\rhoe(t',\x')}
   -\x'_0\intV{'}{\rhoe(t',\x')}
\\
 &= \vd(t',\0) - \x'_0\qe(t')
\end{align}

Clearly there are two cases to consider here. If $\qe(t')\neq0$, it is always
possible to choose a moment point $\x'_0$ so that ${\vd(t',\x'_0)=\0}$. But
if $\qe(t')=0$, then the dipole moment is independent of the choice of
the moment point $\x'_0$. Furthermore, since $\x'_0$ is a fix point,
\begin{align}
 \pdd{t'}{[\vd(t',\x'_0)]}
&= \dd{t'}{[\vd(t',\x'_0)]}
 = \dd{t'}{[\vd(t',\0)]}
 - \x'_0\dd{t'}{\qe(t')}
\end{align}
and, according to the equation of continuity for electric charge,
\begin{align}
 \dd{t'}{\qe(t')} 
 = \intV{'}{\pdd{t'}{\rhoe(t',\x')}}
 = -\ointSvecdot{'}{\je(t',\x')}
\end{align}
If there is no net flow of electric current across the surface $S'$
that encloses $V'$, it therefore follows that
\begin{subequations}
\begin{align}
 \pdd{{t'}}{[\vd(t',\x'_0)]} &= \pdd{{t'}}{[\vd(t',\0)]}
\\
 \pdd{{t'}^2}{^2[\vd(t',\x'_0)]} &= \pdd{{t'}^2}{^2[\vd(t',\0)]}
\end{align}
\end{subequations}

The electric and magnetic fields from an electric Hertzian dipole,
oscillating at the angular frequency $\w=ck$, where ${k=\lambda/(2\pi)}$
is the wave number, $\lambda$ being the wavelength in free space, and
located at the origin $\x'=\0$, which is also the moment point chosen,
\begin{align}
 \vd(t',\0) = \vd_\w(\0)\e^{-\im(\w{t'}+\delta)}
\end{align}
where ${\delta=\w\Delta{t'}}$ is an arbitrary constant phase,
can be obtained from Eqns.~\eqref{eq:E&B}.

We apply the decomposition \eqref{eq:v_decomposition} on $\vd_\w(\0)$,
resulting in a parallel ($\parallel$), and a perpendicular ($\perp$)
component 
\begin{gather}
 \vd_\parallel(\0) =  [\vd_\w(\0)\bdot\nunit']\nunit'
\\
 \vd_\perp(\0)
 = \vd_\w(\0) - \vd_\parallel
 = \vd_\w - [\vd_\w(\0)\bdot\nunit']\nunit'
\end{gather}
Using a spherical coordinate system $(r,\th,\ph)$ where
$\runit\equiv\nunit'$ and, from Eqn.~\eqref{eq:tret}, $t'=t-r/c$, the
fields generated by $\vd$ can be written [\cf\ \textcite[Eqn.~(16),
p.~73]{Cohen-Tannoudji&al:Book:1997}]
\begin{subequations}
\begin{align}
\begin{split}
 \E(t,\x) ={}&
 \frac{\vd_\parallel(\0)}{2\pi\epz}
 \big[\fn{2}(\delta) + \fn{3}(\delta)\big]
\\
 &+\frac{\vd_\perp(\0)}{4\pi\epz}
 \big[\fn{1}(\delta) - \fn{2}(\delta) - \fn{3}(\delta)\big]
\end{split}
\\
 \B(t,\x) ={}& \frac{\runit\cross\vd_\perp(\0)}{4\pi\epz c}
  \big[\fn{1}(\delta) - \fn{2}(\delta)\big]
\end{align}
\end{subequations}
where, for convenience and compactness, we introduced the help quantities
\begin{subequations}
\label{eq:fn}
\begin{align}
 \fn{1}(\delta)
 &\equiv \fn{1}(\w{t'},\delta)
 = \frac{k^2}{r}\cos(kr-\w t + \delta)
\\
 \fn{2}(\delta) 
 &\equiv \fn{2}(\w{t'},\delta)
 = \frac{k}{r^2}\sin(kr-\w t + \delta)
\\
 \fn{3}(\delta)
 &\equiv \fn{3}(\w{t'},\delta)
 = \frac{1}{r^3}\cos(kr-\w t + \delta)
\end{align}
\end{subequations}
with the subscript $n$ denoting how the component $\fn{n}$ in question
behaves as a power of radial distance $r$ from the dipole, \ie,
$\Ordo{\fn{n}}=r^{-n}$.

\subsection{Hertzian dipoles directed along the three Cartesian axes}

Let us now consider three Hertzian dipoles directed along the three
Cartesian axes. We denote their (Fourier) amplitudes 
\begin{align}
 \vd_j(\0) 
 \equiv {\vd_j}_\w(\0)
 = {d_j}_\w(\0)\,\xunit_j\,, j=1,2,3.
\end{align}
where $\xunit_1=\xunit,\xunit_2=\yunit$, and $\xunit_3=\zunit$,

\paragraph{Electric and magnetic fields}

The individual electric Hertzian dipoles $\vd_j(\0),j=1,2,3$, generate
electric and magnetic field vectors $\E_j$ and $\B_j$.  In spherical
polar coordinates they are:
\begin{subequations}
\label{eq:Edipole_fields}
\begin{multline}
\label{eq:Edipole_fields_E1}
 \E_1(t,\x) =
 \frac{d_1(\0)}{2\pi\epz}\big[\fn{2}(\delta_1) + \fn{3}(\delta_1)\big]
  \sin\th\cos\ph\runit
\\
 +\frac{d_1(\0)}{4\pi\epz}\big[\fn{1}(\delta_1)-\fn{2}(\delta_1)-\fn{3}(\delta_1)\big]
  \cos\th\cos\ph\thunit
\\
 -\frac{d_1(\0)}{4\pi\epz}\big[\fn{1}(\delta_1)-\fn{2}(\delta_1)-\fn{3}(\delta_1)\big]
  \sin\ph\phunit
\end{multline}

\begin{multline}
\label{eq:Edipole_fields_B1}
\B_1(t,\x)
 = \frac{d_1(\0)}{4\pi\epz c}\big[\fn{1}(\delta_1)-\fn{2}(\delta_1)\big]
  \sin\ph\thunit
\\
  + \frac{d_1(\0)}{4\pi\epz c}\big[\fn{1}(\delta_1)-\fn{2}(\delta_1)\big]
   \cos\th\cos\ph\phunit
\end{multline}

\begin{multline}
\label{eq:Edipole_fields_E2}
 \E_2(t,\x) =
 \frac{d_2(\0)}{2\pi\epz}\big[\fn{2}(\delta_2)+\fn{3}(\delta_2)\big]
  \sin\th\sin\ph\runit
\\
 +\frac{d_2(\0)}{4\pi\epz}\big[\fn{1}(\delta_2)-\fn{2}(\delta_2)-\fn{3}(\delta_2)\big]
  \cos\th\sin\ph\thunit
\\
 +\frac{d_2(\0)}{4\pi\epz}\big[\fn{1}(\delta_2)-\fn{2}(\delta_2)-\fn{3}(\delta_2)\big]
  \cos\ph\phunit
\end{multline}

\begin{multline}
\label{eq:Edipole_fields_B2}
\B_2(t,\x)
 = -\frac{d_2(\0)}{4\pi\epz c}\big[\fn{1}(\delta_2)-\fn{2}(\delta_2)\big]
  \cos\ph\thunit
\\
  +\frac{d_2(\0)}{4\pi\epz c}\big[\fn{1}(\delta_2)-\fn{2}(\delta_2)\big]
   \cos\th\sin\ph\phunit
\end{multline}

\begin{multline}
\label{eq:Edipole_fields_E3}
 \E_3(t,\x) =
 \frac{d_3(\0)}{2\pi\epz}\big[\fn{2}(\delta_3)+\fn{3}(\delta_3)\big]
  \cos\th\runit
\\
-\frac{d_3(\0)}{4\pi\epz}\big[\fn{1}(\delta_3)-\fn{2}(\delta_3)-\fn{3}(\delta_3)\big]
  \sin\th\thunit
\end{multline}

\begin{multline}
\label{eq:Edipole_fields_B3}
\B_3(t,\x)
 = -\frac{d_3(\0)}{4\pi\epz c}\big[\fn{1}(\delta_3)-\fn{2}(\delta_3)\big]
  \sin\th\phunit\hfill
\end{multline}
\end{subequations}

\paragraph{Vector products of the fields}

When we want to calculate the linear and angular momenta carried by
the fields radiated by Hertzian dipoles directed along the Cartesian
axes, we need to evaluate expressions that involve the vector products
$\E_j\cross\B_k$ for various combinations of $j,k=1,2,3$. Here we list
a few such vector products.
\begin{multline}
\label{eq:E1crossB1}
 \E_1\cross\B_1 =
 \frac{d_1^2(\0)}{16\pi^2\epz^2 c}\big[
  \fn{1}^2(\delta_1)
  - 2\fn{1}(\delta_1)\fn{2}(\delta_1)
\\
  - \fn{1}(\delta_1)\fn{3}(\delta_1)
  + \fn{2}^2(\delta_1)
\\
  + \fn{2}(\delta_1)\fn{3}(\delta_1)
 \big](1-\sin^2\th\cos^2\ph)\,\runit
\\
 - \frac{d_1^2(\0)}{8\pi^2\epz^2 c}\big[
  \fn{1}(\delta_1)\fn{2}(\delta_1)
  + \fn{1}(\delta_1)\fn{3}(\delta_1)
\\
  - \fn{2}^2(\delta_1)
  - \fn{2}(\delta_1)\fn{3}(\delta_1)
 \big]\sin\th\cos\th\cos^2\ph\,\thunit
\\
 + \frac{d_1^2(\0)}{8\pi^2\epz^2 c}\big[
  \fn{1}(\delta_1)\fn{2}(\delta_1)
  + \fn{1}(\delta_1)\fn{3}(\delta_1)
\\
  - \fn{2}^2(\0)(\delta_1)
  - \fn{2}(\delta_1)\fn{3}(\delta_1)
 \big]\sin\th\sin\ph\cos\ph\,\phunit
\end{multline}

\begin{multline}
\label{eq:E2crossB2}
 \E_2\cross\B_2 =
 \frac{d_2^2(\0)}{16\pi^2\epz^2 c}\big[
  \fn{1}^2(\delta_2)
  - 2\fn{1}(\delta_2)\fn{2}(\delta_2)
\\
  - \fn{1}(\delta_2)\fn{3}(\delta_2)
  + \fn{2}^2(\delta_2)
\\
  + \fn{2}(\delta_2)\fn{3}(\delta_2)
 \big](1-\sin^2\th\sin^2\ph)\,\runit
\\
 - \frac{d_2^2(\0)}{8\pi^2\epz^2 c}\big[
  \fn{1}(\delta_2)\fn{2}(\delta_2)
  + \fn{1}(\delta_2)\fn{3}(\delta_2)
\\
  - \fn{2}^2(\delta_2)
  - \fn{2}(\delta_2)\fn{3}(\delta_2)
 \big]\sin\th\cos\th\sin^2\ph\,\thunit
\\
 - \frac{d_2^2(\0)}{8\pi^2\epz^2 c}\big[
  \fn{1}(\delta_2)\fn{2}(\delta_2)
  + \fn{1}(\delta_2)\fn{3}(\delta_2)
\\
  - \fn{2}^2(\delta_2)
  - \fn{2}(\delta_2)\fn{3}(\delta_2)
 \big]\sin\th\sin\ph\cos\ph\,\phunit
\end{multline}

\begin{multline}
\label{eq:E3crossB3}
 \E_3\cross\B_3 =
 \frac{d_3^2(\0)}{16\pi^2\epz^2 c}\big[
  \fn{1}^2(\delta_3)
  - 2\fn{1}(\delta_3)\fn{2}(\delta_3)
  + \fn{2}^2(\delta_3)
\\
  - \fn{1}(\delta_3)\fn{3}(\delta_3)
  + \fn{2}(\delta_3)\fn{3}(\delta_3)
 \big]\sin^2\th\,\runit
\\
 + \frac{d_3^2(\0)}{8\pi^2\epz^2 c}\big[
  \fn{1}(\delta_3)\fn{2}(\delta_3)
  + \fn{1}(\delta_3)\fn{3}(\delta_3)
\\
  - \fn{2}^2(\delta_3)
  - \fn{2}(\delta_3)\fn{3}(\delta_3)
  \big]\sin\th\cos\th\,\thunit
\end{multline}

\begin{multline}
\label{eq:E1crossB2}
 \E_1\cross\B_2 =
 -\frac{d_1(\0)d_2(\0)}{16\pi^2\epz^2 c}\big[
   \fn{1}(\delta_1)\fn{1}(\delta_2)
  -\fn{1}(\delta_1)\fn{2}(\delta_2)
\\
  -\fn{1}(\delta_2)\fn{2}(\delta_1)
  -\fn{1}(\delta_2)\fn{3}(\delta_1)
\\
  +\fn{2}(\delta_1)\fn{2}(\delta_2)
  +\fn{2}(\delta_2)\fn{3}(\delta_1)
 \big](1-\cos^2\th)\sin\ph\cos\ph\,\runit
\\
 -\frac{d_1(\0)d_2(\0)}{8\pi^2\epz^2 c}\big[
   \fn{1}(\delta_2)\fn{2}(\delta_1)
  +\fn{1}(\delta_2)\fn{3}(\delta_1)
\\
  -\fn{2}(\delta_1)\fn{2}(\delta_2)
  -\fn{2}(\delta_2)\fn{3}(\delta_1)
  \big]\sin\th\cos\th\sin\ph\cos\ph\,\thunit
\\
 -\frac{d_1(\0)d_2(\0)}{8\pi^2\epz^2 c}\big[
   \fn{1}(\delta_2)\fn{2}(\delta_1)
  +\fn{1}(\delta_2)\fn{3}(\delta_1)
\\
  -\fn{2}(\delta_1)\fn{2}(\delta_2)
  -\fn{2}(\delta_2)\fn{3}(\delta_1)
  \big]\sin\th\cos^2\ph\,\phunit
\end{multline}

\begin{multline}
\label{eq:E2crossB1}
 \E_2\cross\B_1 =
 -\frac{d_1(\0)d_2(\0)}{16\pi^2\epz^2 c}\big[
   \fn{1}(\delta_1)\fn{1}(\delta_2)
 - \fn{1}(\delta_1)\fn{2}(\delta_2)
\\
 - \fn{1}(\delta_2)\fn{2}(\delta_1)
 - \fn{1}(\delta_1)\fn{3}(\delta_2)
\\
 + \fn{2}(\delta_1)\fn{2}(\delta_2)
 + \fn{2}(\delta_1)\fn{3}(\delta_2)
 \big](1-\cos^2\th)\sin\ph\cos\ph\,\runit
\\
 -\frac{d_1(\0)d_2(\0)}{8\pi^2\epz^2 c}\big[
   \fn{1}(\delta_1)\fn{2}(\delta_2)
  +\fn{1}(\delta_1)\fn{3}(\delta_2)
\\
  -\fn{2}(\delta_1)\fn{2}(\delta_2)
  -\fn{2}(\delta_1)\fn{3}(\delta_2)
  \big]\sin\th\cos\th\sin\ph\cos\ph\,\thunit
\\
 +\frac{d_1(\0)d_2(\0)}{8\pi^2\epz^2 c}\big[
   \fn{1}(\delta_1)\fn{2}(\delta_2)
 + \fn{1}(\delta_1)\fn{3}(\delta_2)
\\
 - \fn{2}(\delta_1)\fn{2}(\delta_2)
 - \fn{2}(\delta_1)\fn{3}(\delta_2)
  \big]\sin\th\sin^2\ph\,\phunit
\end{multline}


\begin{thebibliography}{266}%
\makeatletter
\providecommand \@ifxundefined [1]{%
 \@ifx{#1\undefined}
}%
\providecommand \@ifnum [1]{%
 \ifnum #1\expandafter \@firstoftwo
 \else \expandafter \@secondoftwo
 \fi
}%
\providecommand \@ifx [1]{%
 \ifx #1\expandafter \@firstoftwo
 \else \expandafter \@secondoftwo
 \fi
}%
\providecommand \natexlab [1]{#1}%
\providecommand \enquote  [1]{``#1''}%
\providecommand \bibnamefont  [1]{#1}%
\providecommand \bibfnamefont [1]{#1}%
\providecommand \citenamefont [1]{#1}%
\providecommand \href@noop [0]{\@secondoftwo}%
\providecommand \href [0]{\begingroup \@sanitize@url \@href}%
\providecommand \@href[1]{\@@startlink{#1}\@@href}%
\providecommand \@@href[1]{\endgroup#1\@@endlink}%
\providecommand \@sanitize@url [0]{\catcode `\\12\catcode `\$12\catcode
  `\&12\catcode `\#12\catcode `\^12\catcode `\_12\catcode `\%12\relax}%
\providecommand \@@startlink[1]{}%
\providecommand \@@endlink[0]{}%
\providecommand \url  [0]{\begingroup\@sanitize@url \@url }%
\providecommand \@url [1]{\endgroup\@href {#1}{\urlprefix }}%
\providecommand \urlprefix  [0]{URL }%
\providecommand \Eprint [0]{\href }%
\providecommand \doibase [0]{http://dx.doi.org/}%
\providecommand \selectlanguage [0]{\@gobble}%
\providecommand \bibinfo  [0]{\@secondoftwo}%
\providecommand \bibfield  [0]{\@secondoftwo}%
\providecommand \translation [1]{[#1]}%
\providecommand \BibitemOpen [0]{}%
\providecommand \bibitemStop [0]{}%
\providecommand \bibitemNoStop [0]{.\EOS\space}%
\providecommand \EOS [0]{\spacefactor3000\relax}%
\providecommand \BibitemShut  [1]{\csname bibitem#1\endcsname}%
\let\auto@bib@innerbib\@empty
\bibitem [{\citenamefont {Noether}(1918)}]{Noether:NGWG:1918}%
  \BibitemOpen
  \bibfield  {author} {\bibinfo {author} {\bibfnamefont {Emmy}\ \bibnamefont
  {Noether}},\ }\bibfield  {title} {\enquote {\bibinfo {title} {Invariante
  {V}ariationsprobleme},}\ }\href@noop {} {\bibfield  {journal} {\bibinfo
  {journal} {Nachr.\ Ges.\ Wiss.\ G{\"o}ttingen}\ }\textbf {\bibinfo {volume}
  {1}},\ \bibinfo {pages} {235--257} (\bibinfo {year} {1918})},\ \bibinfo
  {note} {{E}nglish transl.: \emph{Invariant variation problems}, Transp.
  Theor. Stat. Phys., \textbf{1}, 186--207 (1971)}\BibitemShut {NoStop}%
\bibitem [{\citenamefont {Poincar{\'e}}(1906)}]{Poincare:RCMP:1906}%
  \BibitemOpen
  \bibfield  {author} {\bibinfo {author} {\bibfnamefont {M.}~\bibnamefont
  {Poincar{\'e}}},\ }\bibfield  {title} {\enquote {\bibinfo {title} {Sur la
  dynamique de l'{\'e}lectron},}\ }\href {\doibase 10.1007/BF03013466}
  {\bibfield  {journal} {\bibinfo  {journal} {Rendic.\ Circ.\ Matem.\ Palermo
  (1884--1940)}\ }\textbf {\bibinfo {volume} {21}},\ \bibinfo {pages}
  {129--175} (\bibinfo {year} {1906})}\BibitemShut {NoStop}%
\bibitem [{\citenamefont {Fushchich} and \citenamefont
  {Nikitin}(1992)}]{Fushchich&Nikitin:JPA:1992}%
  \BibitemOpen
  \bibfield  {author} {\bibinfo {author} {\bibfnamefont {W.~I.}\ \bibnamefont
  {Fushchich}} and \bibinfo {author} {\bibfnamefont {A.~G.}\ \bibnamefont
  {Nikitin}},\ }\bibfield  {title} {\enquote {\bibinfo {title} {The complete
  sets of conservation laws for the electromagnetic field},}\ }\href {\doibase
  10.1088/0305-4470/25/5/004} {\bibfield  {journal} {\bibinfo  {journal}
  {J.~Phys.~A: Math.\ Gen.}\ }\textbf {\bibinfo {volume} {25}},\ \bibinfo
  {pages} {L231--L233} (\bibinfo {year} {1992})}\BibitemShut {NoStop}%
\bibitem [{\citenamefont {Ibragimov}(2008)}]{Ibragimov:AAM:2008}%
  \BibitemOpen
  \bibfield  {author} {\bibinfo {author} {\bibfnamefont {Nail~H.}\ \bibnamefont
  {Ibragimov}},\ }\bibfield  {title} {\enquote {\bibinfo {title} {Symmetries,
  {Lagrangian} and conservation laws for the {Maxwell} equations},}\ }\href
  {\doibase 10.1007/s10440-008-9270-y} {\bibfield  {journal} {\bibinfo
  {journal} {Acta Applic.\ Math.}\ }\textbf {\bibinfo {volume} {105}},\
  \bibinfo {pages} {157--187} (\bibinfo {year} {2008})}\BibitemShut {NoStop}%
\bibitem [{\citenamefont {Dirac}(1931)}]{Dirac:PRSLA:1931}%
  \BibitemOpen
  \bibfield  {author} {\bibinfo {author} {\bibfnamefont {P.~A.~M.}\
  \bibnamefont {Dirac}},\ }\bibfield  {title} {\enquote {\bibinfo {title}
  {Quantised singularities in the electromagnetic field},}\ }\href {\doibase
  10.1098/rspa.1931.0129} {\bibfield  {journal} {\bibinfo  {journal} {Proc.\
  Roy.\ Soc.\ London Ser.\ A Math.\ Phys.\ Sci.}\ }\textbf {\bibinfo {volume}
  {133}},\ \bibinfo {pages} {60--72} (\bibinfo {year} {1931})}\BibitemShut
  {NoStop}%
\bibitem [{\citenamefont {Schwinger}(1969)}]{Schwinger:S:1969}%
  \BibitemOpen
  \bibfield  {author} {\bibinfo {author} {\bibfnamefont {Julian}\ \bibnamefont
  {Schwinger}},\ }\bibfield  {title} {\enquote {\bibinfo {title} {A magnetic
  model of matter},}\ }\href {\doibase 10.1126/science.165.3895.757} {\bibfield
   {journal} {\bibinfo  {journal} {Science}\ }\textbf {\bibinfo {volume}
  {165}},\ \bibinfo {pages} {757--761} (\bibinfo {year} {1969})}\BibitemShut
  {NoStop}%
\bibitem [{\citenamefont {Schwinger}\ \emph {et~al.}(1998)\citenamefont
  {Schwinger}, \citenamefont {DeRaad}, \citenamefont {Milton}, and
  \citenamefont {Tsai}}]{Schwinger&al:Book:1998}%
  \BibitemOpen
  \bibfield  {author} {\bibinfo {author} {\bibfnamefont {Julian}\ \bibnamefont
  {Schwinger}}, \bibinfo {author} {\bibfnamefont {Lester~L.}\ \bibnamefont
  {DeRaad}, \bibfnamefont {Jr.}}, \bibinfo {author} {\bibfnamefont
  {Kimball~A.}\ \bibnamefont {Milton}},  and \bibinfo {author} {\bibfnamefont
  {{Wu-yang}}\ \bibnamefont {Tsai}},\ }\href@noop {} {\emph {\bibinfo {title}
  {Classical Electrodynamics}}}\ (\bibinfo  {publisher} {Perseus Books},\
  \bibinfo {address} {Reading, MA, USA},\ \bibinfo {year} {1998})\BibitemShut
  {NoStop}%
\bibitem [{\citenamefont {Jackson}(1998)}]{Jackson:Book:1998}%
  \BibitemOpen
  \bibfield  {author} {\bibinfo {author} {\bibfnamefont {John~David}\
  \bibnamefont {Jackson}},\ }\href@noop {} {\emph {\bibinfo {title} {Classical
  Electrodynamics}}},\ \bibinfo {edition} {3rd}\ ed.\ (\bibinfo  {publisher}
  {Wiley \& Sons},\ \bibinfo {address} {New York, NY, USA},\ \bibinfo {year}
  {1998})\BibitemShut {NoStop}%
\bibitem [{\citenamefont {Thid{\'e}}\ \emph {et~al.}(2011)\citenamefont
  {Thid{\'e}}, \citenamefont {Elias}, \citenamefont {Tamburini}, \citenamefont
  {Mohammadi}, and \citenamefont
  {Mendon\c{c}a}}]{Thide&al:Incollection:2011}%
  \BibitemOpen
  \bibfield  {author} {\bibinfo {author} {\bibfnamefont {B.}~\bibnamefont
  {Thid{\'e}}}, \bibinfo {author} {\bibfnamefont {Nicholas~M.}\ \bibnamefont
  {Elias}, \bibfnamefont {II}}, \bibinfo {author} {\bibfnamefont
  {F.}~\bibnamefont {Tamburini}}, \bibinfo {author} {\bibfnamefont {S.~M.}\
  \bibnamefont {Mohammadi}},  and \bibinfo {author} {\bibfnamefont {J.~T.}\
  \bibnamefont {Mendon\c{c}a}},\ }\bibfield  {title} {\enquote {\bibinfo
  {title} {Applications of electromagnetic {OAM} in astrophysics and space
  physics studies},}\ }in\ \href@noop {} {\emph {\bibinfo {booktitle} {Twisted
  Photons: Applications of Light With Orbital Angular Momentum}}},\ \bibinfo
  {editor} {edited by\ \bibinfo {editor} {\bibfnamefont {Juan~P.}\ \bibnamefont
  {Torres}} and \bibinfo {editor} {\bibfnamefont {Lluis}\ \bibnamefont
  {Torner}}}\ (\bibinfo  {publisher} {Wiley-Vch Verlag},\ \bibinfo {address}
  {Weinheim, DE},\ \bibinfo {year} {2011})\ Chap.~\bibinfo {chapter} {9}, pp.\
  \bibinfo {pages} {155--178}\BibitemShut {NoStop}%
\bibitem [{\citenamefont {Thid{\'e}}(2011)}]{Thide:Book:2011}%
  \BibitemOpen
  \bibfield  {author} {\bibinfo {author} {\bibfnamefont {Bo}~\bibnamefont
  {Thid{\'e}}},\ }\href {http://www.plasma.uu.se/CED/Book} {\emph {\bibinfo
  {title} {{E}lectromagnetic {F}ield {T}heory}}},\ \bibinfo {edition} {2nd}\
  ed.\ (\bibinfo  {publisher} {Dover Publications, Inc.},\ \bibinfo {address}
  {Mineola, NY, USA},\ \bibinfo {year} {2011})\ \bibinfo {note} {(In
  press)}\BibitemShut {NoStop}%
\bibitem [{\citenamefont {Weber}(1901)}]{Weber:Book:1901}%
  \BibitemOpen
  \bibfield  {author} {\bibinfo {author} {\bibfnamefont {Heinrich}\
  \bibnamefont {Weber}},\ }\href@noop {} {\emph {\bibinfo {title} {Die
  partiellen {Differential-Gleichungen} der mathematischen {Physik} nach
  {Riemann's Vorlesungen}}}}\ (\bibinfo  {publisher} {Friedrich Vieweg und
  Sohn},\ \bibinfo {address} {Braunschweig},\ \bibinfo {year}
  {1901})\BibitemShut {NoStop}%
\bibitem [{\citenamefont
  {Silberstein}(1907{\natexlab{a}})}]{Silberstein:AP:1907a}%
  \BibitemOpen
  \bibfield  {author} {\bibinfo {author} {\bibfnamefont {Ludwik}\ \bibnamefont
  {Silberstein}},\ }\bibfield  {title} {\enquote {\bibinfo {title}
  {Elektromagnetische {G}rundgleichungen in bivektorieller {B}ehandlung},}\
  }\href {\doibase 10.1002/andp.19073270313} {\bibfield  {journal} {\bibinfo
  {journal} {Ann.\ Phys. (Leipzig)}\ }\textbf {\bibinfo {volume} {327}},\
  \bibinfo {pages} {579--586} (\bibinfo {year}
  {1907}{\natexlab{a}})}\BibitemShut {NoStop}%
\bibitem [{\citenamefont
  {Silberstein}(1907{\natexlab{b}})}]{Silberstein:AP:1907b}%
  \BibitemOpen
  \bibfield  {author} {\bibinfo {author} {\bibfnamefont {Ludwik}\ \bibnamefont
  {Silberstein}},\ }\bibfield  {title} {\enquote {\bibinfo {title} {Nachtrag
  zur {A}bhandlung {\"u}ber elektromagnetische {G}rundgleichungen in
  bivektorieller {B}ehandlung},}\ }\href {\doibase 10.1002/andp.19073291409}
  {\bibfield  {journal} {\bibinfo  {journal} {Ann.\ Phys. (Leipzig)}\ }\textbf
  {\bibinfo {volume} {329}},\ \bibinfo {pages} {783--784} (\bibinfo {year}
  {1907}{\natexlab{b}})}\BibitemShut {NoStop}%
\bibitem [{\citenamefont {Silberstein}(1914)}]{Silberstein:Book:1914}%
  \BibitemOpen
  \bibfield  {author} {\bibinfo {author} {\bibfnamefont {Ludwik}\ \bibnamefont
  {Silberstein}},\ }\href@noop {} {\emph {\bibinfo {title} {The Theory of
  Relativity}}}\ (\bibinfo  {publisher} {MacMillan},\ \bibinfo {address}
  {London},\ \bibinfo {year} {1914})\BibitemShut {NoStop}%
\bibitem [{\citenamefont {Bateman}(1915)}]{Bateman:Book:1915}%
  \BibitemOpen
  \bibfield  {author} {\bibinfo {author} {\bibfnamefont {Harry}\ \bibnamefont
  {Bateman}},\ }\href {http://archive.org/details/mathematicalanal00baterich}
  {\emph {\bibinfo {title} {The Mathematical Analysis of Electrical and Optical
  Wave-Motion on the Basis of {Maxwell's} Equations}}}\ (\bibinfo  {publisher}
  {Cambridge University Press},\ \bibinfo {address} {Cambridge, UK},\ \bibinfo
  {year} {1915})\BibitemShut {NoStop}%
\bibitem [{\citenamefont {Bia{\l}ynicki-Birula} and \citenamefont
  {Bia{\l}ynicka-Birula}(2006)}]{Bialynicki-Birula&Bialynicka-Birula:OC:2006}%
  \BibitemOpen
  \bibfield  {author} {\bibinfo {author} {\bibfnamefont {Iwo}\ \bibnamefont
  {Bia{\l}ynicki-Birula}} and \bibinfo {author} {\bibfnamefont {Zofia}\
  \bibnamefont {Bia{\l}ynicka-Birula}},\ }\bibfield  {title} {\enquote
  {\bibinfo {title} {Beams of electromagnetic radiation carrying angular
  momentum: {The Riemann-Silberstein} vector and the classical-quantum
  correspondence},}\ }\href {\doibase 10.1016/j.optcom.2005.11.071} {\bibfield
  {journal} {\bibinfo  {journal} {Opt. Commun.}\ }\textbf {\bibinfo {volume}
  {264}},\ \bibinfo {pages} {342--351} (\bibinfo {year} {2006})}\BibitemShut
  {NoStop}%
\bibitem [{\citenamefont {Bia{\l}ynicki-Birula} and \citenamefont
  {Bia{\l}ynicka-Birula}(2009)}]{Bialynicki-Birula&Bialynicka-Birula:PRA:2009}%
  \BibitemOpen
  \bibfield  {author} {\bibinfo {author} {\bibfnamefont {Iwo}\ \bibnamefont
  {Bia{\l}ynicki-Birula}} and \bibinfo {author} {\bibfnamefont {Zofia}\
  \bibnamefont {Bia{\l}ynicka-Birula}},\ }\bibfield  {title} {\enquote
  {\bibinfo {title} {Why photons cannot be sharply localized},}\ }\href
  {\doibase 10.1103/PhysRevA.79.032112} {\bibfield  {journal} {\bibinfo
  {journal} {Phys.\ Rev.\ A}\ }\textbf {\bibinfo {volume} {79}},\ \bibinfo
  {pages} {032112} (\bibinfo {year} {2009})}\BibitemShut {NoStop}%
\bibitem [{\citenamefont {Kobe}(1999{\natexlab{a}})}]{Kobe:FP:1999}%
  \BibitemOpen
  \bibfield  {author} {\bibinfo {author} {\bibfnamefont {Donald~H.}\
  \bibnamefont {Kobe}},\ }\bibfield  {title} {\enquote {\bibinfo {title} {A
  relativistic {Schr\"{o}dinger}-like equation for a photon and its second
  quantization},}\ }\href {\doibase 10.1023/A:1018855630724} {\bibfield
  {journal} {\bibinfo  {journal} {Found.\ Phys.}\ }\textbf {\bibinfo {volume}
  {29}},\ \bibinfo {pages} {1203--1231} (\bibinfo {year}
  {1999}{\natexlab{a}})}\BibitemShut {NoStop}%
\bibitem [{\citenamefont {Majorana}(2009)}]{Majorana:ResearchNotes:2009}%
  \BibitemOpen
  \bibfield  {author} {\bibinfo {author} {\bibfnamefont {Ettore}\ \bibnamefont
  {Majorana}},\ }\href@noop {} {\enquote {\bibinfo {title} {Research notes on
  theoretical physics},}\ } (\bibinfo {year} {2009}),\ \bibinfo {note}
  {unpublished, deposited at the `Domus Galileana,' Pisa, quaderno 2, p. 101/1;
  3, p. 11, 160; 15, p. 16; 17, p. 83, 159}\BibitemShut {NoStop}%
\bibitem [{\citenamefont {Oppenheimer}(1931)}]{Oppenheimer:PR:1931}%
  \BibitemOpen
  \bibfield  {author} {\bibinfo {author} {\bibfnamefont {J.~R.}\ \bibnamefont
  {Oppenheimer}},\ }\bibfield  {title} {\enquote {\bibinfo {title} {Note on
  light quanta and the electromagnetic field},}\ }\href {\doibase
  10.1103/PhysRev.38.725} {\bibfield  {journal} {\bibinfo  {journal} {Phys.\
  Rev.}\ }\textbf {\bibinfo {volume} {38}},\ \bibinfo {pages} {725--746}
  (\bibinfo {year} {1931})}\BibitemShut {NoStop}%
\bibitem [{\citenamefont {Laporte} and \citenamefont
  {Uhlenbeck}(1931)}]{Laporte&Uhlenbeck:PR:1931}%
  \BibitemOpen
  \bibfield  {author} {\bibinfo {author} {\bibfnamefont {Otto}\ \bibnamefont
  {Laporte}} and \bibinfo {author} {\bibfnamefont {George~E.}\ \bibnamefont
  {Uhlenbeck}},\ }\bibfield  {title} {\enquote {\bibinfo {title} {Application
  of spinor analysis to the {Maxwell} and {Dirac} equations},}\ }\href
  {\doibase 10.1103/PhysRev.37.1380} {\bibfield  {journal} {\bibinfo  {journal}
  {Phys.\ Rev.}\ }\textbf {\bibinfo {volume} {37}},\ \bibinfo {pages}
  {1380--1397} (\bibinfo {year} {1931})}\BibitemShut {NoStop}%
\bibitem [{\citenamefont {Archibald}(1955)}]{Archibald:CJP:1955}%
  \BibitemOpen
  \bibfield  {author} {\bibinfo {author} {\bibfnamefont {W.~J.}\ \bibnamefont
  {Archibald}},\ }\bibfield  {title} {\enquote {\bibinfo {title} {Field
  equations from particle equations},}\ }\href {\doibase 10.1139/p55-068}
  {\bibfield  {journal} {\bibinfo  {journal} {Canad.\ J.~Phys.}\ }\textbf
  {\bibinfo {volume} {33}},\ \bibinfo {pages} {565--572} (\bibinfo {year}
  {1955})}\BibitemShut {NoStop}%
\bibitem [{\citenamefont {Good}(1957)}]{Good:PR:1957}%
  \BibitemOpen
  \bibfield  {author} {\bibinfo {author} {\bibfnamefont {R.~H.}\ \bibnamefont
  {Good}},\ }\bibfield  {title} {\enquote {\bibinfo {title} {Particle aspect of
  the electromagnetic field equations},}\ }\href {\doibase
  10.1103/PhysRev.105.1914} {\bibfield  {journal} {\bibinfo  {journal} {Phys.\
  Rev.}\ }\textbf {\bibinfo {volume} {105}},\ \bibinfo {pages} {1914--1918}
  (\bibinfo {year} {1957})}\BibitemShut {NoStop}%
\bibitem [{\citenamefont {Hammer} and \citenamefont
  {Good}(1957)}]{Hammer&Good:PR:1957}%
  \BibitemOpen
  \bibfield  {author} {\bibinfo {author} {\bibfnamefont {C.~L.}\ \bibnamefont
  {Hammer}} and \bibinfo {author} {\bibfnamefont {R.~H.}\ \bibnamefont
  {Good}},\ }\bibfield  {title} {\enquote {\bibinfo {title} {Wave equation for
  a massless particle with arbitrary spin},}\ }\href {\doibase
  10.1103/PhysRev.108.882} {\bibfield  {journal} {\bibinfo  {journal} {Phys.\
  Rev.}\ }\textbf {\bibinfo {volume} {108}},\ \bibinfo {pages} {882--886}
  (\bibinfo {year} {1957})}\BibitemShut {NoStop}%
\bibitem [{\citenamefont {Moses}(1959)}]{Moses:PR:1959}%
  \BibitemOpen
  \bibfield  {author} {\bibinfo {author} {\bibfnamefont {H.~E.}\ \bibnamefont
  {Moses}},\ }\bibfield  {title} {\enquote {\bibinfo {title} {Solution of
  {Maxwell's} equations in terms of a spinor notation: the direct and inverse
  problem},}\ }\href {\doibase 10.1103/PhysRev.113.1670} {\bibfield  {journal}
  {\bibinfo  {journal} {Phys.\ Rev.}\ }\textbf {\bibinfo {volume} {113}},\
  \bibinfo {pages} {1670--1679} (\bibinfo {year} {1959})}\BibitemShut {NoStop}%
\bibitem [{\citenamefont {Kur\c{s}uno\u{g}lu}(1961)}]{Kursunoglu:JMP:1961}%
  \BibitemOpen
  \bibfield  {author} {\bibinfo {author} {\bibfnamefont {Behram}\ \bibnamefont
  {Kur\c{s}uno\u{g}lu}},\ }\bibfield  {title} {\enquote {\bibinfo {title}
  {Complex orthogonal and antiorthogonal representation of {Lorentz} group},}\
  }\href {\doibase 10.1063/1.1724209} {\bibfield  {journal} {\bibinfo
  {journal} {J.~Math.\ Phys.}\ }\textbf {\bibinfo {volume} {2}},\ \bibinfo
  {pages} {22--32} (\bibinfo {year} {1961})}\BibitemShut {NoStop}%
\bibitem [{\citenamefont {Good} and \citenamefont
  {Nelson}(1971)}]{Good&Nelson:Book:1971}%
  \BibitemOpen
  \bibfield  {author} {\bibinfo {author} {\bibfnamefont {Roland~H.}\
  \bibnamefont {Good}, \bibfnamefont {Jr.}} and \bibinfo {author}
  {\bibfnamefont {Terence~J.}\ \bibnamefont {Nelson}},\ }\href@noop {} {\emph
  {\bibinfo {title} {Classical Theory of Electric and Magnetic Fields}}}\
  (\bibinfo  {publisher} {Academic Press},\ \bibinfo {address} {London, UK},\
  \bibinfo {year} {1971})\BibitemShut {NoStop}%
\bibitem [{\citenamefont {Mignani}\ \emph {et~al.}(1974)\citenamefont
  {Mignani}, \citenamefont {Recami}, and \citenamefont
  {Baldo}}]{Mignani&al:LNC:1974}%
  \BibitemOpen
  \bibfield  {author} {\bibinfo {author} {\bibfnamefont {R.}~\bibnamefont
  {Mignani}}, \bibinfo {author} {\bibfnamefont {E.}~\bibnamefont {Recami}}, \
  and\ \bibinfo {author} {\bibfnamefont {M.}~\bibnamefont {Baldo}},\ }\bibfield
   {title} {\enquote {\bibinfo {title} {About a {D}irac-like equation for the
  photon according to {Ettore Majorana}},}\ }\href {\doibase
  10.1007/BF02812391} {\bibfield  {journal} {\bibinfo  {journal} {Lett.\ Nuov.\
  Cim.}\ }\textbf {\bibinfo {volume} {11}},\ \bibinfo {pages} {568--572}
  (\bibinfo {year} {1974})}\BibitemShut {NoStop}%
\bibitem [{\citenamefont {Barut}(1980)}]{Barut:Book:1980}%
  \BibitemOpen
  \bibfield  {author} {\bibinfo {author} {\bibfnamefont {Asim~O.}\ \bibnamefont
  {Barut}},\ }\href@noop {} {\emph {\bibinfo {title} {Electrodynamics and
  Classical Theory of Fields and Particles}}}\ (\bibinfo  {publisher} {Dover
  Publications},\ \bibinfo {address} {New York, NY, USA},\ \bibinfo {year}
  {1980})\BibitemShut {NoStop}%
\bibitem [{\citenamefont {Giannetto}(1985)}]{Giannetto:LNC2:1985}%
  \BibitemOpen
  \bibfield  {author} {\bibinfo {author} {\bibfnamefont {E.}~\bibnamefont
  {Giannetto}},\ }\bibfield  {title} {\enquote {\bibinfo {title} {A
  {Majorana-Oppenheimer} formulation of quantum electrodynamics},}\ }\href
  {\doibase 10.1007/BF02746912} {\bibfield  {journal} {\bibinfo  {journal}
  {Lett.\ Nuov.\ Cim.}\ }\bibinfo {series} {2},\ \textbf {\bibinfo {volume}
  {44}},\ \bibinfo {pages} {140--144} (\bibinfo {year} {1985})}\BibitemShut
  {NoStop}%
\bibitem [{\citenamefont {Sipe}(1995)}]{Sipe:PRA:1995}%
  \BibitemOpen
  \bibfield  {author} {\bibinfo {author} {\bibfnamefont {J.~E.}\ \bibnamefont
  {Sipe}},\ }\bibfield  {title} {\enquote {\bibinfo {title} {Photon wave
  functions},}\ }\href {\doibase 10.1103/PhysRevA.52.1875} {\bibfield
  {journal} {\bibinfo  {journal} {Phys.\ Rev.\ A}\ }\textbf {\bibinfo {volume}
  {52}},\ \bibinfo {pages} {1875--1883} (\bibinfo {year} {1995})}\BibitemShut
  {NoStop}%
\bibitem [{\citenamefont
  {Bia{\l}ynicki-Birula}(1994)}]{Bialynicki-Birula:APP:1994}%
  \BibitemOpen
  \bibfield  {author} {\bibinfo {author} {\bibfnamefont {Iwo}\ \bibnamefont
  {Bia{\l}ynicki-Birula}},\ }\bibfield  {title} {\enquote {\bibinfo {title} {On
  the wave function of the photon},}\ }\href@noop {} {\bibfield  {journal}
  {\bibinfo  {journal} {Acta\ Phys.\ Polon.}\ }\textbf {\bibinfo {volume} {A
  86}},\ \bibinfo {pages} {97--116} (\bibinfo {year} {1994})}\BibitemShut
  {NoStop}%
\bibitem [{\citenamefont
  {Bia{\l}ynicki-Birula}(1996)}]{Bialynicki-Birula:PO:1996}%
  \BibitemOpen
  \bibfield  {author} {\bibinfo {author} {\bibfnamefont {I.}~\bibnamefont
  {Bia{\l}ynicki-Birula}},\ }\bibfield  {title} {\enquote {\bibinfo {title}
  {Photon wave function},}\ }in\ \href {\doibase 10.1016/S0079-6638(08)70316-0}
  {\emph {\bibinfo {booktitle} {Prog.\ Opt.}}},\ Vol.\ \bibinfo {volume}
  {XXXVI},\ \bibinfo {editor} {edited by\ \bibinfo {editor} {\bibfnamefont
  {E.}~\bibnamefont {Wolf}}}\ (\bibinfo  {publisher} {Elsevier},\ \bibinfo
  {address} {Amsterdam, Holland},\ \bibinfo {year} {1996})\ pp.\ \bibinfo
  {pages} {245--294}\BibitemShut {NoStop}%
\bibitem [{\citenamefont {Esposito}(1998)}]{Esposito:FP:1998}%
  \BibitemOpen
  \bibfield  {author} {\bibinfo {author} {\bibfnamefont {S.}~\bibnamefont
  {Esposito}},\ }\bibfield  {title} {\enquote {\bibinfo {title} {Covariant
  {Majorana} formulation of electrodynamics},}\ }\href@noop {} {\bibfield
  {journal} {\bibinfo  {journal} {Found.\ Phys.}\ }\textbf {\bibinfo {volume}
  {28}},\ \bibinfo {pages} {231--244} (\bibinfo {year} {1998})}\BibitemShut
  {NoStop}%
\bibitem [{\citenamefont {Gersten}(1999)}]{Gersten:FPL:1999}%
  \BibitemOpen
  \bibfield  {author} {\bibinfo {author} {\bibfnamefont {A.}~\bibnamefont
  {Gersten}},\ }\bibfield  {title} {\enquote {\bibinfo {title} {Maxwell
  equations\textemdash the one-photon quantum equation},}\ }\href {\doibase
  10.1007/BF00670207} {\bibfield  {journal} {\bibinfo  {journal} {Found.\
  Phys.\ Lett.}\ }\textbf {\bibinfo {volume} {12}},\ \bibinfo {pages}
  {291--298} (\bibinfo {year} {1999})}\BibitemShut {NoStop}%
\bibitem [{\citenamefont {Keller}(2005)}]{Keller:PHR:2005}%
  \BibitemOpen
  \bibfield  {author} {\bibinfo {author} {\bibfnamefont {Ole}\ \bibnamefont
  {Keller}},\ }\bibfield  {title} {\enquote {\bibinfo {title} {On the theory of
  spatial localization of photons},}\ }\href {\doibase
  10.1016/j.physrep.2005.01.002} {\bibfield  {journal} {\bibinfo  {journal}
  {Phys.\ Rep.}\ }\textbf {\bibinfo {volume} {411}},\ \bibinfo {pages} {1--232}
  (\bibinfo {year} {2005})}\BibitemShut {NoStop}%
\bibitem [{\citenamefont {Khan}(2005)}]{Khan:PS:2005}%
  \BibitemOpen
  \bibfield  {author} {\bibinfo {author} {\bibfnamefont {Sameen~Ahmed}\
  \bibnamefont {Khan}},\ }\bibfield  {title} {\enquote {\bibinfo {title} {An
  exact matrix representation of {Maxwell's} equations},}\ }\href {\doibase
  10.1238/Physica.Regular.071a00440} {\bibfield  {journal} {\bibinfo  {journal}
  {Phys.\ Scr.}\ }\textbf {\bibinfo {volume} {71}},\ \bibinfo {pages}
  {440--442} (\bibinfo {year} {2005})}\BibitemShut {NoStop}%
\bibitem [{\citenamefont {Dragoman}(2007)}]{Dragoman:JOSAB:2007}%
  \BibitemOpen
  \bibfield  {author} {\bibinfo {author} {\bibfnamefont {Daniela}\ \bibnamefont
  {Dragoman}},\ }\bibfield  {title} {\enquote {\bibinfo {title} {Photon states
  from propagating complex electromagnetic fields},}\ }\href {\doibase
  10.1364/JOSAB.24.000922} {\bibfield  {journal} {\bibinfo  {journal} {J.~Opt.\
  Soc\ Am.\ B}\ }\textbf {\bibinfo {volume} {24}},\ \bibinfo {pages} {922--927}
  (\bibinfo {year} {2007})}\BibitemShut {NoStop}%
\bibitem [{\citenamefont {Smith} and \citenamefont
  {Raymer}(2007)}]{Smith&Raymer:2007}%
  \BibitemOpen
  \bibfield  {author} {\bibinfo {author} {\bibfnamefont {B.~J}\ \bibnamefont
  {Smith}} and \bibinfo {author} {\bibfnamefont {M.~G.}\ \bibnamefont
  {Raymer}},\ }\bibfield  {title} {\enquote {\bibinfo {title} {Photon wave
  functions, wave-packet quantization of light, and coherence theory},}\
  }\href@noop {} {\bibfield  {journal} {\bibinfo  {journal} {New J.~Phys.}\
  }\textbf {\bibinfo {volume} {9}},\ \bibinfo {pages} {414} (\bibinfo {year}
  {2007})}\BibitemShut {NoStop}%
\bibitem [{\citenamefont {Tamburini} and \citenamefont
  {Vicino}(2008)}]{Tamburini&Vicino:PRA:2008}%
  \BibitemOpen
  \bibfield  {author} {\bibinfo {author} {\bibfnamefont {F.}~\bibnamefont
  {Tamburini}} and \bibinfo {author} {\bibfnamefont {D.}~\bibnamefont
  {Vicino}},\ }\bibfield  {title} {\enquote {\bibinfo {title} {Photon wave
  function: A covariant formulation and equivalence with {QED}},}\ }\href
  {\doibase 10.1103/PhysRevA.78.052116} {\bibfield  {journal} {\bibinfo
  {journal} {Phys.\ Rev.\ A}\ }\textbf {\bibinfo {volume} {78}},\ \bibinfo
  {pages} {052116(5)} (\bibinfo {year} {2008})}\BibitemShut {NoStop}%
\bibitem [{\citenamefont {Bogush}\ \emph {et~al.}(2009)\citenamefont {Bogush},
  \citenamefont {Krylov}, \citenamefont {Ovsiyuk}, and \citenamefont
  {Red'kov}}]{Bogush&al:RM:2009}%
  \BibitemOpen
  \bibfield  {author} {\bibinfo {author} {\bibfnamefont {A.~A.}\ \bibnamefont
  {Bogush}}, \bibinfo {author} {\bibfnamefont {G.~G.}\ \bibnamefont {Krylov}},
  \bibinfo {author} {\bibfnamefont {E.~M.}\ \bibnamefont {Ovsiyuk}},  and
  \bibinfo {author} {\bibfnamefont {V.~M.}\ \bibnamefont {Red'kov}},\
  }\bibfield  {title} {\enquote {\bibinfo {title} {Maxwell equations in complex
  form of {Majorana-Oppenheimer}, solutions with cylindric symmetry in {Riemann
  $S_3$} and {Lobachevsky $H_3$} spaces},}\ }\href {\doibase
  10.1007/s11587-009-0067-8} {\bibfield  {journal} {\bibinfo  {journal}
  {Ricerche Mat.}\ }\textbf {\bibinfo {volume} {59}},\ \bibinfo {pages}
  {59--96} (\bibinfo {year} {2009})}\BibitemShut {NoStop}%
\bibitem [{\citenamefont {Wang}\ \emph {et~al.}(2009)\citenamefont {Wang},
  \citenamefont {Xiong}, and \citenamefont {Qiu}}]{Wang&al:PRA:2009}%
  \BibitemOpen
  \bibfield  {author} {\bibinfo {author} {\bibfnamefont {Zhi-Yong}\
  \bibnamefont {Wang}}, \bibinfo {author} {\bibfnamefont {Cai-Dong}\
  \bibnamefont {Xiong}},  and \bibinfo {author} {\bibfnamefont
  {Qi}~\bibnamefont {Qiu}},\ }\bibfield  {title} {\enquote {\bibinfo {title}
  {Photon wave function and \emph{Zitterbewegung}},}\ }\href {\doibase
  10.1103/PhysRevA.80.032118} {\bibfield  {journal} {\bibinfo  {journal}
  {Phys.\ Rev.\ A}\ }\textbf {\bibinfo {volume} {80}},\ \bibinfo {pages}
  {032118} (\bibinfo {year} {2009})}\BibitemShut {NoStop}%
\bibitem [{\citenamefont {Mohr}(2010)}]{Mohr:APNY:2010}%
  \BibitemOpen
  \bibfield  {author} {\bibinfo {author} {\bibfnamefont {Peter~J.}\
  \bibnamefont {Mohr}},\ }\bibfield  {title} {\enquote {\bibinfo {title}
  {Solutions of the {Maxwell} equations and photon wave functions},}\ }\href
  {\doibase 10.1016/j.aop.2009.11.007} {\bibfield  {journal} {\bibinfo
  {journal} {Ann.\ Phys. (N.\,Y.)}\ }\textbf {\bibinfo {volume} {325}},\
  \bibinfo {pages} {607--663} (\bibinfo {year} {2010})}\BibitemShut {NoStop}%
\bibitem [{\citenamefont {Aste}(2012)}]{Aste:JGSP:2012}%
  \BibitemOpen
  \bibfield  {author} {\bibinfo {author} {\bibfnamefont {Andreas}\ \bibnamefont
  {Aste}},\ }\bibfield  {title} {\enquote {\bibinfo {title} {Complex
  representation theory of the electromagnetic field},}\ }\href {\doibase
  10.7546/jgsp-28-2012-47-58} {\bibfield  {journal} {\bibinfo  {journal}
  {J.~Geom.\ Sym.\ Phys.}\ }\textbf {\bibinfo {volume} {28}},\ \bibinfo {pages}
  {47--58} (\bibinfo {year} {2012})}\BibitemShut {NoStop}%
\bibitem [{\citenamefont {Red'kov}\ \emph {et~al.}(2012)\citenamefont
  {Red'kov}, \citenamefont {Tokarevskaya}, and \citenamefont
  {Spix}}]{Redkov&al:AACA:2012}%
  \BibitemOpen
  \bibfield  {author} {\bibinfo {author} {\bibfnamefont {V.~M.}\ \bibnamefont
  {Red'kov}}, \bibinfo {author} {\bibfnamefont {N.~G.}\ \bibnamefont
  {Tokarevskaya}},  and \bibinfo {author} {\bibfnamefont {George~J.}\
  \bibnamefont {Spix}},\ }\bibfield  {title} {\enquote {\bibinfo {title}
  {{Majorana-Oppenheimer} approach to {Maxwell} electrodynamics. {Part I}.
  {Minkowski} space},}\ }\href {\doibase 10.1007/s00006-012-0320-1} {\bibfield
  {journal} {\bibinfo  {journal} {Adv.\ Appl.\ Clifford Algeb.}\ }\textbf
  {\bibinfo {volume} {22}},\ \bibinfo {pages} {1129--1149} (\bibinfo {year}
  {2012})}\BibitemShut {NoStop}%
\bibitem [{\citenamefont {Yamamoto}\ \emph {et~al.}(2012)\citenamefont
  {Yamamoto}, \citenamefont {Yamashita}, and \citenamefont
  {Yajima}}]{Yamamoto&al:JPSJ:2012}%
  \BibitemOpen
  \bibfield  {author} {\bibinfo {author} {\bibfnamefont {Takuo}\ \bibnamefont
  {Yamamoto}}, \bibinfo {author} {\bibfnamefont {Shinji}\ \bibnamefont
  {Yamashita}},  and \bibinfo {author} {\bibfnamefont {Satoshi}\ \bibnamefont
  {Yajima}},\ }\bibfield  {title} {\enquote {\bibinfo {title} {Wave function of
  a photon and the appropriate {Lagrangian}},}\ }\href {\doibase
  10.1143/JPSJ.81.024402} {\bibfield  {journal} {\bibinfo  {journal} {J.~Phys.\
  Soc.\ Japan}\ }\textbf {\bibinfo {volume} {81}},\ \bibinfo {pages} {024402}
  (\bibinfo {year} {2012})}\BibitemShut {NoStop}%
\bibitem [{\citenamefont {Bia{\l}ynicki-Birula} and \citenamefont
  {Bia{\l}ynicka-Birula}(2013)}]{Bialynicki-Birula&Bialynicka-Birula:JPA:2013}%
  \BibitemOpen
  \bibfield  {author} {\bibinfo {author} {\bibfnamefont {Iwo}\ \bibnamefont
  {Bia{\l}ynicki-Birula}} and \bibinfo {author} {\bibfnamefont {Zofia}\
  \bibnamefont {Bia{\l}ynicka-Birula}},\ }\bibfield  {title} {\enquote
  {\bibinfo {title} {The role of the {Riemann-Silberstein} vector in classical
  and quantum theories of electromagnetism},}\ }\href {\doibase
  10.1088/1751-8113/46/5/053001} {\bibfield  {journal} {\bibinfo  {journal}
  {J.~Phys.~A: Math.\ Gen.}\ }\textbf {\bibinfo {volume} {46}},\ \bibinfo
  {pages} {053001} (\bibinfo {year} {2013})}\BibitemShut {NoStop}%
\bibitem [{\citenamefont {Barnett}(2014)}]{Barnett:NJP:2014}%
  \BibitemOpen
  \bibfield  {author} {\bibinfo {author} {\bibfnamefont {Stephen~M.}\
  \bibnamefont {Barnett}},\ }\bibfield  {title} {\enquote {\bibinfo {title}
  {Optical {Dirac} equation},}\ }\href {\doibase 10.1088/1367-2630/16/9/093008}
  {\bibfield  {journal} {\bibinfo  {journal} {New J.~Phys.}\ }\textbf {\bibinfo
  {volume} {16}},\ \bibinfo {pages} {093008} (\bibinfo {year}
  {2014})}\BibitemShut {NoStop}%
\bibitem [{\citenamefont {Dressel}\ \emph {et~al.}(2014)\citenamefont
  {Dressel}, \citenamefont {Bliokh}, and \citenamefont
  {Nori}}]{Dressel&al:ARXIV:2014}%
  \BibitemOpen
  \bibfield  {author} {\bibinfo {author} {\bibfnamefont {Justin}\ \bibnamefont
  {Dressel}}, \bibinfo {author} {\bibfnamefont {Konstantin~Y.}\ \bibnamefont
  {Bliokh}},  and \bibinfo {author} {\bibfnamefont {Franco}\ \bibnamefont
  {Nori}},\ }\href {http://arxiv.org/abs/1411.5002} {\enquote {\bibinfo {title}
  {Spacetime algebra as a powerful tool for electromagnetism},}\ } (\bibinfo
  {year} {2014}),\ \Eprint {http://arxiv.org/abs/1411.5002}
  {arXiv.org:1411.5002 [physics.optics]} \BibitemShut {NoStop}%
\bibitem [{\citenamefont {Cohen-Tannoudji}\ \emph {et~al.}(1997)\citenamefont
  {Cohen-Tannoudji}, \citenamefont {Dupont-Roc}, and \citenamefont
  {Grynberg}}]{Cohen-Tannoudji&al:Book:1997}%
  \BibitemOpen
  \bibfield  {author} {\bibinfo {author} {\bibfnamefont {Claude}\ \bibnamefont
  {Cohen-Tannoudji}}, \bibinfo {author} {\bibfnamefont {Jaques}\ \bibnamefont
  {Dupont-Roc}},  and \bibinfo {author} {\bibfnamefont {Gilbert}\
  \bibnamefont {Grynberg}},\ }\href@noop {} {\emph {\bibinfo {title} {Photons
  and Atoms. Introduction to Quantum Electrodynamics}}}\ (\bibinfo  {publisher}
  {Wiley \& Sons},\ \bibinfo {address} {New York, NY, USA},\ \bibinfo {year}
  {1997})\ \bibinfo {note} {{W}iley Professional Paperback Edition}\BibitemShut
  {NoStop}%
\bibitem [{\citenamefont {Li}(2013)}]{Li:OE:2013}%
  \BibitemOpen
  \bibfield  {author} {\bibinfo {author} {\bibfnamefont {Hongrui}\ \bibnamefont
  {Li}},\ }\bibfield  {title} {\enquote {\bibinfo {title} {Evanescent wave of a
  single photon},}\ }\href {\doibase 10.1117/1.OE.52.7.074103} {\bibfield
  {journal} {\bibinfo  {journal} {Opt. Express}\ }\textbf {\bibinfo {volume}
  {52}},\ \bibinfo {pages} {074103--074103} (\bibinfo {year}
  {2013})}\BibitemShut {NoStop}%
\bibitem [{\citenamefont {Andrews}(2014)}]{Andrews:JN:2014}%
  \BibitemOpen
  \bibfield  {author} {\bibinfo {author} {\bibfnamefont {David~L.}\
  \bibnamefont {Andrews}},\ }\bibfield  {title} {\enquote {\bibinfo {title}
  {Photon-based and classical descriptions in nanophotonics: a review},}\
  }\href {\doibase 10.1117/1.JNP.8.081599} {\bibfield  {journal} {\bibinfo
  {journal} {J.~Nanophot.}\ }\textbf {\bibinfo {volume} {8}},\ \bibinfo {pages}
  {081599} (\bibinfo {year} {2014})}\BibitemShut {NoStop}%
\bibitem [{\citenamefont {Dyson}(1999)}]{Dyson:Maxwell:1999}%
  \BibitemOpen
  \bibfield  {author} {\bibinfo {author} {\bibfnamefont {Freeman~J.}\
  \bibnamefont {Dyson}},\ }\bibfield  {title} {\enquote {\bibinfo {title} {Why
  is {Maxwell's Theory} so hard to understand?}}\ }in\ \href
  {http://www.clerkmaxwellfoundation.org/DysonFreemanArticle.pdf} {\emph
  {\bibinfo {booktitle} {James Clerk Maxwell Commemorative Booklet}}}\
  (\bibinfo  {publisher} {James Clerk Maxwell Foundation},\ \bibinfo {address}
  {Edinburgh, Scotland, UK},\ \bibinfo {year} {1999})\ \bibinfo {note}
  {\url{http://www.clerkmaxwellfoundation.org/DysonFreemanArticle.pdf}}\BibitemShut
  {NoStop}%
\bibitem [{\citenamefont {Siegel}(2003)}]{Siegel:Book:2003}%
  \BibitemOpen
  \bibfield  {author} {\bibinfo {author} {\bibfnamefont {Daniel~M.}\
  \bibnamefont {Siegel}},\ }\href@noop {} {\emph {\bibinfo {title} {Innovation
  in Maxwell's Electromagnetic Theory: Molecular Vortices, Displacement
  Current, and Light}}}\ (\bibinfo  {publisher} {Cambridge University Press},\
  \bibinfo {address} {Cambridge, UK},\ \bibinfo {year} {2003})\BibitemShut
  {NoStop}%
\bibitem [{\citenamefont {Sadovskii}(1899)}]{Sadovskii:1899}%
  \BibitemOpen
  \bibfield  {author} {\bibinfo {author} {\bibfnamefont {A.~I.}\ \bibnamefont
  {Sadovskii}},\ }\bibfield  {title} {\enquote {\bibinfo {title}
  {Ponderomotorniya djstviya elektromagnitnykh' i svtovykh' voln' na
  kristally},}\ }\href@noop {} {\bibfield  {journal} {\bibinfo  {journal}
  {Uchen. Zapis. Imp. Yurev. Univ.}\ }\textbf {\bibinfo {volume} {52}},\
  \bibinfo {pages} {1--126} (\bibinfo {year} {1899})},\ \bibinfo {note}
  {({English} translation: Ponderomotive action of electromagnetic and light
  waves on crystals)}\BibitemShut {NoStop}%
\bibitem [{\citenamefont {Poynting}(1909)}]{Poynting:PRSL:1909}%
  \BibitemOpen
  \bibfield  {author} {\bibinfo {author} {\bibfnamefont {J.~H.}\ \bibnamefont
  {Poynting}},\ }\bibfield  {title} {\enquote {\bibinfo {title} {The wave
  motion of a revolving shaft, and a suggestion as to the angular momentum in a
  beam of circularly polarised light},}\ }\href@noop {} {\bibfield  {journal}
  {\bibinfo  {journal} {Proc.\ Roy.\ Soc.\ London}\ }\textbf {\bibinfo {volume}
  {A 82}},\ \bibinfo {pages} {560--567} (\bibinfo {year} {1909})}\BibitemShut
  {NoStop}%
\bibitem [{\citenamefont {Abraham}(1914)}]{Abraham:PZ:1914}%
  \BibitemOpen
  \bibfield  {author} {\bibinfo {author} {\bibfnamefont {Max}\ \bibnamefont
  {Abraham}},\ }\bibfield  {title} {\enquote {\bibinfo {title} {{D}er
  {D}rehimpuls des {L}ichtes},}\ }\href@noop {} {\bibfield  {journal} {\bibinfo
   {journal} {Physik. Zeitschr.}\ }\textbf {\bibinfo {volume} {XV}},\ \bibinfo
  {pages} {914--918} (\bibinfo {year} {1914})}\BibitemShut {NoStop}%
\bibitem [{\citenamefont {Bateman}(1926)}]{Bateman:PR:1926}%
  \BibitemOpen
  \bibfield  {author} {\bibinfo {author} {\bibfnamefont {H.}~\bibnamefont
  {Bateman}},\ }\bibfield  {title} {\enquote {\bibinfo {title} {The radiation
  of energy and angular momentum},}\ }\href {\doibase 10.1103/PhysRev.27.606}
  {\bibfield  {journal} {\bibinfo  {journal} {Phys.\ Rev.}\ }\textbf {\bibinfo
  {volume} {27}},\ \bibinfo {pages} {606--617} (\bibinfo {year}
  {1926})}\BibitemShut {NoStop}%
\bibitem [{\citenamefont {Heitler}(1954)}]{Heitler:Book:1954}%
  \BibitemOpen
  \bibfield  {author} {\bibinfo {author} {\bibfnamefont {Walter}\ \bibnamefont
  {Heitler}},\ }\href@noop {} {\emph {\bibinfo {title} {The Quantum Theory of
  Radiation}}},\ \bibinfo {edition} {3rd}\ ed.,\ The International Series of
  Monographs on Physics\ (\bibinfo  {publisher} {Clarendon Press},\ \bibinfo
  {address} {Oxford, UK},\ \bibinfo {year} {1954})\BibitemShut {NoStop}%
\bibitem [{\citenamefont {Bogolyubov} and \citenamefont
  {Shirkov}(1959)}]{Bogolyubov&Shirkov:Book:1959}%
  \BibitemOpen
  \bibfield  {author} {\bibinfo {author} {\bibfnamefont {N.~N.}\ \bibnamefont
  {Bogolyubov}} and \bibinfo {author} {\bibfnamefont {D.~V.}\ \bibnamefont
  {Shirkov}},\ }\href@noop {} {\emph {\bibinfo {title} {Introduction to the
  Theory of Quantized Fields}}}\ (\bibinfo  {publisher} {Interscience},\
  \bibinfo {address} {New York},\ \bibinfo {year} {1959})\BibitemShut {NoStop}%
\bibitem [{\citenamefont {Messiah}(1970)}]{Messiah:Book:1970}%
  \BibitemOpen
  \bibfield  {author} {\bibinfo {author} {\bibfnamefont {Albert}\ \bibnamefont
  {Messiah}},\ }\href@noop {} {\emph {\bibinfo {title} {Quantum Mechanics}}}\
  (\bibinfo  {publisher} {North-Holland},\ \bibinfo {address} {Amsterdam, NL},\
  \bibinfo {year} {1970})\BibitemShut {NoStop}%
\bibitem [{\citenamefont {Eyges}(1972)}]{Eyges:Book:1972}%
  \BibitemOpen
  \bibfield  {author} {\bibinfo {author} {\bibfnamefont {Leonard}\ \bibnamefont
  {Eyges}},\ }\href@noop {} {\emph {\bibinfo {title} {The Classical
  Electromagnetic Field}}}\ (\bibinfo  {publisher} {Dover Publications},\
  \bibinfo {address} {New York, NY, USA},\ \bibinfo {year} {1972})\BibitemShut
  {NoStop}%
\bibitem [{\citenamefont {Landau} and \citenamefont
  {Lifshitz}(1975)}]{Landau&Lifshitz:Book:1975}%
  \BibitemOpen
  \bibfield  {author} {\bibinfo {author} {\bibfnamefont {L.~D.}\ \bibnamefont
  {Landau}} and \bibinfo {author} {\bibfnamefont {E.~M.}\ \bibnamefont
  {Lifshitz}},\ }\href@noop {} {\emph {\bibinfo {title} {The Classical Theory
  of Fields}}},\ \bibinfo {series} {Course of Theoretical Physics},
  Vol.~\bibinfo {volume} {2}\ (\bibinfo  {publisher} {Pergamon},\ \bibinfo
  {address} {Oxford, UK},\ \bibinfo {year} {1975})\BibitemShut {NoStop}%
\bibitem [{\citenamefont {Berestetskii}\ \emph {et~al.}(1989)\citenamefont
  {Berestetskii}, \citenamefont {Lifshitz}, and \citenamefont
  {Pitaevskii}}]{Berestetskii&al:Book:1989}%
  \BibitemOpen
  \bibfield  {author} {\bibinfo {author} {\bibfnamefont {V.~B.}\ \bibnamefont
  {Berestetskii}}, \bibinfo {author} {\bibfnamefont {E.~M.}\ \bibnamefont
  {Lifshitz}},  and \bibinfo {author} {\bibfnamefont {L.~P.}\ \bibnamefont
  {Pitaevskii}},\ }\href@noop {} {\emph {\bibinfo {title} {Quantum
  Electrodynamics}}},\ \bibinfo {edition} {2nd}\ ed.,\ \bibinfo {series}
  {Course of Theoretical Physics}, Vol.~\bibinfo {volume} {4}\ (\bibinfo
  {publisher} {Pergamon Press},\ \bibinfo {address} {Oxford, UK},\ \bibinfo
  {year} {1989})\BibitemShut {NoStop}%
\bibitem [{\citenamefont {Mandel} and \citenamefont
  {Wolf}(1995)}]{Mandel&Wolf:Book:1995}%
  \BibitemOpen
  \bibfield  {author} {\bibinfo {author} {\bibfnamefont {Leonard}\ \bibnamefont
  {Mandel}} and \bibinfo {author} {\bibfnamefont {Emil}\ \bibnamefont
  {Wolf}},\ }\href@noop {} {\emph {\bibinfo {title} {Optical Coherence and
  Quantum Optics}}}\ (\bibinfo  {publisher} {Cambridge University Press},\
  \bibinfo {address} {New York, NY, USA},\ \bibinfo {year} {1995})\BibitemShut
  {NoStop}%
\bibitem [{\citenamefont {Griffiths}(1999)}]{Griffiths:Book:1999}%
  \BibitemOpen
  \bibfield  {author} {\bibinfo {author} {\bibfnamefont {David~J.}\
  \bibnamefont {Griffiths}},\ }\href@noop {} {\emph {\bibinfo {title}
  {Introduction to Electrodynamics}}},\ \bibinfo {edition} {3rd}\ ed.\
  (\bibinfo  {publisher} {Prentice-Hall International},\ \bibinfo {address}
  {London, UK},\ \bibinfo {year} {1999})\BibitemShut {NoStop}%
\bibitem [{\citenamefont {Rohrlich}(2007)}]{Rohrlich:Book:2007}%
  \BibitemOpen
  \bibfield  {author} {\bibinfo {author} {\bibfnamefont {Fritz}\ \bibnamefont
  {Rohrlich}},\ }\href@noop {} {\emph {\bibinfo {title} {Classical Charged
  Particles}}},\ \bibinfo {edition} {3rd}\ ed.\ (\bibinfo  {publisher} {World
  Scientific},\ \bibinfo {address} {Singapore},\ \bibinfo {year}
  {2007})\BibitemShut {NoStop}%
\bibitem [{\citenamefont {Gibson}\ \emph
  {et~al.}(2004{\natexlab{a}})\citenamefont {Gibson}, \citenamefont {Courtial},
  \citenamefont {Padgett}, \citenamefont {Vasnetsov}, \citenamefont {Pas{'}ko},
  \citenamefont {Barnett}, and \citenamefont
  {Franke-Arnold}}]{Gibson&al:OE:2004}%
  \BibitemOpen
  \bibfield  {author} {\bibinfo {author} {\bibfnamefont {Graham}\ \bibnamefont
  {Gibson}}, \bibinfo {author} {\bibfnamefont {Johannes}\ \bibnamefont
  {Courtial}}, \bibinfo {author} {\bibfnamefont {Miles~J.}\ \bibnamefont
  {Padgett}}, \bibinfo {author} {\bibfnamefont {Mikhail}\ \bibnamefont
  {Vasnetsov}}, \bibinfo {author} {\bibfnamefont {Valeriy}\ \bibnamefont
  {Pas{'}ko}}, \bibinfo {author} {\bibfnamefont {Stephen~M.}\ \bibnamefont
  {Barnett}},  and \bibinfo {author} {\bibfnamefont {Sonja}\ \bibnamefont
  {Franke-Arnold}},\ }\bibfield  {title} {\enquote {\bibinfo {title}
  {Free-space information transfer using light beams carrying orbital angular
  momentum},}\ }\href {\doibase 10.1364/OPEX.12.005448} {\bibfield  {journal}
  {\bibinfo  {journal} {Opt. Express}\ }\textbf {\bibinfo {volume} {12}},\
  \bibinfo {pages} {5448--5456} (\bibinfo {year}
  {2004}{\natexlab{a}})}\BibitemShut {NoStop}%
\bibitem [{\citenamefont {Thid{\'e}}\ \emph {et~al.}(2007)\citenamefont
  {Thid{\'e}}, \citenamefont {Then}, \citenamefont {Sj{\"o}holm}, \citenamefont
  {Palmer}, \citenamefont {Bergman}, \citenamefont {Carozzi}, \citenamefont
  {Istomin}, \citenamefont {Ibragimov}, and \citenamefont
  {Khamitova}}]{Thide&al:PRL:2007}%
  \BibitemOpen
  \bibfield  {author} {\bibinfo {author} {\bibfnamefont {B.}~\bibnamefont
  {Thid{\'e}}}, \bibinfo {author} {\bibfnamefont {H.}~\bibnamefont {Then}},
  \bibinfo {author} {\bibfnamefont {J.}~\bibnamefont {Sj{\"o}holm}}, \bibinfo
  {author} {\bibfnamefont {K.}~\bibnamefont {Palmer}}, \bibinfo {author}
  {\bibfnamefont {J.}~\bibnamefont {Bergman}}, \bibinfo {author} {\bibfnamefont
  {T.~D.}\ \bibnamefont {Carozzi}}, \bibinfo {author} {\bibfnamefont {Ya.~N.}\
  \bibnamefont {Istomin}}, \bibinfo {author} {\bibfnamefont {N.~H.}\
  \bibnamefont {Ibragimov}},  and \bibinfo {author} {\bibfnamefont
  {R.}~\bibnamefont {Khamitova}},\ }\bibfield  {title} {\enquote {\bibinfo
  {title} {Utilization of photon orbital angular momentum in the low-frequency
  radio domain},}\ }\href {\doibase 10.1103/PhysRevLett.99.087701} {\bibfield
  {journal} {\bibinfo  {journal} {Phys.\ Rev.\ Lett.}\ }\textbf {\bibinfo
  {volume} {99}},\ \bibinfo {pages} {087701(4)} (\bibinfo {year}
  {2007})}\BibitemShut {NoStop}%
\bibitem [{\citenamefont {Franke-Arnold}\ \emph {et~al.}(2008)\citenamefont
  {Franke-Arnold}, \citenamefont {Allen}, and \citenamefont
  {Padgett}}]{Franke-Arnold&al:LPR:2008}%
  \BibitemOpen
  \bibfield  {author} {\bibinfo {author} {\bibfnamefont {S.}~\bibnamefont
  {Franke-Arnold}}, \bibinfo {author} {\bibfnamefont {L.}~\bibnamefont
  {Allen}},  and \bibinfo {author} {\bibfnamefont {M.}~\bibnamefont
  {Padgett}},\ }\bibfield  {title} {\enquote {\bibinfo {title} {Advances in
  optical angular momentum},}\ }\href {\doibase 10.1103/PhysRevA.82.033822}
  {\bibfield  {journal} {\bibinfo  {journal} {Laser \& Photon.\ Rev.}\ }\textbf
  {\bibinfo {volume} {2}},\ \bibinfo {pages} {299--313} (\bibinfo {year}
  {2008})}\BibitemShut {NoStop}%
\bibitem [{\citenamefont {Tamburini}\ \emph
  {et~al.}(2011{\natexlab{a}})\citenamefont {Tamburini}, \citenamefont
  {Thid\'e}, \citenamefont {Molina-Terriza}, and \citenamefont
  {Anzolin}}]{Tamburini&al:NPHY:2011}%
  \BibitemOpen
  \bibfield  {author} {\bibinfo {author} {\bibfnamefont {Fabrizio}\
  \bibnamefont {Tamburini}}, \bibinfo {author} {\bibfnamefont {Bo}~\bibnamefont
  {Thid\'e}}, \bibinfo {author} {\bibfnamefont {Gabriel}\ \bibnamefont
  {Molina-Terriza}},  and \bibinfo {author} {\bibfnamefont {Gabriele}\
  \bibnamefont {Anzolin}},\ }\bibfield  {title} {\enquote {\bibinfo {title}
  {Twisting of light around rotating black holes},}\ }\href {\doibase
  10.1038/NPHYS1907} {\bibfield  {journal} {\bibinfo  {journal} {Nature Phys.}\
  }\textbf {\bibinfo {volume} {7}},\ \bibinfo {pages} {195--197} (\bibinfo
  {year} {2011}{\natexlab{a}})}\BibitemShut {NoStop}%
\bibitem [{\citenamefont {Tamburini}\ \emph
  {et~al.}(2011{\natexlab{b}})\citenamefont {Tamburini}, \citenamefont {Mari},
  \citenamefont {Thid\'e}, \citenamefont {Barbieri}, and \citenamefont
  {Romanato}}]{Tamburini&al:APL:2011}%
  \BibitemOpen
  \bibfield  {author} {\bibinfo {author} {\bibfnamefont {Fabrizio}\
  \bibnamefont {Tamburini}}, \bibinfo {author} {\bibfnamefont {Elettra}\
  \bibnamefont {Mari}}, \bibinfo {author} {\bibfnamefont {Bo}~\bibnamefont
  {Thid\'e}}, \bibinfo {author} {\bibfnamefont {Cesare}\ \bibnamefont
  {Barbieri}},  and \bibinfo {author} {\bibfnamefont {Filippo}\ \bibnamefont
  {Romanato}},\ }\bibfield  {title} {\enquote {\bibinfo {title} {Experimental
  verification of photon angular momentum and vorticity with radio
  techniques},}\ }\href {\doibase 10.1063/1.3659466} {\bibfield  {journal}
  {\bibinfo  {journal} {Appl.\ Phys.\ Lett.}\ }\textbf {\bibinfo {volume}
  {99}},\ \bibinfo {pages} {204102} (\bibinfo {year}
  {2011}{\natexlab{b}})}\BibitemShut {NoStop}%
\bibitem [{\citenamefont {Tamburini}\ \emph
  {et~al.}(2012{\natexlab{a}})\citenamefont {Tamburini}, \citenamefont {Mari},
  \citenamefont {Sponselli}, \citenamefont {Thid\'e}, \citenamefont
  {Bianchini}, and \citenamefont {Romanato}}]{Tamburini&al:NJP:2012}%
  \BibitemOpen
  \bibfield  {author} {\bibinfo {author} {\bibfnamefont {Fabrizio}\
  \bibnamefont {Tamburini}}, \bibinfo {author} {\bibfnamefont {Elettra}\
  \bibnamefont {Mari}}, \bibinfo {author} {\bibfnamefont {Anna}\ \bibnamefont
  {Sponselli}}, \bibinfo {author} {\bibfnamefont {Bo}~\bibnamefont {Thid\'e}},
  \bibinfo {author} {\bibfnamefont {Antonio}\ \bibnamefont {Bianchini}},  and
  \bibinfo {author} {\bibfnamefont {Filippo}\ \bibnamefont {Romanato}},\
  }\bibfield  {title} {\enquote {\bibinfo {title} {Encoding many channels on
  the same frequency through radio vorticity: first experimental test},}\
  }\href {\doibase 10.1088/1367-2630/14/3/033001} {\bibfield  {journal}
  {\bibinfo  {journal} {New J.~Phys.}\ }\textbf {\bibinfo {volume} {14}},\
  \bibinfo {pages} {03301} (\bibinfo {year} {2012}{\natexlab{a}})}\BibitemShut
  {NoStop}%
\bibitem [{\citenamefont {Tamburini}\ \emph
  {et~al.}(2012{\natexlab{b}})\citenamefont {Tamburini}, \citenamefont
  {Thid{\'e}}, \citenamefont {Mari}, \citenamefont {Sponselli}, \citenamefont
  {Bianchini}, and \citenamefont {Romanato}}]{Tamburini&al:NJP:2012a}%
  \BibitemOpen
  \bibfield  {author} {\bibinfo {author} {\bibfnamefont {F.}~\bibnamefont
  {Tamburini}}, \bibinfo {author} {\bibfnamefont {B.}~\bibnamefont
  {Thid{\'e}}}, \bibinfo {author} {\bibfnamefont {E.}~\bibnamefont {Mari}},
  \bibinfo {author} {\bibfnamefont {A.}~\bibnamefont {Sponselli}}, \bibinfo
  {author} {\bibfnamefont {A.}~\bibnamefont {Bianchini}},  and \bibinfo
  {author} {\bibfnamefont {F.}~\bibnamefont {Romanato}},\ }\bibfield  {title}
  {\enquote {\bibinfo {title} {Reply to comment on {`Encoding} many channels on
  the same frequency through radio vorticity: first experimental test'},}\
  }\href {\doibase 10.1088/1367-2630/14/11/118002} {\bibfield  {journal}
  {\bibinfo  {journal} {New J.~Phys.}\ }\textbf {\bibinfo {volume} {14}},\
  \bibinfo {pages} {118002} (\bibinfo {year} {2012}{\natexlab{b}})}\BibitemShut
  {NoStop}%
\bibitem [{\citenamefont {Tamburini}\ \emph {et~al.}(2015)\citenamefont
  {Tamburini}, \citenamefont {Mari}, \citenamefont {Parisi}, \citenamefont
  {Spinello}, \citenamefont {Oldoni}, \citenamefont {Ravanelli}, \citenamefont
  {Coassini}, \citenamefont {Someda}, \citenamefont {Thid{\'e}}, and
  \citenamefont {Romanato}}]{Tamburini&al:RS:2015}%
  \BibitemOpen
  \bibfield  {author} {\bibinfo {author} {\bibfnamefont {F.}~\bibnamefont
  {Tamburini}}, \bibinfo {author} {\bibfnamefont {E.}~\bibnamefont {Mari}},
  \bibinfo {author} {\bibfnamefont {G.}~\bibnamefont {Parisi}}, \bibinfo
  {author} {\bibfnamefont {F.}~\bibnamefont {Spinello}}, \bibinfo {author}
  {\bibfnamefont {M.}~\bibnamefont {Oldoni}}, \bibinfo {author} {\bibfnamefont
  {R.~A.}\ \bibnamefont {Ravanelli}}, \bibinfo {author} {\bibfnamefont
  {P.}~\bibnamefont {Coassini}}, \bibinfo {author} {\bibfnamefont {Carlo~G}\
  \bibnamefont {Someda}}, \bibinfo {author} {\bibfnamefont {B.}~\bibnamefont
  {Thid{\'e}}},  and \bibinfo {author} {\bibfnamefont {F.}~\bibnamefont
  {Romanato}},\ }\bibfield  {title} {\enquote {\bibinfo {title} {Tripling the
  capacity of a point-to-point radio link by using electromagnetic vortices},}\
  }\href {\doibase 10.1002/2015RS005662} {\bibfield  {journal} {\bibinfo
  {journal} {Radio Sci.}\ }\textbf {\bibinfo {volume} {50}} (\bibinfo {year}
  {2015}),\ 10.1002/2015RS005662}\BibitemShut {NoStop}%
\bibitem [{\citenamefont {Gibson}\ \emph
  {et~al.}(2004{\natexlab{b}})\citenamefont {Gibson}, \citenamefont {Courtial},
  \citenamefont {Barnett}, and \citenamefont
  {Franke-Arnold}}]{Gibson&al:SPIE:2004}%
  \BibitemOpen
  \bibfield  {author} {\bibinfo {author} {\bibfnamefont {Graham}\ \bibnamefont
  {Gibson}}, \bibinfo {author} {\bibfnamefont {Johannes}\ \bibnamefont
  {Courtial}}, \bibinfo {author} {\bibfnamefont {Mikhail Vasnetsov~Stephen}\
  \bibnamefont {Barnett}},  and \bibinfo {author} {\bibfnamefont {Sonja}\
  \bibnamefont {Franke-Arnold}},\ }\bibfield  {title} {\enquote {\bibinfo
  {title} {Increasing the data density of free-space optical communications
  using orbital angular momentum},}\ }in\ \href {\doibase 10.1117/12.557176}
  {\emph {\bibinfo {booktitle} {Free-Space Laser Communications IV}}},\
  \bibinfo {series} {Proceedings of {SPIE}}, Vol.\ \bibinfo {volume} {5550},\
  \bibinfo {editor} {edited by\ \bibinfo {editor} {\bibfnamefont {Jennifer~C.}\
  \bibnamefont {Rickling}} and \bibinfo {editor} {\bibfnamefont {David~G.}\
  \bibnamefont {Voetz}}}\ (\bibinfo {address} {Denver, {CO,} {USA}},\ \bibinfo
  {year} {2004})\ pp.\ \bibinfo {pages} {367--373}\BibitemShut {NoStop}%
\bibitem [{\citenamefont {Wang}\ \emph {et~al.}(2012)\citenamefont {Wang},
  \citenamefont {Yang}, \citenamefont {Fazal}, \citenamefont {Ahmed},
  \citenamefont {Yan}, \citenamefont {Huang}, \citenamefont {Ren},
  \citenamefont {Yue}, \citenamefont {Dolinar}, \citenamefont {Tur}, and
  \citenamefont {Willner}}]{Wang&al:NPHO:2012}%
  \BibitemOpen
  \bibfield  {author} {\bibinfo {author} {\bibfnamefont {Jian}\ \bibnamefont
  {Wang}}, \bibinfo {author} {\bibfnamefont {Jeng-Yuan}\ \bibnamefont {Yang}},
  \bibinfo {author} {\bibfnamefont {Irfan~M.}\ \bibnamefont {Fazal}}, \bibinfo
  {author} {\bibfnamefont {Nisar}\ \bibnamefont {Ahmed}}, \bibinfo {author}
  {\bibfnamefont {Yan}\ \bibnamefont {Yan}}, \bibinfo {author} {\bibfnamefont
  {Hao}\ \bibnamefont {Huang}}, \bibinfo {author} {\bibfnamefont {Yongxiong}\
  \bibnamefont {Ren}}, \bibinfo {author} {\bibfnamefont {Yang}\ \bibnamefont
  {Yue}}, \bibinfo {author} {\bibfnamefont {Samuel}\ \bibnamefont {Dolinar}},
  \bibinfo {author} {\bibfnamefont {Moshe}\ \bibnamefont {Tur}},  and
  \bibinfo {author} {\bibfnamefont {Alan~E.}\ \bibnamefont {Willner}},\
  }\bibfield  {title} {\enquote {\bibinfo {title} {Terabit free-space data
  transmission employing orbital angular momentum multiplexing},}\ }\href
  {\doibase 10.1038/nphoton.2012.138} {\bibfield  {journal} {\bibinfo
  {journal} {Nature Photon.}\ }\textbf {\bibinfo {volume} {6}},\ \bibinfo
  {pages} {488--496} (\bibinfo {year} {2012})}\BibitemShut {NoStop}%
\bibitem [{\citenamefont {Litchinitser}(2012)}]{Litchinitser:S:2012}%
  \BibitemOpen
  \bibfield  {author} {\bibinfo {author} {\bibfnamefont {Natalia~M.}\
  \bibnamefont {Litchinitser}},\ }\bibfield  {title} {\enquote {\bibinfo
  {title} {Structured light meets structured matter},}\ }\href {\doibase
  10.1126/science.1226204} {\bibfield  {journal} {\bibinfo  {journal}
  {Science}\ }\textbf {\bibinfo {volume} {337}},\ \bibinfo {pages} {1054--1055}
  (\bibinfo {year} {2012})}\BibitemShut {NoStop}%
\bibitem [{\citenamefont {Bozinovic}\ \emph {et~al.}(2013)\citenamefont
  {Bozinovic}, \citenamefont {Yue}, \citenamefont {Ren}, \citenamefont {Tur},
  \citenamefont {Kristensen}, \citenamefont {Huang}, \citenamefont {Willner},\
  and\ \citenamefont {Ramachandran}}]{Bozinovic&al:S:2013}%
  \BibitemOpen
  \bibfield  {author} {\bibinfo {author} {\bibfnamefont {Nenad}\ \bibnamefont
  {Bozinovic}}, \bibinfo {author} {\bibfnamefont {Yang}\ \bibnamefont {Yue}},
  \bibinfo {author} {\bibfnamefont {Yongxiong}\ \bibnamefont {Ren}}, \bibinfo
  {author} {\bibfnamefont {Moshe}\ \bibnamefont {Tur}}, \bibinfo {author}
  {\bibfnamefont {Poul}\ \bibnamefont {Kristensen}}, \bibinfo {author}
  {\bibfnamefont {Hao}\ \bibnamefont {Huang}}, \bibinfo {author} {\bibfnamefont
  {Alan~E.}\ \bibnamefont {Willner}},  and \bibinfo {author} {\bibfnamefont
  {Siddharth}\ \bibnamefont {Ramachandran}},\ }\bibfield  {title} {\enquote
  {\bibinfo {title} {Terabit-scale orbital angular momentum mode division
  multiplexing in fibers},}\ }\href {\doibase 10.1126/science.1237861}
  {\bibfield  {journal} {\bibinfo  {journal} {Science}\ }\textbf {\bibinfo
  {volume} {340}},\ \bibinfo {pages} {1545--1548} (\bibinfo {year}
  {2013})}\BibitemShut {NoStop}%
\bibitem [{\citenamefont {Mahmouli} and \citenamefont
  {Walker}(2013)}]{Mahmouli&Walker:IEEEWCL:2013}%
  \BibitemOpen
  \bibfield  {author} {\bibinfo {author} {\bibfnamefont {F.~E.}\ \bibnamefont
  {Mahmouli}} and \bibinfo {author} {\bibfnamefont {S.~D.}\ \bibnamefont
  {Walker}},\ }\bibfield  {title} {\enquote {\bibinfo {title} {4-{Gbps}
  uncompressed video transmission over a 60-{GHz} orbital angular momentum
  wireless channel},}\ }\href {\doibase 10.1109/WCL.2013.012513.120686}
  {\bibfield  {journal} {\bibinfo  {journal} {IEEE Wirel.\ Comm.\ Lett.}\
  }\textbf {\bibinfo {volume} {2}},\ \bibinfo {pages} {223--226} (\bibinfo
  {year} {2013})}\BibitemShut {NoStop}%
\bibitem [{\citenamefont {Willner}(2013)}]{Willner:MOC:2013}%
  \BibitemOpen
  \bibfield  {author} {\bibinfo {author} {\bibfnamefont {A.E.}\ \bibnamefont
  {Willner}},\ }\bibfield  {title} {\enquote {\bibinfo {title} {Tb/s optical
  communications using orbital angular momentum},}\ }in\ \href@noop {} {\emph
  {\bibinfo {booktitle} {Microoptics Conference ({MOC}), 2013 18th}}}\
  (\bibinfo {year} {2013})\ pp.\ \bibinfo {pages} {1--2}\BibitemShut {NoStop}%
\bibitem [{\citenamefont {Ren}\ \emph {et~al.}(2014)\citenamefont {Ren},
  \citenamefont {Li}, \citenamefont {Xie}, \citenamefont {Yan}, \citenamefont
  {Cao}, \citenamefont {Huang}, \citenamefont {Ahemd}, \citenamefont {Lavery},
  \citenamefont {Zhao}, \citenamefont {Zhang}, \citenamefont {Tur},
  \citenamefont {Padgett}, \citenamefont {Caire}, \citenamefont {Molisch},\
  and\ \citenamefont {Willner}}]{Ren&al:GLOBECOM:2014}%
  \BibitemOpen
  \bibfield  {author} {\bibinfo {author} {\bibfnamefont {Yongxiong}\
  \bibnamefont {Ren}}, \bibinfo {author} {\bibfnamefont {Long}\ \bibnamefont
  {Li}}, \bibinfo {author} {\bibfnamefont {Guodong}\ \bibnamefont {Xie}},
  \bibinfo {author} {\bibfnamefont {Yan}\ \bibnamefont {Yan}}, \bibinfo
  {author} {\bibfnamefont {Yinwen}\ \bibnamefont {Cao}}, \bibinfo {author}
  {\bibfnamefont {Hao}\ \bibnamefont {Huang}}, \bibinfo {author} {\bibfnamefont
  {N.}~\bibnamefont {Ahemd}}, \bibinfo {author} {\bibfnamefont {M.J.}\
  \bibnamefont {Lavery}}, \bibinfo {author} {\bibfnamefont {Zhe}\ \bibnamefont
  {Zhao}}, \bibinfo {author} {\bibfnamefont {Chongfu}\ \bibnamefont {Zhang}},
  \bibinfo {author} {\bibfnamefont {M.}~\bibnamefont {Tur}}, \bibinfo {author}
  {\bibfnamefont {M.}~\bibnamefont {Padgett}}, \bibinfo {author} {\bibfnamefont
  {G.}~\bibnamefont {Caire}}, \bibinfo {author} {\bibfnamefont {A.F.}\
  \bibnamefont {Molisch}},  and \bibinfo {author} {\bibfnamefont {A.E.}\
  \bibnamefont {Willner}},\ }\bibfield  {title} {\enquote {\bibinfo {title}
  {Experimental demonstration of 16 gbit/s millimeter-wave communications using
  {MIMO} processing of 2 {OAM} modes on each of two transmitter/receiver
  antenna apertures},}\ }in\ \href {\doibase 10.1109/GLOCOM.2014.7037403}
  {\emph {\bibinfo {booktitle} {2014 {IEEE} Global Communications Conference
  ({GLOBECOM})}}}\ (\bibinfo {year} {2014})\ pp.\ \bibinfo {pages}
  {3821--3826}\BibitemShut {NoStop}%
\bibitem [{\citenamefont {Xie}\ \emph {et~al.}(2014)\citenamefont {Xie},
  \citenamefont {Li}, \citenamefont {Ren}, \citenamefont {Huang}, \citenamefont
  {Yan}, \citenamefont {Ahmed}, \citenamefont {Zhao}, \citenamefont {Lavery},
  \citenamefont {Ashrafi}, \citenamefont {Ashrafi}, \citenamefont {Tur},
  \citenamefont {Molisch}, and \citenamefont {Willner}}]{Xie&al:GC:2014}%
  \BibitemOpen
  \bibfield  {author} {\bibinfo {author} {\bibfnamefont {Guodong}\ \bibnamefont
  {Xie}}, \bibinfo {author} {\bibfnamefont {Long}\ \bibnamefont {Li}}, \bibinfo
  {author} {\bibfnamefont {Yongxiong}\ \bibnamefont {Ren}}, \bibinfo {author}
  {\bibfnamefont {Hao}\ \bibnamefont {Huang}}, \bibinfo {author} {\bibfnamefont
  {Yan}\ \bibnamefont {Yan}}, \bibinfo {author} {\bibfnamefont
  {N.}~\bibnamefont {Ahmed}}, \bibinfo {author} {\bibfnamefont {Zhe}\
  \bibnamefont {Zhao}}, \bibinfo {author} {\bibfnamefont {M.P.J.}\ \bibnamefont
  {Lavery}}, \bibinfo {author} {\bibfnamefont {N.}~\bibnamefont {Ashrafi}},
  \bibinfo {author} {\bibfnamefont {S.}~\bibnamefont {Ashrafi}}, \bibinfo
  {author} {\bibfnamefont {M.}~\bibnamefont {Tur}}, \bibinfo {author}
  {\bibfnamefont {A.F.}\ \bibnamefont {Molisch}},  and \bibinfo {author}
  {\bibfnamefont {A.E.}\ \bibnamefont {Willner}},\ }\bibfield  {title}
  {\enquote {\bibinfo {title} {Performance metrics and design parameters for an
  {FSO} communications link based on multiplexing of multiple
  orbital-angular-momentum beams},}\ }in\ \href {\doibase
  10.1109/GLOCOMW.2014.7063478} {\emph {\bibinfo {booktitle} {Globecom
  Workshops ({GC} Wkshps), 2014}}}\ (\bibinfo {year} {2014})\ pp.\ \bibinfo
  {pages} {481--486}\BibitemShut {NoStop}%
\bibitem [{\citenamefont {Yan}\ \emph {et~al.}(2014{\natexlab{a}})\citenamefont
  {Yan}, \citenamefont {Xie}, \citenamefont {Huang}, \citenamefont {Lavery},
  \citenamefont {Ahemd}, \citenamefont {Bao}, \citenamefont {Ren},
  \citenamefont {Molisch}, \citenamefont {Tur}, \citenamefont {Padgett}, and
  \citenamefont {Willner}}]{Yan&al:ICC:2014}%
  \BibitemOpen
  \bibfield  {author} {\bibinfo {author} {\bibfnamefont {Yan}\ \bibnamefont
  {Yan}}, \bibinfo {author} {\bibfnamefont {Guodong}\ \bibnamefont {Xie}},
  \bibinfo {author} {\bibfnamefont {Hao}\ \bibnamefont {Huang}}, \bibinfo
  {author} {\bibfnamefont {M.J.}\ \bibnamefont {Lavery}}, \bibinfo {author}
  {\bibfnamefont {N.}~\bibnamefont {Ahemd}}, \bibinfo {author} {\bibfnamefont
  {Changjing}\ \bibnamefont {Bao}}, \bibinfo {author} {\bibfnamefont
  {Yongxiong}\ \bibnamefont {Ren}}, \bibinfo {author} {\bibfnamefont {A.F.}\
  \bibnamefont {Molisch}}, \bibinfo {author} {\bibfnamefont {M.}~\bibnamefont
  {Tur}}, \bibinfo {author} {\bibfnamefont {M.}~\bibnamefont {Padgett}},  and
  \bibinfo {author} {\bibfnamefont {A.E.}\ \bibnamefont {Willner}},\ }\bibfield
   {title} {\enquote {\bibinfo {title} {Demonstration of 8-mode 32-gbit/s
  millimeter-wave free-space communication link using 4
  orbital-angular-momentum modes on 2 polarizations},}\ }in\ \href {\doibase
  10.1109/ICC.2014.6884088} {\emph {\bibinfo {booktitle} {2014 {IEEE}
  International Conference on Communications ({ICC})}}}\ (\bibinfo {year}
  {2014})\ pp.\ \bibinfo {pages} {4850--4855}\BibitemShut {NoStop}%
\bibitem [{\citenamefont {Yan}\ \emph {et~al.}(2014{\natexlab{b}})\citenamefont
  {Yan}, \citenamefont {Xie}, \citenamefont {Lavery}, \citenamefont {Huang},
  \citenamefont {Ahmed}, \citenamefont {Bao}, \citenamefont {Ren},
  \citenamefont {Cao}, \citenamefont {Li}, \citenamefont {Zhao}, \citenamefont
  {Molisch}, \citenamefont {Tur}, \citenamefont {Padgett}, and \citenamefont
  {Willner}}]{Yan&al:NCOM:2014}%
  \BibitemOpen
  \bibfield  {author} {\bibinfo {author} {\bibfnamefont {Yan}\ \bibnamefont
  {Yan}}, \bibinfo {author} {\bibfnamefont {Guodong}\ \bibnamefont {Xie}},
  \bibinfo {author} {\bibfnamefont {Martin P.~J.}\ \bibnamefont {Lavery}},
  \bibinfo {author} {\bibfnamefont {Hao}\ \bibnamefont {Huang}}, \bibinfo
  {author} {\bibfnamefont {Nisar}\ \bibnamefont {Ahmed}}, \bibinfo {author}
  {\bibfnamefont {Changjing}\ \bibnamefont {Bao}}, \bibinfo {author}
  {\bibfnamefont {Yongxiong}\ \bibnamefont {Ren}}, \bibinfo {author}
  {\bibfnamefont {Yinwen}\ \bibnamefont {Cao}}, \bibinfo {author}
  {\bibfnamefont {Long}\ \bibnamefont {Li}}, \bibinfo {author} {\bibfnamefont
  {Zhe}\ \bibnamefont {Zhao}}, \bibinfo {author} {\bibfnamefont {Andreas~F.}\
  \bibnamefont {Molisch}}, \bibinfo {author} {\bibfnamefont {Moshe}\
  \bibnamefont {Tur}}, \bibinfo {author} {\bibfnamefont {Miles~J.}\
  \bibnamefont {Padgett}},  and \bibinfo {author} {\bibfnamefont {Alan~E.}\
  \bibnamefont {Willner}},\ }\bibfield  {title} {\enquote {\bibinfo {title}
  {High-capacity millimetre-wave communications with orbital angular momentum
  multiplexing},}\ }\href {\doibase 10.1038/ncomms5876} {\bibfield  {journal}
  {\bibinfo  {journal} {Nature Commun.}\ }\textbf {\bibinfo {volume} {5}},\
  \bibinfo {pages} {1--9} (\bibinfo {year} {2014}{\natexlab{b}})}\BibitemShut
  {NoStop}%
\bibitem [{\citenamefont {Willner}\ \emph {et~al.}(2015)\citenamefont
  {Willner}, \citenamefont {Huang}, \citenamefont {Yan}, \citenamefont {Ren},
  \citenamefont {Ahmed}, \citenamefont {Xie}, \citenamefont {Bao},
  \citenamefont {Li}, \citenamefont {Cao}, \citenamefont {Zhao}, \citenamefont
  {Wang}, \citenamefont {Lavery}, \citenamefont {Tur}, \citenamefont
  {Ramachandran}, \citenamefont {Molisch}, \citenamefont {Ashrafi}, and
  \citenamefont {Ashrafi}}]{Willner&al:AOP:2015}%
  \BibitemOpen
  \bibfield  {author} {\bibinfo {author} {\bibfnamefont {A.~E.}\ \bibnamefont
  {Willner}}, \bibinfo {author} {\bibfnamefont {H.}~\bibnamefont {Huang}},
  \bibinfo {author} {\bibfnamefont {Y.}~\bibnamefont {Yan}}, \bibinfo {author}
  {\bibfnamefont {Y.}~\bibnamefont {Ren}}, \bibinfo {author} {\bibfnamefont
  {N.}~\bibnamefont {Ahmed}}, \bibinfo {author} {\bibfnamefont
  {G.}~\bibnamefont {Xie}}, \bibinfo {author} {\bibfnamefont {C.}~\bibnamefont
  {Bao}}, \bibinfo {author} {\bibfnamefont {L.}~\bibnamefont {Li}}, \bibinfo
  {author} {\bibfnamefont {Y.}~\bibnamefont {Cao}}, \bibinfo {author}
  {\bibfnamefont {Z.}~\bibnamefont {Zhao}}, \bibinfo {author} {\bibfnamefont
  {J.}~\bibnamefont {Wang}}, \bibinfo {author} {\bibfnamefont {M.~P.~J.}\
  \bibnamefont {Lavery}}, \bibinfo {author} {\bibfnamefont {M.}~\bibnamefont
  {Tur}}, \bibinfo {author} {\bibfnamefont {S.}~\bibnamefont {Ramachandran}},
  \bibinfo {author} {\bibfnamefont {A.~F.}\ \bibnamefont {Molisch}}, \bibinfo
  {author} {\bibfnamefont {N.}~\bibnamefont {Ashrafi}},  and \bibinfo
  {author} {\bibfnamefont {S.}~\bibnamefont {Ashrafi}},\ }\bibfield  {title}
  {\enquote {\bibinfo {title} {Optical communications using orbital angular
  momentum beams},}\ }\href {http://aop.osa.org/abstract.cfm?URI=aop-7-1-66}
  {\bibfield  {journal} {\bibinfo  {journal} {Adv.\ Opt.\ Photon.}\ }\textbf
  {\bibinfo {volume} {7}},\ \bibinfo {pages} {66--106} (\bibinfo {year}
  {2015})}\BibitemShut {NoStop}%
\bibitem [{\citenamefont {Papanicolaou} and \citenamefont
  {Tomaras}(1991)}]{Papanicolaou&Tomaras:NPB:1991}%
  \BibitemOpen
  \bibfield  {author} {\bibinfo {author} {\bibfnamefont {N.}~\bibnamefont
  {Papanicolaou}} and \bibinfo {author} {\bibfnamefont {T.~N.}\ \bibnamefont
  {Tomaras}},\ }\bibfield  {title} {\enquote {\bibinfo {title} {Dynamics of
  magnetic vortices},}\ }\href {\doibase 10.1016/0550-3213(91)90410-Y}
  {\bibfield  {journal} {\bibinfo  {journal} {Nucl.\ Phys.~B}\ }\textbf
  {\bibinfo {volume} {360}},\ \bibinfo {pages} {425--462} (\bibinfo {year}
  {1991})}\BibitemShut {NoStop}%
\bibitem [{\citenamefont {Zagrodzinski}(2002)}]{Zagrodzinski:PC:2002}%
  \BibitemOpen
  \bibfield  {author} {\bibinfo {author} {\bibfnamefont {J.}~\bibnamefont
  {Zagrodzinski}},\ }\bibfield  {title} {\enquote {\bibinfo {title} {Vortices
  in different branches of physics},}\ }\href {\doibase
  10.1016/S0921-4534(01)01219-9} {\bibfield  {journal} {\bibinfo  {journal}
  {Physica C}\ }\textbf {\bibinfo {volume} {369}},\ \bibinfo {pages} {45--54}
  (\bibinfo {year} {2002})}\BibitemShut {NoStop}%
\bibitem [{\citenamefont {Yao}\ \emph {et~al.}(2006)\citenamefont {Yao},
  \citenamefont {Franke-Arnold}, \citenamefont {Courtial}, \citenamefont
  {Barnett}, and \citenamefont {Padgett}}]{Yao&al:OE:2006}%
  \BibitemOpen
  \bibfield  {author} {\bibinfo {author} {\bibfnamefont {Eric}\ \bibnamefont
  {Yao}}, \bibinfo {author} {\bibfnamefont {Sonja}\ \bibnamefont
  {Franke-Arnold}}, \bibinfo {author} {\bibfnamefont {Johannes}\ \bibnamefont
  {Courtial}}, \bibinfo {author} {\bibfnamefont {Stephen}\ \bibnamefont
  {Barnett}},  and \bibinfo {author} {\bibfnamefont {Miles}\ \bibnamefont
  {Padgett}},\ }\bibfield  {title} {\enquote {\bibinfo {title} {{Fourier}
  relationship between angular position and optical orbital angular
  momentum},}\ }\href {\doibase 10.1364/OE.14.009071} {\bibfield  {journal}
  {\bibinfo  {journal} {Opt. Express}\ }\textbf {\bibinfo {volume} {14}},\
  \bibinfo {pages} {9071--9076} (\bibinfo {year} {2006})}\BibitemShut {NoStop}%
\bibitem [{\citenamefont {Jha}\ \emph {et~al.}(2008)\citenamefont {Jha},
  \citenamefont {Jack}, \citenamefont {Yao}, \citenamefont {Leach},
  \citenamefont {Boyd}, \citenamefont {Buller}, \citenamefont {Barnett},
  \citenamefont {Franke-Arnold}, and \citenamefont
  {Padgett}}]{Jha&al:PRA:2008}%
  \BibitemOpen
  \bibfield  {author} {\bibinfo {author} {\bibfnamefont {A.~K.}\ \bibnamefont
  {Jha}}, \bibinfo {author} {\bibfnamefont {B.}~\bibnamefont {Jack}}, \bibinfo
  {author} {\bibfnamefont {E.}~\bibnamefont {Yao}}, \bibinfo {author}
  {\bibfnamefont {J.}~\bibnamefont {Leach}}, \bibinfo {author} {\bibfnamefont
  {R.~W.}\ \bibnamefont {Boyd}}, \bibinfo {author} {\bibfnamefont {G.~S.}\
  \bibnamefont {Buller}}, \bibinfo {author} {\bibfnamefont {S.~M.}\
  \bibnamefont {Barnett}}, \bibinfo {author} {\bibfnamefont {S.}~\bibnamefont
  {Franke-Arnold}},  and \bibinfo {author} {\bibfnamefont {M.~J.}\
  \bibnamefont {Padgett}},\ }\bibfield  {title} {\enquote {\bibinfo {title}
  {{Fourier} relationship between the angle and angular momentum of entangled
  photons},}\ }\href {\doibase 10.1103/PhysRevA.78.043810} {\bibfield
  {journal} {\bibinfo  {journal} {Phys.\ Rev.\ A}\ }\textbf {\bibinfo {volume}
  {78}},\ \bibinfo {pages} {043810} (\bibinfo {year} {2008})}\BibitemShut
  {NoStop}%
\bibitem [{\citenamefont {Jack}\ \emph {et~al.}(2008)\citenamefont {Jack},
  \citenamefont {Padgett}, and \citenamefont
  {Franke-Arnold}}]{Jack&al:NJP:2008}%
  \BibitemOpen
  \bibfield  {author} {\bibinfo {author} {\bibfnamefont {B.}~\bibnamefont
  {Jack}}, \bibinfo {author} {\bibfnamefont {M.~J.}\ \bibnamefont {Padgett}}, \
  and\ \bibinfo {author} {\bibfnamefont {S.}~\bibnamefont {Franke-Arnold}},\
  }\bibfield  {title} {\enquote {\bibinfo {title} {Angular diffraction},}\
  }\href@noop {} {\bibfield  {journal} {\bibinfo  {journal} {New J.~Phys.}\
  }\textbf {\bibinfo {volume} {10}},\ \bibinfo {pages} {103013} (\bibinfo
  {year} {2008})}\BibitemShut {NoStop}%
\bibitem [{\citenamefont {Padgett}\ \emph {et~al.}(2015)\citenamefont
  {Padgett}, \citenamefont {Miatto}, \citenamefont {Lavery}, \citenamefont
  {Zeilinger}, and \citenamefont {Boyd}}]{Padgett&al:NJP:2015}%
  \BibitemOpen
  \bibfield  {author} {\bibinfo {author} {\bibfnamefont {Miles~J.}\
  \bibnamefont {Padgett}}, \bibinfo {author} {\bibfnamefont {Filippo~M.}\
  \bibnamefont {Miatto}}, \bibinfo {author} {\bibfnamefont {Martin P.~J.}\
  \bibnamefont {Lavery}}, \bibinfo {author} {\bibfnamefont {Anton}\
  \bibnamefont {Zeilinger}},  and \bibinfo {author} {\bibfnamefont
  {Robert~W.}\ \bibnamefont {Boyd}},\ }\bibfield  {title} {\enquote {\bibinfo
  {title} {Divergence of an orbital-angular-momentum-carrying beam upon
  propagation},}\ }\href {\doibase 10.1088/1367-2630/17/2/023011} {\bibfield
  {journal} {\bibinfo  {journal} {New J.~Phys.}\ }\textbf {\bibinfo {volume}
  {17}},\ \bibinfo {pages} {023011} (\bibinfo {year} {2015})}\BibitemShut
  {NoStop}%
\bibitem [{\citenamefont {Chen}\ \emph {et~al.}(2013)\citenamefont {Chen},
  \citenamefont {Mazilu}, \citenamefont {Arita}, \citenamefont {Wright}, and
  \citenamefont {Dholakia}}]{Chen&al:OL:2013}%
  \BibitemOpen
  \bibfield  {author} {\bibinfo {author} {\bibfnamefont {Mingzhou}\
  \bibnamefont {Chen}}, \bibinfo {author} {\bibfnamefont {Michael}\
  \bibnamefont {Mazilu}}, \bibinfo {author} {\bibfnamefont {Yoshihiko}\
  \bibnamefont {Arita}}, \bibinfo {author} {\bibfnamefont {Ewan~M.}\
  \bibnamefont {Wright}},  and \bibinfo {author} {\bibfnamefont {Kishan}\
  \bibnamefont {Dholakia}},\ }\bibfield  {title} {\enquote {\bibinfo {title}
  {Dynamics of microparticles trapped in a perfect vortex beam},}\ }\href
  {\doibase 10.1364/OL.38.004919} {\bibfield  {journal} {\bibinfo  {journal}
  {Opt. Lett.}\ }\textbf {\bibinfo {volume} {38}},\ \bibinfo {pages} {4919}
  (\bibinfo {year} {2013})}\BibitemShut {NoStop}%
\bibitem [{\citenamefont {Ostrovsky}\ \emph {et~al.}(2013)\citenamefont
  {Ostrovsky}, \citenamefont {Rickenstorff-Parrao}, and \citenamefont
  {Arriz{\'o}n}}]{Ostrovsky&al:OL:2013}%
  \BibitemOpen
  \bibfield  {author} {\bibinfo {author} {\bibfnamefont {Andrey~S.}\
  \bibnamefont {Ostrovsky}}, \bibinfo {author} {\bibfnamefont {Carolina}\
  \bibnamefont {Rickenstorff-Parrao}},  and \bibinfo {author} {\bibfnamefont
  {V{\'\i}ctor}\ \bibnamefont {Arriz{\'o}n}},\ }\bibfield  {title} {\enquote
  {\bibinfo {title} {Generation of the ``perfect'' optical vortex using a
  liquid-crystal spatial light modulator},}\ }\href {\doibase
  10.1364/OL.38.000534} {\bibfield  {journal} {\bibinfo  {journal} {Opt.
  Lett.}\ }\textbf {\bibinfo {volume} {38}},\ \bibinfo {pages} {534--536}
  (\bibinfo {year} {2013})}\BibitemShut {NoStop}%
\bibitem [{\citenamefont {Garc{\'\i}a-Garc{\'\i}a}\ \emph
  {et~al.}(2014)\citenamefont {Garc{\'\i}a-Garc{\'\i}a}, \citenamefont
  {Rickenstorff-Parrao}, \citenamefont {Ramos-Garc{\'\i}a}, \citenamefont
  {Arriz{\'o}n}, and \citenamefont {Ostrovsky}}]{Garcia-Garcia&al:OL:2014}%
  \BibitemOpen
  \bibfield  {author} {\bibinfo {author} {\bibfnamefont {Joaqu{\'\i}n}\
  \bibnamefont {Garc{\'\i}a-Garc{\'\i}a}}, \bibinfo {author} {\bibfnamefont
  {Carolina}\ \bibnamefont {Rickenstorff-Parrao}}, \bibinfo {author}
  {\bibfnamefont {Rub{\'e}n}\ \bibnamefont {Ramos-Garc{\'\i}a}}, \bibinfo
  {author} {\bibfnamefont {V{\'\i}ctor}\ \bibnamefont {Arriz{\'o}n}},  and
  \bibinfo {author} {\bibfnamefont {Andrey~S.}\ \bibnamefont {Ostrovsky}},\
  }\bibfield  {title} {\enquote {\bibinfo {title} {Simple technique for
  generating the perfect optical vortex},}\ }\href {\doibase
  10.1364/OL.39.005305} {\bibfield  {journal} {\bibinfo  {journal} {Opt.
  Lett.}\ }\textbf {\bibinfo {volume} {39}},\ \bibinfo {pages} {5305--5308}
  (\bibinfo {year} {2014})}\BibitemShut {NoStop}%
\bibitem [{\citenamefont {Vaity} and \citenamefont
  {Rusch}(2015)}]{Vaity&Rusch:OL:2015}%
  \BibitemOpen
  \bibfield  {author} {\bibinfo {author} {\bibfnamefont {Pravin}\ \bibnamefont
  {Vaity}} and \bibinfo {author} {\bibfnamefont {Leslie}\ \bibnamefont
  {Rusch}},\ }\bibfield  {title} {\enquote {\bibinfo {title} {Perfect vortex
  beam: {Fourier} transformation of a {Bessel} beam},}\ }\href {\doibase
  10.1364/OL.40.000597} {\bibfield  {journal} {\bibinfo  {journal} {Opt.
  Lett.}\ }\textbf {\bibinfo {volume} {40}},\ \bibinfo {pages} {597--600}
  (\bibinfo {year} {2015})}\BibitemShut {NoStop}%
\bibitem [{\citenamefont {Li}\ \emph {et~al.}(2015)\citenamefont {Li},
  \citenamefont {Xie}, \citenamefont {Ren}, \citenamefont {Ahmed},
  \citenamefont {Huang}, \citenamefont {Zhao}, \citenamefont {Liao},
  \citenamefont {Lavery}, \citenamefont {Yan}, \citenamefont {Bao},
  \citenamefont {Wang}, \citenamefont {Ashrafi}, \citenamefont {Ashrafi},
  \citenamefont {Linquist}, \citenamefont {Tur}, and \citenamefont
  {Willner}}]{Li&al:OSA:2015}%
  \BibitemOpen
  \bibfield  {author} {\bibinfo {author} {\bibfnamefont {Long}\ \bibnamefont
  {Li}}, \bibinfo {author} {\bibfnamefont {Guodong}\ \bibnamefont {Xie}},
  \bibinfo {author} {\bibfnamefont {Yongxiong}\ \bibnamefont {Ren}}, \bibinfo
  {author} {\bibfnamefont {Nisar}\ \bibnamefont {Ahmed}}, \bibinfo {author}
  {\bibfnamefont {Hao}\ \bibnamefont {Huang}}, \bibinfo {author} {\bibfnamefont
  {Zhe}\ \bibnamefont {Zhao}}, \bibinfo {author} {\bibfnamefont {Peicheng}\
  \bibnamefont {Liao}}, \bibinfo {author} {\bibfnamefont {Martin}\ \bibnamefont
  {Lavery}}, \bibinfo {author} {\bibfnamefont {Yan}\ \bibnamefont {Yan}},
  \bibinfo {author} {\bibfnamefont {Bhangjing}\ \bibnamefont {Bao}}, \bibinfo
  {author} {\bibfnamefont {Zhe}\ \bibnamefont {Wang}}, \bibinfo {author}
  {\bibfnamefont {Nima}\ \bibnamefont {Ashrafi}}, \bibinfo {author}
  {\bibfnamefont {Solyman}\ \bibnamefont {Ashrafi}}, \bibinfo {author}
  {\bibfnamefont {Roger}\ \bibnamefont {Linquist}}, \bibinfo {author}
  {\bibfnamefont {Moshe}\ \bibnamefont {Tur}},  and \bibinfo {author}
  {\bibfnamefont {Alan}\ \bibnamefont {Willner}},\ }\bibfield  {title}
  {\enquote {\bibinfo {title} {Performance enhancement of an
  orbital-angular-momentum-based free-space optical communication link through
  beam divergence controlling},}\ }in\ \href {\doibase 10.1364/OFC.2015.M2F.6}
  {\emph {\bibinfo {booktitle} {Optical Fiber Communication Conference}}},\
  \bibinfo {series and number} {{OSA} Technical Digest (online)}\ (\bibinfo
  {publisher} {Optical Society of America},\ \bibinfo {year} {2015})\ p.\
  \bibinfo {pages} {M2F.6}\BibitemShut {NoStop}%
\bibitem [{\citenamefont {Bohr}(1913)}]{Bohr:PM:1913}%
  \BibitemOpen
  \bibfield  {author} {\bibinfo {author} {\bibfnamefont {N.}~\bibnamefont
  {Bohr}},\ }\bibfield  {title} {\enquote {\bibinfo {title} {On the
  constitution of atoms and molecules},}\ }\href {\doibase
  10.1080/14786441308634955} {\bibfield  {journal} {\bibinfo  {journal}
  {Philos.\ Mag.}\ }\bibinfo {series} {6},\ \textbf {\bibinfo {volume} {26}},\
  \bibinfo {pages} {1--25} (\bibinfo {year} {1913})}\BibitemShut {NoStop}%
\bibitem [{\citenamefont {Sommerfeld}(1921)}]{Sommerfeld:Book:1921}%
  \BibitemOpen
  \bibfield  {author} {\bibinfo {author} {\bibfnamefont {Arnold}\ \bibnamefont
  {Sommerfeld}},\ }\href@noop {} {\emph {\bibinfo {title} {Atombau und
  Spektrallinien}}}\ (\bibinfo  {publisher} {F. Vieweg \& Sohn},\ \bibinfo
  {year} {1921})\BibitemShut {NoStop}%
\bibitem [{\citenamefont {Bohm}(1989)}]{Bohm:Book:1989}%
  \BibitemOpen
  \bibfield  {author} {\bibinfo {author} {\bibfnamefont {David}\ \bibnamefont
  {Bohm}},\ }\href@noop {} {\emph {\bibinfo {title} {Quantum Theory}}}\
  (\bibinfo  {publisher} {Dover Publications Inc.},\ \bibinfo {address} {New
  York},\ \bibinfo {year} {1989})\BibitemShut {NoStop}%
\bibitem [{\citenamefont {Nienhuis}(2008)}]{Nienhuis:Incollection:2008}%
  \BibitemOpen
  \bibfield  {author} {\bibinfo {author} {\bibfnamefont {Gerard}\ \bibnamefont
  {Nienhuis}},\ }\bibfield  {title} {\enquote {\bibinfo {title} {Angular
  momentum and vortices in optics},}\ }in\ \href@noop {} {\emph {\bibinfo
  {booktitle} {Structured Light and its Applications: An Introduction to
  Phase-Structured Beams and Nanoscale Optical Forces}}}\ (\bibinfo
  {publisher} {Elsevier Inc.},\ \bibinfo {address} {Amsterdam, NL},\ \bibinfo
  {year} {2008})\ pp.\ \bibinfo {pages} {19--61}\BibitemShut {NoStop}%
\bibitem [{\citenamefont {Berry}(2004)}]{Berry:JOA:2004c}%
  \BibitemOpen
  \bibfield  {author} {\bibinfo {author} {\bibfnamefont {M.~V.}\ \bibnamefont
  {Berry}},\ }\bibfield  {title} {\enquote {\bibinfo {title} {Optical vortices
  evolving from helicoidal integer and fractional phase steps},}\ }\href
  {\doibase 10.1088/1464-4258/6/2/018} {\bibfield  {journal} {\bibinfo
  {journal} {J.~Opt.\ A: Pure Appl.\ Opt.}\ }\textbf {\bibinfo {volume} {6}},\
  \bibinfo {pages} {259--268} (\bibinfo {year} {2004})}\BibitemShut {NoStop}%
\bibitem [{\citenamefont {\v{C}elechovsk\'{y}} and \citenamefont
  {Bouchal}(2007)}]{Celechovsky&Bouchal:NJP:2007}%
  \BibitemOpen
  \bibfield  {author} {\bibinfo {author} {\bibfnamefont {R.}~\bibnamefont
  {\v{C}elechovsk\'{y}}} and \bibinfo {author} {\bibfnamefont
  {Z.}~\bibnamefont {Bouchal}},\ }\bibfield  {title} {\enquote {\bibinfo
  {title} {Optical implementation of the vortex information channel},}\ }\href
  {\doibase 10.1088/1367-2630/9/9/328} {\bibfield  {journal} {\bibinfo
  {journal} {New J.~Phys.}\ }\textbf {\bibinfo {volume} {9}},\ \bibinfo {pages}
  {328} (\bibinfo {year} {2007})}\BibitemShut {NoStop}%
\bibitem [{\citenamefont {Barreiro}\ \emph {et~al.}(2008)\citenamefont
  {Barreiro}, \citenamefont {Wei}, and \citenamefont
  {Kwiat}}]{Barreiro&al:NPHY:2008}%
  \BibitemOpen
  \bibfield  {author} {\bibinfo {author} {\bibfnamefont {Julio~T.}\
  \bibnamefont {Barreiro}}, \bibinfo {author} {\bibfnamefont {Tzu-Chieh}\
  \bibnamefont {Wei}},  and \bibinfo {author} {\bibfnamefont {Paul~G.}\
  \bibnamefont {Kwiat}},\ }\bibfield  {title} {\enquote {\bibinfo {title}
  {Beating the channel capacity limit for linear photonic superdense coding},}\
  }\href {\doibase 10.1038/nphys919} {\bibfield  {journal} {\bibinfo  {journal}
  {Nature Phys.}\ }\textbf {\bibinfo {volume} {4}},\ \bibinfo {pages}
  {282--286} (\bibinfo {year} {2008})}\BibitemShut {NoStop}%
\bibitem [{\citenamefont {Pors}\ \emph {et~al.}(2008)\citenamefont {Pors},
  \citenamefont {Oemrawsingh}, \citenamefont {Aiello}, \citenamefont {van
  Exter}, \citenamefont {Eliel}, \citenamefont {'t~Hooft}, and \citenamefont
  {Woerdman}}]{Pors&al:PRL:2008}%
  \BibitemOpen
  \bibfield  {author} {\bibinfo {author} {\bibfnamefont {J.~B.}\ \bibnamefont
  {Pors}}, \bibinfo {author} {\bibfnamefont {S.~S.~R.}\ \bibnamefont
  {Oemrawsingh}}, \bibinfo {author} {\bibfnamefont {A.}~\bibnamefont {Aiello}},
  \bibinfo {author} {\bibfnamefont {M.~P.}\ \bibnamefont {van Exter}}, \bibinfo
  {author} {\bibfnamefont {E.~R.}\ \bibnamefont {Eliel}}, \bibinfo {author}
  {\bibfnamefont {G.~W.}\ \bibnamefont {'t~Hooft}},  and \bibinfo {author}
  {\bibfnamefont {J.~P.}\ \bibnamefont {Woerdman}},\ }\bibfield  {title}
  {\enquote {\bibinfo {title} {Shannon dimensionality of quantum channels and
  its application to photon entanglement},}\ }\href {\doibase
  10.1103/PhysRevLett.101.120502} {\bibfield  {journal} {\bibinfo  {journal}
  {Phys.\ Rev.\ Lett.}\ }\textbf {\bibinfo {volume} {101}},\ \bibinfo {pages}
  {120502} (\bibinfo {year} {2008})}\BibitemShut {NoStop}%
\bibitem [{\citenamefont {Martelli}\ \emph {et~al.}(2011)\citenamefont
  {Martelli}, \citenamefont {Gatto}, \citenamefont {Boffi}, and \citenamefont
  {Martinelli}}]{Martelli&al:EL:2011}%
  \BibitemOpen
  \bibfield  {author} {\bibinfo {author} {\bibfnamefont {P.}~\bibnamefont
  {Martelli}}, \bibinfo {author} {\bibfnamefont {A.}~\bibnamefont {Gatto}},
  \bibinfo {author} {\bibfnamefont {P.}~\bibnamefont {Boffi}},  and \bibinfo
  {author} {\bibfnamefont {M.}~\bibnamefont {Martinelli}},\ }\bibfield  {title}
  {\enquote {\bibinfo {title} {Free-space optical transmission with orbital
  angular momentum division multiplexing},}\ }\href {\doibase
  10.1049/el.2011.1766} {\bibfield  {journal} {\bibinfo  {journal} {Electron.
  Lett.}\ }\textbf {\bibinfo {volume} {47}},\ \bibinfo {pages} {972--973}
  (\bibinfo {year} {2011})}\BibitemShut {NoStop}%
\bibitem [{\citenamefont {Kumar}\ \emph {et~al.}(2011)\citenamefont {Kumar},
  \citenamefont {Prabhakar}, \citenamefont {Vaity}, and \citenamefont
  {Singh}}]{Kumar&al:OL:2011}%
  \BibitemOpen
  \bibfield  {author} {\bibinfo {author} {\bibfnamefont {Ashok}\ \bibnamefont
  {Kumar}}, \bibinfo {author} {\bibfnamefont {Shashi}\ \bibnamefont
  {Prabhakar}}, \bibinfo {author} {\bibfnamefont {Pravin}\ \bibnamefont
  {Vaity}},  and \bibinfo {author} {\bibfnamefont {R.~P.}\ \bibnamefont
  {Singh}},\ }\bibfield  {title} {\enquote {\bibinfo {title} {Information
  content of optical vortex fields},}\ }\href {\doibase 10.1364/OL.36.001161}
  {\bibfield  {journal} {\bibinfo  {journal} {Opt. Lett.}\ }\textbf {\bibinfo
  {volume} {36}},\ \bibinfo {pages} {1161--1163} (\bibinfo {year}
  {2011})}\BibitemShut {NoStop}%
\bibitem [{\citenamefont {Mair}\ \emph {et~al.}(2001)\citenamefont {Mair},
  \citenamefont {Vaziri}, \citenamefont {Weihs}, and \citenamefont
  {Zeilinger}}]{Mair&al:N:2001}%
  \BibitemOpen
  \bibfield  {author} {\bibinfo {author} {\bibfnamefont {A.}~\bibnamefont
  {Mair}}, \bibinfo {author} {\bibfnamefont {A.}~\bibnamefont {Vaziri}},
  \bibinfo {author} {\bibfnamefont {G.}~\bibnamefont {Weihs}},  and \bibinfo
  {author} {\bibfnamefont {A.}~\bibnamefont {Zeilinger}},\ }\bibfield  {title}
  {\enquote {\bibinfo {title} {Entanglement of the orbital angular momentum
  states of photons},}\ }\href {\doibase 10.1038/35085529} {\bibfield
  {journal} {\bibinfo  {journal} {Nature}\ }\textbf {\bibinfo {volume} {412}},\
  \bibinfo {pages} {313--316} (\bibinfo {year} {2001})}\BibitemShut {NoStop}%
\bibitem [{\citenamefont {Watson}(2002)}]{Watson:S:2002}%
  \BibitemOpen
  \bibfield  {author} {\bibinfo {author} {\bibfnamefont {Andrew}\ \bibnamefont
  {Watson}},\ }\bibfield  {title} {\enquote {\bibinfo {title} {New twist could
  pack photons with data},}\ }\href {\doibase 10.1126/science.296.5577.2316b}
  {\bibfield  {journal} {\bibinfo  {journal} {Science}\ }\textbf {\bibinfo
  {volume} {296}},\ \bibinfo {pages} {2316--2317} (\bibinfo {year}
  {2002})}\BibitemShut {NoStop}%
\bibitem [{\citenamefont {G{\"o}tte}\ \emph {et~al.}(2008)\citenamefont
  {G{\"o}tte}, \citenamefont {{O'Holleran}}, \citenamefont {Preece},
  \citenamefont {Flossmann}, \citenamefont {Franke-Arnold}, \citenamefont
  {Barnett}, and \citenamefont {Padgett}}]{Gotte&al:OE:2008}%
  \BibitemOpen
  \bibfield  {author} {\bibinfo {author} {\bibfnamefont {J.~B.}\ \bibnamefont
  {G{\"o}tte}}, \bibinfo {author} {\bibfnamefont {K.}~\bibnamefont
  {{O'Holleran}}}, \bibinfo {author} {\bibfnamefont {Daryl}\ \bibnamefont
  {Preece}}, \bibinfo {author} {\bibfnamefont {Florian}\ \bibnamefont
  {Flossmann}}, \bibinfo {author} {\bibfnamefont {Sonja}\ \bibnamefont
  {Franke-Arnold}}, \bibinfo {author} {\bibfnamefont {Stephen~M.}\ \bibnamefont
  {Barnett}},  and \bibinfo {author} {\bibfnamefont {Miles~J.}\ \bibnamefont
  {Padgett}},\ }\bibfield  {title} {\enquote {\bibinfo {title} {Light beams
  with fractional orbital angular momentum and their vortex structure},}\
  }\href {\doibase 10.1364/OE.16.000993} {\bibfield  {journal} {\bibinfo
  {journal} {Opt. Express}\ }\textbf {\bibinfo {volume} {16}},\ \bibinfo
  {pages} {993--1006} (\bibinfo {year} {2008})}\BibitemShut {NoStop}%
\bibitem [{\citenamefont {Huang}\ \emph
  {et~al.}(2012{\natexlab{a}})\citenamefont {Huang}, \citenamefont {Lin}, and
  \citenamefont {Shih}}]{Huang&al:OC:2012}%
  \BibitemOpen
  \bibfield  {author} {\bibinfo {author} {\bibfnamefont {Hsiao-Chih}\
  \bibnamefont {Huang}}, \bibinfo {author} {\bibfnamefont {Yen-Ting}\
  \bibnamefont {Lin}},  and \bibinfo {author} {\bibfnamefont {Ming-Feng}\
  \bibnamefont {Shih}},\ }\bibfield  {title} {\enquote {\bibinfo {title}
  {Measuring the fractional orbital angular momentum of a vortex light beam by
  cascaded {Mach-Zehnder} interferometers},}\ }\href {\doibase
  10.1016/j.optcom.2011.09.063} {\bibfield  {journal} {\bibinfo  {journal}
  {Opt. Commun.}\ }\textbf {\bibinfo {volume} {285}},\ \bibinfo {pages}
  {383--388} (\bibinfo {year} {2012}{\natexlab{a}})}\BibitemShut {NoStop}%
\bibitem [{\citenamefont {Fadeyeva}\ \emph {et~al.}(2014)\citenamefont
  {Fadeyeva}, \citenamefont {Rubass}, \citenamefont {Aleksandrov}, and
  \citenamefont {Volyar}}]{Fadeyeva&al:JOSAB:2014}%
  \BibitemOpen
  \bibfield  {author} {\bibinfo {author} {\bibfnamefont {Tatyana~A.}\
  \bibnamefont {Fadeyeva}}, \bibinfo {author} {\bibfnamefont {Alexander~F.}\
  \bibnamefont {Rubass}}, \bibinfo {author} {\bibfnamefont {Rodion~V.}\
  \bibnamefont {Aleksandrov}},  and \bibinfo {author} {\bibfnamefont
  {Aleksander~V.}\ \bibnamefont {Volyar}},\ }\bibfield  {title} {\enquote
  {\bibinfo {title} {Does the optical angular momentum change smoothly in
  fractional-charged vortex beams?}}\ }\href {\doibase 10.1364/JOSAB.31.000798}
  {\bibfield  {journal} {\bibinfo  {journal} {J.~Opt.\ Soc\ Am.\ B}\ }\textbf
  {\bibinfo {volume} {31}},\ \bibinfo {pages} {798} (\bibinfo {year}
  {2014})}\BibitemShut {NoStop}%
\bibitem [{\citenamefont {Zernike}(1934)}]{Zernike:P:1934}%
  \BibitemOpen
  \bibfield  {author} {\bibinfo {author} {\bibfnamefont {F.}~\bibnamefont
  {Zernike}, \bibfnamefont {von}},\ }\bibfield  {title} {\enquote {\bibinfo
  {title} {Beugungstheorie des {S}chneidenverfahrens und seiner verbesserten
  {F}orm, der {P}hasenkontrastmethode},}\ }\href {\doibase
  10.1016/S0031-8914(34)80259-5} {\bibfield  {journal} {\bibinfo  {journal}
  {Physica}\ }\textbf {\bibinfo {volume} {1}},\ \bibinfo {pages} {689--704}
  (\bibinfo {year} {1934})}\BibitemShut {NoStop}%
\bibitem [{\citenamefont {Dholakia} and \citenamefont {{\v C}i{\v
  z}m{\'a}r}(2011)}]{Dholakia&Cizmar:NPHO:2011}%
  \BibitemOpen
  \bibfield  {author} {\bibinfo {author} {\bibfnamefont {K.}~\bibnamefont
  {Dholakia}} and \bibinfo {author} {\bibfnamefont {T.}~\bibnamefont {{\v
  C}i{\v z}m{\'a}r}},\ }\bibfield  {title} {\enquote {\bibinfo {title} {Shaping
  the future of manipulation},}\ }\href {\doibase 10.1038/nphoton.2011.80}
  {\bibfield  {journal} {\bibinfo  {journal} {Nature Photon.}\ }\textbf
  {\bibinfo {volume} {5}},\ \bibinfo {pages} {335--342} (\bibinfo {year}
  {2011})}\BibitemShut {NoStop}%
\bibitem [{\citenamefont {Rogel-Salazar}\ \emph {et~al.}(2014)\citenamefont
  {Rogel-Salazar}, \citenamefont {Trevi{\~n}o}, and \citenamefont
  {Ch{\'a}vez-Cerda}}]{Rogel-Salazar&al:JOSAB:2014}%
  \BibitemOpen
  \bibfield  {author} {\bibinfo {author} {\bibfnamefont {Jes{\'u}s}\
  \bibnamefont {Rogel-Salazar}}, \bibinfo {author} {\bibfnamefont {Juan~Pablo}\
  \bibnamefont {Trevi{\~n}o}},  and \bibinfo {author} {\bibfnamefont
  {Sabino}\ \bibnamefont {Ch{\'a}vez-Cerda}},\ }\bibfield  {title} {\enquote
  {\bibinfo {title} {Engineering structured light with optical vortices},}\
  }\href {\doibase 10.1364/JOSAB.31.000A46} {\bibfield  {journal} {\bibinfo
  {journal} {J.~Opt.\ Soc\ Am.\ B}\ }\textbf {\bibinfo {volume} {31}},\
  \bibinfo {pages} {A46--A50} (\bibinfo {year} {2014})}\BibitemShut {NoStop}%
\bibitem [{\citenamefont {Molina-Terriza}\ \emph {et~al.}(2002)\citenamefont
  {Molina-Terriza}, \citenamefont {Torres}, and \citenamefont
  {Torner}}]{Molina-Terriza&al:PRL:2002}%
  \BibitemOpen
  \bibfield  {author} {\bibinfo {author} {\bibfnamefont {Gabriel}\ \bibnamefont
  {Molina-Terriza}}, \bibinfo {author} {\bibfnamefont {Juan~P.}\ \bibnamefont
  {Torres}},  and \bibinfo {author} {\bibfnamefont {Lluis}\ \bibnamefont
  {Torner}},\ }\bibfield  {title} {\enquote {\bibinfo {title} {Management of
  the angular momentum of light: Preparation of photons in multidimensional
  vector states of angular momentum},}\ }\href {\doibase
  10.1103/PhysRevLett.88.013601} {\bibfield  {journal} {\bibinfo  {journal}
  {Phys.\ Rev.\ Lett.}\ }\textbf {\bibinfo {volume} {88}},\ \bibinfo {pages}
  {013601(4)} (\bibinfo {year} {2002})}\BibitemShut {NoStop}%
\bibitem [{\citenamefont {Torner}\ \emph {et~al.}(2005)\citenamefont {Torner},
  \citenamefont {Torres}, and \citenamefont {Carrasco}}]{Torner&al:OE:2005}%
  \BibitemOpen
  \bibfield  {author} {\bibinfo {author} {\bibfnamefont {Lluis}\ \bibnamefont
  {Torner}}, \bibinfo {author} {\bibfnamefont {Juan~P.}\ \bibnamefont
  {Torres}},  and \bibinfo {author} {\bibfnamefont {Silvia}\ \bibnamefont
  {Carrasco}},\ }\bibfield  {title} {\enquote {\bibinfo {title} {Digital spiral
  imaging},}\ }\href {\doibase 10.1364/OPEX.13.000873} {\bibfield  {journal}
  {\bibinfo  {journal} {Opt. Express}\ }\textbf {\bibinfo {volume} {13}},\
  \bibinfo {pages} {873--881} (\bibinfo {year} {2005})}\BibitemShut {NoStop}%
\bibitem [{\citenamefont {Leach}\ \emph {et~al.}(2004)\citenamefont {Leach},
  \citenamefont {Yao}, and \citenamefont {Padgett}}]{Leach&al:NJP:2004}%
  \BibitemOpen
  \bibfield  {author} {\bibinfo {author} {\bibfnamefont {Jonathan}\
  \bibnamefont {Leach}}, \bibinfo {author} {\bibfnamefont {Eric}\ \bibnamefont
  {Yao}},  and \bibinfo {author} {\bibfnamefont {Miles~J.}\ \bibnamefont
  {Padgett}},\ }\bibfield  {title} {\enquote {\bibinfo {title} {Observation of
  the vortex structure of a non-integer vortex beam},}\ }\href {\doibase
  10.1088/1367-2630/6/1/071} {\bibfield  {journal} {\bibinfo  {journal} {New
  J.~Phys.}\ }\textbf {\bibinfo {volume} {6}},\ \bibinfo {pages} {71} (\bibinfo
  {year} {2004})}\BibitemShut {NoStop}%
\bibitem [{\citenamefont {Molina-Terriza}\ \emph
  {et~al.}(2007{\natexlab{a}})\citenamefont {Molina-Terriza}, \citenamefont
  {Rebane}, \citenamefont {Torres}, \citenamefont {Torner}, and \citenamefont
  {Carrasco}}]{Molina-Terriza&al:JEOS:2007}%
  \BibitemOpen
  \bibfield  {author} {\bibinfo {author} {\bibfnamefont {Gabriel}\ \bibnamefont
  {Molina-Terriza}}, \bibinfo {author} {\bibfnamefont {Liis}\ \bibnamefont
  {Rebane}}, \bibinfo {author} {\bibfnamefont {Juan~P.}\ \bibnamefont
  {Torres}}, \bibinfo {author} {\bibfnamefont {Lluis}\ \bibnamefont {Torner}},
   and \bibinfo {author} {\bibfnamefont {Silvia}\ \bibnamefont {Carrasco}},\
  }\bibfield  {title} {\enquote {\bibinfo {title} {Probing canonical
  geometrical objects by digital spiral imaging},}\ }\href {\doibase
  10.2971/jeos.2007.07014} {\bibfield  {journal} {\bibinfo  {journal} {J.~Eur.\
  Opt.\ Soc.}\ }\textbf {\bibinfo {volume} {2}},\ \bibinfo {pages} {07014(6)}
  (\bibinfo {year} {2007}{\natexlab{a}})}\BibitemShut {NoStop}%
\bibitem [{\citenamefont {Lavery}\ \emph {et~al.}(2011)\citenamefont {Lavery},
  \citenamefont {Berkhout}, \citenamefont {Courtial}, and \citenamefont
  {Padgett}}]{Lavery&al:JO:2011}%
  \BibitemOpen
  \bibfield  {author} {\bibinfo {author} {\bibfnamefont {Martin P.~J.}\
  \bibnamefont {Lavery}}, \bibinfo {author} {\bibfnamefont {Gregorius C.~G.}\
  \bibnamefont {Berkhout}}, \bibinfo {author} {\bibfnamefont {Johannes}\
  \bibnamefont {Courtial}},  and \bibinfo {author} {\bibfnamefont {Miles~J.}\
  \bibnamefont {Padgett}},\ }\bibfield  {title} {\enquote {\bibinfo {title}
  {Measurement of the light orbital angular momentum spectrum using an optical
  geometric transformation},}\ }\href {\doibase 10.1088/2040-8978/13/6/064006}
  {\bibfield  {journal} {\bibinfo  {journal} {J.~Opt.}\ }\textbf {\bibinfo
  {volume} {13}},\ \bibinfo {pages} {064006} (\bibinfo {year}
  {2011})}\BibitemShut {NoStop}%
\bibitem [{\citenamefont {Vucetic} and \citenamefont
  {Yuan}(2003)}]{Vucetic&Yuan:Book:2003}%
  \BibitemOpen
  \bibfield  {author} {\bibinfo {author} {\bibfnamefont {Branka}\ \bibnamefont
  {Vucetic}} and \bibinfo {author} {\bibfnamefont {Jinhong}\ \bibnamefont
  {Yuan}},\ }\href@noop {} {\emph {\bibinfo {title} {Space-Time Coding}}}\
  (\bibinfo  {publisher} {John Wiley \& Sons},\ \bibinfo {address} {UK},\
  \bibinfo {year} {2003})\BibitemShut {NoStop}%
\bibitem [{\citenamefont {Boffi}\ \emph {et~al.}(2013)\citenamefont {Boffi},
  \citenamefont {Martelli}, \citenamefont {Gatto}, and \citenamefont
  {Martinelli}}]{Boffi&al:SPIE:2013}%
  \BibitemOpen
  \bibfield  {author} {\bibinfo {author} {\bibfnamefont {Pierpaolo}\
  \bibnamefont {Boffi}}, \bibinfo {author} {\bibfnamefont {P.}~\bibnamefont
  {Martelli}}, \bibinfo {author} {\bibfnamefont {A.}~\bibnamefont {Gatto}}, \
  and\ \bibinfo {author} {\bibfnamefont {M.}~\bibnamefont {Martinelli}},\
  }\bibfield  {title} {\enquote {\bibinfo {title} {Optical vortices: an
  innovative approach to increase spectral efficiency by fiber mode-division
  multiplexing},}\ }in\ \href {\doibase 10.1117/12.2003588} {\emph {\bibinfo
  {booktitle} {Next-Generation Optical Communication: Components, Sub-Systems,
  and Systems I}}},\ \bibinfo {series} {Proc. of SPIE}, Vol.\ \bibinfo {volume}
  {8647}\ (\bibinfo {address} {San Francisco, CA, USA},\ \bibinfo {year}
  {2013})\ p.\ \bibinfo {pages} {864705}\BibitemShut {NoStop}%
\bibitem [{\citenamefont {Yu}(2015)}]{Yu:OE:2015}%
  \BibitemOpen
  \bibfield  {author} {\bibinfo {author} {\bibfnamefont {Siyuan}\ \bibnamefont
  {Yu}},\ }\bibfield  {title} {\enquote {\bibinfo {title} {Potentials and
  challenges of using orbital angular momentum communications in optical
  interconnects},}\ }\href {\doibase 10.1364/OE.23.003075} {\bibfield
  {journal} {\bibinfo  {journal} {Opt. Express}\ }\textbf {\bibinfo {volume}
  {23}},\ \bibinfo {pages} {3075--3087} (\bibinfo {year} {2015})}\BibitemShut
  {NoStop}%
\bibitem [{\citenamefont {Sasaki} and \citenamefont
  {{McNulty}}(2008)}]{Sasaki&al:PRL:2008}%
  \BibitemOpen
  \bibfield  {author} {\bibinfo {author} {\bibfnamefont {Shigemi}\ \bibnamefont
  {Sasaki}} and \bibinfo {author} {\bibfnamefont {Ian}\ \bibnamefont
  {{McNulty}}},\ }\bibfield  {title} {\enquote {\bibinfo {title} {Proposal for
  generating brilliant {X}-ray beams carrying orbital angular momentum},}\
  }\href {\doibase 10.1103/PhysRevLett.100.124801} {\bibfield  {journal}
  {\bibinfo  {journal} {Phys.\ Rev.\ Lett.}\ }\textbf {\bibinfo {volume}
  {100}},\ \bibinfo {pages} {124801} (\bibinfo {year} {2008})}\BibitemShut
  {NoStop}%
\bibitem [{\citenamefont {Courtial}\ \emph {et~al.}(1997)\citenamefont
  {Courtial}, \citenamefont {Dholakia}, \citenamefont {Allen}, and
  \citenamefont {Padgett}}]{Courtial&al:OC::1997}%
  \BibitemOpen
  \bibfield  {author} {\bibinfo {author} {\bibfnamefont {J.}~\bibnamefont
  {Courtial}}, \bibinfo {author} {\bibfnamefont {K.}~\bibnamefont {Dholakia}},
  \bibinfo {author} {\bibfnamefont {L.}~\bibnamefont {Allen}},  and \bibinfo
  {author} {\bibfnamefont {M.J.}\ \bibnamefont {Padgett}},\ }\bibfield  {title}
  {\enquote {\bibinfo {title} {Gaussian beams with very high orbital angular
  momentum},}\ }\href {\doibase 10.1016/S0030-4018(97)00376-3} {\bibfield
  {journal} {\bibinfo  {journal} {Opt. Commun.}\ }\textbf {\bibinfo {volume}
  {144}},\ \bibinfo {pages} {210--213} (\bibinfo {year} {1997})}\BibitemShut
  {NoStop}%
\bibitem [{\citenamefont {Belinfante}(1940)}]{Belinfante:P:1940}%
  \BibitemOpen
  \bibfield  {author} {\bibinfo {author} {\bibfnamefont {F.~J.}\ \bibnamefont
  {Belinfante}},\ }\bibfield  {title} {\enquote {\bibinfo {title} {On the
  current and the density of the electric charge, the energy, the linear
  momentum and the angular momentum of arbitrary fields},}\ }\href {\doibase
  10.1016/S0031-8914(40)90091-X} {\bibfield  {journal} {\bibinfo  {journal}
  {Physica}\ }\textbf {\bibinfo {volume} {7}},\ \bibinfo {pages} {449--474}
  (\bibinfo {year} {1940})}\BibitemShut {NoStop}%
\bibitem [{\citenamefont {Soper}(1976)}]{Soper:Book:1976}%
  \BibitemOpen
  \bibfield  {author} {\bibinfo {author} {\bibfnamefont {Davison~E.}\
  \bibnamefont {Soper}},\ }\href@noop {} {\emph {\bibinfo {title} {Classical
  Field Theory}}}\ (\bibinfo  {publisher} {John Wiley \&~Sons,~Inc.},\ \bibinfo
  {address} {New York, NY, USA},\ \bibinfo {year} {1976})\BibitemShut {NoStop}%
\bibitem [{\citenamefont {Truesdell}(1968)}]{Truesdell:Book:1968}%
  \BibitemOpen
  \bibfield  {author} {\bibinfo {author} {\bibfnamefont {Clifford}\
  \bibnamefont {Truesdell}},\ }\href@noop {} {\emph {\bibinfo {title} {Essays
  in the History of Mechanics}}}\ (\bibinfo  {publisher} {Springer-Verlag},\
  \bibinfo {address} {Berlin, Heidelberg, New York},\ \bibinfo {year}
  {1968})\BibitemShut {NoStop}%
\bibitem [{\citenamefont {Ribari{\v c}} and \citenamefont {{\v S}u{\v
  s}ter{\v s}i{\v c}}(1990)}]{Ribaric&Sustersic:Book:1990}%
  \BibitemOpen
  \bibfield  {author} {\bibinfo {author} {\bibfnamefont {Marijan}\ \bibnamefont
  {Ribari{\v c}}} and \bibinfo {author} {\bibfnamefont {Luka}\ \bibnamefont
  {{\v S}u{\v s}ter{\v s}i{\v c}}},\ }\href@noop {} {\emph {\bibinfo {title}
  {Conservation Laws and Open Questions of Classical Electrodynamics}}}\
  (\bibinfo  {publisher} {World Scientific},\ \bibinfo {address} {Singapore,
  New Jersey, London, Hong Kong},\ \bibinfo {year} {1990})\BibitemShut
  {NoStop}%
\bibitem [{\citenamefont {Allen}\ \emph {et~al.}(2003)\citenamefont {Allen},
  \citenamefont {Barnett}, and \citenamefont {Padgett}}]{Allen&al:Book:2003}%
  \BibitemOpen
  \bibfield  {author} {\bibinfo {author} {\bibfnamefont {L.}~\bibnamefont
  {Allen}}, \bibinfo {author} {\bibfnamefont {S.~M.}\ \bibnamefont {Barnett}},
   and \bibinfo {author} {\bibfnamefont {M.~J.}\ \bibnamefont {Padgett}},\
  }\href@noop {} {\emph {\bibinfo {title} {Optical Angular Momentum}}}\
  (\bibinfo  {publisher} {IOP},\ \bibinfo {address} {Bristol, UK},\ \bibinfo
  {year} {2003})\BibitemShut {NoStop}%
\bibitem [{\citenamefont {Andrews}(2008)}]{Andrews:Book:2008}%
  \BibitemOpen
  \bibfield  {author} {\bibinfo {author} {\bibfnamefont {David~L.}\
  \bibnamefont {Andrews}},\ }\href@noop {} {\emph {\bibinfo {title} {Structured
  Light and Its Applications: An Introduction to Phase-Structured Beams and
  Nanoscale Optical Forces}}}\ (\bibinfo  {publisher} {Academic Press},\
  \bibinfo {address} {Amsterdam, NL},\ \bibinfo {year} {2008})\BibitemShut
  {NoStop}%
\bibitem [{\citenamefont {Torres} and \citenamefont
  {Torner}(2011)}]{Torres&Torner:Book:2011}%
  \BibitemOpen
  \bibfield  {author} {\bibinfo {author} {\bibfnamefont {Juan~P.}\ \bibnamefont
  {Torres}} and \bibinfo {author} {\bibfnamefont {Lluis}\ \bibnamefont
  {Torner}},\ }\href@noop {} {\emph {\bibinfo {title} {Twisted Photons:
  Applications of Light With Orbital Angular Momentum}}}\ (\bibinfo
  {publisher} {Wiley-Vch Verlag, John Wiley and Sons},\ \bibinfo {address}
  {Weinheim, DE},\ \bibinfo {year} {2011})\BibitemShut {NoStop}%
\bibitem [{\citenamefont {Yao} and \citenamefont
  {Padgett}(2011)}]{Yao&Padgett:AOP:2011}%
  \BibitemOpen
  \bibfield  {author} {\bibinfo {author} {\bibfnamefont {Alison~M.}\
  \bibnamefont {Yao}} and \bibinfo {author} {\bibfnamefont {Miles~J.}\
  \bibnamefont {Padgett}},\ }\bibfield  {title} {\enquote {\bibinfo {title}
  {Orbital angular momentum: origins, behavior and applications},}\ }\href
  {\doibase 10.1364/AOP.3.000161} {\bibfield  {journal} {\bibinfo  {journal}
  {Adv.\ Opt.\ Photon.}\ }\textbf {\bibinfo {volume} {3}},\ \bibinfo {pages}
  {161--204} (\bibinfo {year} {2011})}\BibitemShut {NoStop}%
\bibitem [{\citenamefont {Beth}(1935)}]{Beth:PR:1935}%
  \BibitemOpen
  \bibfield  {author} {\bibinfo {author} {\bibfnamefont {Richard~A.}\
  \bibnamefont {Beth}},\ }\bibfield  {title} {\enquote {\bibinfo {title}
  {Direct detection of the angular momentum of light},}\ }\href {\doibase
  10.1103/PhysRev.48.471} {\bibfield  {journal} {\bibinfo  {journal} {Phys.\
  Rev.}\ }\textbf {\bibinfo {volume} {48}},\ \bibinfo {pages} {471} (\bibinfo
  {year} {1935})}\BibitemShut {NoStop}%
\bibitem [{\citenamefont {Beth}(1936)}]{Beth:PR:1936}%
  \BibitemOpen
  \bibfield  {author} {\bibinfo {author} {\bibfnamefont {Richard~A.}\
  \bibnamefont {Beth}},\ }\bibfield  {title} {\enquote {\bibinfo {title}
  {Mechanical detection and measurement of the angular momentum of light},}\
  }\href {\doibase 10.1103/PhysRev.50.115} {\bibfield  {journal} {\bibinfo
  {journal} {Phys.\ Rev.}\ }\textbf {\bibinfo {volume} {50}},\ \bibinfo {pages}
  {115--125} (\bibinfo {year} {1936})}\BibitemShut {NoStop}%
\bibitem [{\citenamefont {Holbourn}(1936)}]{Holbourn:N:1936}%
  \BibitemOpen
  \bibfield  {author} {\bibinfo {author} {\bibfnamefont {A.~H.~S.}\
  \bibnamefont {Holbourn}},\ }\bibfield  {title} {\enquote {\bibinfo {title}
  {Angular momentum of circularly polarised light},}\ }\href {\doibase
  10.1038/137031a0} {\bibfield  {journal} {\bibinfo  {journal} {Nature}\
  }\textbf {\bibinfo {volume} {137}},\ \bibinfo {pages} {31--31} (\bibinfo
  {year} {1936})}\BibitemShut {NoStop}%
\bibitem [{\citenamefont {Carrara}(1949)}]{Carrara:N:1949}%
  \BibitemOpen
  \bibfield  {author} {\bibinfo {author} {\bibfnamefont {Nello}\ \bibnamefont
  {Carrara}},\ }\bibfield  {title} {\enquote {\bibinfo {title} {Torque and
  angular momentum of centimetre electromagnetic waves},}\ }\href {\doibase
  10.1038/164882c0} {\bibfield  {journal} {\bibinfo  {journal} {Nature}\
  }\textbf {\bibinfo {volume} {164}},\ \bibinfo {pages} {882--884} (\bibinfo
  {year} {1949})}\BibitemShut {NoStop}%
\bibitem [{\citenamefont {Allen}(1966)}]{Allen:AJP:1966}%
  \BibitemOpen
  \bibfield  {author} {\bibinfo {author} {\bibfnamefont {P.~J.}\ \bibnamefont
  {Allen}},\ }\bibfield  {title} {\enquote {\bibinfo {title} {A radiation
  torque experiment},}\ }\href {\doibase 10.1119/1.1972585} {\bibfield
  {journal} {\bibinfo  {journal} {Am.\ J.~Phys.}\ }\textbf {\bibinfo {volume}
  {34}},\ \bibinfo {pages} {1185--1192} (\bibinfo {year} {1966})}\BibitemShut
  {NoStop}%
\bibitem [{\citenamefont {Carusotto}\ \emph {et~al.}(1968)\citenamefont
  {Carusotto}, \citenamefont {Fornaca}, and \citenamefont
  {Polacco}}]{Carusotto&al:NC:1968}%
  \BibitemOpen
  \bibfield  {author} {\bibinfo {author} {\bibfnamefont {S.}~\bibnamefont
  {Carusotto}}, \bibinfo {author} {\bibfnamefont {G.}~\bibnamefont {Fornaca}},
   and \bibinfo {author} {\bibfnamefont {E.}~\bibnamefont {Polacco}},\
  }\bibfield  {title} {\enquote {\bibinfo {title} {Radiation beats and rotating
  systems},}\ }\href {\doibase 10.1007/BF0271096} {\bibfield  {journal}
  {\bibinfo  {journal} {Nuov.\ Cim.}\ }\textbf {\bibinfo {volume} {53}},\
  \bibinfo {pages} {87--97} (\bibinfo {year} {1968})}\BibitemShut {NoStop}%
\bibitem [{\citenamefont {Chang} and \citenamefont
  {Lee}(1985)}]{Chang&Lee:JOSAB:1985}%
  \BibitemOpen
  \bibfield  {author} {\bibinfo {author} {\bibfnamefont {Soo}\ \bibnamefont
  {Chang}} and \bibinfo {author} {\bibfnamefont {Sang~Soo}\ \bibnamefont
  {Lee}},\ }\bibfield  {title} {\enquote {\bibinfo {title} {Optical torque
  exerted on a homogeneous sphere levitated in the circularly polarized
  fundamental-mode laser beam},}\ }\href {\doibase 10.1364/JOSAB.2.001853}
  {\bibfield  {journal} {\bibinfo  {journal} {J.~Opt.\ Soc\ Am.\ B}\ }\textbf
  {\bibinfo {volume} {2}},\ \bibinfo {pages} {1853--1860} (\bibinfo {year}
  {1985})}\BibitemShut {NoStop}%
\bibitem [{\citenamefont {Vul'fson}(1987)}]{Vulfson:USP:1987}%
  \BibitemOpen
  \bibfield  {author} {\bibinfo {author} {\bibfnamefont {K.~S.}\ \bibnamefont
  {Vul'fson}},\ }\bibfield  {title} {\enquote {\bibinfo {title} {Angular
  momentum of electromagnetic waves},}\ }\href {\doibase
  10.1070/PU1987v030n08ABEH002940} {\bibfield  {journal} {\bibinfo  {journal}
  {Sov.\ Phys.\ Usp.}\ }\textbf {\bibinfo {volume} {30}},\ \bibinfo {pages}
  {667--674} (\bibinfo {year} {1987})}\BibitemShut {NoStop}%
\bibitem [{\citenamefont {Kristensen}\ \emph {et~al.}(1994)\citenamefont
  {Kristensen}, \citenamefont {Beijersbergen}, and \citenamefont
  {Woerdman}}]{Kristensen&al:OC:1994}%
  \BibitemOpen
  \bibfield  {author} {\bibinfo {author} {\bibfnamefont {M.}~\bibnamefont
  {Kristensen}}, \bibinfo {author} {\bibfnamefont {M.~W.}\ \bibnamefont
  {Beijersbergen}},  and \bibinfo {author} {\bibfnamefont {J.~P.}\
  \bibnamefont {Woerdman}},\ }\bibfield  {title} {\enquote {\bibinfo {title}
  {Angular momentum and spin-orbit coupling for microwave photons},}\ }\href
  {\doibase 10.1016/0030-4018(94)90547-9} {\bibfield  {journal} {\bibinfo
  {journal} {Opt. Commun.}\ }\textbf {\bibinfo {volume} {104}},\ \bibinfo
  {pages} {229--233} (\bibinfo {year} {1994})}\BibitemShut {NoStop}%
\bibitem [{\citenamefont {He}\ \emph {et~al.}(1995)\citenamefont {He},
  \citenamefont {Friese}, \citenamefont {Heckenberg}, and \citenamefont
  {Rubinsztein-Dunlop}}]{He&al:PRL:1995}%
  \BibitemOpen
  \bibfield  {author} {\bibinfo {author} {\bibfnamefont {H.}~\bibnamefont
  {He}}, \bibinfo {author} {\bibfnamefont {M.~E.~J.}\ \bibnamefont {Friese}},
  \bibinfo {author} {\bibfnamefont {N.~R.}\ \bibnamefont {Heckenberg}},  and
  \bibinfo {author} {\bibfnamefont {H.}~\bibnamefont {Rubinsztein-Dunlop}},\
  }\bibfield  {title} {\enquote {\bibinfo {title} {Direct observation of
  transfer of angular momentum to absorptive particles from a laser beam with a
  phase singularity},}\ }\href {\doibase 10.1103/PhysRevLett.75.826} {\bibfield
   {journal} {\bibinfo  {journal} {Phys.\ Rev.\ Lett.}\ }\textbf {\bibinfo
  {volume} {75}},\ \bibinfo {pages} {826--829} (\bibinfo {year}
  {1995})}\BibitemShut {NoStop}%
\bibitem [{\citenamefont {Friese}\ \emph {et~al.}(1996)\citenamefont {Friese},
  \citenamefont {Enger}, \citenamefont {Rubinsztein-Dunlop}, and
  \citenamefont {Heckenberg}}]{Friese&al:PRA:1996}%
  \BibitemOpen
  \bibfield  {author} {\bibinfo {author} {\bibfnamefont {M.~E.~J.}\
  \bibnamefont {Friese}}, \bibinfo {author} {\bibfnamefont {J.}~\bibnamefont
  {Enger}}, \bibinfo {author} {\bibfnamefont {H.}~\bibnamefont
  {Rubinsztein-Dunlop}},  and \bibinfo {author} {\bibfnamefont {N.~R.}\
  \bibnamefont {Heckenberg}},\ }\bibfield  {title} {\enquote {\bibinfo {title}
  {Optical angular-momentum transfer to trapped absorbing particles},}\ }\href
  {\doibase 10.1103/PhysRevA.54.1593} {\bibfield  {journal} {\bibinfo
  {journal} {Phys.\ Rev.\ A}\ }\textbf {\bibinfo {volume} {54}},\ \bibinfo
  {pages} {1593--1596} (\bibinfo {year} {1996})}\BibitemShut {NoStop}%
\bibitem [{\citenamefont {Then} and \citenamefont
  {Thid\'e}(2008)}]{Then&Thide:ARXIV:2008}%
  \BibitemOpen
  \bibfield  {author} {\bibinfo {author} {\bibfnamefont {H.}~\bibnamefont
  {Then}} and \bibinfo {author} {\bibfnamefont {B.}~\bibnamefont {Thid\'e}},\
  }\href {http://arxiv.org/abs/0803.0200} {\enquote {\bibinfo {title}
  {Mechanical properties of the radio frequency field emitted by an antenna
  array},}\ } (\bibinfo {year} {2008}),\ \Eprint
  {http://arxiv.org/abs/0803.0200} {arXiv.org:0803.0200 [physics.class-ph]}
  \BibitemShut {NoStop}%
\bibitem [{\citenamefont {Helmerson}\ \emph {et~al.}(2009)\citenamefont
  {Helmerson}, \citenamefont {Andersen}, \citenamefont {Clad\'e}, \citenamefont
  {Natarajan}, \citenamefont {Phillips}, \citenamefont {Ramanathan},
  \citenamefont {Ryu}, and \citenamefont
  {Vaziri}}]{Helmerson&al:Topologica:2009}%
  \BibitemOpen
  \bibfield  {author} {\bibinfo {author} {\bibfnamefont {K.}~\bibnamefont
  {Helmerson}}, \bibinfo {author} {\bibfnamefont {M.~F.}\ \bibnamefont
  {Andersen}}, \bibinfo {author} {\bibfnamefont {P.}~\bibnamefont {Clad\'e}},
  \bibinfo {author} {\bibfnamefont {V.}~\bibnamefont {Natarajan}}, \bibinfo
  {author} {\bibfnamefont {W.~D.}\ \bibnamefont {Phillips}}, \bibinfo {author}
  {\bibfnamefont {A.}~\bibnamefont {Ramanathan}}, \bibinfo {author}
  {\bibfnamefont {C.}~\bibnamefont {Ryu}},  and \bibinfo {author}
  {\bibfnamefont {A.}~\bibnamefont {Vaziri}},\ }\bibfield  {title} {\enquote
  {\bibinfo {title} {Vortices and persistent currents: Rotating a
  {Bose-Einstein} condensate using photons with orbital angular momentum},}\
  }\href {\doibase 10.3731/topologica.2.002} {\bibfield  {journal} {\bibinfo
  {journal} {Topologica}\ }\textbf {\bibinfo {volume} {2}},\ \bibinfo {pages}
  {002--1--002--12} (\bibinfo {year} {2009})}\BibitemShut {NoStop}%
\bibitem [{\citenamefont {Padgett} and \citenamefont
  {Bowman}(2011)}]{Padgett&Bowman:NPHY:2011}%
  \BibitemOpen
  \bibfield  {author} {\bibinfo {author} {\bibfnamefont {Miles}\ \bibnamefont
  {Padgett}} and \bibinfo {author} {\bibfnamefont {Richard}\ \bibnamefont
  {Bowman}},\ }\bibfield  {title} {\enquote {\bibinfo {title} {Tweezers with a
  twist},}\ }\href {\doibase 10.1038/nphoton.2011.81} {\bibfield  {journal}
  {\bibinfo  {journal} {Nature Phys.}\ }\textbf {\bibinfo {volume} {5}},\
  \bibinfo {pages} {343--348} (\bibinfo {year} {2011})}\BibitemShut {NoStop}%
\bibitem [{\citenamefont {Ramanathan}\ \emph {et~al.}(2011)\citenamefont
  {Ramanathan}, \citenamefont {Wright}, \citenamefont {Muniz}, \citenamefont
  {Zelan}, \citenamefont {Hill}, \citenamefont {Lobb}, \citenamefont
  {Helmerson}, \citenamefont {Phillips}, and \citenamefont
  {Campbell}}]{Ramanathan&al:PRL:2011}%
  \BibitemOpen
  \bibfield  {author} {\bibinfo {author} {\bibfnamefont {A.}~\bibnamefont
  {Ramanathan}}, \bibinfo {author} {\bibfnamefont {K.~C.}\ \bibnamefont
  {Wright}}, \bibinfo {author} {\bibfnamefont {S.~R.}\ \bibnamefont {Muniz}},
  \bibinfo {author} {\bibfnamefont {M.}~\bibnamefont {Zelan}}, \bibinfo
  {author} {\bibfnamefont {W.~T.}\ \bibnamefont {Hill}}, \bibinfo {author}
  {\bibfnamefont {C.~J.}\ \bibnamefont {Lobb}}, \bibinfo {author}
  {\bibfnamefont {K.}~\bibnamefont {Helmerson}}, \bibinfo {author}
  {\bibfnamefont {W.~D.}\ \bibnamefont {Phillips}},  and \bibinfo {author}
  {\bibfnamefont {G.~K.}\ \bibnamefont {Campbell}},\ }\bibfield  {title}
  {\enquote {\bibinfo {title} {Superflow in a toroidal {Bose-Einstein}
  condensate: An atom circuit with a tunable weak link},}\ }\href {\doibase
  10.1103/PhysRevLett.106.130401} {\bibfield  {journal} {\bibinfo  {journal}
  {Phys.\ Rev.\ Lett.}\ }\textbf {\bibinfo {volume} {106}},\ \bibinfo {pages}
  {130401} (\bibinfo {year} {2011})}\BibitemShut {NoStop}%
\bibitem [{\citenamefont {Elias}(2012)}]{Elias:AA:2012}%
  \BibitemOpen
  \bibfield  {author} {\bibinfo {author} {\bibfnamefont {Nicholas~M.}\
  \bibnamefont {Elias}, \bibfnamefont {II}},\ }\bibfield  {title} {\enquote
  {\bibinfo {title} {Photon orbital angular momentum and torque metrics for
  single telescopes and interferometers},}\ }\href {\doibase
  10.1051/0004-6361/201218774} {\bibfield  {journal} {\bibinfo  {journal}
  {Astron.\ Astrophys.}\ }\textbf {\bibinfo {volume} {541}},\ \bibinfo {pages}
  {A101} (\bibinfo {year} {2012})}\BibitemShut {NoStop}%
\bibitem [{\citenamefont {Emile}\ \emph {et~al.}(2014)\citenamefont {Emile},
  \citenamefont {Brousseau}, \citenamefont {Emile}, \citenamefont {Niemiec},
  \citenamefont {Madhjoubi}, and \citenamefont
  {Thid\'{e}}}]{Emile&al:PRL:2014}%
  \BibitemOpen
  \bibfield  {author} {\bibinfo {author} {\bibfnamefont {Olivier}\ \bibnamefont
  {Emile}}, \bibinfo {author} {\bibfnamefont {Christian}\ \bibnamefont
  {Brousseau}}, \bibinfo {author} {\bibfnamefont {Janine}\ \bibnamefont
  {Emile}}, \bibinfo {author} {\bibfnamefont {Ronan}\ \bibnamefont {Niemiec}},
  \bibinfo {author} {\bibfnamefont {Kouroch}\ \bibnamefont {Madhjoubi}},  and
  \bibinfo {author} {\bibfnamefont {Bo}~\bibnamefont {Thid\'{e}}},\ }\bibfield
  {title} {\enquote {\bibinfo {title} {Electromagnetically induced torque on a
  large ring in the microwave range},}\ }\href {\doibase
  10.1103/PhysRevLett.112.053902} {\bibfield  {journal} {\bibinfo  {journal}
  {Phys.\ Rev.\ Lett.}\ }\textbf {\bibinfo {volume} {112}},\ \bibinfo {pages}
  {053902(4)} (\bibinfo {year} {2014})}\BibitemShut {NoStop}%
\bibitem [{\citenamefont {Thid{\'e}}(2007)}]{Thide:PPCF:2007}%
  \BibitemOpen
  \bibfield  {author} {\bibinfo {author} {\bibfnamefont {B.}~\bibnamefont
  {Thid{\'e}}},\ }\bibfield  {title} {\enquote {\bibinfo {title} {Nonlinear
  physics of the ionosphere and {LOIS}/{LOFAR}},}\ }\href {\doibase
  10.1088/0741-3335/49/12B/S09} {\bibfield  {journal} {\bibinfo  {journal}
  {Plasma Phys.\ Contr.\ Fusion}\ }\textbf {\bibinfo {volume} {49}},\ \bibinfo
  {pages} {B103--B107} (\bibinfo {year} {2007})}\BibitemShut {NoStop}%
\bibitem [{\citenamefont {Mendon\c{c}a}\ \emph {et~al.}(2008)\citenamefont
  {Mendon\c{c}a}, \citenamefont {Thid{\'e}}, \citenamefont {Bergman},
  \citenamefont {Mohammadi}, \citenamefont {Eliasson}, \citenamefont {Baan},\
  and\ \citenamefont {Then}}]{Mendonca&al:ARXIV:2008}%
  \BibitemOpen
  \bibfield  {author} {\bibinfo {author} {\bibfnamefont {J.~T.}\ \bibnamefont
  {Mendon\c{c}a}}, \bibinfo {author} {\bibfnamefont {B.}~\bibnamefont
  {Thid{\'e}}}, \bibinfo {author} {\bibfnamefont {J.~E.~S.}\ \bibnamefont
  {Bergman}}, \bibinfo {author} {\bibfnamefont {S.~M.}\ \bibnamefont
  {Mohammadi}}, \bibinfo {author} {\bibfnamefont {B.}~\bibnamefont {Eliasson}},
  \bibinfo {author} {\bibfnamefont {W.}~\bibnamefont {Baan}},  and \bibinfo
  {author} {\bibfnamefont {H.}~\bibnamefont {Then}},\ }\href@noop {} {\enquote
  {\bibinfo {title} {Photon orbital angular momentum in a plasma vortex},}\ }
  (\bibinfo {year} {2008}),\ \Eprint {http://arxiv.org/abs/0804.3221}
  {arXiv.org:0804.3221 [physics.plasm-ph]} \BibitemShut {NoStop}%
\bibitem [{\citenamefont {Mendon\c{c}a}\ \emph {et~al.}(2009)\citenamefont
  {Mendon\c{c}a}, \citenamefont {Thid{\'e}}, and \citenamefont
  {Then}}]{Mendonca&al:PRL:2009}%
  \BibitemOpen
  \bibfield  {author} {\bibinfo {author} {\bibfnamefont {J.~T.}\ \bibnamefont
  {Mendon\c{c}a}}, \bibinfo {author} {\bibfnamefont {B.}~\bibnamefont
  {Thid{\'e}}},  and \bibinfo {author} {\bibfnamefont {H.}~\bibnamefont
  {Then}},\ }\bibfield  {title} {\enquote {\bibinfo {title} {Stimulated {R}aman
  and {B}rillouin backscattering of collimated beams carrying orbital angular
  momentum},}\ }\href {\doibase 10.1103/PhysRevLett.102.185005} {\bibfield
  {journal} {\bibinfo  {journal} {Phys.\ Rev.\ Lett.}\ }\textbf {\bibinfo
  {volume} {102}},\ \bibinfo {pages} {185005(4)} (\bibinfo {year}
  {2009})}\BibitemShut {NoStop}%
\bibitem [{\citenamefont {Tamburini}\ \emph {et~al.}(2010)\citenamefont
  {Tamburini}, \citenamefont {Sponselli}, \citenamefont {Thid\'{e}}, and
  \citenamefont {Mendon\c{c}a}}]{Tamburini&al:EPL:2010}%
  \BibitemOpen
  \bibfield  {author} {\bibinfo {author} {\bibfnamefont {F.}~\bibnamefont
  {Tamburini}}, \bibinfo {author} {\bibfnamefont {A.}~\bibnamefont
  {Sponselli}}, \bibinfo {author} {\bibfnamefont {B.}~\bibnamefont
  {Thid\'{e}}},  and \bibinfo {author} {\bibfnamefont {J.~T.}\ \bibnamefont
  {Mendon\c{c}a}},\ }\bibfield  {title} {\enquote {\bibinfo {title} {Photon
  orbital angular momentum and mass in a plasma vortex},}\ }\href {\doibase
  10.1209/0295-5075/90/45001} {\bibfield  {journal} {\bibinfo  {journal}
  {Europhys.\ Lett.}\ }\textbf {\bibinfo {volume} {90}},\ \bibinfo {pages}
  {45001} (\bibinfo {year} {2010})}\BibitemShut {NoStop}%
\bibitem [{\citenamefont {Tamburini} and \citenamefont
  {Thid\'{e}}(2011)}]{Tamburini&Thide:EPL:2011}%
  \BibitemOpen
  \bibfield  {author} {\bibinfo {author} {\bibfnamefont {F.}~\bibnamefont
  {Tamburini}} and \bibinfo {author} {\bibfnamefont {B.}~\bibnamefont
  {Thid\'{e}}},\ }\bibfield  {title} {\enquote {\bibinfo {title} {Storming
  {Majorana's Tower} with {OAM} states of light in a plasma},}\ }\href
  {\doibase 10.1209/0295-5075/96/64005} {\bibfield  {journal} {\bibinfo
  {journal} {Europhys.\ Lett.}\ }\textbf {\bibinfo {volume} {96}},\ \bibinfo
  {pages} {64005} (\bibinfo {year} {2011})}\BibitemShut {NoStop}%
\bibitem [{\citenamefont {Mendon\c{c}a}(2012)}]{Mendonca:PPCF:2012}%
  \BibitemOpen
  \bibfield  {author} {\bibinfo {author} {\bibfnamefont {J.~T.}\ \bibnamefont
  {Mendon\c{c}a}},\ }\bibfield  {title} {\enquote {\bibinfo {title} {Twisted
  waves in a plasma},}\ }\href {\doibase 10.1088/0741-3335/54/12/124031}
  {\bibfield  {journal} {\bibinfo  {journal} {Plasma Phys.\ Contr.\ Fusion}\
  }\textbf {\bibinfo {volume} {54}},\ \bibinfo {pages} {124031} (\bibinfo
  {year} {2012})}\BibitemShut {NoStop}%
\bibitem [{\citenamefont {Ibanescu}\ \emph {et~al.}(2000)\citenamefont
  {Ibanescu}, \citenamefont {Fink}, \citenamefont {Fan}, \citenamefont
  {Thomas}, and \citenamefont {Joannopoulos}}]{Ibanescu&al:S:2000}%
  \BibitemOpen
  \bibfield  {author} {\bibinfo {author} {\bibfnamefont {M.}~\bibnamefont
  {Ibanescu}}, \bibinfo {author} {\bibfnamefont {Y.}~\bibnamefont {Fink}},
  \bibinfo {author} {\bibfnamefont {S.}~\bibnamefont {Fan}}, \bibinfo {author}
  {\bibfnamefont {E.~L.}\ \bibnamefont {Thomas}},  and \bibinfo {author}
  {\bibfnamefont {J.~D.}\ \bibnamefont {Joannopoulos}},\ }\bibfield  {title}
  {\enquote {\bibinfo {title} {An all-dielectric coaxial waveguide},}\ }\href
  {\doibase 10.1126/science.289.5478.415} {\bibfield  {journal} {\bibinfo
  {journal} {Science}\ }\textbf {\bibinfo {volume} {289}},\ \bibinfo {pages}
  {415--419} (\bibinfo {year} {2000})}\BibitemShut {NoStop}%
\bibitem [{\citenamefont {Alexeyev}\ \emph {et~al.}(1998)\citenamefont
  {Alexeyev}, \citenamefont {Fadeyeva}, \citenamefont {Volyar}, and
  \citenamefont {Soskin}}]{Alexeyev&al:SPQEOE:1998}%
  \BibitemOpen
  \bibfield  {author} {\bibinfo {author} {\bibfnamefont {A.~N.}\ \bibnamefont
  {Alexeyev}}, \bibinfo {author} {\bibfnamefont {T.}~\bibnamefont {Fadeyeva}},
  \bibinfo {author} {\bibfnamefont {A.~V.}\ \bibnamefont {Volyar}},  and
  \bibinfo {author} {\bibfnamefont {M.~S.}\ \bibnamefont {Soskin}},\ }\bibfield
   {title} {\enquote {\bibinfo {title} {Optical vortices and the flow of their
  angular momentum in a multimode fiber},}\ }\href
  {http://opto.org.ua/users/pdf/n1_98/082_198.pdf} {\bibfield  {journal}
  {\bibinfo  {journal} {Semicond.\ Phys.\ Quant.\ Electron.\ Optoelectron.}\
  }\textbf {\bibinfo {volume} {1}},\ \bibinfo {pages} {82--89} (\bibinfo {year}
  {1998})}\BibitemShut {NoStop}%
\bibitem [{\citenamefont {Brunet}\ \emph {et~al.}(2014)\citenamefont {Brunet},
  \citenamefont {Vaity}, \citenamefont {Messaddeq}, \citenamefont
  {{LaRochelle}}, and \citenamefont {Rusch}}]{Brunet&al:OE:2014}%
  \BibitemOpen
  \bibfield  {author} {\bibinfo {author} {\bibfnamefont {Charles}\ \bibnamefont
  {Brunet}}, \bibinfo {author} {\bibfnamefont {Pravin}\ \bibnamefont {Vaity}},
  \bibinfo {author} {\bibfnamefont {Youn{\`e}s}\ \bibnamefont {Messaddeq}},
  \bibinfo {author} {\bibfnamefont {Sophie}\ \bibnamefont {{LaRochelle}}}, \
  and\ \bibinfo {author} {\bibfnamefont {Leslie~A.}\ \bibnamefont {Rusch}},\
  }\bibfield  {title} {\enquote {\bibinfo {title} {Design, fabrication and
  validation of an {OAM} fiber supporting 36 states},}\ }\href
  {http://www.opticsinfobase.org/abstract.cfm?uri=oe-22-21-26117} {\bibfield
  {journal} {\bibinfo  {journal} {Opt. Express}\ }\textbf {\bibinfo {volume}
  {22}},\ \bibinfo {pages} {26117--26127} (\bibinfo {year} {2014})}\BibitemShut
  {NoStop}%
\bibitem [{\citenamefont {de~Melo~e Souza}\ \emph {et~al.}(2009)\citenamefont
  {de~Melo~e Souza}, \citenamefont {Cougo-Pinto}, \citenamefont {Farina}, and
  \citenamefont {Moriconi}}]{Souza&al:AJP:2009}%
  \BibitemOpen
  \bibfield  {author} {\bibinfo {author} {\bibfnamefont {R.}~\bibnamefont
  {de~Melo~e Souza}}, \bibinfo {author} {\bibfnamefont {M.~V.}\ \bibnamefont
  {Cougo-Pinto}}, \bibinfo {author} {\bibfnamefont {C.}~\bibnamefont {Farina}},
   and \bibinfo {author} {\bibfnamefont {M.}~\bibnamefont {Moriconi}},\
  }\bibfield  {title} {\enquote {\bibinfo {title} {Multipole radiation fields
  from the {Jefimenko} equation for the magnetic field and the
  {Panofsky-Phillips} equation for the electric field},}\ }\href@noop {}
  {\bibfield  {journal} {\bibinfo  {journal} {Am.\ J.~Phys.}\ }\textbf
  {\bibinfo {volume} {77}},\ \bibinfo {pages} {67--72} (\bibinfo {year}
  {2009})}\BibitemShut {NoStop}%
\bibitem [{\citenamefont {Stratton}(1941)}]{Stratton:Book:1941}%
  \BibitemOpen
  \bibfield  {author} {\bibinfo {author} {\bibfnamefont {Julius~Adams}\
  \bibnamefont {Stratton}},\ }\href@noop {} {\emph {\bibinfo {title}
  {Electromagnetic Theory}}}\ (\bibinfo  {publisher} {McGraw-Hill Book Co.},\
  \bibinfo {address} {New York, NY, USA},\ \bibinfo {year} {1941})\BibitemShut
  {NoStop}%
\bibitem [{\citenamefont {Panofsky} and \citenamefont
  {Phillips}(1962)}]{Panofsky&Phillips:Book:1962}%
  \BibitemOpen
  \bibfield  {author} {\bibinfo {author} {\bibfnamefont {Wolfgang K.~H.}\
  \bibnamefont {Panofsky}} and \bibinfo {author} {\bibfnamefont {Melba}\
  \bibnamefont {Phillips}},\ }\href@noop {} {\emph {\bibinfo {title} {Classical
  Electricity and Magnetism}}},\ \bibinfo {edition} {2nd}\ ed.\ (\bibinfo
  {publisher} {Addison-Wesley Publishing Company},\ \bibinfo {address}
  {Reading, MA, USA},\ \bibinfo {year} {1962})\BibitemShut {NoStop}%
\bibitem [{\citenamefont {Jefimenko}(1966)}]{Jefimenko:Book:1966}%
  \BibitemOpen
  \bibfield  {author} {\bibinfo {author} {\bibfnamefont {Oleg~D.}\ \bibnamefont
  {Jefimenko}},\ }\href@noop {} {\emph {\bibinfo {title} {Electricity and
  Magnetism. An introduction to the theory of electric and magnetic fields}}}\
  (\bibinfo  {publisher} {Appleton-Century-Crofts},\ \bibinfo {address} {New
  York, NY, USA},\ \bibinfo {year} {1966})\BibitemShut {NoStop}%
\bibitem [{\citenamefont {Heald} and \citenamefont
  {Marion}(1995)}]{Heald&Marion:Book:1995}%
  \BibitemOpen
  \bibfield  {author} {\bibinfo {author} {\bibfnamefont {Mark~A.}\ \bibnamefont
  {Heald}} and \bibinfo {author} {\bibfnamefont {Jerry~B.}\ \bibnamefont
  {Marion}},\ }\href@noop {} {\emph {\bibinfo {title} {Classical
  Electromagnetic Radiation}}},\ \bibinfo {edition} {3rd}\ ed.\ (\bibinfo
  {publisher} {Saunders College Publishing},\ \bibinfo {address} {Fort Worth,
  TX, USA},\ \bibinfo {year} {1995})\BibitemShut {NoStop}%
\bibitem [{\citenamefont {Thid{\'e}}\ \emph {et~al.}(2010)\citenamefont
  {Thid{\'e}}, \citenamefont {Lindberg}, \citenamefont {Then}, and
  \citenamefont {Tamburini}}]{Thide&al:ARXIV:2010}%
  \BibitemOpen
  \bibfield  {author} {\bibinfo {author} {\bibfnamefont {B.}~\bibnamefont
  {Thid{\'e}}}, \bibinfo {author} {\bibfnamefont {J.}~\bibnamefont {Lindberg}},
  \bibinfo {author} {\bibfnamefont {H.}~\bibnamefont {Then}},  and \bibinfo
  {author} {\bibfnamefont {F.}~\bibnamefont {Tamburini}},\ }\href
  {http://arxiv.org/abs/1001.0954} {\enquote {\bibinfo {title} {Linear and
  angular momentum of electromagnetic fields generated by an arbitrary
  distribution of charge and current densities at rest},}\ } (\bibinfo {year}
  {2010}),\ \Eprint {http://arxiv.org/abs/1001.0954} {arXiv.org:1001.0954
  [physics.class-ph]} \BibitemShut {NoStop}%
\bibitem [{\citenamefont {Thid\'{e}}\ \emph {et~al.}(2014)\citenamefont
  {Thid\'{e}}, \citenamefont {Tamburini}, \citenamefont {Then}, \citenamefont
  {Someda}, \citenamefont {Mari}, \citenamefont {Parisi}, \citenamefont
  {Spinello}, and \citenamefont {Romanato}}]{Thide&al:SPIE:2014}%
  \BibitemOpen
  \bibfield  {author} {\bibinfo {author} {\bibfnamefont {B.}~\bibnamefont
  {Thid\'{e}}}, \bibinfo {author} {\bibfnamefont {F.}~\bibnamefont
  {Tamburini}}, \bibinfo {author} {\bibfnamefont {H.}~\bibnamefont {Then}},
  \bibinfo {author} {\bibfnamefont {C.~G.}\ \bibnamefont {Someda}}, \bibinfo
  {author} {\bibfnamefont {E.}~\bibnamefont {Mari}}, \bibinfo {author}
  {\bibfnamefont {G.}~\bibnamefont {Parisi}}, \bibinfo {author} {\bibfnamefont
  {F.}~\bibnamefont {Spinello}},  and \bibinfo {author} {\bibfnamefont
  {F.}~\bibnamefont {Romanato}},\ }\bibfield  {title} {\enquote {\bibinfo
  {title} {Angular momentum radio},}\ }in\ \href {\doibase 10.1117/12.2041797}
  {\emph {\bibinfo {booktitle} {Complex Light and Optical Forces VIII}}},\
  \bibinfo {series} {Proceedings of {SPIE}}, Vol.\ \bibinfo {volume} {8999},\
  \bibinfo {organization} {The International Society for Optical Engineering}\
  (\bibinfo  {publisher} {SPIE},\ \bibinfo {address} {San Francisco, CA, USA},\
  \bibinfo {year} {2014})\ pp.\ \bibinfo {pages}
  {89990B--89990B--11}\BibitemShut {NoStop}%
\bibitem [{\citenamefont {Heras}(1994)}]{Heras:AJP:1994}%
  \BibitemOpen
  \bibfield  {author} {\bibinfo {author} {\bibfnamefont {Jos\'{e}~A.}\
  \bibnamefont {Heras}},\ }\bibfield  {title} {\enquote {\bibinfo {title}
  {Radiation fields of a dipole in arbitrary motion},}\ }\href {\doibase
  10.1119/1.17759} {\bibfield  {journal} {\bibinfo  {journal} {Am.\ J.~Phys.}\
  }\textbf {\bibinfo {volume} {62}},\ \bibinfo {pages} {1109--1115} (\bibinfo
  {year} {1994})}\BibitemShut {NoStop}%
\bibitem [{\citenamefont {Bliokh}\ \emph {et~al.}(2013)\citenamefont {Bliokh},
  \citenamefont {Bekshaev}, and \citenamefont {Nori}}]{Bliokh&al:NJP:2013}%
  \BibitemOpen
  \bibfield  {author} {\bibinfo {author} {\bibfnamefont {Konstantin~Y.}\
  \bibnamefont {Bliokh}}, \bibinfo {author} {\bibfnamefont {Aleksandr~Y.}\
  \bibnamefont {Bekshaev}},  and \bibinfo {author} {\bibfnamefont {Franco}\
  \bibnamefont {Nori}},\ }\bibfield  {title} {\enquote {\bibinfo {title} {Dual
  electromagnetism: helicity, spin, momentum and angular momentum},}\ }\href
  {\doibase 10.1088/1367-2630/15/3/033026} {\bibfield  {journal} {\bibinfo
  {journal} {New J.~Phys.}\ }\textbf {\bibinfo {volume} {15}},\ \bibinfo
  {pages} {033026} (\bibinfo {year} {2013})}\BibitemShut {NoStop}%
\bibitem [{\citenamefont {Eddington}(2010)}]{Eddington:Book:2010}%
  \BibitemOpen
  \bibfield  {author} {\bibinfo {author} {\bibfnamefont {Arthur~Stanley}\
  \bibnamefont {Eddington}},\ }\href@noop {} {\emph {\bibinfo {title} {The
  Nature of the Physical World}}}\ (\bibinfo  {publisher} {Kessinger
  Publishing},\ \bibinfo {year} {2010})\BibitemShut {NoStop}%
\bibitem [{\citenamefont {Zeh}(2001)}]{Zeh:Book:2001}%
  \BibitemOpen
  \bibfield  {author} {\bibinfo {author} {\bibfnamefont {H.~Dieter}\
  \bibnamefont {Zeh}},\ }\href@noop {} {\emph {\bibinfo {title} {The Physical
  Basis of The Direction of Time}}},\ \bibinfo {edition} {4th}\ ed.\ (\bibinfo
  {publisher} {Springer},\ \bibinfo {year} {2001})\BibitemShut {NoStop}%
\bibitem [{\citenamefont {Wheeler} and \citenamefont
  {Zurek}(2014)}]{Wheeler&Zurek:Book:2014}%
  \BibitemOpen
  \bibinfo {editor} {\bibfnamefont {John~Archibald}\ \bibnamefont {Wheeler}}\
  and\ \bibinfo {editor} {\bibfnamefont {Wojciech~Hubert}\ \bibnamefont
  {Zurek}},\ eds.,\ \href@noop {} {\emph {\bibinfo {title} {Quantum Theory and
  Measurement}}},\ Princton Legacy Library\ (\bibinfo  {publisher} {Princeton
  University Press},\ \bibinfo {address} {Princeton, NJ, USA},\ \bibinfo {year}
  {2014})\BibitemShut {NoStop}%
\bibitem [{\citenamefont {Brown}(1936)}]{Brown:E:1936}%
  \BibitemOpen
  \bibfield  {author} {\bibinfo {author} {\bibfnamefont {G.~H.}\ \bibnamefont
  {Brown}},\ }\bibfield  {title} {\enquote {\bibinfo {title} {Turnstile
  aerials},}\ }\href@noop {} {\bibfield  {journal} {\bibinfo  {journal}
  {Electronics}\ }\textbf {\bibinfo {volume} {9}},\ \bibinfo {pages} {14--17}
  (\bibinfo {year} {1936})}\BibitemShut {NoStop}%
\bibitem [{\citenamefont {Someda}(2006)}]{Someda:Book:2006}%
  \BibitemOpen
  \bibfield  {author} {\bibinfo {author} {\bibfnamefont {Carlo~G.}\
  \bibnamefont {Someda}},\ }\href@noop {} {\emph {\bibinfo {title}
  {Electromagnetic Waves}}},\ \bibinfo {edition} {2nd}\ ed.\ (\bibinfo
  {publisher} {{CRC}},\ \bibinfo {address} {Boca Raton, {FL}, {USA}},\ \bibinfo
  {year} {2006})\BibitemShut {NoStop}%
\bibitem [{\citenamefont {Coles}\ \emph {et~al.}(2013)\citenamefont {Coles},
  \citenamefont {Williams}, \citenamefont {Saadi}, \citenamefont {Bradshaw},\
  and\ \citenamefont {Andrews}}]{Coles&al:LPR:2013}%
  \BibitemOpen
  \bibfield  {author} {\bibinfo {author} {\bibfnamefont {Matt~M.}\ \bibnamefont
  {Coles}}, \bibinfo {author} {\bibfnamefont {Mathew~D.}\ \bibnamefont
  {Williams}}, \bibinfo {author} {\bibfnamefont {Kamel}\ \bibnamefont {Saadi}},
  \bibinfo {author} {\bibfnamefont {David~S.}\ \bibnamefont {Bradshaw}},  and
  \bibinfo {author} {\bibfnamefont {David~L.}\ \bibnamefont {Andrews}},\
  }\bibfield  {title} {\enquote {\bibinfo {title} {Chiral nanoemitter array: A
  launchpad for optical vortices},}\ }\href {\doibase 10.1002/lpor.201300117}
  {\bibfield  {journal} {\bibinfo  {journal} {Laser \& Photon.\ Rev.}\ }\textbf
  {\bibinfo {volume} {7}},\ \bibinfo {pages} {1088--1092} (\bibinfo {year}
  {2013})}\BibitemShut {NoStop}%
\bibitem [{\citenamefont {Williams}\ \emph {et~al.}(2014)\citenamefont
  {Williams}, \citenamefont {Coles}, \citenamefont {Bradshaw}, and
  \citenamefont {Andrews}}]{Williams&al:PRL:2014}%
  \BibitemOpen
  \bibfield  {author} {\bibinfo {author} {\bibfnamefont {Mathew~D.}\
  \bibnamefont {Williams}}, \bibinfo {author} {\bibfnamefont {Matt~M.}\
  \bibnamefont {Coles}}, \bibinfo {author} {\bibfnamefont {David~S.}\
  \bibnamefont {Bradshaw}},  and \bibinfo {author} {\bibfnamefont {David~L.}\
  \bibnamefont {Andrews}},\ }\bibfield  {title} {\enquote {\bibinfo {title}
  {Direct generation of optical vortices},}\ }\href {\doibase
  10.1103/PhysRevA.89.033837} {\bibfield  {journal} {\bibinfo  {journal}
  {Phys.\ Rev.\ A}\ }\textbf {\bibinfo {volume} {89}},\ \bibinfo {pages}
  {033837} (\bibinfo {year} {2014})}\BibitemShut {NoStop}%
\bibitem [{\citenamefont {Gagnon}\ \emph {et~al.}(2010)\citenamefont {Gagnon},
  \citenamefont {Petosa}, and \citenamefont
  {{McNamara}}}]{Gagnon&al:IEEETAP:2010}%
  \BibitemOpen
  \bibfield  {author} {\bibinfo {author} {\bibfnamefont {N.}~\bibnamefont
  {Gagnon}}, \bibinfo {author} {\bibfnamefont {A.}~\bibnamefont {Petosa}}, \
  and\ \bibinfo {author} {\bibfnamefont {D.A.}\ \bibnamefont {{McNamara}}},\
  }\bibfield  {title} {\enquote {\bibinfo {title} {Thin microwave
  quasi-transparent phase-shifting surface ({PSS})},}\ }\href {\doibase
  10.1109/TAP.2010.2041150} {\bibfield  {journal} {\bibinfo  {journal} {IEEE
  Trans.\ Ant.\ Prop.}\ }\textbf {\bibinfo {volume} {58}},\ \bibinfo {pages}
  {1193--1201} (\bibinfo {year} {2010})}\BibitemShut {NoStop}%
\bibitem [{\citenamefont {Salem} and \citenamefont
  {Caloz}(2013)}]{Salem&Caloz:METAMATERIALS:2013}%
  \BibitemOpen
  \bibfield  {author} {\bibinfo {author} {\bibfnamefont {M.A.}\ \bibnamefont
  {Salem}} and \bibinfo {author} {\bibfnamefont {C.}~\bibnamefont {Caloz}},\
  }\bibfield  {title} {\enquote {\bibinfo {title} {Precision orbital angular
  momentum ({OAM}) multiplexing communication using a metasurface},}\ }in\
  \href {\doibase 10.1109/MetaMaterials.2013.6808964} {\emph {\bibinfo
  {booktitle} {2013 7th International Congress on Advanced Electromagnetic
  Materials in Microwaves and Optics ({METAMATERIALS})}}}\ (\bibinfo {year}
  {2013})\ pp.\ \bibinfo {pages} {94--96}\BibitemShut {NoStop}%
\bibitem [{\citenamefont {Cheng}\ \emph {et~al.}(2014)\citenamefont {Cheng},
  \citenamefont {Hong}, and \citenamefont {Hao}}]{Cheng&al:SR:2014}%
  \BibitemOpen
  \bibfield  {author} {\bibinfo {author} {\bibfnamefont {Li}~\bibnamefont
  {Cheng}}, \bibinfo {author} {\bibfnamefont {Wei}\ \bibnamefont {Hong}}, \
  and\ \bibinfo {author} {\bibfnamefont {Zhang-Cheng}\ \bibnamefont {Hao}},\
  }\bibfield  {title} {\enquote {\bibinfo {title} {Generation of
  electromagnetic waves with arbitrary orbital angular momentum modes},}\
  }\href {\doibase 10.1038/srep04814} {\bibfield  {journal} {\bibinfo
  {journal} {Sci. Rep.}\ }\textbf {\bibinfo {volume} {4}} (\bibinfo {year}
  {2014}),\ 10.1038/srep04814}\BibitemShut {NoStop}%
\bibitem [{\citenamefont {Gao}\ \emph {et~al.}(2014)\citenamefont {Gao},
  \citenamefont {Huang}, \citenamefont {Wei}, \citenamefont {Zhai},
  \citenamefont {Xu}, \citenamefont {Yin}, \citenamefont {Zhou}, and
  \citenamefont {Gu}}]{Gao&al:APL:2014}%
  \BibitemOpen
  \bibfield  {author} {\bibinfo {author} {\bibfnamefont {Xinlu}\ \bibnamefont
  {Gao}}, \bibinfo {author} {\bibfnamefont {Shanguo}\ \bibnamefont {Huang}},
  \bibinfo {author} {\bibfnamefont {Yongfeng}\ \bibnamefont {Wei}}, \bibinfo
  {author} {\bibfnamefont {Wensheng}\ \bibnamefont {Zhai}}, \bibinfo {author}
  {\bibfnamefont {Wenjing}\ \bibnamefont {Xu}}, \bibinfo {author}
  {\bibfnamefont {Shan}\ \bibnamefont {Yin}}, \bibinfo {author} {\bibfnamefont
  {Jing}\ \bibnamefont {Zhou}},  and \bibinfo {author} {\bibfnamefont
  {Wanyi}\ \bibnamefont {Gu}},\ }\bibfield  {title} {\enquote {\bibinfo {title}
  {An orbital angular momentum radio communication system optimized by
  intensity controlled masks effectively: Theoretical design and experimental
  verification},}\ }\href {\doibase 10.1063/1.4904090} {\bibfield  {journal}
  {\bibinfo  {journal} {Appl.\ Phys.\ Lett.}\ }\textbf {\bibinfo {volume}
  {105}},\ \bibinfo {pages} {241109} (\bibinfo {year} {2014})}\BibitemShut
  {NoStop}%
\bibitem [{\citenamefont {Schemmel}\ \emph {et~al.}(2014)\citenamefont
  {Schemmel}, \citenamefont {Pisano}, and \citenamefont
  {Maffei}}]{Schemmel&al:OE:2014}%
  \BibitemOpen
  \bibfield  {author} {\bibinfo {author} {\bibfnamefont {Peter}\ \bibnamefont
  {Schemmel}}, \bibinfo {author} {\bibfnamefont {Giampaolo}\ \bibnamefont
  {Pisano}},  and \bibinfo {author} {\bibfnamefont {Bruno}\ \bibnamefont
  {Maffei}},\ }\bibfield  {title} {\enquote {\bibinfo {title} {Modular spiral
  phase plate design for orbital angular momentum generation at millimetre
  wavelengths},}\ }\href {\doibase 10.1364/OE.22.014712} {\bibfield  {journal}
  {\bibinfo  {journal} {Opt. Express}\ }\textbf {\bibinfo {volume} {22}},\
  \bibinfo {pages} {14712--14726} (\bibinfo {year} {2014})}\BibitemShut
  {NoStop}%
\bibitem [{\citenamefont {Yu} and \citenamefont
  {Capasso}(2014)}]{Yu&Capasso:NMAT:2014}%
  \BibitemOpen
  \bibfield  {author} {\bibinfo {author} {\bibfnamefont {Nanfang}\ \bibnamefont
  {Yu}} and \bibinfo {author} {\bibfnamefont {Federico}\ \bibnamefont
  {Capasso}},\ }\bibfield  {title} {\enquote {\bibinfo {title} {Flat optics
  with designer metasurfaces},}\ }\href {\doibase 10.1038/nmat3839} {\bibfield
  {journal} {\bibinfo  {journal} {Nature Mater.}\ }\textbf {\bibinfo {volume}
  {13}},\ \bibinfo {pages} {139--150} (\bibinfo {year} {2014})}\BibitemShut
  {NoStop}%
\bibitem [{\citenamefont {Hui}\ \emph {et~al.}(2015)\citenamefont {Hui},
  \citenamefont {Zheng}, \citenamefont {Hu}, \citenamefont {Xu}, \citenamefont
  {Jin}, \citenamefont {Chi}, and \citenamefont
  {Zhang}}]{Hui&al:IEEEAWPL:2015}%
  \BibitemOpen
  \bibfield  {author} {\bibinfo {author} {\bibfnamefont {X.}~\bibnamefont
  {Hui}}, \bibinfo {author} {\bibfnamefont {S.}~\bibnamefont {Zheng}}, \bibinfo
  {author} {\bibfnamefont {Y.}~\bibnamefont {Hu}}, \bibinfo {author}
  {\bibfnamefont {C.}~\bibnamefont {Xu}}, \bibinfo {author} {\bibfnamefont
  {X.}~\bibnamefont {Jin}}, \bibinfo {author} {\bibfnamefont {H.}~\bibnamefont
  {Chi}},  and \bibinfo {author} {\bibfnamefont {X.}~\bibnamefont {Zhang}},\
  }\bibfield  {title} {\enquote {\bibinfo {title} {Ultralow reflectivity spiral
  phase plate for generation of millimeter-wave {OAM} beam},}\ }\href {\doibase
  10.1109/LAWP.2014.2387431} {\bibfield  {journal} {\bibinfo  {journal} {IEEE
  Ant.\ Wirel.\ Prop.\ Letter.}\ }\textbf {\bibinfo {volume} {14}},\ \bibinfo
  {pages} {966--969} (\bibinfo {year} {2015})}\BibitemShut {NoStop}%
\bibitem [{\citenamefont {Trevi\~{n}o}\ \emph {et~al.}(2013)\citenamefont
  {Trevi\~{n}o}, \citenamefont {L\'{o}pez-Cruz}, and \citenamefont
  {Ch\'{a}vez-Cerda}}]{Trevino&al:OE:2013}%
  \BibitemOpen
  \bibfield  {author} {\bibinfo {author} {\bibfnamefont {Juan~P.}\ \bibnamefont
  {Trevi\~{n}o}}, \bibinfo {author} {\bibfnamefont {Omar}\ \bibnamefont
  {L\'{o}pez-Cruz}},  and \bibinfo {author} {\bibfnamefont {Sabino}\
  \bibnamefont {Ch\'{a}vez-Cerda}},\ }\bibfield  {title} {\enquote {\bibinfo
  {title} {Segmented vortex telescope and its tolerance to diffraction effects
  and primary aberrations},}\ }\href {\doibase 10.1117/1.OE.52.8.081605}
  {\bibfield  {journal} {\bibinfo  {journal} {Opt. Express}\ }\textbf {\bibinfo
  {volume} {52}},\ \bibinfo {pages} {081605--081605} (\bibinfo {year}
  {2013})}\BibitemShut {NoStop}%
\bibitem [{\citenamefont {Bennis}\ \emph {et~al.}(2013)\citenamefont {Bennis},
  \citenamefont {Niemiec}, \citenamefont {Brousseau}, \citenamefont
  {Mahdjoubi}, and \citenamefont {Emile}}]{Bennis&al:EuCAP:2013}%
  \BibitemOpen
  \bibfield  {author} {\bibinfo {author} {\bibfnamefont {A.}~\bibnamefont
  {Bennis}}, \bibinfo {author} {\bibfnamefont {R.}~\bibnamefont {Niemiec}},
  \bibinfo {author} {\bibfnamefont {C.}~\bibnamefont {Brousseau}}, \bibinfo
  {author} {\bibfnamefont {K.}~\bibnamefont {Mahdjoubi}},  and \bibinfo
  {author} {\bibfnamefont {O.}~\bibnamefont {Emile}},\ }\bibfield  {title}
  {\enquote {\bibinfo {title} {Flat plate for {OAM} generation in the
  millimeter band},}\ }in\ \href@noop {} {\emph {\bibinfo {booktitle} {7th
  European Conference on Antennas and Propagation (EuCAP)}}}\ (\bibinfo {year}
  {2013})\ pp.\ \bibinfo {pages} {3203--3207}\BibitemShut {NoStop}%
\bibitem [{\citenamefont {Krenn}\ \emph {et~al.}(2014)\citenamefont {Krenn},
  \citenamefont {Fickler}, \citenamefont {Fink}, \citenamefont {Handsteiner},
  \citenamefont {Malik}, \citenamefont {Scheidl}, \citenamefont {Ursin}, and
  \citenamefont {Zeilinger}}]{Krenn&al:NJP:2014}%
  \BibitemOpen
  \bibfield  {author} {\bibinfo {author} {\bibfnamefont {Mario}\ \bibnamefont
  {Krenn}}, \bibinfo {author} {\bibfnamefont {Robert}\ \bibnamefont {Fickler}},
  \bibinfo {author} {\bibfnamefont {Matthias}\ \bibnamefont {Fink}}, \bibinfo
  {author} {\bibfnamefont {Johannes}\ \bibnamefont {Handsteiner}}, \bibinfo
  {author} {\bibfnamefont {Mehul}\ \bibnamefont {Malik}}, \bibinfo {author}
  {\bibfnamefont {Thomas}\ \bibnamefont {Scheidl}}, \bibinfo {author}
  {\bibfnamefont {Rupert}\ \bibnamefont {Ursin}},  and \bibinfo {author}
  {\bibfnamefont {Anton}\ \bibnamefont {Zeilinger}},\ }\bibfield  {title}
  {\enquote {\bibinfo {title} {Communication with spatially modulated light
  through turbulent air across {Vienna}},}\ }\href {\doibase
  10.1088/1367-2630/16/11/113028} {\bibfield  {journal} {\bibinfo  {journal}
  {New J.~Phys.}\ }\textbf {\bibinfo {volume} {16}},\ \bibinfo {pages} {113028}
  (\bibinfo {year} {2014})}\BibitemShut {NoStop}%
\bibitem [{\citenamefont {Bochmann}\ \emph {et~al.}(2013)\citenamefont
  {Bochmann}, \citenamefont {Vainsencher}, \citenamefont {Awschalom}, and
  \citenamefont {Cleland}}]{Bochmann&al:NPHY:2013}%
  \BibitemOpen
  \bibfield  {author} {\bibinfo {author} {\bibfnamefont {Joerg}\ \bibnamefont
  {Bochmann}}, \bibinfo {author} {\bibfnamefont {Amit}\ \bibnamefont
  {Vainsencher}}, \bibinfo {author} {\bibfnamefont {David~D.}\ \bibnamefont
  {Awschalom}},  and \bibinfo {author} {\bibfnamefont {Andrew~N.}\
  \bibnamefont {Cleland}},\ }\bibfield  {title} {\enquote {\bibinfo {title}
  {Nanomechanical coupling between microwave and optical photons},}\ }\href
  {\doibase 10.1038/nphys2748} {\bibfield  {journal} {\bibinfo  {journal}
  {Nature Phys.}\ }\textbf {\bibinfo {volume} {9}},\ \bibinfo {pages}
  {712--716} (\bibinfo {year} {2013})}\BibitemShut {NoStop}%
\bibitem [{\citenamefont {Shi} and \citenamefont
  {Bhattacharya}(2013{\natexlab{a}})}]{Shi&Bhattacharya:JPB:2013}%
  \BibitemOpen
  \bibfield  {author} {\bibinfo {author} {\bibfnamefont {H.}~\bibnamefont
  {Shi}} and \bibinfo {author} {\bibfnamefont {M.}~\bibnamefont
  {Bhattacharya}},\ }\bibfield  {title} {\enquote {\bibinfo {title} {Mechanical
  memory for photons with orbital angular momentum},}\ }\href {\doibase
  10.1088/0953-4075/46/15/151001} {\bibfield  {journal} {\bibinfo  {journal}
  {J.~Phys.~B: Atom.\ Mol.\ Opt.\ Phys.}\ }\textbf {\bibinfo {volume} {46}},\
  \bibinfo {pages} {151001} (\bibinfo {year} {2013}{\natexlab{a}})}\BibitemShut
  {NoStop}%
\bibitem [{\citenamefont {Shi} and \citenamefont
  {Bhattacharya}(2013{\natexlab{b}})}]{Shi&Bhattacharya:JMO:2013}%
  \BibitemOpen
  \bibfield  {author} {\bibinfo {author} {\bibfnamefont {H.}~\bibnamefont
  {Shi}} and \bibinfo {author} {\bibfnamefont {M.}~\bibnamefont
  {Bhattacharya}},\ }\bibfield  {title} {\enquote {\bibinfo {title} {Coupling a
  small torsional oscillator to large optical angular momentum},}\ }\href
  {\doibase 10.1080/09500340.2013.778341} {\bibfield  {journal} {\bibinfo
  {journal} {J.~Mod.\ Opt.}\ }\textbf {\bibinfo {volume} {60}},\ \bibinfo
  {pages} {382--386} (\bibinfo {year} {2013}{\natexlab{b}})}\BibitemShut
  {NoStop}%
\bibitem [{\citenamefont {Andrews}\ \emph {et~al.}(2014)\citenamefont
  {Andrews}, \citenamefont {Peterson}, \citenamefont {Purdy}, \citenamefont
  {Cicak}, \citenamefont {Simmonds}, \citenamefont {Regal}, and \citenamefont
  {Lehnert}}]{Andrews&al:NPHY:2014}%
  \BibitemOpen
  \bibfield  {author} {\bibinfo {author} {\bibfnamefont {R.~W.}\ \bibnamefont
  {Andrews}}, \bibinfo {author} {\bibfnamefont {R.~W.}\ \bibnamefont
  {Peterson}}, \bibinfo {author} {\bibfnamefont {T.~P.}\ \bibnamefont {Purdy}},
  \bibinfo {author} {\bibfnamefont {K.}~\bibnamefont {Cicak}}, \bibinfo
  {author} {\bibfnamefont {R.~W.}\ \bibnamefont {Simmonds}}, \bibinfo {author}
  {\bibfnamefont {C.~A.}\ \bibnamefont {Regal}},  and \bibinfo {author}
  {\bibfnamefont {K.~W.}\ \bibnamefont {Lehnert}},\ }\bibfield  {title}
  {\enquote {\bibinfo {title} {Bidirectional and efficient conversion between
  microwave and optical light},}\ }\href {\doibase 10.1038/nphys2911}
  {\bibfield  {journal} {\bibinfo  {journal} {Nature Phys.}\ }\textbf {\bibinfo
  {volume} {10}},\ \bibinfo {pages} {321--326} (\bibinfo {year}
  {2014})}\BibitemShut {NoStop}%
\bibitem [{\citenamefont {Aspelmeyer}\ \emph {et~al.}(2010)\citenamefont
  {Aspelmeyer}, \citenamefont {Gr{\"o}blacher}, \citenamefont {Hammerer}, and
  \citenamefont {Kiesel}}]{Aspelmeyer&al:JOSAB:2010}%
  \BibitemOpen
  \bibfield  {author} {\bibinfo {author} {\bibfnamefont {M.}~\bibnamefont
  {Aspelmeyer}}, \bibinfo {author} {\bibfnamefont {S.}~\bibnamefont
  {Gr{\"o}blacher}}, \bibinfo {author} {\bibfnamefont {K.}~\bibnamefont
  {Hammerer}},  and \bibinfo {author} {\bibfnamefont {N.}~\bibnamefont
  {Kiesel}},\ }\bibfield  {title} {\enquote {\bibinfo {title} {Quantum
  optomechanics---throwing a glance},}\ }\href {\doibase
  10.1364/JOSAB.27.00A189} {\bibfield  {journal} {\bibinfo  {journal} {J.~Opt.\
  Soc\ Am.\ B}\ }\textbf {\bibinfo {volume} {27}},\ \bibinfo {pages} {A189}
  (\bibinfo {year} {2010})}\BibitemShut {NoStop}%
\bibitem [{\citenamefont {Aspelmeyer}\ \emph {et~al.}(2014)\citenamefont
  {Aspelmeyer}, \citenamefont {Kippenberg}, and \citenamefont
  {Marquardt}}]{Aspelmeyer&al:RMP:2014}%
  \BibitemOpen
  \bibfield  {author} {\bibinfo {author} {\bibfnamefont {Markus}\ \bibnamefont
  {Aspelmeyer}}, \bibinfo {author} {\bibfnamefont {Tobias~J.}\ \bibnamefont
  {Kippenberg}},  and \bibinfo {author} {\bibfnamefont {Florian}\
  \bibnamefont {Marquardt}},\ }\bibfield  {title} {\enquote {\bibinfo {title}
  {Cavity optomechanics},}\ }\href {\doibase 10.1103/RevModPhys.86.1391}
  {\bibfield  {journal} {\bibinfo  {journal} {Rev.\ Mod.\ Phys.}\ }\textbf
  {\bibinfo {volume} {86}},\ \bibinfo {pages} {1391--1452} (\bibinfo {year}
  {2014})}\BibitemShut {NoStop}%
\bibitem [{\citenamefont {Bagci}\ \emph {et~al.}(2014)\citenamefont {Bagci},
  \citenamefont {Simonsen}, \citenamefont {Schmid}, \citenamefont {Villanueva},
  \citenamefont {Zeuthen}, \citenamefont {Appel}, \citenamefont {Taylor},
  \citenamefont {S{\o}rensen}, \citenamefont {Usami}, \citenamefont
  {Schliesser}, and \citenamefont {Polzik}}]{Bagci&al:N:2014}%
  \BibitemOpen
  \bibfield  {author} {\bibinfo {author} {\bibfnamefont {T.}~\bibnamefont
  {Bagci}}, \bibinfo {author} {\bibfnamefont {A.}~\bibnamefont {Simonsen}},
  \bibinfo {author} {\bibfnamefont {S.}~\bibnamefont {Schmid}}, \bibinfo
  {author} {\bibfnamefont {L.~G.}\ \bibnamefont {Villanueva}}, \bibinfo
  {author} {\bibfnamefont {E.}~\bibnamefont {Zeuthen}}, \bibinfo {author}
  {\bibfnamefont {J.}~\bibnamefont {Appel}}, \bibinfo {author} {\bibfnamefont
  {J.~M.}\ \bibnamefont {Taylor}}, \bibinfo {author} {\bibfnamefont
  {A.}~\bibnamefont {S{\o}rensen}}, \bibinfo {author} {\bibfnamefont
  {K.}~\bibnamefont {Usami}}, \bibinfo {author} {\bibfnamefont
  {A.}~\bibnamefont {Schliesser}},  and \bibinfo {author} {\bibfnamefont
  {E.~S.}\ \bibnamefont {Polzik}},\ }\bibfield  {title} {\enquote {\bibinfo
  {title} {Optical detection of radio waves through a nanomechanical
  transducer},}\ }\href {\doibase 10.1038/nature13029} {\bibfield  {journal}
  {\bibinfo  {journal} {Nature}\ }\textbf {\bibinfo {volume} {507}},\ \bibinfo
  {pages} {81--85} (\bibinfo {year} {2014})}\BibitemShut {NoStop}%
\bibitem [{\citenamefont {Zhang}\ \emph {et~al.}(2015)\citenamefont {Zhang},
  \citenamefont {Bariani}, \citenamefont {Dong}, \citenamefont {Zhang}, and
  \citenamefont {Meystre}}]{Zhang&al:PRL:2015}%
  \BibitemOpen
  \bibfield  {author} {\bibinfo {author} {\bibfnamefont {Keye}\ \bibnamefont
  {Zhang}}, \bibinfo {author} {\bibfnamefont {Francesco}\ \bibnamefont
  {Bariani}}, \bibinfo {author} {\bibfnamefont {Ying}\ \bibnamefont {Dong}},
  \bibinfo {author} {\bibfnamefont {Weiping}\ \bibnamefont {Zhang}},  and
  \bibinfo {author} {\bibfnamefont {Pierre}\ \bibnamefont {Meystre}},\
  }\bibfield  {title} {\enquote {\bibinfo {title} {Proposal for an
  optomechanical microwave sensor at the subphoton level},}\ }\href {\doibase
  10.1103/PhysRevLett.114.113601} {\bibfield  {journal} {\bibinfo  {journal}
  {Phys.\ Rev.\ Lett.}\ }\textbf {\bibinfo {volume} {114}},\ \bibinfo {pages}
  {113601} (\bibinfo {year} {2015})}\BibitemShut {NoStop}%
\bibitem [{\citenamefont {Mondal}\ \emph {et~al.}(2014)\citenamefont {Mondal},
  \citenamefont {Deb}, and \citenamefont {Majumder}}]{Mondal&al:PRA:2014}%
  \BibitemOpen
  \bibfield  {author} {\bibinfo {author} {\bibfnamefont {Pradip~Kumar}\
  \bibnamefont {Mondal}}, \bibinfo {author} {\bibfnamefont {Bimalendu}\
  \bibnamefont {Deb}},  and \bibinfo {author} {\bibfnamefont {Sonjoy}\
  \bibnamefont {Majumder}},\ }\bibfield  {title} {\enquote {\bibinfo {title}
  {Angular momentum transfer in interaction of {Laguerre-Gaussian} beams with
  atoms and molecules},}\ }\href {\doibase 10.1103/PhysRevA.89.063418}
  {\bibfield  {journal} {\bibinfo  {journal} {Phys.\ Rev.\ A}\ }\textbf
  {\bibinfo {volume} {89}},\ \bibinfo {pages} {063418} (\bibinfo {year}
  {2014})}\BibitemShut {NoStop}%
\bibitem [{\citenamefont {Wolf}\ \emph {et~al.}(2001)\citenamefont {Wolf},
  \citenamefont {Awschalom}, \citenamefont {Buhrman}, \citenamefont {Daughton},
  \citenamefont {Moln{\'a}r}, \citenamefont {Roukes}, \citenamefont
  {Chtchelkanova}, and \citenamefont {Treger}}]{Wolf&al:S:2001}%
  \BibitemOpen
  \bibfield  {author} {\bibinfo {author} {\bibfnamefont {S.~A.}\ \bibnamefont
  {Wolf}}, \bibinfo {author} {\bibfnamefont {D.~D.}\ \bibnamefont {Awschalom}},
  \bibinfo {author} {\bibfnamefont {R.~A.}\ \bibnamefont {Buhrman}}, \bibinfo
  {author} {\bibfnamefont {J.~M.}\ \bibnamefont {Daughton}}, \bibinfo {author}
  {\bibfnamefont {S.~von}\ \bibnamefont {Moln{\'a}r}}, \bibinfo {author}
  {\bibfnamefont {M.~L.}\ \bibnamefont {Roukes}}, \bibinfo {author}
  {\bibfnamefont {A.~Y.}\ \bibnamefont {Chtchelkanova}},  and \bibinfo
  {author} {\bibfnamefont {D.~M.}\ \bibnamefont {Treger}},\ }\bibfield  {title}
  {\enquote {\bibinfo {title} {Spintronics: A spin-based electronics vision for
  the future},}\ }\href {\doibase 10.1126/science.1065389} {\bibfield
  {journal} {\bibinfo  {journal} {Science}\ }\textbf {\bibinfo {volume}
  {294}},\ \bibinfo {pages} {1488--1495} (\bibinfo {year} {2001})}\BibitemShut
  {NoStop}%
\bibitem [{\citenamefont {Murakami}\ \emph {et~al.}(2003)\citenamefont
  {Murakami}, \citenamefont {Nagaosa}, and \citenamefont
  {Zhang}}]{Murakami&al:S:2003}%
  \BibitemOpen
  \bibfield  {author} {\bibinfo {author} {\bibfnamefont {Shuichi}\ \bibnamefont
  {Murakami}}, \bibinfo {author} {\bibfnamefont {Naoto}\ \bibnamefont
  {Nagaosa}},  and \bibinfo {author} {\bibfnamefont {Shou-Cheng}\
  \bibnamefont {Zhang}},\ }\bibfield  {title} {\enquote {\bibinfo {title}
  {Dissipationless quantum spin current at room temperature},}\ }\href
  {\doibase 10.1126/science.1087128} {\bibfield  {journal} {\bibinfo  {journal}
  {Science}\ }\textbf {\bibinfo {volume} {301}},\ \bibinfo {pages} {1348--1351}
  (\bibinfo {year} {2003})}\BibitemShut {NoStop}%
\bibitem [{\citenamefont {Sinova}\ \emph {et~al.}(2004)\citenamefont {Sinova},
  \citenamefont {Culcer}, \citenamefont {Niu}, \citenamefont {Sinitsyn},
  \citenamefont {Jungwirth}, and \citenamefont
  {{MacDonald}}}]{Sinova&al:PRL:2004}%
  \BibitemOpen
  \bibfield  {author} {\bibinfo {author} {\bibfnamefont {Jairo}\ \bibnamefont
  {Sinova}}, \bibinfo {author} {\bibfnamefont {Dimitrie}\ \bibnamefont
  {Culcer}}, \bibinfo {author} {\bibfnamefont {Q.}~\bibnamefont {Niu}},
  \bibinfo {author} {\bibfnamefont {N.~A.}\ \bibnamefont {Sinitsyn}}, \bibinfo
  {author} {\bibfnamefont {T.}~\bibnamefont {Jungwirth}},  and \bibinfo
  {author} {\bibfnamefont {A.~H.}\ \bibnamefont {{MacDonald}}},\ }\bibfield
  {title} {\enquote {\bibinfo {title} {Universal intrinsic spin hall effect},}\
  }\href {\doibase 10.1103/PhysRevLett.92.126603} {\bibfield  {journal}
  {\bibinfo  {journal} {Phys.\ Rev.\ Lett.}\ }\textbf {\bibinfo {volume}
  {92}},\ \bibinfo {pages} {126603} (\bibinfo {year} {2004})}\BibitemShut
  {NoStop}%
\bibitem [{\citenamefont {\v{Z}uti\'{c}}\ \emph {et~al.}(2004)\citenamefont
  {\v{Z}uti\'{c}}, \citenamefont {Fabian}, and \citenamefont
  {Das~Sarma}}]{Zutic&al:RMP:2004}%
  \BibitemOpen
  \bibfield  {author} {\bibinfo {author} {\bibfnamefont {Igor}\ \bibnamefont
  {\v{Z}uti\'{c}}}, \bibinfo {author} {\bibfnamefont {Jaroslav}\ \bibnamefont
  {Fabian}},  and \bibinfo {author} {\bibfnamefont {S.}~\bibnamefont
  {Das~Sarma}},\ }\bibfield  {title} {\enquote {\bibinfo {title} {Spintronics:
  Fundamentals and applications},}\ }\href {\doibase 10.1103/RevModPhys.76.323}
  {\bibfield  {journal} {\bibinfo  {journal} {Rev.\ Mod.\ Phys.}\ }\textbf
  {\bibinfo {volume} {76}},\ \bibinfo {pages} {323--410} (\bibinfo {year}
  {2004})}\BibitemShut {NoStop}%
\bibitem [{\citenamefont {Deac}\ \emph {et~al.}(2008)\citenamefont {Deac},
  \citenamefont {Fukushima}, \citenamefont {Kubota}, \citenamefont {Maehara},
  \citenamefont {Suzuki}, \citenamefont {Yuasa}, \citenamefont {Nagamine},
  \citenamefont {Tsunekawa}, \citenamefont {Djayaprawira}, and \citenamefont
  {Watanabe}}]{Deac&al:NPHY:2008}%
  \BibitemOpen
  \bibfield  {author} {\bibinfo {author} {\bibfnamefont {Alina~M.}\
  \bibnamefont {Deac}}, \bibinfo {author} {\bibfnamefont {Akio}\ \bibnamefont
  {Fukushima}}, \bibinfo {author} {\bibfnamefont {Hitoshi}\ \bibnamefont
  {Kubota}}, \bibinfo {author} {\bibfnamefont {Hiroki}\ \bibnamefont
  {Maehara}}, \bibinfo {author} {\bibfnamefont {Yoshishige}\ \bibnamefont
  {Suzuki}}, \bibinfo {author} {\bibfnamefont {Shinji}\ \bibnamefont {Yuasa}},
  \bibinfo {author} {\bibfnamefont {Yoshinori}\ \bibnamefont {Nagamine}},
  \bibinfo {author} {\bibfnamefont {Koji}\ \bibnamefont {Tsunekawa}}, \bibinfo
  {author} {\bibfnamefont {David~D.}\ \bibnamefont {Djayaprawira}},  and
  \bibinfo {author} {\bibfnamefont {Naoki}\ \bibnamefont {Watanabe}},\
  }\bibfield  {title} {\enquote {\bibinfo {title} {Bias-driven high-power
  microwave emission from {MgO}-based tunnel magnetoresistance devices},}\
  }\href {\doibase 10.1038/nphys1036} {\bibfield  {journal} {\bibinfo
  {journal} {Nature Phys.}\ }\textbf {\bibinfo {volume} {4}},\ \bibinfo {pages}
  {803--809} (\bibinfo {year} {2008})}\BibitemShut {NoStop}%
\bibitem [{\citenamefont {Bernevig}\ \emph {et~al.}(2005)\citenamefont
  {Bernevig}, \citenamefont {Hughes}, and \citenamefont
  {Zhang}}]{Bernevig&al:PRL:2005}%
  \BibitemOpen
  \bibfield  {author} {\bibinfo {author} {\bibfnamefont {B.~Andrei}\
  \bibnamefont {Bernevig}}, \bibinfo {author} {\bibfnamefont {Taylor~L.}\
  \bibnamefont {Hughes}},  and \bibinfo {author} {\bibfnamefont {Shou-Cheng}\
  \bibnamefont {Zhang}},\ }\bibfield  {title} {\enquote {\bibinfo {title}
  {Orbitronics: The intrinsic orbital current in \emph{p}-doped silicon},}\
  }\href {\doibase 10.1103/PhysRevLett.95.066601} {\bibfield  {journal}
  {\bibinfo  {journal} {Phys.\ Rev.\ Lett.}\ }\textbf {\bibinfo {volume}
  {95}},\ \bibinfo {pages} {066601} (\bibinfo {year} {2005})}\BibitemShut
  {NoStop}%
\bibitem [{\citenamefont {Dussaux}\ \emph {et~al.}(2010)\citenamefont
  {Dussaux}, \citenamefont {Georges}, \citenamefont {Grollier}, \citenamefont
  {Cros}, \citenamefont {Khvalkovskiy}, \citenamefont {Fukushima},
  \citenamefont {Konoto}, \citenamefont {Kubota}, \citenamefont {Yakushiji},
  \citenamefont {Yuasa}, \citenamefont {Zvezdin}, \citenamefont {Ando}, and
  \citenamefont {Fert}}]{Dussaux&al:NCOM:2010}%
  \BibitemOpen
  \bibfield  {author} {\bibinfo {author} {\bibfnamefont {A.}~\bibnamefont
  {Dussaux}}, \bibinfo {author} {\bibfnamefont {B.}~\bibnamefont {Georges}},
  \bibinfo {author} {\bibfnamefont {J.}~\bibnamefont {Grollier}}, \bibinfo
  {author} {\bibfnamefont {V.}~\bibnamefont {Cros}}, \bibinfo {author}
  {\bibfnamefont {A.~V.}\ \bibnamefont {Khvalkovskiy}}, \bibinfo {author}
  {\bibfnamefont {A.}~\bibnamefont {Fukushima}}, \bibinfo {author}
  {\bibfnamefont {M.}~\bibnamefont {Konoto}}, \bibinfo {author} {\bibfnamefont
  {H.}~\bibnamefont {Kubota}}, \bibinfo {author} {\bibfnamefont
  {K.}~\bibnamefont {Yakushiji}}, \bibinfo {author} {\bibfnamefont
  {S.}~\bibnamefont {Yuasa}}, \bibinfo {author} {\bibfnamefont {K.~A.i}\
  \bibnamefont {Zvezdin}}, \bibinfo {author} {\bibfnamefont {K.}~\bibnamefont
  {Ando}},  and \bibinfo {author} {\bibfnamefont {A.}~\bibnamefont {Fert}},\
  }\bibfield  {title} {\enquote {\bibinfo {title} {Large microwave generation
  from current-driven magnetic vortex oscillators in magnetic tunnel
  junctions},}\ }\href {\doibase 10.1038/ncomms1006} {\bibfield  {journal}
  {\bibinfo  {journal} {Nature Commun.}\ }\textbf {\bibinfo {volume} {1}},\
  \bibinfo {pages} {8} (\bibinfo {year} {2010})}\BibitemShut {NoStop}%
\bibitem [{\citenamefont {Kirilyuk}\ \emph {et~al.}(2010)\citenamefont
  {Kirilyuk}, \citenamefont {Kimel}, and \citenamefont
  {Rasing}}]{Kirilyuk&al:RMP:2010}%
  \BibitemOpen
  \bibfield  {author} {\bibinfo {author} {\bibfnamefont {Andrei}\ \bibnamefont
  {Kirilyuk}}, \bibinfo {author} {\bibfnamefont {Alexey~V.}\ \bibnamefont
  {Kimel}},  and \bibinfo {author} {\bibfnamefont {Theo}\ \bibnamefont
  {Rasing}},\ }\bibfield  {title} {\enquote {\bibinfo {title} {Ultrafast
  optical manipulation of magnetic order},}\ }\href {\doibase
  10.1103/RevModPhys.82.2731} {\bibfield  {journal} {\bibinfo  {journal}
  {Reviews of Modern Physics}\ }\textbf {\bibinfo {volume} {82}},\ \bibinfo
  {pages} {2731--2784} (\bibinfo {year} {2010})}\BibitemShut {NoStop}%
\bibitem [{\citenamefont {Tokatly}(2010)}]{Tokatly:PRB:2010}%
  \BibitemOpen
  \bibfield  {author} {\bibinfo {author} {\bibfnamefont {I.~V.}\ \bibnamefont
  {Tokatly}},\ }\bibfield  {title} {\enquote {\bibinfo {title} {Orbital
  momentum {Hall} effect in \emph{p}-doped graphane},}\ }\href {\doibase
  10.1103/PhysRevB.82.161404} {\bibfield  {journal} {\bibinfo  {journal}
  {Phys.\ Rev.\ B}\ }\textbf {\bibinfo {volume} {82}},\ \bibinfo {pages}
  {161404} (\bibinfo {year} {2010})}\BibitemShut {NoStop}%
\bibitem [{\citenamefont {Verma}(2011)}]{Verma:EFY:2011}%
  \BibitemOpen
  \bibfield  {author} {\bibinfo {author} {\bibfnamefont {S.~S.}\ \bibnamefont
  {Verma}},\ }\bibfield  {title} {\enquote {\bibinfo {title} {Atomtronics
  trying to replace {Electronics}},}\ }\href {www.efymag.com} {\bibfield
  {journal} {\bibinfo  {journal} {Electronics For You}\ }\textbf {\bibinfo
  {volume} {43}},\ \bibinfo {pages} {60--64} (\bibinfo {year}
  {2011})}\BibitemShut {NoStop}%
\bibitem [{\citenamefont {Maekawa}\ \emph {et~al.}(2013)\citenamefont
  {Maekawa}, \citenamefont {Adachi}, \citenamefont {Uchida}, \citenamefont
  {Ieda}, and \citenamefont {Saitoh}}]{Maekawa&al:2013}%
  \BibitemOpen
  \bibfield  {author} {\bibinfo {author} {\bibfnamefont {Sadamichi}\
  \bibnamefont {Maekawa}}, \bibinfo {author} {\bibfnamefont {Hiroto}\
  \bibnamefont {Adachi}}, \bibinfo {author} {\bibfnamefont {Ken-ichi}\
  \bibnamefont {Uchida}}, \bibinfo {author} {\bibfnamefont {Jun'ichi}\
  \bibnamefont {Ieda}},  and \bibinfo {author} {\bibfnamefont {Eiji}\
  \bibnamefont {Saitoh}},\ }\bibfield  {title} {\enquote {\bibinfo {title}
  {Spin current: Experimental and theoretical aspects},}\ }\href {\doibase
  10.7566/JPSJ.82.102002} {\bibfield  {journal} {\bibinfo  {journal} {J.~Phys.\
  Soc.\ Japan}\ }\textbf {\bibinfo {volume} {82}},\ \bibinfo {pages} {102002}
  (\bibinfo {year} {2013})}\BibitemShut {NoStop}%
\bibitem [{\citenamefont {Vannucchi}\ \emph {et~al.}(2013)\citenamefont
  {Vannucchi}, \citenamefont {Vasconcellos}, and \citenamefont
  {Luzzi}}]{Vannucchi&al:APL:2013}%
  \BibitemOpen
  \bibfield  {author} {\bibinfo {author} {\bibfnamefont {Fabio~Stucchi}\
  \bibnamefont {Vannucchi}}, \bibinfo {author} {\bibfnamefont
  {{\'A}urea~Rosas}\ \bibnamefont {Vasconcellos}},  and \bibinfo {author}
  {\bibfnamefont {Roberto}\ \bibnamefont {Luzzi}},\ }\bibfield  {title}
  {\enquote {\bibinfo {title} {Emission of monochromatic microwave radiation
  from a nonequilibrium condensation of excited magnons},}\ }\href {\doibase
  10.1063/1.4818312} {\bibfield  {journal} {\bibinfo  {journal} {Appl.\ Phys.\
  Lett.}\ }\textbf {\bibinfo {volume} {103}},\ \bibinfo {pages} {072401}
  (\bibinfo {year} {2013})}\BibitemShut {NoStop}%
\bibitem [{\citenamefont {Yan}\ \emph {et~al.}(2013)\citenamefont {Yan},
  \citenamefont {Kamra}, \citenamefont {Cao}, and \citenamefont
  {Bauer}}]{Yan&al:PRB:2013}%
  \BibitemOpen
  \bibfield  {author} {\bibinfo {author} {\bibfnamefont {Peng}\ \bibnamefont
  {Yan}}, \bibinfo {author} {\bibfnamefont {Akashdeep}\ \bibnamefont {Kamra}},
  \bibinfo {author} {\bibfnamefont {Yunshan}\ \bibnamefont {Cao}},  and
  \bibinfo {author} {\bibfnamefont {Gerrit E.~W.}\ \bibnamefont {Bauer}},\
  }\bibfield  {title} {\enquote {\bibinfo {title} {Angular and linear momentum
  of excited ferromagnets},}\ }\href {\doibase 10.1103/PhysRevB.88.144413}
  {\bibfield  {journal} {\bibinfo  {journal} {Phys.\ Rev.\ B}\ }\textbf
  {\bibinfo {volume} {88}},\ \bibinfo {pages} {144413} (\bibinfo {year}
  {2013})}\BibitemShut {NoStop}%
\bibitem [{\citenamefont {Brataas} and \citenamefont
  {Hals}(2014)}]{Brataas&Hals:NNAN:2014}%
  \BibitemOpen
  \bibfield  {author} {\bibinfo {author} {\bibfnamefont {Arne}\ \bibnamefont
  {Brataas}} and \bibinfo {author} {\bibfnamefont {Kjetil M.~D.}\
  \bibnamefont {Hals}},\ }\bibfield  {title} {\enquote {\bibinfo {title}
  {Spin-orbit torques in action},}\ }\href {\doibase 10.1038/nnano.2014.8}
  {\bibfield  {journal} {\bibinfo  {journal} {Nature Nanotech.}\ }\textbf
  {\bibinfo {volume} {9}},\ \bibinfo {pages} {86--88} (\bibinfo {year}
  {2014})}\BibitemShut {NoStop}%
\bibitem [{\citenamefont {Kuschel} and \citenamefont
  {Reiss}(2015)}]{Kuschel&Reiss:NNAN:2015}%
  \BibitemOpen
  \bibfield  {author} {\bibinfo {author} {\bibfnamefont {Timo}\ \bibnamefont
  {Kuschel}} and \bibinfo {author} {\bibfnamefont {G{\"u}nter}\ \bibnamefont
  {Reiss}},\ }\bibfield  {title} {\enquote {\bibinfo {title} {Spin orbitronics:
  Charges ride the spin wave},}\ }\href {\doibase 10.1038/nnano.2014.279}
  {\bibfield  {journal} {\bibinfo  {journal} {Nature Nanotech.}\ }\textbf
  {\bibinfo {volume} {10}},\ \bibinfo {pages} {22--24} (\bibinfo {year}
  {2015})}\BibitemShut {NoStop}%
\bibitem [{\citenamefont {Vogel}\ \emph {et~al.}(2011)\citenamefont {Vogel},
  \citenamefont {Martens}, \citenamefont {Weigand}, and \citenamefont
  {Meier}}]{Vogel&al:APL:2011}%
  \BibitemOpen
  \bibfield  {author} {\bibinfo {author} {\bibfnamefont {Andreas}\ \bibnamefont
  {Vogel}}, \bibinfo {author} {\bibfnamefont {Michael}\ \bibnamefont
  {Martens}}, \bibinfo {author} {\bibfnamefont {Markus}\ \bibnamefont
  {Weigand}},  and \bibinfo {author} {\bibfnamefont {Guido}\ \bibnamefont
  {Meier}},\ }\bibfield  {title} {\enquote {\bibinfo {title} {Signal transfer
  in a chain of stray-field coupled ferromagnetic squares},}\ }\href {\doibase
  10.1063/1.3614551} {\bibfield  {journal} {\bibinfo  {journal} {Appl.\ Phys.\
  Lett.}\ }\textbf {\bibinfo {volume} {99}},\ \bibinfo {pages} {042506}
  (\bibinfo {year} {2011})}\BibitemShut {NoStop}%
\bibitem [{\citenamefont {Mochizuki}(2012)}]{Mochizuki:PRL:2012}%
  \BibitemOpen
  \bibfield  {author} {\bibinfo {author} {\bibfnamefont {Masahito}\
  \bibnamefont {Mochizuki}},\ }\bibfield  {title} {\enquote {\bibinfo {title}
  {Spin-wave modes and their intense excitation effects in skyrmion
  crystals},}\ }\href {\doibase 10.1103/PhysRevLett.108.017601} {\bibfield
  {journal} {\bibinfo  {journal} {Phys.\ Rev.\ Lett.}\ }\textbf {\bibinfo
  {volume} {108}},\ \bibinfo {pages} {017601} (\bibinfo {year}
  {2012})}\BibitemShut {NoStop}%
\bibitem [{\citenamefont {Pulecio}\ \emph {et~al.}(2014)\citenamefont
  {Pulecio}, \citenamefont {Warnicke}, \citenamefont {Pollard}, \citenamefont
  {Arena}, and \citenamefont {Zhu}}]{Pulecio&al:NCOM:2014}%
  \BibitemOpen
  \bibfield  {author} {\bibinfo {author} {\bibfnamefont {J.~F.}\ \bibnamefont
  {Pulecio}}, \bibinfo {author} {\bibfnamefont {P.}~\bibnamefont {Warnicke}},
  \bibinfo {author} {\bibfnamefont {S.~D.}\ \bibnamefont {Pollard}}, \bibinfo
  {author} {\bibfnamefont {D.~A.}\ \bibnamefont {Arena}},  and \bibinfo
  {author} {\bibfnamefont {Y.}~\bibnamefont {Zhu}},\ }\bibfield  {title}
  {\enquote {\bibinfo {title} {Coherence and modality of driven
  interlayer-coupled magnetic vortices},}\ }\href {\doibase 10.1038/ncomms4760}
  {\bibfield  {journal} {\bibinfo  {journal} {Nature Commun.}\ }\textbf
  {\bibinfo {volume} {5}} (\bibinfo {year} {2014}),\
  10.1038/ncomms4760}\BibitemShut {NoStop}%
\bibitem [{\citenamefont {Quinteiro} and \citenamefont
  {Berakdar}(2009)}]{Quinteiro&Beradkar:OE:2009}%
  \BibitemOpen
  \bibfield  {author} {\bibinfo {author} {\bibfnamefont {G.~F.}\ \bibnamefont
  {Quinteiro}} and \bibinfo {author} {\bibfnamefont {J.}~\bibnamefont
  {Berakdar}},\ }\bibfield  {title} {\enquote {\bibinfo {title} {Electric
  currents induced by twisted light in quantum rings},}\ }\href {\doibase
  10.1364/OE.17.020465} {\bibfield  {journal} {\bibinfo  {journal} {Opt.
  Express}\ }\textbf {\bibinfo {volume} {17}},\ \bibinfo {pages} {20465--20475}
  (\bibinfo {year} {2009})}\BibitemShut {NoStop}%
\bibitem [{\citenamefont {K{\"o}ksal} and \citenamefont
  {Berakdar}(2012)}]{Koksal&Berakdar:PRA:2012}%
  \BibitemOpen
  \bibfield  {author} {\bibinfo {author} {\bibfnamefont {K.}~\bibnamefont
  {K{\"o}ksal}} and \bibinfo {author} {\bibfnamefont {J.}~\bibnamefont
  {Berakdar}},\ }\bibfield  {title} {\enquote {\bibinfo {title} {Charge-current
  generation in atomic systems induced by optical vortices},}\ }\href {\doibase
  10.1103/PhysRevA.86.063812} {\bibfield  {journal} {\bibinfo  {journal}
  {Phys.\ Rev.\ A}\ }\textbf {\bibinfo {volume} {86}},\ \bibinfo {pages}
  {063812} (\bibinfo {year} {2012})}\BibitemShut {NoStop}%
\bibitem [{\citenamefont {Nieminen}\ \emph {et~al.}(2015)\citenamefont
  {Nieminen}, \citenamefont {V{\"a}nsk{\"a}}, \citenamefont {Tittonen},
  \citenamefont {Koch}, and \citenamefont {Kira}}]{Nieminen&al:NJP:2015}%
  \BibitemOpen
  \bibfield  {author} {\bibinfo {author} {\bibfnamefont {J.~V.}\ \bibnamefont
  {Nieminen}}, \bibinfo {author} {\bibfnamefont {O.}~\bibnamefont
  {V{\"a}nsk{\"a}}}, \bibinfo {author} {\bibfnamefont {I.}~\bibnamefont
  {Tittonen}}, \bibinfo {author} {\bibfnamefont {S.~W.}\ \bibnamefont {Koch}},
   and \bibinfo {author} {\bibfnamefont {M.}~\bibnamefont {Kira}},\
  }\bibfield  {title} {\enquote {\bibinfo {title} {Accessing orbital angular
  momentum of quantum-ring excitons via directional semiconductor
  luminescence},}\ }\href {\doibase 10.1088/1367-2630/17/3/033046} {\bibfield
  {journal} {\bibinfo  {journal} {New J.~Phys.}\ }\textbf {\bibinfo {volume}
  {17}},\ \bibinfo {pages} {033046} (\bibinfo {year} {2015})}\BibitemShut
  {NoStop}%
\bibitem [{\citenamefont {Nyquist}(1928)}]{Nyquist:AIEET:1928}%
  \BibitemOpen
  \bibfield  {author} {\bibinfo {author} {\bibfnamefont {H.}~\bibnamefont
  {Nyquist}},\ }\bibfield  {title} {\enquote {\bibinfo {title} {Certain topics
  in telegraph transmission theory},}\ }\href {\doibase
  10.1109/T-AIEE.1928.5055024} {\bibfield  {journal} {\bibinfo  {journal} {AIEE
  Trans.}\ }\textbf {\bibinfo {volume} {47}},\ \bibinfo {pages} {617--644}
  (\bibinfo {year} {1928})}\BibitemShut {NoStop}%
\bibitem [{\citenamefont {Shannon}(1949)}]{Shannon:PIRE:1949}%
  \BibitemOpen
  \bibfield  {author} {\bibinfo {author} {\bibfnamefont {C.E.}\ \bibnamefont
  {Shannon}},\ }\bibfield  {title} {\enquote {\bibinfo {title} {Communication
  in the presence of noise},}\ }\href {\doibase 10.1109/JRPROC.1949.232969}
  {\bibfield  {journal} {\bibinfo  {journal} {Proceedings of the {IRE}}\
  }\textbf {\bibinfo {volume} {37}},\ \bibinfo {pages} {10--21} (\bibinfo
  {year} {1949})},\ \bibinfo {note} {05728}\BibitemShut {NoStop}%
\bibitem [{\citenamefont {Jerri}(1977)}]{Jerri:PIEEE:1977}%
  \BibitemOpen
  \bibfield  {author} {\bibinfo {author} {\bibfnamefont {A.J.}\ \bibnamefont
  {Jerri}},\ }\bibfield  {title} {\enquote {\bibinfo {title} {The {Shannon}
  sampling theorem---its various extensions and applications: A tutorial
  review},}\ }\href {\doibase 10.1109/PROC.1977.10771} {\bibfield  {journal}
  {\bibinfo  {journal} {Proc.\ IEEE}\ }\textbf {\bibinfo {volume} {65}},\
  \bibinfo {pages} {1565--1596} (\bibinfo {year} {1977})}\BibitemShut {NoStop}%
\bibitem [{\citenamefont {Then}\ \emph {et~al.}(2008)\citenamefont {Then},
  \citenamefont {Thid\'e}, \citenamefont {Mendon\c{c}a}, \citenamefont
  {Carozzi}, \citenamefont {Bergman}, \citenamefont {Baan}, \citenamefont
  {Mohammadi}, and \citenamefont {Eliasson}}]{Then&al:ARXIV:2008}%
  \BibitemOpen
  \bibfield  {author} {\bibinfo {author} {\bibfnamefont {H.}~\bibnamefont
  {Then}}, \bibinfo {author} {\bibfnamefont {B.}~\bibnamefont {Thid\'e}},
  \bibinfo {author} {\bibfnamefont {J.~T.}\ \bibnamefont {Mendon\c{c}a}},
  \bibinfo {author} {\bibfnamefont {T.~D.}\ \bibnamefont {Carozzi}}, \bibinfo
  {author} {\bibfnamefont {J.~E.~S.}\ \bibnamefont {Bergman}}, \bibinfo
  {author} {\bibfnamefont {W.~A.}\ \bibnamefont {Baan}}, \bibinfo {author}
  {\bibfnamefont {S.}~\bibnamefont {Mohammadi}},  and \bibinfo {author}
  {\bibfnamefont {B.}~\bibnamefont {Eliasson}},\ }\href
  {http://arxiv.org/abs/0805.2735} {\enquote {\bibinfo {title} {Detecting
  orbital angular momentum in radio signals},}\ } (\bibinfo {year} {2008}),\
  \Eprint {http://arxiv.org/abs/0805.2735} {arXiv.org:0805.2735 [astro-ph]}
  \BibitemShut {NoStop}%
\bibitem [{\citenamefont {Sj{\"o}holm} and \citenamefont
  {Palmer}(2009)}]{Sjoholm&Palmer:MSc:2009}%
  \BibitemOpen
  \bibfield  {author} {\bibinfo {author} {\bibfnamefont {J.}~\bibnamefont
  {Sj{\"o}holm}} and \bibinfo {author} {\bibfnamefont {K.}~\bibnamefont
  {Palmer}},\ }\emph {\bibinfo {title} {Angular Momentum of Electromagnetic
  Radiation. Fundamental physics applied to the radio domain for innovative
  studies of space and development of new concepts in wireless
  communications}},\ \href@noop {} {\bibinfo {type} {Master's thesis}},\
  \bibinfo  {school} {Uppsala University}, \bibinfo {address} {Uppsala, SE}
  (\bibinfo {year} {2009}),\ \Eprint {http://arxiv.org/abs/0905.0190}
  {arXiv.org:0905.0190 [physics.class-ph]} \BibitemShut {NoStop}%
\bibitem [{\citenamefont {Mohammadi}\ \emph
  {et~al.}(2010{\natexlab{a}})\citenamefont {Mohammadi}, \citenamefont
  {Daldorff}, \citenamefont {Bergman}, \citenamefont {Karlsson}, \citenamefont
  {Thid{\'e}}, \citenamefont {Forozesh}, and \citenamefont
  {Carozzi}}]{Mohammadi&al:IEEETAP:2010}%
  \BibitemOpen
  \bibfield  {author} {\bibinfo {author} {\bibfnamefont {Siavoush~M.}\
  \bibnamefont {Mohammadi}}, \bibinfo {author} {\bibfnamefont {Lars K.~S.}\
  \bibnamefont {Daldorff}}, \bibinfo {author} {\bibfnamefont {Jan E.~S.}\
  \bibnamefont {Bergman}}, \bibinfo {author} {\bibfnamefont {Roger~L.}\
  \bibnamefont {Karlsson}}, \bibinfo {author} {\bibfnamefont {Bo}~\bibnamefont
  {Thid{\'e}}}, \bibinfo {author} {\bibfnamefont {Kamyar}\ \bibnamefont
  {Forozesh}},  and \bibinfo {author} {\bibfnamefont {Tobia~D.}\ \bibnamefont
  {Carozzi}},\ }\bibfield  {title} {\enquote {\bibinfo {title} {Orbital angular
  momentum in radio---a system study},}\ }\href@noop {} {\bibfield  {journal}
  {\bibinfo  {journal} {IEEE Trans.\ Ant.\ Prop.}\ }\textbf {\bibinfo {volume}
  {58}},\ \bibinfo {pages} {565--572} (\bibinfo {year}
  {2010}{\natexlab{a}})}\BibitemShut {NoStop}%
\bibitem [{\citenamefont {Edfors} and \citenamefont
  {Johansson}(2012)}]{Edfors&Johansson:IEEETAP:2012}%
  \BibitemOpen
  \bibfield  {author} {\bibinfo {author} {\bibfnamefont {O.}~\bibnamefont
  {Edfors}} and \bibinfo {author} {\bibfnamefont {A.J.}\ \bibnamefont
  {Johansson}},\ }\bibfield  {title} {\enquote {\bibinfo {title} {Is orbital
  angular momentum ({OAM}) based radio communication an unexploited area?}}\
  }\href {\doibase 10.1109/TAP.2011.2173142} {\bibfield  {journal} {\bibinfo
  {journal} {IEEE Trans.\ Ant.\ Prop.}\ }\textbf {\bibinfo {volume} {60}},\
  \bibinfo {pages} {1126--1131} (\bibinfo {year} {2012})}\BibitemShut {NoStop}%
\bibitem [{\citenamefont {Huang}\ \emph
  {et~al.}(2012{\natexlab{b}})\citenamefont {Huang}, \citenamefont {Chen},
  \citenamefont {M{\"u}hlenbernd}, \citenamefont {Li}, \citenamefont {Bai},
  \citenamefont {Tan}, \citenamefont {Jin}, \citenamefont {Zentgraf}, and
  \citenamefont {Zhang}}]{Huang&al:NL:2012}%
  \BibitemOpen
  \bibfield  {author} {\bibinfo {author} {\bibfnamefont {Lingling}\
  \bibnamefont {Huang}}, \bibinfo {author} {\bibfnamefont {Xianzhong}\
  \bibnamefont {Chen}}, \bibinfo {author} {\bibfnamefont {Holger}\ \bibnamefont
  {M{\"u}hlenbernd}}, \bibinfo {author} {\bibfnamefont {Guixin}\ \bibnamefont
  {Li}}, \bibinfo {author} {\bibfnamefont {Benfeng}\ \bibnamefont {Bai}},
  \bibinfo {author} {\bibfnamefont {Qiaofeng}\ \bibnamefont {Tan}}, \bibinfo
  {author} {\bibfnamefont {Guofan}\ \bibnamefont {Jin}}, \bibinfo {author}
  {\bibfnamefont {Thomas}\ \bibnamefont {Zentgraf}},  and \bibinfo {author}
  {\bibfnamefont {Shuang}\ \bibnamefont {Zhang}},\ }\bibfield  {title}
  {\enquote {\bibinfo {title} {Dispersionless phase discontinuities for
  controlling light propagation},}\ }\href {\doibase 10.1021/nl303031j}
  {\bibfield  {journal} {\bibinfo  {journal} {Nano Lett.}\ }\textbf {\bibinfo
  {volume} {12}},\ \bibinfo {pages} {5750--5755} (\bibinfo {year}
  {2012}{\natexlab{b}})}\BibitemShut {NoStop}%
\bibitem [{\citenamefont {Tennant} and \citenamefont
  {Allen}(2012)}]{Tennant&Allen:EL:2012}%
  \BibitemOpen
  \bibfield  {author} {\bibinfo {author} {\bibfnamefont {A.}~\bibnamefont
  {Tennant}} and \bibinfo {author} {\bibfnamefont {B.}~\bibnamefont
  {Allen}},\ }\bibfield  {title} {\enquote {\bibinfo {title} {Generation of
  {OAM} radio waves using circular time-switched array antenna},}\ }\href
  {\doibase 10.1049/el.2012.2664} {\bibfield  {journal} {\bibinfo  {journal}
  {Electron. Lett.}\ }\textbf {\bibinfo {volume} {48}},\ \bibinfo {pages}
  {1365--1366} (\bibinfo {year} {2012})}\BibitemShut {NoStop}%
\bibitem [{\citenamefont {Bai}\ \emph {et~al.}(2013)\citenamefont {Bai},
  \citenamefont {Tennant}, \citenamefont {Allen}, and \citenamefont
  {Rehman}}]{Bai&al:LAPC:2013}%
  \BibitemOpen
  \bibfield  {author} {\bibinfo {author} {\bibfnamefont {Qiang}\ \bibnamefont
  {Bai}}, \bibinfo {author} {\bibfnamefont {A.}~\bibnamefont {Tennant}},
  \bibinfo {author} {\bibfnamefont {B.}~\bibnamefont {Allen}},  and \bibinfo
  {author} {\bibfnamefont {M.U.}\ \bibnamefont {Rehman}},\ }\bibfield  {title}
  {\enquote {\bibinfo {title} {Generation of orbital angular momentum ({OAM})
  radio beams with phased patch array},}\ }in\ \href {\doibase
  10.1109/LAPC.2013.6711931} {\emph {\bibinfo {booktitle} {Antennas and
  Propagation Conference ({LAPC}), 2013 Loughborough}}}\ (\bibinfo {year}
  {2013})\ pp.\ \bibinfo {pages} {410--413}\BibitemShut {NoStop}%
\bibitem [{\citenamefont {Barbuto}\ \emph {et~al.}(2013)\citenamefont
  {Barbuto}, \citenamefont {Toscano}, and \citenamefont
  {Bilotti}}]{Barbuto&al:APSURSI:2013}%
  \BibitemOpen
  \bibfield  {author} {\bibinfo {author} {\bibfnamefont {M.}~\bibnamefont
  {Barbuto}}, \bibinfo {author} {\bibfnamefont {A.}~\bibnamefont {Toscano}}, \
  and\ \bibinfo {author} {\bibfnamefont {F.}~\bibnamefont {Bilotti}},\
  }\bibfield  {title} {\enquote {\bibinfo {title} {Single patch antenna
  generating electromagnetic field with orbital angular momentum},}\ }in\ \href
  {\doibase 10.1109/APS.2013.6711591} {\emph {\bibinfo {booktitle} {2013 {IEEE}
  Antennas and Propagation Society International Symposium ({APSURSI})}}}\
  (\bibinfo {year} {2013})\ pp.\ \bibinfo {pages} {1866--1867}\BibitemShut
  {NoStop}%
\bibitem [{\citenamefont {Deng}\ \emph {et~al.}(2013)\citenamefont {Deng},
  \citenamefont {Chen}, \citenamefont {Zhang}, \citenamefont {Li}, and
  \citenamefont {Feng}}]{Deng&al:IJAP:2013}%
  \BibitemOpen
  \bibfield  {author} {\bibinfo {author} {\bibfnamefont {Changjiang}\
  \bibnamefont {Deng}}, \bibinfo {author} {\bibfnamefont {Wenhua}\ \bibnamefont
  {Chen}}, \bibinfo {author} {\bibfnamefont {Zhijun}\ \bibnamefont {Zhang}},
  \bibinfo {author} {\bibfnamefont {Yue}\ \bibnamefont {Li}},  and \bibinfo
  {author} {\bibfnamefont {Zhenghe}\ \bibnamefont {Feng}},\ }\bibfield  {title}
  {\enquote {\bibinfo {title} {Generation of {OAM} radio waves using circular
  {Vivaldi} antenna array},}\ }\href {\doibase 10.1155/2013/847859} {\bibfield
  {journal} {\bibinfo  {journal} {Int.\ J.~Ant.\ Prop.}\ }\textbf {\bibinfo
  {volume} {2013}} (\bibinfo {year} {2013}),\ 10.1155/2013/847859}\BibitemShut
  {NoStop}%
\bibitem [{\citenamefont {Gao}\ \emph {et~al.}(2013)\citenamefont {Gao},
  \citenamefont {Huang}, \citenamefont {Zhou}, \citenamefont {Wei},
  \citenamefont {Gao}, \citenamefont {Zhang}, and \citenamefont
  {Gu}}]{Gao&al:JO:2013}%
  \BibitemOpen
  \bibfield  {author} {\bibinfo {author} {\bibfnamefont {Xinlu}\ \bibnamefont
  {Gao}}, \bibinfo {author} {\bibfnamefont {Shanguo}\ \bibnamefont {Huang}},
  \bibinfo {author} {\bibfnamefont {Jing}\ \bibnamefont {Zhou}}, \bibinfo
  {author} {\bibfnamefont {Yongfeng}\ \bibnamefont {Wei}}, \bibinfo {author}
  {\bibfnamefont {Chao}\ \bibnamefont {Gao}}, \bibinfo {author} {\bibfnamefont
  {Xukai}\ \bibnamefont {Zhang}},  and \bibinfo {author} {\bibfnamefont
  {Wanyi}\ \bibnamefont {Gu}},\ }\bibfield  {title} {\enquote {\bibinfo {title}
  {Generating, multiplexing/demultiplexing and receiving the orbital angular
  momentum of radio frequency signals using an optical true time delay unit},}\
  }\href {\doibase 10.1088/2040-8978/15/10/105401} {\bibfield  {journal}
  {\bibinfo  {journal} {J.~Opt.}\ }\textbf {\bibinfo {volume} {15}},\ \bibinfo
  {pages} {105401} (\bibinfo {year} {2013})}\BibitemShut {NoStop}%
\bibitem [{\citenamefont {Li}\ \emph {et~al.}(2013)\citenamefont {Li},
  \citenamefont {Ohashi}, and \citenamefont {Kasai}}]{Li&al:EuMC:2013}%
  \BibitemOpen
  \bibfield  {author} {\bibinfo {author} {\bibfnamefont {Zhengyi}\ \bibnamefont
  {Li}}, \bibinfo {author} {\bibfnamefont {Y.}~\bibnamefont {Ohashi}},  and
  \bibinfo {author} {\bibfnamefont {K.}~\bibnamefont {Kasai}},\ }\bibfield
  {title} {\enquote {\bibinfo {title} {Quasi-{LoS} {MIMO} wireless
  communication with twisted radio wave},}\ }in\ \href@noop {} {\emph {\bibinfo
  {booktitle} {Microwave Conference ({EuMC}), 2013 European}}}\ (\bibinfo
  {year} {2013})\ pp.\ \bibinfo {pages} {577--580}\BibitemShut {NoStop}%
\bibitem [{\citenamefont {Weber}(2013)}]{Weber:OPTIK:2013}%
  \BibitemOpen
  \bibfield  {author} {\bibinfo {author} {\bibfnamefont {H.}~\bibnamefont
  {Weber}},\ }\bibfield  {title} {\enquote {\bibinfo {title} {Angular momentum
  of low frequency electromagnetic fields},}\ }\href {\doibase
  10.1016/j.ijleo.2013.02.009} {\bibfield  {journal} {\bibinfo  {journal}
  {Optik---Intern.\ J.~Light Electron\ Opt.}\ }\textbf {\bibinfo {volume}
  {124}},\ \bibinfo {pages} {4313--4314} (\bibinfo {year} {2013})}\BibitemShut
  {NoStop}%
\bibitem [{\citenamefont {Bai}\ \emph {et~al.}(2014)\citenamefont {Bai},
  \citenamefont {Tennant}, and \citenamefont {Allen}}]{Bai&al:EL:2014}%
  \BibitemOpen
  \bibfield  {author} {\bibinfo {author} {\bibfnamefont {Q.}~\bibnamefont
  {Bai}}, \bibinfo {author} {\bibfnamefont {A.}~\bibnamefont {Tennant}},  and
  \bibinfo {author} {\bibfnamefont {B.}~\bibnamefont {Allen}},\ }\bibfield
  {title} {\enquote {\bibinfo {title} {Experimental circular phased array for
  generating {OAM} radio beams},}\ }\href {\doibase 10.1049/el.2014.2860}
  {\bibfield  {journal} {\bibinfo  {journal} {Electron. Lett.}\ }\textbf
  {\bibinfo {volume} {50}},\ \bibinfo {pages} {1414--1415} (\bibinfo {year}
  {2014})}\BibitemShut {NoStop}%
\bibitem [{\citenamefont {Barbuto}\ \emph {et~al.}(2014)\citenamefont
  {Barbuto}, \citenamefont {Trotta}, \citenamefont {Bilotti}, and
  \citenamefont {Toscano}}]{Barbuto&al:PIER:2014}%
  \BibitemOpen
  \bibfield  {author} {\bibinfo {author} {\bibfnamefont {Mirko}\ \bibnamefont
  {Barbuto}}, \bibinfo {author} {\bibfnamefont {Fabrizio}\ \bibnamefont
  {Trotta}}, \bibinfo {author} {\bibfnamefont {Filiberto}\ \bibnamefont
  {Bilotti}},  and \bibinfo {author} {\bibfnamefont {Alessandro}\
  \bibnamefont {Toscano}},\ }\bibfield  {title} {\enquote {\bibinfo {title}
  {Circular polarized patch antenna generating orbital angular momentum},}\
  }\href {http://www.jpier.org/PIER/pier.php?paper=14050204} {\bibfield
  {journal} {\bibinfo  {journal} {Progr.\ Electromagn.\ Res.}\ }\textbf
  {\bibinfo {volume} {148}},\ \bibinfo {pages} {23--30} (\bibinfo {year}
  {2014})}\BibitemShut {NoStop}%
\bibitem [{\citenamefont {Cano}\ \emph {et~al.}(2014)\citenamefont {Cano},
  \citenamefont {Allen}, \citenamefont {Bai}, and \citenamefont
  {Tennant}}]{Cano&al:LAPC:2014}%
  \BibitemOpen
  \bibfield  {author} {\bibinfo {author} {\bibfnamefont {E.}~\bibnamefont
  {Cano}}, \bibinfo {author} {\bibfnamefont {B.}~\bibnamefont {Allen}},
  \bibinfo {author} {\bibfnamefont {Qiang}\ \bibnamefont {Bai}},  and
  \bibinfo {author} {\bibfnamefont {A.}~\bibnamefont {Tennant}},\ }\bibfield
  {title} {\enquote {\bibinfo {title} {Generation and detection of {OAM}
  signals for radio communications},}\ }in\ \href {\doibase
  10.1109/LAPC.2014.6996473} {\emph {\bibinfo {booktitle} {Antennas and
  Propagation Conference ({LAPC}), 2014 Loughborough}}}\ (\bibinfo {year}
  {2014})\ pp.\ \bibinfo {pages} {637--640}\BibitemShut {NoStop}%
\bibitem [{\citenamefont {Opare} and \citenamefont
  {Kuang}(2014)}]{Opare&Kuan:2014}%
  \BibitemOpen
  \bibfield  {author} {\bibinfo {author} {\bibfnamefont {K.A.}\ \bibnamefont
  {Opare}} and \bibinfo {author} {\bibfnamefont {Yujun}\ \bibnamefont
  {Kuang}},\ }\bibfield  {title} {\enquote {\bibinfo {title} {Performance of an
  ideal wireless orbital angular momentum communication system using
  multiple-input multiple-output techniques},}\ }in\ \href {\doibase
  10.1109/TEMU.2014.6917751} {\emph {\bibinfo {booktitle} {2014 International
  Conference on Telecommunications and Multimedia ({TEMU})}}}\ (\bibinfo {year}
  {2014})\ pp.\ \bibinfo {pages} {144--149}\BibitemShut {NoStop}%
\bibitem [{\citenamefont {Bai}\ \emph {et~al.}(2015)\citenamefont {Bai},
  \citenamefont {Jin}, \citenamefont {Liu}, \citenamefont {Geng}, and
  \citenamefont {Liang}}]{Bai&al:IJAP:2015}%
  \BibitemOpen
  \bibfield  {author} {\bibinfo {author} {\bibfnamefont {Xudong}\ \bibnamefont
  {Bai}}, \bibinfo {author} {\bibfnamefont {Ronghong}\ \bibnamefont {Jin}},
  \bibinfo {author} {\bibfnamefont {Liang}\ \bibnamefont {Liu}}, \bibinfo
  {author} {\bibfnamefont {Junping}\ \bibnamefont {Geng}},  and \bibinfo
  {author} {\bibfnamefont {Xianling}\ \bibnamefont {Liang}},\ }\bibfield
  {title} {\enquote {\bibinfo {title} {Generation of {OAM} radio waves with
  three polarizations using circular horn antenna array},}\ }\href {\doibase
  10.1155/2015/132549} {\bibfield  {journal} {\bibinfo  {journal} {Int.\
  J.~Ant.\ Prop.}\ }\textbf {\bibinfo {volume} {2015}},\ \bibinfo {pages}
  {e132549} (\bibinfo {year} {2015})}\BibitemShut {NoStop}%
\bibitem [{\citenamefont {Opare}\ \emph {et~al.}(2015)\citenamefont {Opare},
  \citenamefont {Kuang}, \citenamefont {Kponyo}, \citenamefont {Nwizege}, and
  \citenamefont {Enzhan}}]{Opare&al:ACCT:2015}%
  \BibitemOpen
  \bibfield  {author} {\bibinfo {author} {\bibfnamefont {Kwasi~A.}\
  \bibnamefont {Opare}}, \bibinfo {author} {\bibfnamefont {Yujun}\ \bibnamefont
  {Kuang}}, \bibinfo {author} {\bibfnamefont {Jerry~J.}\ \bibnamefont
  {Kponyo}}, \bibinfo {author} {\bibfnamefont {Kenneth~S.}\ \bibnamefont
  {Nwizege}},  and \bibinfo {author} {\bibfnamefont {Zhang}\ \bibnamefont
  {Enzhan}},\ }\bibfield  {title} {\enquote {\bibinfo {title} {The degrees of
  freedom in wireless line-of-sight {OAM} multiplexing systems using a circular
  array of receiving antennas},}\ }in\ \href {\doibase 10.1109/ACCT.2015.52}
  {\emph {\bibinfo {booktitle} {2015 Fifth International Conference on Advanced
  Computing Communication Technologies ({ACCT})}}}\ (\bibinfo {year} {2015})\
  pp.\ \bibinfo {pages} {608--613}\BibitemShut {NoStop}%
\bibitem [{\citenamefont {Liu}\ \emph {et~al.}(2015)\citenamefont {Liu},
  \citenamefont {Cheng}, \citenamefont {Yang}, \citenamefont {Wang},
  \citenamefont {Qin}, and \citenamefont {Li}}]{Liu&al:IEEEAWPL:2015}%
  \BibitemOpen
  \bibfield  {author} {\bibinfo {author} {\bibfnamefont {Kang}\ \bibnamefont
  {Liu}}, \bibinfo {author} {\bibfnamefont {Yongqiang}\ \bibnamefont {Cheng}},
  \bibinfo {author} {\bibfnamefont {Zhaocheng}\ \bibnamefont {Yang}}, \bibinfo
  {author} {\bibfnamefont {Hongqiang}\ \bibnamefont {Wang}}, \bibinfo {author}
  {\bibfnamefont {Yuliang}\ \bibnamefont {Qin}},  and \bibinfo {author}
  {\bibfnamefont {Xiang}\ \bibnamefont {Li}},\ }\bibfield  {title} {\enquote
  {\bibinfo {title} {Orbital-angular-momentum-based electromagnetic vortex
  imaging},}\ }\href {\doibase 10.1109/LAWP.2014.2376970} {\bibfield  {journal}
  {\bibinfo  {journal} {IEEE Ant.\ Wirel.\ Prop.\ Letter.}\ }\textbf {\bibinfo
  {volume} {14}},\ \bibinfo {pages} {711--714} (\bibinfo {year}
  {2015})}\BibitemShut {NoStop}%
\bibitem [{\citenamefont {Roichman}\ \emph {et~al.}(2008)\citenamefont
  {Roichman}, \citenamefont {Sun}, \citenamefont {Roichman}, \citenamefont
  {Amato-Grill}, and \citenamefont {Grier}}]{Roichman&al:PRL:2008}%
  \BibitemOpen
  \bibfield  {author} {\bibinfo {author} {\bibfnamefont {Yohai}\ \bibnamefont
  {Roichman}}, \bibinfo {author} {\bibfnamefont {Bo}~\bibnamefont {Sun}},
  \bibinfo {author} {\bibfnamefont {Yael}\ \bibnamefont {Roichman}}, \bibinfo
  {author} {\bibfnamefont {Jesse}\ \bibnamefont {Amato-Grill}},  and \bibinfo
  {author} {\bibfnamefont {David~G.}\ \bibnamefont {Grier}},\ }\bibfield
  {title} {\enquote {\bibinfo {title} {Optical forces arising from phase
  gradients},}\ }\href {\doibase 10.1103/PhysRevLett.100.013602} {\bibfield
  {journal} {\bibinfo  {journal} {Phys.\ Rev.\ Lett.}\ }\textbf {\bibinfo
  {volume} {100}},\ \bibinfo {pages} {013602(4)} (\bibinfo {year}
  {2008})}\BibitemShut {NoStop}%
\bibitem [{\citenamefont {Berkhout}\ \emph {et~al.}(2010)\citenamefont
  {Berkhout}, \citenamefont {Lavery}, \citenamefont {Courtial}, \citenamefont
  {Beijersbergen}, and \citenamefont {Padgett}}]{Berkhout&al:PRL:2010}%
  \BibitemOpen
  \bibfield  {author} {\bibinfo {author} {\bibfnamefont {Gregorius C.~G.}\
  \bibnamefont {Berkhout}}, \bibinfo {author} {\bibfnamefont {Martin P.~J.}\
  \bibnamefont {Lavery}}, \bibinfo {author} {\bibfnamefont {Johannes}\
  \bibnamefont {Courtial}}, \bibinfo {author} {\bibfnamefont {Marco~W.}\
  \bibnamefont {Beijersbergen}},  and \bibinfo {author} {\bibfnamefont
  {Miles~J.}\ \bibnamefont {Padgett}},\ }\bibfield  {title} {\enquote {\bibinfo
  {title} {Efficient sorting of orbital angular momentum states of light},}\
  }\href {\doibase 10.1103/PhysRevLett.105.153601} {\bibfield  {journal}
  {\bibinfo  {journal} {Phys.\ Rev.\ Lett.}\ }\textbf {\bibinfo {volume}
  {105}},\ \bibinfo {pages} {153601(4)} (\bibinfo {year} {2010})}\BibitemShut
  {NoStop}%
\bibitem [{\citenamefont {Mohammadi}\ \emph
  {et~al.}(2010{\natexlab{b}})\citenamefont {Mohammadi}, \citenamefont
  {Daldorff}, \citenamefont {Forozesh}, \citenamefont {Thid{\'e}},
  \citenamefont {Bergman}, \citenamefont {Isham}, \citenamefont {Karlsson},\
  and\ \citenamefont {Carozzi}}]{Mohammadi&al:RS:2010}%
  \BibitemOpen
  \bibfield  {author} {\bibinfo {author} {\bibfnamefont {Siavoush~M.}\
  \bibnamefont {Mohammadi}}, \bibinfo {author} {\bibfnamefont {Lars K.~S.}\
  \bibnamefont {Daldorff}}, \bibinfo {author} {\bibfnamefont {Kamyar}\
  \bibnamefont {Forozesh}}, \bibinfo {author} {\bibfnamefont {Bo}~\bibnamefont
  {Thid{\'e}}}, \bibinfo {author} {\bibfnamefont {Jan E.~S.}\ \bibnamefont
  {Bergman}}, \bibinfo {author} {\bibfnamefont {Brett}\ \bibnamefont {Isham}},
  \bibinfo {author} {\bibfnamefont {Roger}\ \bibnamefont {Karlsson}},  and
  \bibinfo {author} {\bibfnamefont {T.~D.}\ \bibnamefont {Carozzi}},\
  }\bibfield  {title} {\enquote {\bibinfo {title} {Orbital angular momentum in
  radio: Measurement methods},}\ }\href {\doibase 10.1029/2009RS004299}
  {\bibfield  {journal} {\bibinfo  {journal} {Radio Sci.}\ }\textbf {\bibinfo
  {volume} {45}},\ \bibinfo {pages} {RS4007} (\bibinfo {year}
  {2010}{\natexlab{b}})}\BibitemShut {NoStop}%
\bibitem [{\citenamefont {Lavery}\ \emph {et~al.}(2013)\citenamefont {Lavery},
  \citenamefont {Robertson}, \citenamefont {Sponselli}, \citenamefont
  {Courtial}, \citenamefont {Steinhoff}, \citenamefont {Tyler}, \citenamefont
  {Willner}, and \citenamefont {Padgett}}]{Lavery&al:NJP:2013}%
  \BibitemOpen
  \bibfield  {author} {\bibinfo {author} {\bibfnamefont {Martin P.~J.}\
  \bibnamefont {Lavery}}, \bibinfo {author} {\bibfnamefont {David~J.}\
  \bibnamefont {Robertson}}, \bibinfo {author} {\bibfnamefont {Anna}\
  \bibnamefont {Sponselli}}, \bibinfo {author} {\bibfnamefont {Johannes}\
  \bibnamefont {Courtial}}, \bibinfo {author} {\bibfnamefont {Nicholas~K.}\
  \bibnamefont {Steinhoff}}, \bibinfo {author} {\bibfnamefont {Glenn~A.}\
  \bibnamefont {Tyler}}, \bibinfo {author} {\bibfnamefont {Alan~E.}\
  \bibnamefont {Willner}},  and \bibinfo {author} {\bibfnamefont {Miles~J.}\
  \bibnamefont {Padgett}},\ }\bibfield  {title} {\enquote {\bibinfo {title}
  {Efficient measurement of an optical orbital-angular-momentum spectrum
  comprising more than 50 states},}\ }\href {\doibase
  10.1088/1367-2630/15/1/013024} {\bibfield  {journal} {\bibinfo  {journal}
  {New J.~Phys.}\ }\textbf {\bibinfo {volume} {15}},\ \bibinfo {pages}
  {013024(7)} (\bibinfo {year} {2013})}\BibitemShut {NoStop}%
\bibitem [{\citenamefont {Wang}\ \emph {et~al.}(2011)\citenamefont {Wang},
  \citenamefont {Yang}, \citenamefont {Fazal}, \citenamefont {Ahmed},
  \citenamefont {Yan}, \citenamefont {Shamee}, \citenamefont {Willner},
  \citenamefont {Birnbaum}, \citenamefont {Choi}, and \citenamefont
  {Erkmen}}]{Wang&al:ECOC:2011}%
  \BibitemOpen
  \bibfield  {author} {\bibinfo {author} {\bibfnamefont {J.}~\bibnamefont
  {Wang}}, \bibinfo {author} {\bibfnamefont {J.~Y.}\ \bibnamefont {Yang}},
  \bibinfo {author} {\bibfnamefont {I.~M.}\ \bibnamefont {Fazal}}, \bibinfo
  {author} {\bibfnamefont {N.}~\bibnamefont {Ahmed}}, \bibinfo {author}
  {\bibfnamefont {Y.}~\bibnamefont {Yan}}, \bibinfo {author} {\bibfnamefont
  {B.}~\bibnamefont {Shamee}}, \bibinfo {author} {\bibfnamefont {A.~E.}\
  \bibnamefont {Willner}}, \bibinfo {author} {\bibfnamefont {K.}~\bibnamefont
  {Birnbaum}}, \bibinfo {author} {\bibfnamefont {J.}~\bibnamefont {Choi}}, \
  and\ \bibinfo {author} {\bibfnamefont {B.}~\bibnamefont {Erkmen}},\
  }\bibfield  {title} {\enquote {\bibinfo {title} {Demonstration of
  12.8-bit/s/hz spectral efficiency using 16-{QAM} signals over multiple
  orbital-angular-momentum modes},}\ }in\ \href
  {http://ieeexplore.ieee.org/xpls/abs_all.jsp?arnumber=6066145} {\emph
  {\bibinfo {booktitle} {Optical Communication ({ECOC}), 2011 37th European
  Conference and Exhibition on}}}\ (\bibinfo {year} {2011})\ pp.\ \bibinfo
  {pages} {1--3}\BibitemShut {NoStop}%
\bibitem [{\citenamefont {Tamburini}\ \emph
  {et~al.}(2013{\natexlab{a}})\citenamefont {Tamburini}, \citenamefont
  {Thid\'{e}}, \citenamefont {Mari}, \citenamefont {Parisi}, \citenamefont
  {Spinello}, \citenamefont {Oldoni}, \citenamefont {Ravanelli}, \citenamefont
  {Coassini}, \citenamefont {Someda}, and \citenamefont
  {Romanato}}]{Tamburini&al:ARXIV:2013b}%
  \BibitemOpen
  \bibfield  {author} {\bibinfo {author} {\bibfnamefont {Fabrizio}\
  \bibnamefont {Tamburini}}, \bibinfo {author} {\bibfnamefont {Bo}~\bibnamefont
  {Thid\'{e}}}, \bibinfo {author} {\bibfnamefont {Elettra}\ \bibnamefont
  {Mari}}, \bibinfo {author} {\bibfnamefont {Giuseppe}\ \bibnamefont {Parisi}},
  \bibinfo {author} {\bibfnamefont {Fabio}\ \bibnamefont {Spinello}}, \bibinfo
  {author} {\bibfnamefont {Matteo}\ \bibnamefont {Oldoni}}, \bibinfo {author}
  {\bibfnamefont {Roberto~A.}\ \bibnamefont {Ravanelli}}, \bibinfo {author}
  {\bibfnamefont {Piero}\ \bibnamefont {Coassini}}, \bibinfo {author}
  {\bibfnamefont {Carlo~G.}\ \bibnamefont {Someda}},  and \bibinfo {author}
  {\bibfnamefont {Filippo}\ \bibnamefont {Romanato}},\ }\href
  {http://arxiv.org/abs/1307.5569} {\enquote {\bibinfo {title}
  {\emph{N}-tupling the capacity of each polarization state in radio links by
  using electromagnetic vorticity},}\ } (\bibinfo {year}
  {2013}{\natexlab{a}}),\ \Eprint {http://arxiv.org/abs/1307.5569}
  {arXiv.org:1307.5569 [physics.optics]} \BibitemShut {NoStop}%
\bibitem [{\citenamefont {Low}(1997)}]{Low:Book:1997}%
  \BibitemOpen
  \bibfield  {author} {\bibinfo {author} {\bibfnamefont {Francis~E.}\
  \bibnamefont {Low}},\ }\href@noop {} {\emph {\bibinfo {title} {Classical
  Field Theory. Electromagnetism and Gravitation}}}\ (\bibinfo  {publisher}
  {John Wiley \& Sons},\ \bibinfo {address} {New York, NY, USA},\ \bibinfo
  {year} {1997})\BibitemShut {NoStop}%
\bibitem [{\citenamefont {{\"U}nal}(1997)}]{Unal:FP:1997}%
  \BibitemOpen
  \bibfield  {author} {\bibinfo {author} {\bibfnamefont {Nuri}\ \bibnamefont
  {{\"U}nal}},\ }\bibfield  {title} {\enquote {\bibinfo {title} {A simple model
  of the classical \emph{Zitterbewegung}: Photon wave function},}\ }\href
  {\doibase 10.1007/BF02550173} {\bibfield  {journal} {\bibinfo  {journal}
  {Found.\ Phys.}\ }\textbf {\bibinfo {volume} {27}},\ \bibinfo {pages}
  {731--746} (\bibinfo {year} {1997})}\BibitemShut {NoStop}%
\bibitem [{\citenamefont {Kobe}(1999{\natexlab{b}})}]{Kobe:PLA:1999}%
  \BibitemOpen
  \bibfield  {author} {\bibinfo {author} {\bibfnamefont {Donald~H.}\
  \bibnamefont {Kobe}},\ }\bibfield  {title} {\enquote {\bibinfo {title}
  {\emph{Zitterbewegung} of a photon},}\ }\href {\doibase
  10.1016/S0375-9601(99)00011-0} {\bibfield  {journal} {\bibinfo  {journal}
  {Phys.\ Lett.\ A}\ }\textbf {\bibinfo {volume} {253}},\ \bibinfo {pages}
  {7--11} (\bibinfo {year} {1999}{\natexlab{b}})}\BibitemShut {NoStop}%
\bibitem [{\citenamefont {Leary} and \citenamefont
  {Smith}(2014)}]{Leary&Smith:PRA:2014}%
  \BibitemOpen
  \bibfield  {author} {\bibinfo {author} {\bibfnamefont {C.~C.}\ \bibnamefont
  {Leary}} and \bibinfo {author} {\bibfnamefont {Karl~H.}\ \bibnamefont
  {Smith}},\ }\bibfield  {title} {\enquote {\bibinfo {title} {Unified dynamics
  of electrons and photons via \emph{Zitterbewegung} and spin-orbit
  interaction},}\ }\href {\doibase 10.1103/PhysRevA.89.023831} {\bibfield
  {journal} {\bibinfo  {journal} {Phys.\ Rev.\ A}\ }\textbf {\bibinfo {volume}
  {89}},\ \bibinfo {pages} {023831--1--11} (\bibinfo {year}
  {2014})}\BibitemShut {NoStop}%
\bibitem [{\citenamefont {Einstein} and \citenamefont
  {deHaas}(1915)}]{Einstein&deHaas:PKAWA:1915}%
  \BibitemOpen
  \bibfield  {author} {\bibinfo {author} {\bibfnamefont {A.}~\bibnamefont
  {Einstein}} and \bibinfo {author} {\bibfnamefont {W.~J.}\ \bibnamefont
  {deHaas}},\ }\bibfield  {title} {\enquote {\bibinfo {title} {Experimental
  proof of the existence of {Amp{\`e}re's} molecular currents},}\ }\href@noop
  {} {\bibfield  {journal} {\bibinfo  {journal} {Proc. Kon. Akad. Wetensch.
  Amsterdam}\ }\textbf {\bibinfo {volume} {18}},\ \bibinfo {pages} {696--708}
  (\bibinfo {year} {1915})}\BibitemShut {NoStop}%
\bibitem [{\citenamefont {Andrews}\ \emph {et~al.}(2001)\citenamefont
  {Andrews}, \citenamefont {Mitra}, and \citenamefont
  {deCarvalho}}]{Andrews&al:N:2001}%
  \BibitemOpen
  \bibfield  {author} {\bibinfo {author} {\bibfnamefont {Michael~R.}\
  \bibnamefont {Andrews}}, \bibinfo {author} {\bibfnamefont {Partha~P.}\
  \bibnamefont {Mitra}},  and \bibinfo {author} {\bibfnamefont {Robert}\
  \bibnamefont {deCarvalho}},\ }\bibfield  {title} {\enquote {\bibinfo {title}
  {Tripling the capacity of wireless communications using electromagnetic
  polarization},}\ }\href {\doibase 10.1038/35053015} {\bibfield  {journal}
  {\bibinfo  {journal} {Nature}\ }\textbf {\bibinfo {volume} {409}},\ \bibinfo
  {pages} {316--318} (\bibinfo {year} {2001})}\BibitemShut {NoStop}%
\bibitem [{\citenamefont {Essiambre} and \citenamefont
  {Tkach}(2012)}]{Essiambre&Tkach:PIEEE:2012}%
  \BibitemOpen
  \bibfield  {author} {\bibinfo {author} {\bibfnamefont {R.}~\bibnamefont
  {Essiambre}} and \bibinfo {author} {\bibfnamefont {R.~W.}\ \bibnamefont
  {Tkach}},\ }\bibfield  {title} {\enquote {\bibinfo {title} {Capacity trends
  and limits of optical communication networks},}\ }\href {\doibase
  10.1109/JPROC.2012.2182970} {\bibfield  {journal} {\bibinfo  {journal}
  {Proc.\ IEEE}\ }\textbf {\bibinfo {volume} {100}},\ \bibinfo {pages}
  {1035--1055} (\bibinfo {year} {2012})}\BibitemShut {NoStop}%
\bibitem [{\citenamefont {Arik}\ \emph {et~al.}(2013)\citenamefont {Arik},
  \citenamefont {Askarov}, and \citenamefont {Kahn}}]{Arik&al:JLWT:2013}%
  \BibitemOpen
  \bibfield  {author} {\bibinfo {author} {\bibfnamefont {S.~O.}\ \bibnamefont
  {Arik}}, \bibinfo {author} {\bibfnamefont {D.}~\bibnamefont {Askarov}}, \
  and\ \bibinfo {author} {\bibfnamefont {J.~M.}\ \bibnamefont {Kahn}},\
  }\bibfield  {title} {\enquote {\bibinfo {title} {Effect of mode coupling on
  signal processing complexity in mode-division multiplexing},}\ }\href
  {\doibase 10.1109/JLT.2012.2234083} {\bibfield  {journal} {\bibinfo
  {journal} {J.~Lightw.\ Technol.}\ }\textbf {\bibinfo {volume} {31}},\
  \bibinfo {pages} {423--431} (\bibinfo {year} {2013})}\BibitemShut {NoStop}%
\bibitem [{\citenamefont {Tamburini}\ \emph
  {et~al.}(2013{\natexlab{b}})\citenamefont {Tamburini}, \citenamefont
  {Thid\'{e}}, \citenamefont {Boaga}, \citenamefont {Carraro}, \citenamefont
  {del Pup}, \citenamefont {Bianchini}, \citenamefont {Someda}, and
  \citenamefont {Romanato}}]{Tamburini&al:ARXIV:2013}%
  \BibitemOpen
  \bibfield  {author} {\bibinfo {author} {\bibfnamefont {F.}~\bibnamefont
  {Tamburini}}, \bibinfo {author} {\bibfnamefont {B.}~\bibnamefont
  {Thid\'{e}}}, \bibinfo {author} {\bibfnamefont {V.}~\bibnamefont {Boaga}},
  \bibinfo {author} {\bibfnamefont {F.}~\bibnamefont {Carraro}}, \bibinfo
  {author} {\bibfnamefont {M.}~\bibnamefont {del Pup}}, \bibinfo {author}
  {\bibfnamefont {A.}~\bibnamefont {Bianchini}}, \bibinfo {author}
  {\bibfnamefont {C.~G.}\ \bibnamefont {Someda}},  and \bibinfo {author}
  {\bibfnamefont {F.}~\bibnamefont {Romanato}},\ }\href
  {http://arxiv.org/abs/1302.2990} {\enquote {\bibinfo {title} {Experimental
  demonstration of free-space information transfer using phase modulated
  orbital angular momentum radio},}\ } (\bibinfo {year} {2013}{\natexlab{b}}),\
  \Eprint {http://arxiv.org/abs/1302.2990} {arXiv.org:1302.2990
  [physics.class-ph]]} \BibitemShut {NoStop}%
\bibitem [{\citenamefont {Fernandez-Corbaton}\ \emph
  {et~al.}(2012)\citenamefont {Fernandez-Corbaton}, \citenamefont
  {Zambrana-Puyalto}, and \citenamefont
  {Molina-Terriza}}]{Fernandez-Corbaton&al:PRA:2012}%
  \BibitemOpen
  \bibfield  {author} {\bibinfo {author} {\bibfnamefont {Ivan}\ \bibnamefont
  {Fernandez-Corbaton}}, \bibinfo {author} {\bibfnamefont {Xavier}\
  \bibnamefont {Zambrana-Puyalto}},  and \bibinfo {author} {\bibfnamefont
  {Gabriel}\ \bibnamefont {Molina-Terriza}},\ }\bibfield  {title} {\enquote
  {\bibinfo {title} {Helicity and angular momentum: A symmetry-based framework
  for the study of light-matter interactions},}\ }\href {\doibase
  10.1103/PhysRevA.86.042103} {\bibfield  {journal} {\bibinfo  {journal}
  {Phys.\ Rev.\ A}\ }\textbf {\bibinfo {volume} {86}},\ \bibinfo {pages}
  {042103} (\bibinfo {year} {2012})}\BibitemShut {NoStop}%
\bibitem [{\citenamefont {Philbin}(2013)}]{Philbin:PRA:2013}%
  \BibitemOpen
  \bibfield  {author} {\bibinfo {author} {\bibfnamefont {T.~G.}\ \bibnamefont
  {Philbin}},\ }\bibfield  {title} {\enquote {\bibinfo {title} {Lipkin's
  conservation law, {Noether's} theorem, and the relation to optical
  helicity},}\ }\href {\doibase 10.1103/PhysRevA.87.043843} {\bibfield
  {journal} {\bibinfo  {journal} {Phys.\ Rev.\ A}\ }\textbf {\bibinfo {volume}
  {87}},\ \bibinfo {pages} {043843} (\bibinfo {year} {2013})}\BibitemShut
  {NoStop}%
\bibitem [{Note1()}]{Note1}%
  \BibitemOpen
  \bibinfo {note} {In certain disciplines a non-standard convention is
  sometimes used in which the term Lorentz force density denotes only the
  second term in the RHS of Eqn.~\protect \textup {\hbox {\mathsurround \z@
  \protect \normalfont (\ignorespaces \ref {eq:Lorentz_force_density}\unskip
  \@@italiccorr )}}. We follow the standard convention.}\BibitemShut {Stop}%
\bibitem [{\citenamefont {van Enk} and \citenamefont
  {Nienhuis}(1994)}]{vanEnk&Nienhuis:JMO:1994}%
  \BibitemOpen
  \bibfield  {author} {\bibinfo {author} {\bibfnamefont {S.~J.}\ \bibnamefont
  {van Enk}} and \bibinfo {author} {\bibfnamefont {G.}~\bibnamefont
  {Nienhuis}},\ }\bibfield  {title} {\enquote {\bibinfo {title} {Commutation
  rules and eigenvalues of spin and orbital angular momentum of radiation
  fields},}\ }\href@noop {} {\bibfield  {journal} {\bibinfo  {journal}
  {J.~Mod.\ Opt.}\ }\textbf {\bibinfo {volume} {41}},\ \bibinfo {pages}
  {963--977} (\bibinfo {year} {1994})}\BibitemShut {NoStop}%
\bibitem [{Note2()}]{Note2}%
  \BibitemOpen
  \bibinfo {note} {In fact, the surface integral may also be a vector that is
  (at most) a function of time $t$.}\BibitemShut {Stop}%
\bibitem [{Note3()}]{Note3}%
  \BibitemOpen
  \bibinfo {note} {See in particular Section V.5: ``Daniel Bernoulli and Euler
  on the Dependence or Independence of the Law of Moment of Momentum in
  1744.''}\BibitemShut {NoStop}%
\bibitem [{\citenamefont {Neuenschwander}(2011)}]{Neuenschwander:Book:2011}%
  \BibitemOpen
  \bibfield  {author} {\bibinfo {author} {\bibfnamefont {Dwight~E.}\
  \bibnamefont {Neuenschwander}},\ }\href@noop {} {\emph {\bibinfo {title}
  {{Emmy Noether's} Wonderful Theorem}}}\ (\bibinfo  {publisher} {Johns Hopkins
  University Press},\ \bibinfo {address} {Baltimore, MD},\ \bibinfo {year}
  {2011})\BibitemShut {NoStop}%
\bibitem [{\citenamefont {Bliokh}\ \emph {et~al.}(2014)\citenamefont {Bliokh},
  \citenamefont {Dressel}, and \citenamefont {Nori}}]{Bliokh&al:NJP:2014}%
  \BibitemOpen
  \bibfield  {author} {\bibinfo {author} {\bibfnamefont {Konstantin~Y.}\
  \bibnamefont {Bliokh}}, \bibinfo {author} {\bibfnamefont {Justin}\
  \bibnamefont {Dressel}},  and \bibinfo {author} {\bibfnamefont {Franco}\
  \bibnamefont {Nori}},\ }\bibfield  {title} {\enquote {\bibinfo {title}
  {Conservation of the spin and orbital angular momenta in electromagnetism},}\
  }\href {\doibase 10.1088/1367-2630/16/9/093037} {\bibfield  {journal}
  {\bibinfo  {journal} {New J.~Phys.}\ }\textbf {\bibinfo {volume} {16}},\
  \bibinfo {pages} {093037} (\bibinfo {year} {2014})}\BibitemShut {NoStop}%
\bibitem [{\citenamefont {Leader}(2013)}]{Leader:PPN:2013}%
  \BibitemOpen
  \bibfield  {author} {\bibinfo {author} {\bibfnamefont {E.}~\bibnamefont
  {Leader}},\ }\bibfield  {title} {\enquote {\bibinfo {title} {The angular
  momentum controversy: What's it all about and does it matter?}}\ }\href
  {http://download.springer.com/static/pdf/665/art\%253A10.1134\%252FS1063779613060142.pdf?auth66=1397049607\_86f50f6cfee5f8c80da18adbc6a3ac48\&ext=.pdf}
  {\bibfield  {journal} {\bibinfo  {journal} {Phys.\ Part.\ Nucl.}\ }\textbf
  {\bibinfo {volume} {44}},\ \bibinfo {pages} {926--929} (\bibinfo {year}
  {2013})}\BibitemShut {NoStop}%
\bibitem [{\citenamefont {Bia{\l}ynicki-Birula} and \citenamefont
  {Bia{\l}ynicka-Birula}(2011)}]{Bialynicki-Birula&Bialynicka-Birula:JO:2011}%
  \BibitemOpen
  \bibfield  {author} {\bibinfo {author} {\bibfnamefont {Iwo}\ \bibnamefont
  {Bia{\l}ynicki-Birula}} and \bibinfo {author} {\bibfnamefont {Zofia}\
  \bibnamefont {Bia{\l}ynicka-Birula}},\ }\bibfield  {title} {\enquote
  {\bibinfo {title} {Canonical separation of angular momentum of light into its
  orbital and spin parts},}\ }\href {\doibase 10.1088/2040-8978/13/6/064014}
  {\bibfield  {journal} {\bibinfo  {journal} {J.~Opt.}\ }\textbf {\bibinfo
  {volume} {13}},\ \bibinfo {pages} {064014} (\bibinfo {year}
  {2011})}\BibitemShut {NoStop}%
\bibitem [{\citenamefont {Allen}\ \emph {et~al.}(1999)\citenamefont {Allen},
  \citenamefont {Padgett}, and \citenamefont {Babiker}}]{Allen&al:PO:1999}%
  \BibitemOpen
  \bibfield  {author} {\bibinfo {author} {\bibfnamefont {L.}~\bibnamefont
  {Allen}}, \bibinfo {author} {\bibfnamefont {M.~J.}\ \bibnamefont {Padgett}},
   and \bibinfo {author} {\bibfnamefont {M.}~\bibnamefont {Babiker}},\
  }\bibfield  {title} {\enquote {\bibinfo {title} {The orbital angular momentum
  of light},}\ }in\ \href@noop {} {\emph {\bibinfo {booktitle} {Prog.\
  Opt.}}},\ Vol.\ \bibinfo {volume} {XXXIX},\ \bibinfo {editor} {edited by\
  \bibinfo {editor} {\bibfnamefont {E.}~\bibnamefont {Wolf}}}\ (\bibinfo
  {publisher} {Elsevier},\ \bibinfo {address} {Amsterdam, Holland},\ \bibinfo
  {year} {1999})\ pp.\ \bibinfo {pages} {291--372}\BibitemShut {NoStop}%
\bibitem [{\citenamefont {Franke-Arnold}\ \emph {et~al.}(2004)\citenamefont
  {Franke-Arnold}, \citenamefont {Barnett}, \citenamefont {Yao}, \citenamefont
  {Leach}, \citenamefont {Courtial}, and \citenamefont
  {Padgett}}]{Franke-Arnold&al:NJP:2004}%
  \BibitemOpen
  \bibfield  {author} {\bibinfo {author} {\bibfnamefont {Sonja}\ \bibnamefont
  {Franke-Arnold}}, \bibinfo {author} {\bibfnamefont {Stephen~M.}\ \bibnamefont
  {Barnett}}, \bibinfo {author} {\bibfnamefont {Eric}\ \bibnamefont {Yao}},
  \bibinfo {author} {\bibfnamefont {Jonathan}\ \bibnamefont {Leach}}, \bibinfo
  {author} {\bibfnamefont {Johannes}\ \bibnamefont {Courtial}},  and \bibinfo
  {author} {\bibfnamefont {Miles}\ \bibnamefont {Padgett}},\ }\bibfield
  {title} {\enquote {\bibinfo {title} {Uncertainty principle for angular
  position and angular momentum},}\ }\href@noop {} {\bibfield  {journal}
  {\bibinfo  {journal} {New J.~Phys.}\ }\textbf {\bibinfo {volume} {6}},\
  \bibinfo {pages} {1--8} (\bibinfo {year} {2004})}\BibitemShut {NoStop}%
\bibitem [{\citenamefont {Molina-Terriza}\ \emph
  {et~al.}(2007{\natexlab{b}})\citenamefont {Molina-Terriza}, \citenamefont
  {Torres}, and \citenamefont {Torner}}]{Molina-Terriza&al:NPHY:2007}%
  \BibitemOpen
  \bibfield  {author} {\bibinfo {author} {\bibfnamefont {Gabriel}\ \bibnamefont
  {Molina-Terriza}}, \bibinfo {author} {\bibfnamefont {Juan~P.}\ \bibnamefont
  {Torres}},  and \bibinfo {author} {\bibfnamefont {Lluis}\ \bibnamefont
  {Torner}},\ }\bibfield  {title} {\enquote {\bibinfo {title} {Twisted
  photons},}\ }\href@noop {} {\bibfield  {journal} {\bibinfo  {journal} {Nature
  Phys.}\ }\textbf {\bibinfo {volume} {3}},\ \bibinfo {pages} {305--310}
  (\bibinfo {year} {2007}{\natexlab{b}})}\BibitemShut {NoStop}%
\bibitem [{\citenamefont {Boyer}(2005)}]{Boyer:AJP:2005}%
  \BibitemOpen
  \bibfield  {author} {\bibinfo {author} {\bibfnamefont {Timothy~H.}\
  \bibnamefont {Boyer}},\ }\bibfield  {title} {\enquote {\bibinfo {title}
  {Illustrations of the relativistic conservation law for the center of
  energy},}\ }\href {\doibase 10.1119/1.1900101} {\bibfield  {journal}
  {\bibinfo  {journal} {Am.\ J.~Phys.}\ }\textbf {\bibinfo {volume} {73}},\
  \bibinfo {pages} {953--961} (\bibinfo {year} {2005})}\BibitemShut {NoStop}%
\end{thebibliography}

%

\end{document}